\documentclass[a4paper,fleqn]{cas-sc}

\usepackage[authoryear,longnamesfirst]{natbib}
\usepackage{bm}
\usepackage{relsize}
\usepackage{amsthm}

\def\tsc#1{\csdef{#1}{\textsc{\lowercase{#1}}\xspace}}
\tsc{WGM}
\tsc{QE}
\tsc{EP}
\tsc{PMS}
\tsc{BEC}
\tsc{DE}
\DeclareMathOperator*{\A}{ \mathlarger{\mathlarger{\mathlarger{\boldsymbol{\mathsf{A}}}}} }

\begin{document}
\let\WriteBookmarks\relax
\def\floatpagepagefraction{1}
\def\textpagefraction{.001}
\shorttitle{Leveraging social media news}
\shortauthors{Y. Xie et~al.}

\title [mode = title]{A GPU-Accelerated Three-Dimensional Crack Element Method for Transient Dynamic Fracture Simulation}                      



\author[1]{Yuxi Xie}[type=editor,
                        auid=000,bioid=1,
                        orcid=0000-0001-8681-4053]
\cormark[1]
\ead{yuxi_xie2017@outlook.com}
\ead[url]{https://www.linkedin.com/in/yuxi-x-142917134/}

\credit{Conceptualization of this study, Methodology, Code Implementation, Writing - original draft preparation and revision}

\affiliation[1]{organization={Computational and Multiscale Mechanics                                    Group, ANSYS Inc},
                addressline={7374 Las Positas Rd}, 
                city={Livermore},
                postcode={94551}, 
                state={C.A.},
                country={USA}}


\author[1]{C.T. Wu}[%
   ]

\credit{Conceptualization of this study, Methodology, Writing - revision}

\author[1]{Wei Hu}

\credit{GPU Implementation}

\author[2]{Lu Xu}

\credit{Project Administration}

\affiliation[2]{organization={Intel},
                addressline={2501 NE Century Blvd}, 
                postcode={97124}, 
                postcodesep={}, 
                city={Hillsboro},
                state={O.R.},
                country={USA}}

\author[3]{Tinh Q. Bui}

\credit{Technical Discussion}

\affiliation[3]{organization={Duy Tan university},
                city={Ho Chi Minh City},
                country={Vietnam}}

\author[4]{Shaofan Li}

\credit{Project Administration}

\affiliation[4]{organization={The University of California at Berkeley},
                city={Berkeley},
                postcode={94720}, 
                state={C.A.}, 
                country={USA}}

\cortext[cor1]{Corresponding author}


\begin{abstract}
This work presents a novel three-dimensional Crack Element Method (CEM) designed to model transient dynamic crack propagation in quasi-brittle materials efficiently. CEM introduces an advanced element-splitting algorithm that enables element-wise crack growth, including crack branching. Based on the evolving topology of split elements, an original formulation for computing the fracture energy release rate in three dimensions is derived. A series of benchmark examples is conducted to demonstrate that the proposed 3D CEM accurately simulates both single crack propagation and complex crack branching scenarios. Furthermore, all three-dimensional simulations are GPU-accelerated, achieving high levels of computational efficiency, consistency, and accuracy.
\end{abstract}

\begin{graphicalabstract}
\includegraphics[height=2.0in]{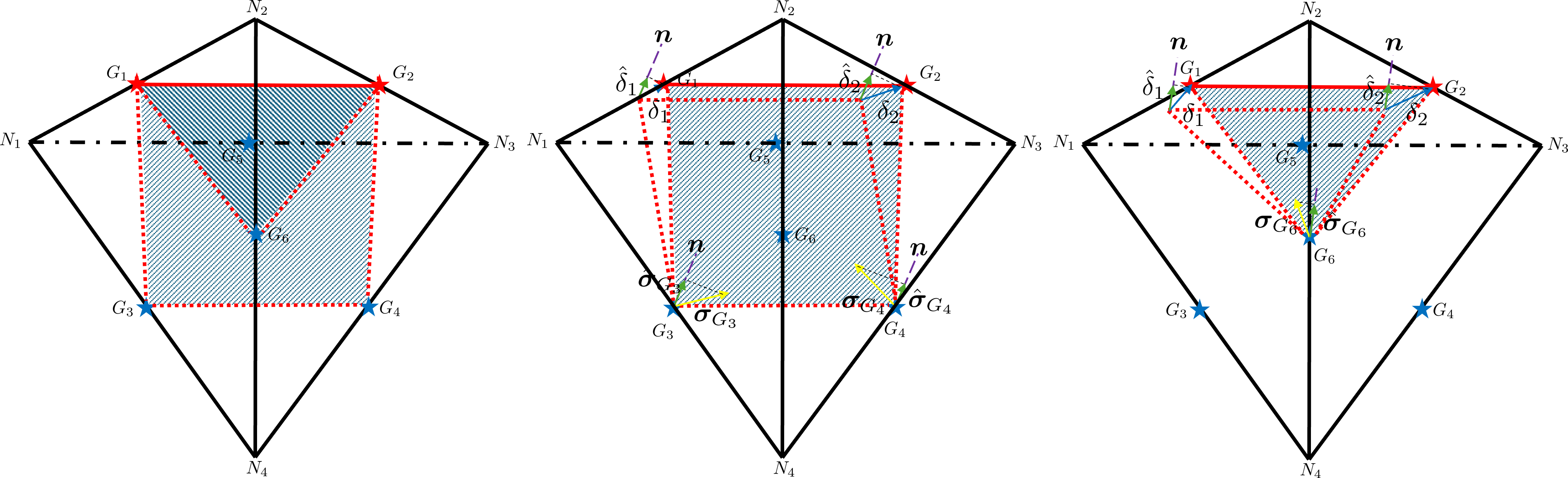} \\
\hfill \\
\includegraphics[height=2.4in]{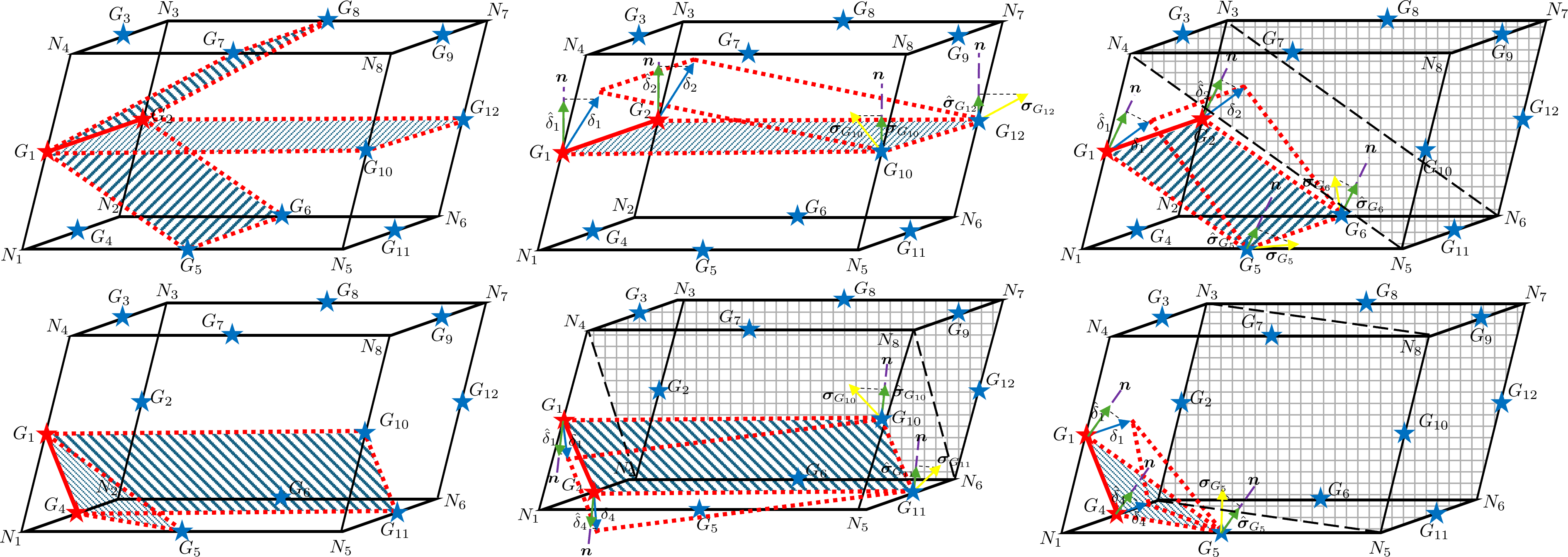}
\hfill \\
\includegraphics[height=2.5in]{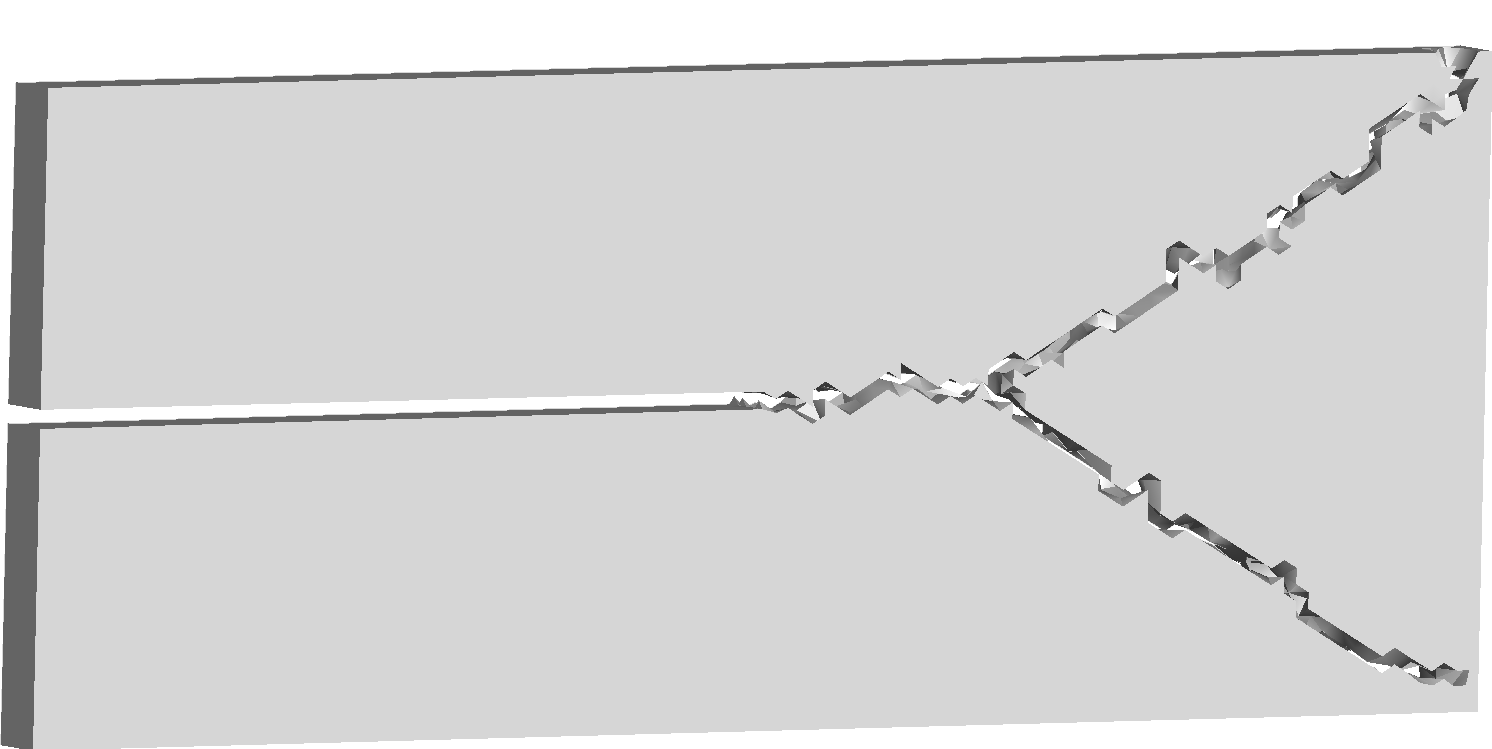}
\end{graphicalabstract}

\begin{highlights}
\item A novel three-dimensional Crack Element Method (CEM) is developed, incorporating an advanced element-splitting algorithm and a new formulation for computing the fracture energy release rate.
\item The 3D element-splitting algorithm enables element-wise crack growth, including crack branching, and approximates the displacement field of fractured elements using the Edge-based Smoothed Finite Element Method (ES-FEM).
\item The fracture energy release rate is evaluated based on the evolving topology of split elements, allowing for accurate and efficient tracking of crack propagation in 3D problems.
\item All 3D benchmark simulations leverage NVIDIA GPU acceleration, highlighting the method's suitability for high-performance computing (HPC) in industrial settings.
\end{highlights}


\maketitle

\section{Introduction}
Fatigue and cracking remain critical challenges across various engineering disciplines, as they represent primary causes of structural failure. In safety-critical applications, accurate prediction of crack propagation lifespan is essential for maintaining structural integrity. In practice, the reliability of components and systems must be verified over a defined operational period before service deployment (\cite{ren2017three}). To comply with such requirements, manufacturers must perform and submit risk assessments aligned with the designated service life. Achieving compliance necessitates estimating crack propagation life, defined as the period from crack initiation to its growth to a critical size under an acceptable risk threshold. Under sufficiently high cyclic loading, cracks will propagate, a process that can be effectively analyzed through the principles of fracture mechanics.

Dynamic crack propagation involves the rapid advancement of cracks in materials under time-dependent loading, wherein both the crack tip velocity and the associated stress fields evolve continuously. Unlike quasi-static crack growth, which assumes a state of equilibrium at each incremental step, dynamic fracture is characterized by the pronounced influence of inertial effects, stress wave emissions, and transient interactions. This phenomenon is particularly relevant for analyzing material failure under high-rate loading conditions such as impact, blast, and seismic events.
Numerical methods play a pivotal role in studying dynamic crack propagation by bridging the gap between analytical models and experimental observations. While analytical solutions offer theoretical insights, they are often restricted to simplified geometries and boundary conditions. Experimental studies, although helpful in observing real-world crack behavior, tend to be costly and offer limited detail on internal stress and displacement fields. Numerical approaches, in contrast, offer the flexibility to simulate complex phenomena such as crack initiation, propagation, branching, arrest, and interactions with material heterogeneities and interfaces, especially under dynamic and non-linear loading conditions. The primary numerical techniques for modeling dynamic fracture include the eXtended Finite Element Method (XFEM, \cite{belytschko1999elastic}, \cite{wang20173}), Cohesive Zone Models (CZM, \cite{xu1994numerical}, \cite{rumi2025polygen}), Peridynamics (PD, \cite{silling2000reformulation}), and Meshfree and Particle Methods (\cite{liu2003smoothed}). These methods have significantly advanced the ability to predict and analyze fracture behavior in various engineering applications.

\begin{figure}[htp]
        \centering
        \begin{minipage}{0.45\linewidth}
            \begin{center}
            \includegraphics[height=2.4in]{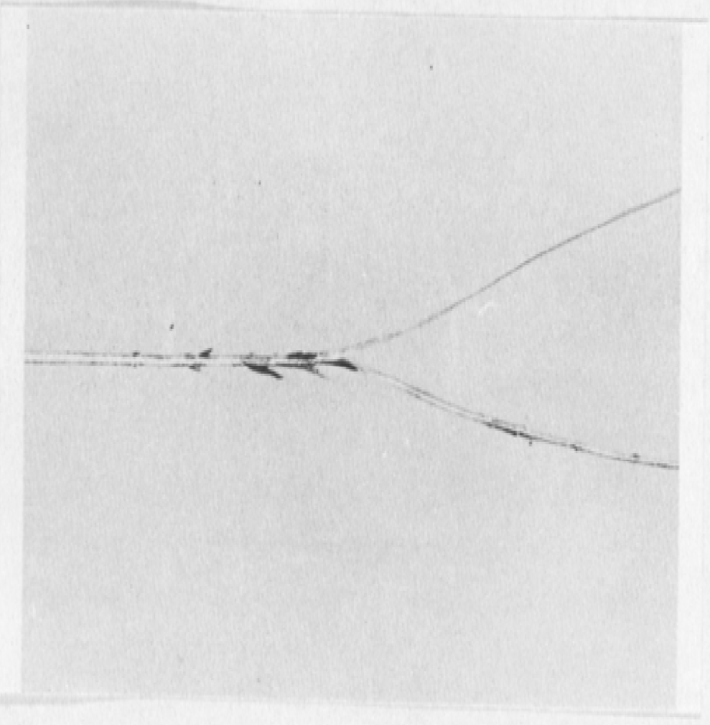}
            \end{center}
            \begin{center}
            (a)
            \end{center}
        \end{minipage}
        \begin{minipage}{0.45\linewidth}
            \begin{center}
            \includegraphics[height=2.4in]{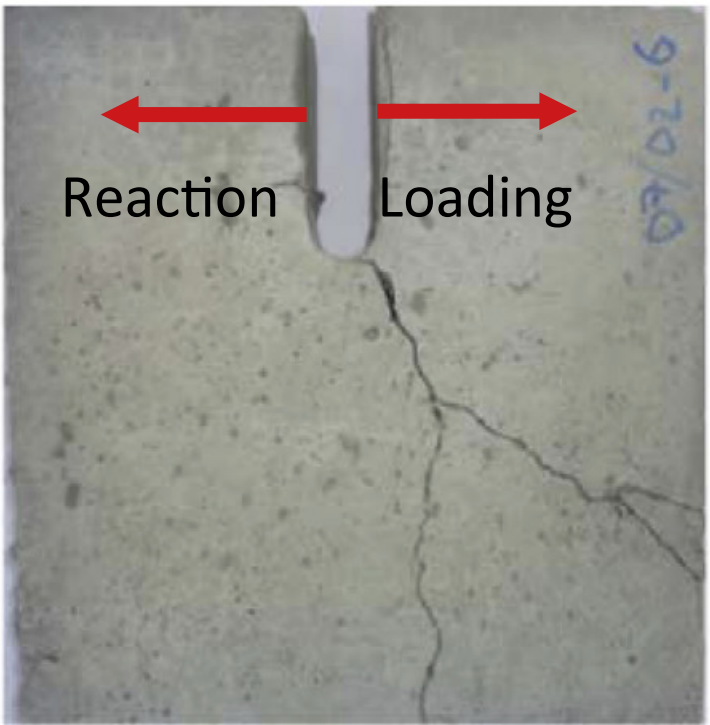}
            \end{center}
            \begin{center}
            (b)
            \end{center}
        \end{minipage}   
        \caption{(a). Symmetric branching in single edge notch specimen (\cite{ramulu1985mechanics}); (b). Asymmetric crack branching in concrete specimen (\cite{ovzbolt2013dynamic}).}
        \label{fig32: crack-branching-concept}
\end{figure}

Among various cracking phenomena, crack branching is frequently observed across numerous engineering applications, particularly in brittle materials and metallic alloys susceptible to stress corrosion cracking (\cite{kalthoff1973propagation}). Branching may occur in symmetric or asymmetric patterns (see Figure.\ref{fig32: crack-branching-concept}). A comprehensive investigation of the crack branching process, including its underlying mechanisms, branching criteria, experimental characterization, and numerical modeling, is critical for accurate and reliable prediction of crack growth behavior. Developing a deep understanding of crack initiation and its subsequent propagation is essential for improving the structural integrity and durability of engineering components.

Numerous crack models and computational techniques have been developed to simulate crack initiation and propagation, including complex behaviors such as branching, coalescence, and intersection, while maintaining computational efficiency. Broadly, these modeling approaches can be classified into two categories: discrete crack models and smeared crack models. Discrete crack models explicitly represent crack geometry and include methods such as remeshing, element deletion, enrichment techniques (e.g., XFEM), cracking particle methods, and cohesive zone models. These approaches provide detailed descriptions of crack evolution but often require complex implementations and mesh manipulation strategies. In contrast, smeared crack models conceptualize the effect of cracking as a phenomenon distributed across a region, avoiding explicit crack surface representation. Examples include nonlocal damage models, gradient-enhanced formulations, viscous regularization techniques, and phase-field models. While these approaches often offer enhanced robustness in handling complex crack topologies, they are generally computationally intensive and may compromise the precise geometric representation of discontinuities.

Furthermore, numerical methods for simulating crack branching can also be categorized based on the presence of spatial derivatives in their governing equations: continuum methods and discontinuous methods. Continuum-based techniques, which model materials as continuous media and are formulated using partial differential equations, include the Finite Element Method, Extended Finite Element Method, Boundary Element Method, and various meshfree methods. Discontinuous methods, such as peridynamics and the discrete element method, eliminate spatial derivatives and are well-suited for capturing discontinuities, including crack branching and fragmentation, without requiring remeshing or enrichment. Each of these numerical approaches offers specific strengths and limitations, depending on the application context, material behavior, and loading conditions. Despite significant advances, a universally accepted numerical framework capable of accurately and efficiently simulating crack branching under general conditions remains an open challenge in computational fracture mechanics.

Unlike two-dimensional crack propagation, where the crack path remains confined to a plane, three-dimensional crack branching exhibits complex morphological features such as crack front twisting, surface roughening, and out-of-plane kinking. These complexities significantly increase the difficulty of accurate prediction and numerical simulation. As a result, research on three-dimensional transient dynamic crack propagation and branching remains relatively limited. The primary challenges and obstacles associated with advancing this study area can be summarized as follows,
\begin{itemize}
\item \textbf{Complex Crack Morphologies}: In three-dimensional settings, crack propagation and branching can occur within and out of the original fracture plane, resulting in a wide range of intricate crack morphologies. These include upward or downward kinks, twisting of the crack front, and branched structures resembling "palm tree" or "fan" shapes. Furthermore, phenomena such as branch merging, crack nucleation, and the formation of closed loops further complicate the geometric representation, making clear visualization and interpretation exceedingly challenging.
\item \textbf{Crack Front Instability and Branching Criteria}: Unlike two-dimensional crack branching, which can often be predicted using dynamic criteria based on crack tip velocity, no universally accepted criterion exists to fully describe the onset and evolution of crack bifurcation in three-dimensional settings. While several approaches, such as Gibbs free energy minimization, stress intensity factor (SIF) analysis, and linear stability analysis, have been proposed, their applicability is generally limited to specific material systems, geometries, or loading conditions. As a result, the prediction of crack front instability and branching in three dimensions remains an open and complex challenge.
\item \textbf{High Computational Costs}: The shift from two-dimensional to three-dimensional crack modeling significantly amplifies computational complexity. For instance, a crack tip in two dimensions becomes a crack front in three dimensions, characterized by intricate curved geometries that demand substantially greater memory and processing resources. Additionally, the tight coupling of spatial and temporal resolution exacerbates computational demands: complex three-dimensional morphologies necessitate smaller element sizes, which in turn require finer timesteps for stability and accuracy. These combined effects render traditional CPU-based parallel computing approaches insufficient, highlighting the need for high-performance computing solutions such as GPU acceleration.
\end{itemize}

This study introduces a novel three-dimensional Crack Element Method (CEM) to address the aforementioned challenges. The proposed method incorporates an easily implementable element failure criterion, grounded in a newly developed computational formulation for fracture energy release rate. This criterion, along with the associated energy release rate computation, has been previously validated through two-dimensional numerical studies (\cite{xie2025practicalfiniteelementapproach}). In this framework, when an element's computed fracture energy release rate exceeds a critical threshold (i.e., $\mathcal{G}$ > $\mathcal{G}_c$), the element is deactivated and excluded from subsequent computations. This mechanism enables the natural and straightforward representation of complex three-dimensional crack patterns, eliminating the need for additional morphological processing. All three-dimensional benchmark simulations are executed using Graphics Processing Unit (GPU) parallelization, effectively addressing the high computational demands and demonstrating the model’s suitability for large-scale, high-fidelity fracture simulations.

The paper is organized as follows to provide a clear and comprehensive introduction to the proposed three-dimensional CEM. Section 2 reviews the variational principles governing three-dimensional fracture in solids and presents the CEM-based weak formulation for continuum media with discontinuities induced by cracks. Section 3 details the novel three-dimensional CEM and introduces the corresponding computational approach for evaluating the fracture energy release rate. Section 4 showcases a series of transient dynamic benchmark simulations, including examples of three-dimensional crack branching under Neumann and Dirichlet boundary conditions. Finally, Section 5 summarizes the key findings and outlines future research directions to address more complex challenges in fracture mechanics.

\section{Variational Formulation for Dynamic Fracture Problems}
This section presents the variational formulation for the three-dimensional transient dynamic fracture problem in quasi-brittle materials. Subsequently, the Edge-based Smoothed Finite Element Method (ES-FEM) is introduced as the discretization framework, employing both four-node tetrahedral and eight-node hexahedral elements to approximate the solution of the variational form in three-dimensional fracture simulations. The notation and indices in the study are introduced. 

\subsection{Variational three-dimensional fractures}
In continuum mechanics, fracture is characterized by a strong discontinuity in the displacement field occurring within certain regions of the domain. Accurately capturing the intricate cracking behavior across various materials and structural systems requires a robust framework capable of handling complex crack geometries, as well as the processes of crack nucleation and growth. These phenomena can often exhibit irregular, non-smooth patterns in both spatial and temporal dimensions.

The simplest fracture propagation criterion was erected by Irwin et al., which is formulated as follows,
\begin{eqnarray}
\dot{\ell} = \begin{cases} 
=0, \text{ if } K < K_c \text{ (no propagation) } \\
\ge 0, \text{ if } K=K_c \text{ (propagation) }
\end{cases}
\nonumber
\end{eqnarray}
in which, $\ell$ indicates length of a crack path, $K_c$ indicates the fracture toughness of the material and $K$ indicates a stress intensity factor. An improved criterion was introduced by Griffth (\cite{griffith1921vi}), which formulates as,
\begin{eqnarray}
\dot{\ell} = \begin{cases}
=0,\text{ if }\mathcal{G} < \mathcal{G}_c \text{ (no propagation) }\\
\ge0, \text{ if } \mathcal{G} \ge \mathcal{G}_c \text{ (propagation) }
\nonumber
\end{cases}
\end{eqnarray}
in which, $\mathcal{G}_c$ denotes the critical fracture energy release rate. A well-recognized limitation of Griffith’s theory is its inability to account for crack initiation. This shortcoming can be addressed by adopting a cohesive fracture model, in which the surface fracture energy $\varphi$ is based on the magnitude of the crack opening, typically expressed as $\varphi = \varphi\left( [\![ u ]\!]\right)$.

\begin{figure}[htp]
	\centering
            \begin{center}
            \includegraphics[height=2.5in]{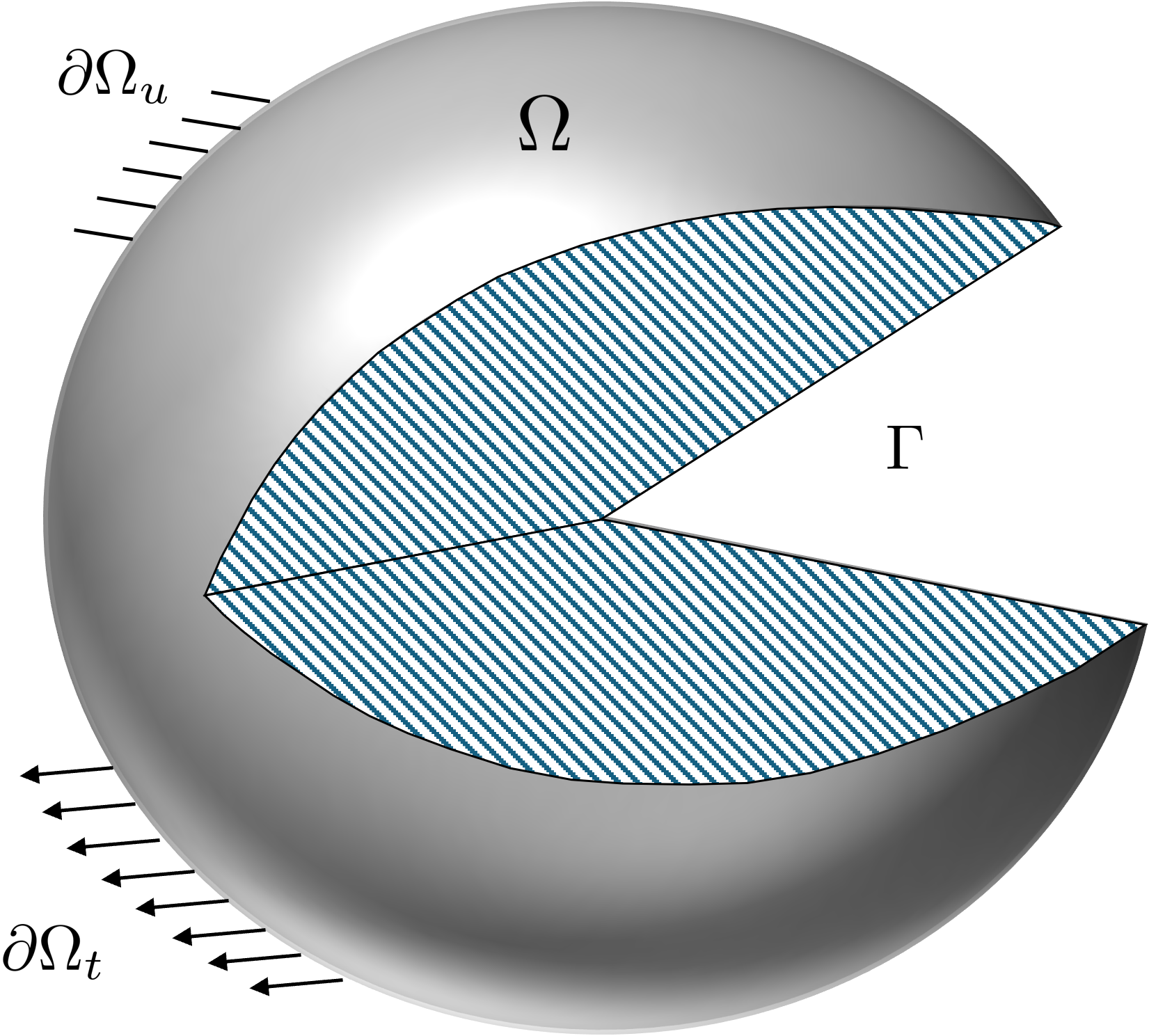}
            \end{center}
        \caption{Illustrative representation of a solid domain $\Omega$ having internal discontinuity $\Gamma$.}
        \label{fig33: weak-form-illustration}
\end{figure}

The reformulation of the brittle fracture problem within a variational framework was introduced by Francfort et al. (\cite{francfort1998revisiting}). A key advantage over Griffth's theory is that the variational approach shows both facture initiation and evolution to emerge naturally - without presupposing any initial defects, predefined crack paths, or imposing ad hoc evolution rules. At the heart of this method is its ability to regularize the fracture problem, which is fundamentally a mathematical free - discontinuity problem, into a more tractable form. This is achieved by introducing a new damage-like variable that evolves smoothly, avoiding explicit discontinuities. The main steps leading to this regularized formulation are outlined below (\cite{bourdin2000numerical}, \cite{bourdin2008variational}, \cite{ambrosio1990approximation}),

An arbitrary domain $\Omega \in \mathbb{R}^d$ ($d\in\left\{ 2,3 \right\}$) having external boundary $\partial \Omega$ and internal discontinuity boundary $\Gamma$, as shown in Figure.\ref{fig33: weak-form-illustration}.
The displacement of a material point $\bm{x} \in \Omega$ at time $t \in \left[0, T \right]$ is represented by $\bm{u}\left(\bm{x}, t \right) \in \mathbb{R}^d$. The spatial conponents of vectors and tensors are indexed using $i,j = 1,2,\cdots, d$. Beisdes, the a portion of displacement field is subjected to either Dirichlet boundary condition, $\bm{u}\left(\bm{x}, t \right) = \bar{\bm{u}}\left( \bm{x}, t \right)$, on $\partial \Omega_{u} \in \partial \Omega$, and Neumann boundary condition, $\bm{t}\left(\bm{x},t \right) =\bar{\bm{t}} \left(\bm{x}, t \right) $, on $\partial \Omega_{t} \in \partial \Omega$. The small deformation is assumed so that the infinitesimal strain tensor is defined as,
\begin{eqnarray}
\bm{\varepsilon} = \nabla^{s} \bm{u} = \frac{1}{2} \left(\frac{\partial \bm{u}}{\partial \bm{x}} + \left( \frac{\partial \bm{u}}{\partial \bm{x}} \right)^T \right)
\nonumber
\end{eqnarray}
in which, $\nabla^s$ denotes the symmetric gradient operator.
The isotropic elasticity of constitutive model is provided in the study, therefore the elastic energy density of the domain $\Omega$ is give as,
\begin{eqnarray}
\psi \left( \bm{\varepsilon} \right) = \frac{1}{2} \lambda \bm{\varepsilon} : \bm{\varepsilon} + \mu \bm{\varepsilon}:\bm{\varepsilon}
\nonumber
\end{eqnarray}
where $\lambda$ and $\mu$ are Lame constants. 

The time-dependent internal discontinuity boundary, $\Gamma \left(t \right)$, denotes a collection of discrete cracks. According to Griffith's theory of brittle fracture, the formation of a unit area of crack surface consumes energy equal to the critical fracture energy density, $\mathcal{G}_c$. The total energy of the domain, $\mathcal{W}_{potential}$, is given as follows,
\begin{eqnarray}
\mathcal{W}_{potential} \left(\bm{u}, \Gamma \right) = -\int_{\Omega} \psi \left(\nabla^s \bm{u} \right) d \Omega + \int_{\Gamma} \mathcal{G}_c d \Gamma + \int_{\Omega} \bm{u} \cdot \bm{d} d\Omega + \int_{\partial \Omega_{t} }\bm{u} \cdot \bar{\bm{t}} d \ell  ~.
\end{eqnarray}
Under the assumption of brittle fracture, the total fracture energy is given by the integral of the critical fracture energy density over the entire crack surface. Furthermore, the irreversibility of fracture implies that $\Gamma\left(t \right) \in \Gamma \left(t + \Delta t \right)$ for all $\Delta t > 0$. As a result, cracks cannot retract or translate within the body, but are allowed to grow, branch, or coalesce over time. 

Since the we consider the transient-dynamic case, the kinetic of the domain $\Omega$ must to be included and is given as
\begin{eqnarray}
\mathcal{T}_{kinetic}\left(\dot{\bm{u}} \right) = \frac{1}{2} \int_{\Omega} \rho \dot{\bm{u}} \cdot \dot{\bm{u}} d\Omega
\nonumber
\end{eqnarray}
in which, $\dot{\Box}$ denotes the time derivative of a specific variable and $\rho$ is material density. Accordingly, the temporal-dependent Lagrangian formulae of the studied domain with discontinuity fracture problem is shown as,
\begin{eqnarray}
\mathcal{L} \left(\bm{u}, \dot{\bm{u}}, \Gamma \right) &=& \mathcal{T}_{kinetic} - \mathcal{W}_{potential} \nonumber \\
&=& \frac{1}{2} \int_{\Omega} \rho \dot{\bm{u}} \cdot \dot{\bm{u}} d\Omega + \int_{\Omega} \psi \left(\nabla^s \bm{u} \right) d \Omega - \int_{\Gamma} \mathcal{G}_c d \Gamma - \int_{\Omega} \bm{u} \cdot \bm{b} d\Omega - \int_{\partial \Omega_{t} }\bm{u} \cdot \bar{\bm{t}} d \ell
\label{eq:weak_form}
\end{eqnarray}
Numerically tracking the evolution of the discontinuity boundary $\Gamma$ can be computationally intensive and challenging, especially when managing interactions among multiple cracks in two dimensions, or when modeling cracks with intricate geometries in three dimensions. This difficulty is further amplified in dynamic fracture simulation, where the ability to accurately and robustly capture phenomena such as crack branching is especially importance.

\subsection{ES-FEM discretization in three-dimension}
In the proposed three-dimensional CEM, the Edge-based Smooth Tetrahedron FEM (ES-T-FEM, \cite{he2013edge}) and the Edge-based Smooth Hexahedron FEM (ES-H-FEM) are utilized to discretize the three-dimensional variational formulation of the discontinuity fracture problem, i.e., Eq.(\ref{eq:weak_form}). It is worth emphasizing that the Edge-based Smooth Hexahedron FEM (ES-H-FEM) is NOT the Simplified FEM based on 8-node Hexahedron (S-FEM-H8, \cite{li2019highly}). The latter uses node-shared and face-shared neighboring elements while the former only uses edge-shared neighboring elements. 

\begin{figure}[htp]
        \centering
        \begin{minipage}{0.45\linewidth}
            \begin{center}
            \includegraphics[height=2.4in]{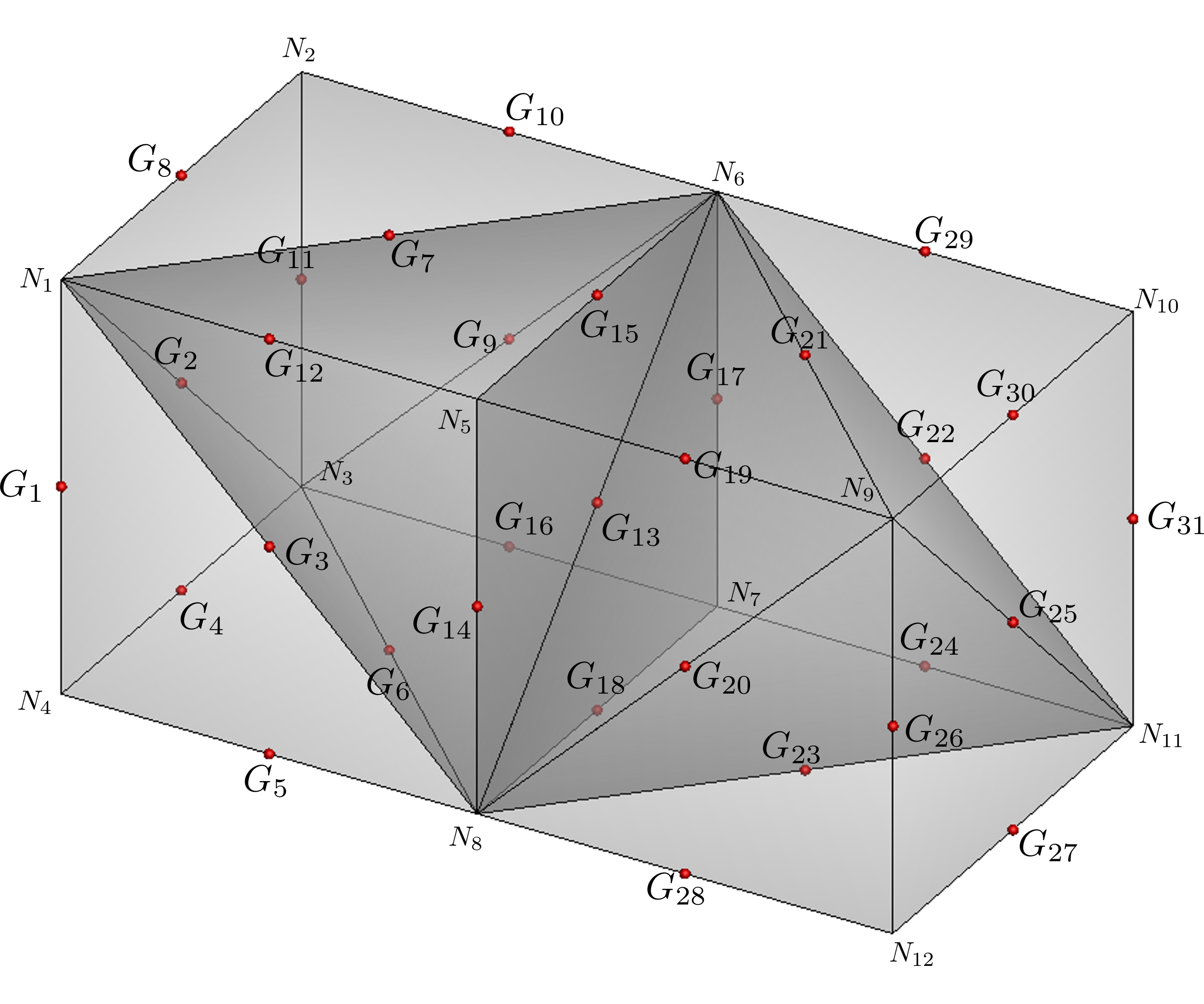}
            \end{center}
            \begin{center}
            (a)
            \end{center}
        \end{minipage}
        \hfill
        \begin{minipage}{0.45\linewidth}
            \begin{center}
            \includegraphics[height=2.4in]{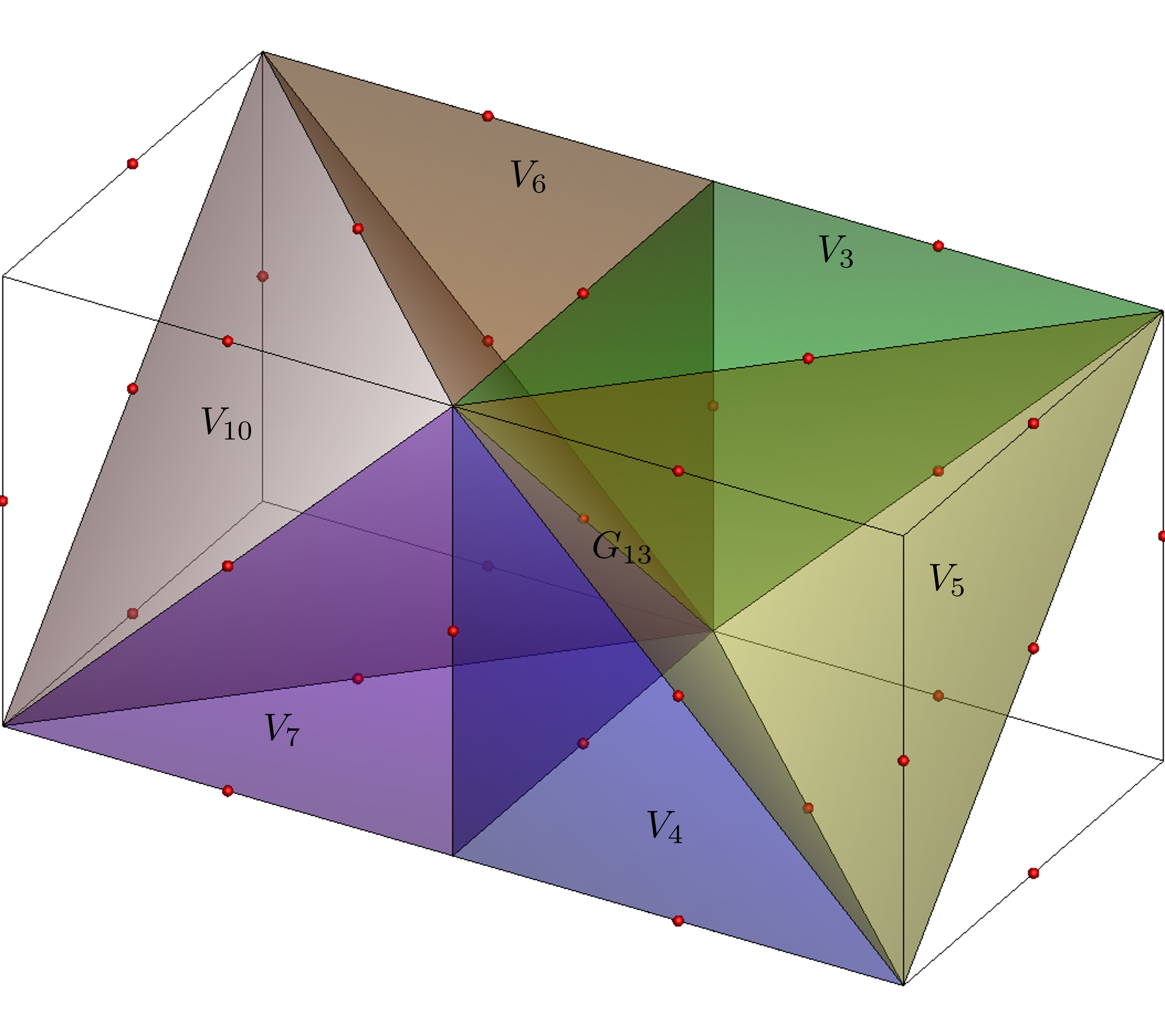}
            \end{center}
            \begin{center}
            (b)
            \end{center}
        \end{minipage}   
        \hfill
        \begin{minipage}{0.45\linewidth}
            \begin{center}
            \includegraphics[height=2.4in]{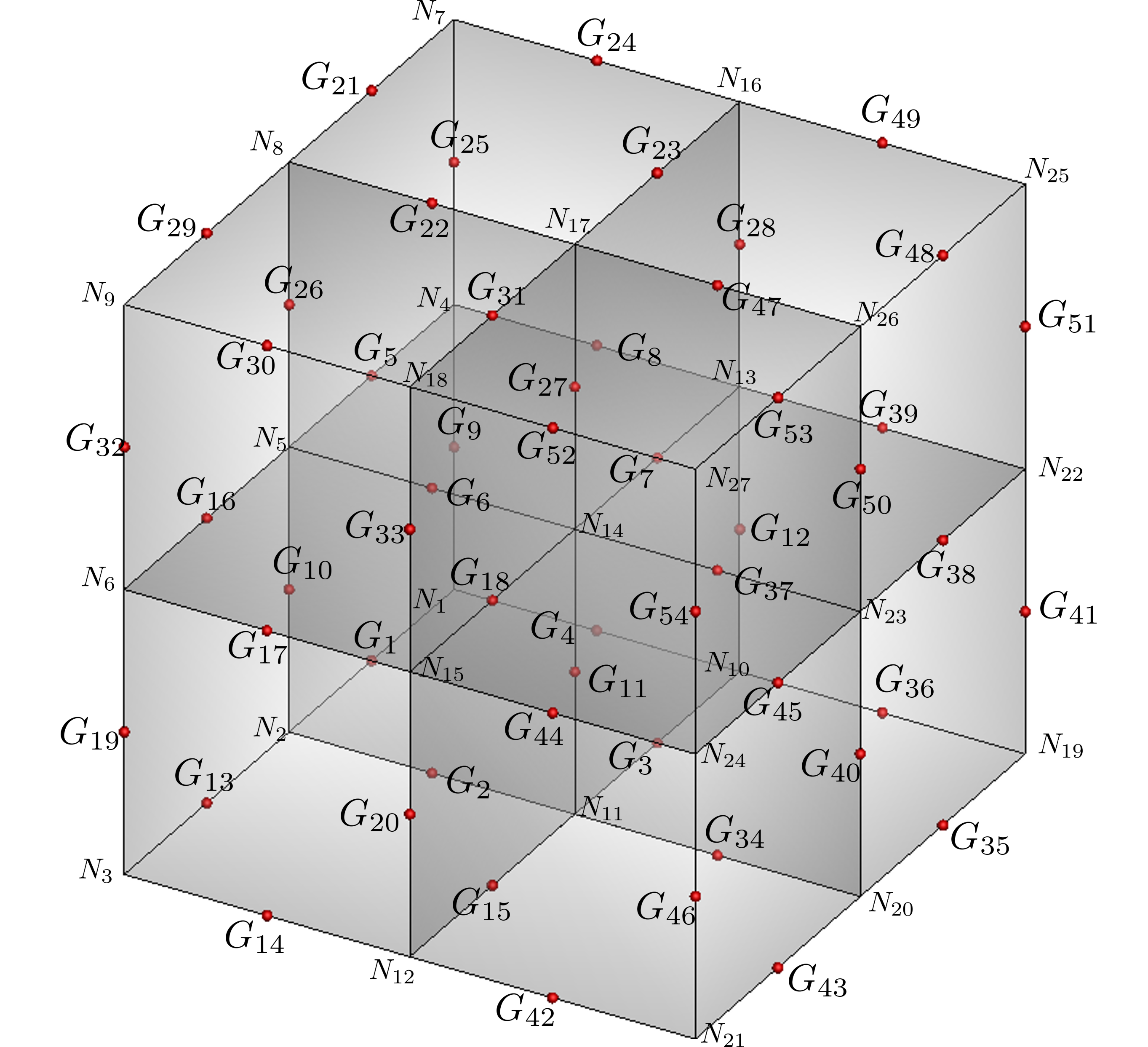}
            \end{center}
            \begin{center}
            (c)
            \end{center}
        \end{minipage} 
        \hfill
        \begin{minipage}{0.45\linewidth}
            \begin{center}
            \includegraphics[height=2.4in]{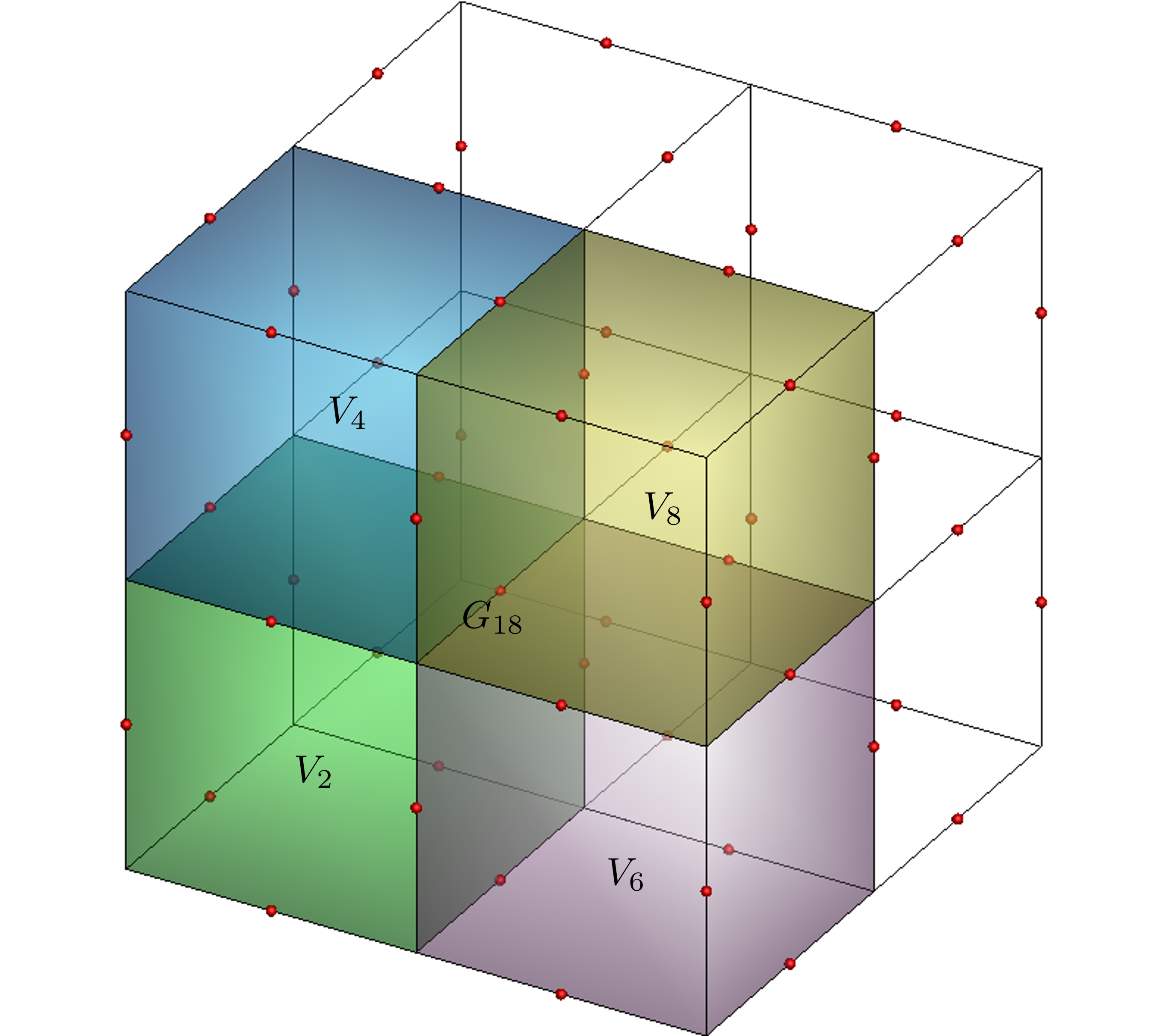}
            \end{center}
            \begin{center}
            (d)
            \end{center}
        \end{minipage} 
        \caption{(a). The studied domain is discretized by tetrahedron elements. Notation $V_i$ denotes element ID, $G_j$ denotes edge quadrature ID and $N_k$ denotes node ID: (a). ES-T-FEM formulation of Constant Strain Tetrahedron element, totally $10$ elements, $12$ nodes and $31$ edge quadrature points; (b). neighboring elements of edge quadrature point $G_{13}$, i.e., $V_3, V_4, V_5, V_6, V_7, V_{13}$. other elements are blank; (c). ES-H-FEM formulation of Trilinear Strain Hexahedron element, totally $4$ elements, $27$ nodes and $54$ edge quadrature points; (d). neighboring elements of edge quadrature point $G_{18}$, i.e., $V_2, V_4, V_6, V_8$. other elements are blank.}
        \label{fig34: ES-T-FEM_CSTet_Hex_formulation}
\end{figure}

Similar to two-dimensional ES-FEM, the studied domain is partitioned by edge quadrature points of elements, i.e., $\Omega = \bigcap_{i=1}^{N_{edges}}$, where $N_{edges}$ denotes the total number of edges of all elements taking up of the entire domain. Contrary to a definite number of neighboring elements sharing one edge in two-dimension, the number of elements sharing one edge in three-dimension is NOT determinate. For example, in Figure.\ref{fig34: ES-T-FEM_CSTet_Hex_formulation}(b) the neighboring elements of edge quadrature $G_{13}$ include $V_3$ (green), $V_4$ (blue), $V_5$ (yellow), $V_6$ (brown), $E_7$ (purple) and $E_{13}$ (pink); in Figure.\ref{fig34: ES-T-FEM_CSTet_Hex_formulation}(d) the neighboring elements of edge quadrature $G_{18}$ include $V_2$ (green), $V_4$ (blue), $V_6$ (pink) and $V_8$ (yellow).

The displacement field at an edge quadrature point $G_i$ can be interpolated using the following expression,
\begin{eqnarray}
\bm{u}\left(G_i \right) = \sum_{j \in \mathcal{M}_e^{G_i}} \left( \sum_{k \in \mathcal{L}_n^{V_j}} N_k^{V_j} \left(G_i \right) \bm{u}_k \times \frac{V_j}{\sum_{m \in \mathcal{N}_e^{G_i}}V_m} \right)
\end{eqnarray}
where $V_j$ denotes neighboring element ID, $G_i$ denotes edge quadrature point ID, both $\mathcal{M}_e^{G_i}$ and $\mathcal{N}_e^{G_i}$ denote the neighboring elements set of edge quadrature point $G_i$, $\mathcal{L}_n^{V_j}$ denotes nodes set of the neighboring element $V_j$, $N_k^{V_j}\left(G_i \right)$ denotes the $k$-th nodal shape function of element $V_j$ at edge quadrature point $G_i$ location, $\bm{u}_k$ denotes the displacement of node $k$. Accordingly, the infinitesimal strain field can be derived by differentiating the displacement $\bm{u}\left(G_i \right)$ with respect to the current spatial coordinates $\bm{x}$, as illustrated below,
\begin{eqnarray}
\bm{\varepsilon} = \frac{\partial \bm{u}\left(G_i \right)}{\partial \bm{x}} = \sum_{j \in \mathcal{M}_e^{G_i}} \left( \sum_{k \in \mathcal{L}_n^{V_j}} \frac{\partial N_k^{V_j} \left(G_i \right)} {\partial \bm{x}} \bm{u}_k \times \frac{V_j}{\sum_{m \in \mathcal{N}_e^{G_i}}V_m} \right)
\end{eqnarray}
therefore the discretized formulation of matrix is available,
\begin{eqnarray}
\bm{\varepsilon} = \left( \varepsilon_{11}, \varepsilon_{22}, \varepsilon_{33}, \gamma_{12}, \gamma_{23}, \gamma_{13} \right)^T = \A_j \left( \A_k \left[ B_k^{V_j} \right] \rm{w}_{V_j} \right) \cdot \left\{\bm{u} \right\}
\label{eq:strain_discretization}
\end{eqnarray}
in which, $j$ denotes the neighboring element ID, $k$ denotes nodal ID of a neighboring element, $\rm{w}_{V_j} = \frac{V_j}{\sum_n V_n}$ denotes the volume weight of each neighboring element, $\left[B_k^{V_j} \right]$ denotes stress-strain matrix of node $k$ in element $V_j$ and $\A$ denotes element assemble operator in Finite Element Method.

In the Constant Tetrahedron element formulation, Eq.(\ref{eq:strain_discretization}) can be expanded as follows,
\begin{eqnarray}
&\bm{\varepsilon} = \begin{Bmatrix}
\varepsilon_{11} \\
\varepsilon_{22} \\
\varepsilon_{33} \\
\gamma_{12} \\
\gamma_{23} \\
\gamma_{13}
\end{Bmatrix} = \left[ \begin{array}{ccc}
\frac{\partial N_i^{e_j}}{\partial x} \rm {w}_{V_j}, & 0, & 0 \\
0, & \frac{\partial N_i^{e_j}}{\partial y} \rm{w}_{V_j}, & 0 \\
0, & 0, & \frac{\partial N_i^{e_j}}{\partial z} \rm{w}_{V_j} \\
\frac{\partial N_i^{e_j}}{\partial y} \rm{w}_{V_j}, & \frac{\partial N_i^{e_j}}{\partial x} \rm{w}_{V_j}, & 0 \\
0, & \frac{\partial N_i^{e_j}}{\partial z} \rm{w}_{V_j}, & \frac{\partial N_i^{e_j}}{\partial y} \rm{w}_{V_j} \\
\frac{\partial N_i^{e_j}}{\partial z} \rm{w}_{V_j}, & 0, & \frac{\partial N_i^{e_j}}{\partial x} \rm{w}_{V_j}
\end{array} \right]_{{i=1\sim 4}, V_j=1\sim \mathcal{N}_e} \begin{Bmatrix}
u_i^{1} \\
v_i^{1} \\
w_i^{1} \\
\vdots \\
u_i^{\mathcal{N}_e} \\
v_i^{\mathcal{N}_e} \\
w_i^{\mathcal{N}_e}
\end{Bmatrix}_{{i=1\sim 4}, V_j=1\sim \mathcal{N}_e}  ~.
\label{eq:CSTet_strain_discretization}
\end{eqnarray}
in which, $N_i^{e_j}$ denotes $i$-th nodal shape function of neighboring element $e_j$, $\rm{w}_{V_j}=\frac{V_j}{\sum_k V_k}$ denotes the volume weight of each neighboring element, $u_i^{e_j}, v_i^{e_j}, w_i^{e_j}$ denote $i$-th nodal displacement of neighboring element $e_j$ along $x,y,z$ direction, respectively. $\mathcal{N}_e$ denotes the number of neighboring elements, so that the subscript $i=1\sim4, V_j=1\sim \mathcal{N}_e$ indicate the dimensions of matrix $B$ are $6 \times 12\mathcal{N}_e$, the dimensions of nodal displacement vector are $12\mathcal{N}_e\times 1$.\\
In trilinear Hexahedron element formulation, Eq.(\ref{eq:strain_discretization}) can be expanded as follows,
\begin{eqnarray}
&\bm{\varepsilon} = \begin{Bmatrix}
\varepsilon_{11} \\
\varepsilon_{22} \\
\varepsilon_{33} \\
\gamma_{12} \\
\gamma_{23} \\
\gamma_{13}
\end{Bmatrix} = \left[ \begin{array}{ccc}
\frac{\partial N_i^{e_j}}{\partial x} \rm {w}_{V_j}, & 0, & 0 \\
0, & \frac{\partial N_i^{e_j}}{\partial y} \rm{w}_{V_j}, & 0 \\
0, & 0, & \frac{\partial N_i^{e_j}}{\partial z} \rm{w}_{V_j} \\
\frac{\partial N_i^{e_j}}{\partial y} \rm{w}_{V_j}, & \frac{\partial N_i^{e_j}}{\partial x} \rm{w}_{V_j}, & 0 \\
0, & \frac{\partial N_i^{e_j}}{\partial z} \rm{w}_{V_j}, & \frac{\partial N_i^{e_j}}{\partial y} \rm{w}_{V_j} \\
\frac{\partial N_i^{e_j}}{\partial z} \rm{w}_{V_j}, & 0, & \frac{\partial N_i^{e_j}}{\partial x} \rm{w}_{V_j}
\end{array} \right]_{{i=1\sim 8}, V_j=1\sim \mathcal{N}_e} \begin{Bmatrix}
u_i^{1} \\
v_i^{1} \\
w_i^{1} \\
\vdots \\
u_i^{\mathcal{N}_e} \\
v_i^{\mathcal{N}_e} \\
w_i^{\mathcal{N}_e}
\end{Bmatrix}_{{i=1\sim 8}, V_j=1\sim \mathcal{N}_e}  ~.
\label{eq:hex_strain_discretization}
\end{eqnarray}
in which, the subscript $i=1\sim8, V_j=1\sim \mathcal{N}_e$ indicate the dimensions of $B$ matrix are $6\times 24\mathcal{N}_e$, the dimensions of nodal displacement vector are $24\mathcal{N}_e\times 1$. Other notations follow the Eq.(\ref{eq:CSTet_strain_discretization}).

Different from standard three-dimensional FEM, the internal force $f_{int}$ assembly process from $\int_{\Omega} \psi\left(\nabla^s \bm{u} \right) d\Omega - \int_{\Gamma}\mathcal{G}_c d\Gamma$ in Eq.(\ref{eq:weak_form}) can be formulated as following equation,
\begin{eqnarray}
f_{int} = \int_{\Omega \backslash \Gamma} \nabla^s \bm{N} : \bm{\sigma} d\Omega = \A_j \left( \A_k \left[ B_k^{V_j} \right]^T \rm{w}_{V_j} \cdot \bm{\sigma} \right) \cdot V_j
\label{eq:fint_discretization}
\end{eqnarray}
in which, $j$ denotes the neighboring element ID, $k$ denotes nodal ID of a neighboring element, $\rm{w}_{V_j} = \frac{V_j}{\sum_n V_n}$ denotes the volume weight of each neighboring element and $V_j$ denotes the volume of a neighboring element. It is worth noting that the energy consumed by unit area of the crack surface is implemented through domain removal, i.e., $\Omega \backslash \Gamma$, rather than the explicit formulation. Accordingly, in the Constant Tetrahedron element formulation, Eq.(\ref{eq:fint_discretization}) can be expanded as,
\begin{eqnarray}
\bm{f}_{int} = \begin{Bmatrix}
f_{xi}^{e_j} \\
f_{yi}^{e_j} \\
f_{wi}^{e_j} 
\end{Bmatrix}_{i= 1\sim 4} = \left[ \begin{array}{ccc}
\frac{\partial N_i^{e_j}}{\partial x} \rm {w}_{V_j}, & 0, & 0 \\
0, & \frac{\partial N_i^{e_j}}{\partial y} \rm{w}_{V_j}, & 0 \\
0, & 0, & \frac{\partial N_i^{e_j}}{\partial z} \rm{w}_{V_j} \\
\frac{\partial N_i^{e_j}}{\partial y} \rm{w}_{V_j}, & \frac{\partial N_i^{e_j}}{\partial x} \rm{w}_{V_j}, & 0 \\
0, & \frac{\partial N_i^{e_j}}{\partial z} \rm{w}_{V_j}, & \frac{\partial N_i^{e_j}}{\partial y} \rm{w}_{V_j} \\
\frac{\partial N_i^{e_j}}{\partial z} \rm{w}_{V_j}, & 0, & \frac{\partial N_i^{e_j}}{\partial x} \rm{w}_{V_j}
\end{array} \right]_{{i=1\sim 4}, V_j=1\sim \mathcal{N}_e} ^T \begin{Bmatrix}
\sigma_{11} \\
\sigma_{22} \\
\sigma_{33} \\
\sigma_{12} \\
\sigma_{23} \\
\sigma_{13}
\end{Bmatrix} \cdot \frac{1}{6}\sum_j V_j
\label{eq:CSTet_fint_discretization}
\end{eqnarray}
in which, $f_{xi}^{e_j}, f_{yi}^{e_j}, f_{zi}^{e_j}$ denote $i$-th nodal force of neighboring element $e_j$ in $x$-, $y$- and $z-$ directions, respectively. $\left[\cdot \right]^T$ denotes transpose of a matrix. In the trilinear Hexahedron element formulation, Eq.(\ref{eq:fint_discretization}) can be expanded as follows,
\begin{eqnarray}
\bm{f}_{int} = \begin{Bmatrix}
f_{xi}^{e_j} \\
f_{yi}^{e_j} \\
f_{wi}^{e_j} 
\end{Bmatrix}_{i= 1\sim 8} = \left[ \begin{array}{ccc}
\frac{\partial N_i^{e_j}}{\partial x} \rm {w}_{V_j}, & 0, & 0 \\
0, & \frac{\partial N_i^{e_j}}{\partial y} \rm{w}_{V_j}, & 0 \\
0, & 0, & \frac{\partial N_i^{e_j}}{\partial z} \rm{w}_{V_j} \\
\frac{\partial N_i^{e_j}}{\partial y} \rm{w}_{V_j}, & \frac{\partial N_i^{e_j}}{\partial x} \rm{w}_{V_j}, & 0 \\
0, & \frac{\partial N_i^{e_j}}{\partial z} \rm{w}_{V_j}, & \frac{\partial N_i^{e_j}}{\partial y} \rm{w}_{V_j} \\
\frac{\partial N_i^{e_j}}{\partial z} \rm{w}_{V_j}, & 0, & \frac{\partial N_i^{e_j}}{\partial x} \rm{w}_{V_j}
\end{array} \right]_{{i=1\sim 8}, V_j=1\sim \mathcal{N}_e} ^T \begin{Bmatrix}
\sigma_{11} \\
\sigma_{22} \\
\sigma_{33} \\
\sigma_{12} \\
\sigma_{23} \\
\sigma_{13}
\end{Bmatrix} \cdot \frac{1}{12}\sum_j V_j
\label{eq:hex_fint_discretization}
\end{eqnarray}

The mass matrix from $\frac{1}{2} \int_{\Omega} \rho \dot{\bm{u}} \cdot \dot{\bm{u}}d\Omega$ in Eq.(\ref{eq:weak_form}) is discretized as a diagonal matrix:
\begin{eqnarray}
\bm{M} =  \begin{bmatrix}
m_{1}, & 0, & \cdots, & 0 \\
0, & m_{2},  & \cdots, & 0 \\
\vdots & \vdots & \ddots & \vdots \\
0 & 0 & \cdots & m_{N}
\end{bmatrix} = \A_e \begin{bmatrix}
N_1^e \\
N_2^e \\
\vdots \\
N_k^e
\end{bmatrix}^T \cdot \begin{bmatrix} N_1^e, N_2^e, \cdots, N_k^e
\end{bmatrix} \rho A_e
\label{eq:mass_discretization}
\end{eqnarray}
which follows the standard FEM mass matrix assemble operation. And the external force from $\int_{\Omega}\bm{u} \cdot \bm{b}d\Omega + \int_{\partial \Omega_t} \bm{u} \cdot \bar{\bm{t}}d\ell$ in Eq.(\ref{eq:weak_form}) is discretized as follows,
\begin{eqnarray}
\bm{f}_{ext} = \A_e \left[N^e\right] \cdot \bm{b} A_e + \A_e \left[ \hat{N}^e \right] \cdot \bar{\bm{t}} \ell_e
\end{eqnarray}
in which, $\left[N^e \right]$ denotes the standard three-dimensional element shape function, $\left[\hat{N}^e \right]$ denotes the degenerated element shape function matrix, $\ell_e$ denotes the integral line segment.

Similar to two-dimensional temporal discretization, an explicit \textit{Newmark} with $\beta=0$ is utilized, i.e.,
\begin{eqnarray}
 \rm \bm{a}_{n+1} &=& \frac{1}{\bm{M}} \left( \bm{f}^{ext}_{n+1} - \bm{f}^{int} \left( \rm \bm{u}_{n+1} \right) \right) = \frac{1}{\bm{M}} \left[ \bm{f}^{ext}_{\rm n+1} - \bm{f}^{int} \left( \rm \bm{u}_{n} + \rm \bm{v}_{n} \Delta t_n + \frac{1}{2} \rm \bm{a}_n \Delta t_n^2 \right) \right] \\
\rm \bm{v}_{n+1} &=& \rm \bm{v}_n + \left[ \left(1-\gamma \right) \rm \bm{a}_n + \gamma \rm \bm{a}_{n+1} \right] \Delta t_n \\
\rm \bm{ u }_{n+1} &=& \rm \bm{u}_n + \rm \bm{v}_n \Delta t_n + \frac{1}{2} \rm \bm{a}_n \Delta t_n^2 ~.
\end{eqnarray}

\section{A Novel Three-dimensional Crack Element Method}
In the Lagrangian formulation for solids containing discontinuities, Eq.(\ref{eq:weak_form}) represents the energy dissipated along fracture surfaces, i.e., $\int_{\Gamma} \mathcal{G}_c d\Gamma$. One of the principal innovations of the proposed three-dimensional CEM lies in developing a novel crack element formulation that mimics the crack growth, dissipates the fracture energy, and functions as a natural mechanism for tracking the evolution of three-dimensional fracture surfaces. Building upon the three-dimensional Edge-based Smoothed Finite Element Method (ES-FEM), the proposed CEM framework incorporates two element-specific formulations: the Edge-based Smoothed Tetrahedral Finite Element Method (ES-T-FEM) and the Edge-based Smoothed Hexahedral Finite Element Method (ES-H-FEM). These formulations enable the systematic construction of crack element models tailored to tetrahedral and hexahedral meshes, respectively, ensuring accurate and robust simulation of three-dimensional dynamic crack propagation.

\begin{figure}[htp]
	\centering
        \begin{minipage}{0.3\linewidth}
            \begin{center}
            \includegraphics[height=2.0in]{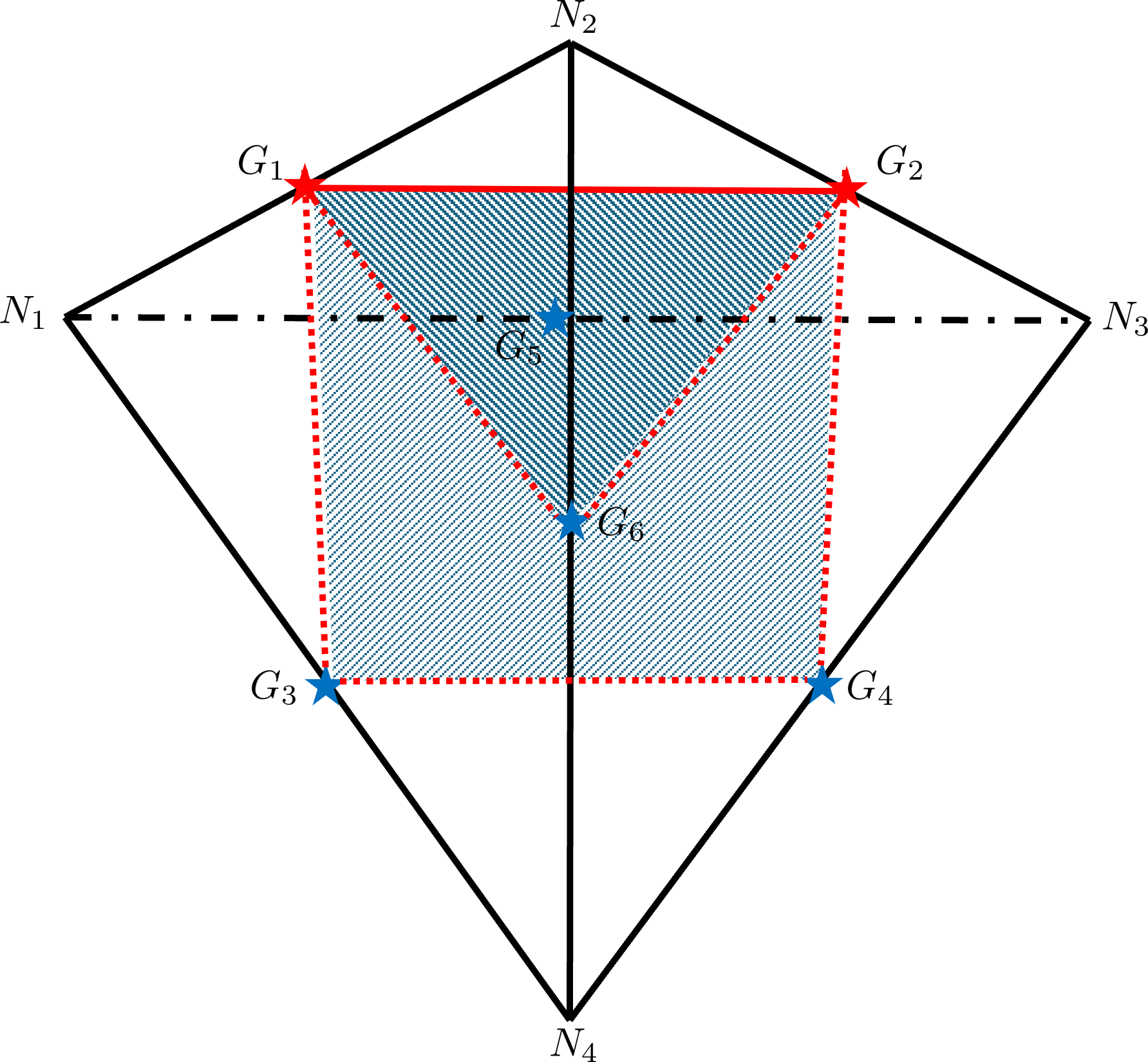}
            \end{center}
            \begin{center}
            (a)
            \end{center}
        \end{minipage}
        \hfill
        \begin{minipage}{0.3\linewidth}
            \begin{center}
            \includegraphics[height=2.0in]{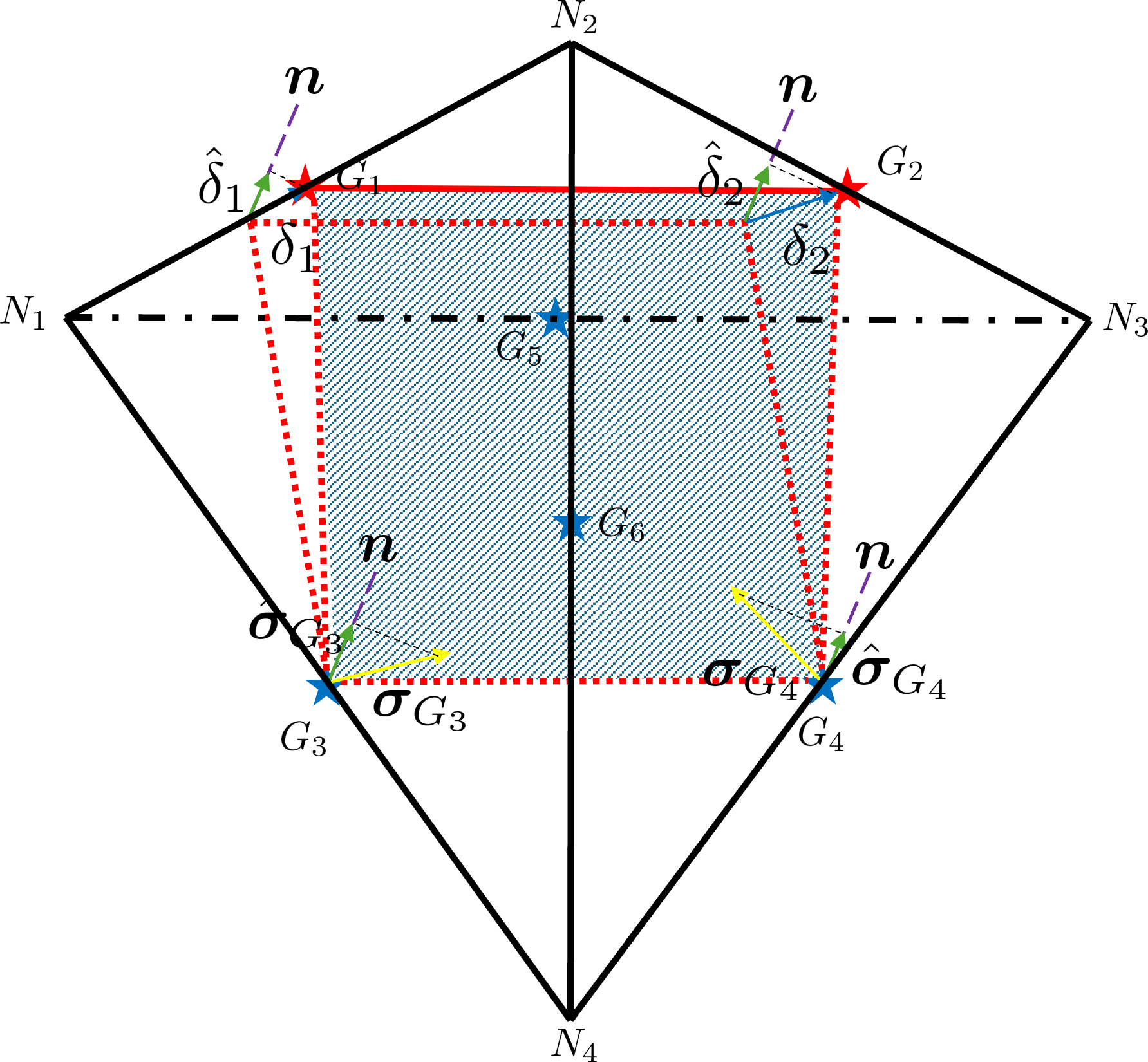}
            \end{center}
            \begin{center}
            (b)
            \end{center}
        \end{minipage}   
        \hfill
        \begin{minipage}{0.3\linewidth}
            \begin{center}
            \includegraphics[height=2.0in]{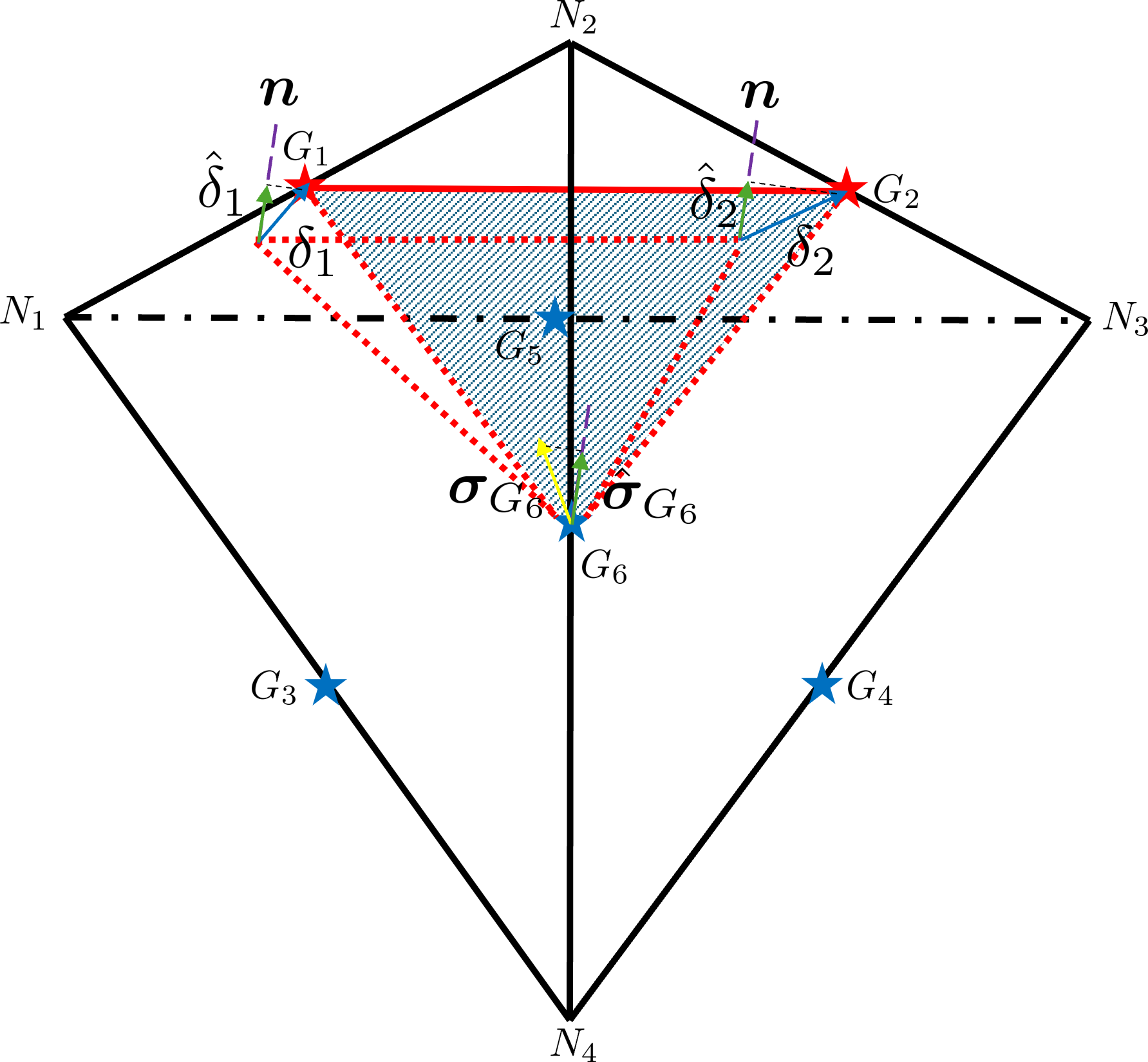}
            \end{center}
            \begin{center}
            (c)
            \end{center}
        \end{minipage}
        \caption{Three-dimensional crack types of Constant Strain Tetrahedron element, where $N_i$ denotes node ID, $G_j$ denotes edge quadrature point ID, $\delta_i$ denotes stretch of an edge, $\bm{n}$ denotes the normal vector of crack surface, $\hat{\bm{\delta}}_i = \left(\bm{\delta} \cdot \bm{n} \right) \bm{n}$ denotes the projection of $\delta_i$ on normal vector $\bm{n}$, $\bm{\sigma}_{G}$ denotes the maximum principal stress at edge quadrature $G$, $\hat{\bm{\sigma}}_G = \left(\bm{\sigma}_G \cdot \bm{n} \right)\bm{n}$ denotes the projection of $\bm{\sigma}_G$ on normal vector $\bm{n}$, and blue grids demonstrate possible crack surface: (a). possible three-dimensional crack patterns; (b). three-dimensional crack pattern $I$ and corresponding calculation of fracture energy release rate $\mathcal{G}_I$; (c). three-dimensional crack pattern $II$ and correspinding computation of fracture energy release rate $\mathcal{G}_{II}$.}
        \label{fig35: Gc_compute_CSTet}
\end{figure}

In Constant Strain Tetrahedron element type, as shown in Figure.\ref{fig35: Gc_compute_CSTet}, there are two options of three-dimensional crack patterns (shadow portions): one is a quadrilateral plane formed by edge quadrature points $G_1$, $G_2$, $G_4$, $G_3$ (See Appendix-\ref{Appendix-I: coplanarity}); another is a triangle formed by edge quadrature points $G_1$, $G_2$, $G_6$ (see Figure.\ref{fig35: Gc_compute_CSTet}(a)). It is noted that $N_i$ denotes node ID, $G_j$ denotes edge quadrature point ID. The solid red line connected by the two red star symbols, $G_1$ and $G_2$, denotes the current crack front line since a crack-tip in two dimensions becomes a crack front line in three dimensions. 

In the first possible crack pattern in three-dimension (see Figure.\ref{fig35: Gc_compute_CSTet}(b)), a fracture opening displacement is formed along the crack front line, whose tensile stretches at two edges, $G_1$ and $G_2$, are $\bm{\delta}_1$ and $\bm{\delta}_2$, respectively. They are marked by two blue arrows and computed as following equation:
\begin{eqnarray}
\bm{\delta} = \left( \bm{u}_{N_i} - \bm{u}_{N_j} \right) \cdot \mathcal{H}\left(\frac{\left\| \bm{x}_{N_j} - \bm{x}_{N_i} \right\|}{\left\| \bm{X}_{N_j} - \bm{X}_{N_i} \right\|} - 1 \right)
\label{eq:stretch-delta}
\end{eqnarray}
where $N_i$ and $N_j$ denote node ID at the two sides of the edge, $\bm{x}$ and $\bm{X}$ denote deformed and undeformed configuration of nodes, $\bm{u}$ denotes displacement of nodes, $\mathcal{H}$ is the Heaviside function, $\left\|\cdot \right\|$ is Euclidean norm of a vector.
Meanwhile, the maximum principal stress at opposite edge formed by $G_3$ and $G_4$ are described as $\bm{\sigma}_{G_3}$ and $\bm{\sigma}_{G_4}$, which are marked by yellow arrows. The unit normal vector to the quadrilateral plane $G_1G_2G_4G_3$ is denoted by $\bm{n}$, marked by purple dashed line.  Accordingly, the projection of edge stretch $\bm{\delta}$ on normal vector $\bm{n}$ is denoted by $\hat{\bm{\delta}}$ marked by green arrow, and projection of edge maximum principal stress $\bm{\sigma}$ on unit normal vector $\bm{n}$ is denoted by $\hat{\bm{\sigma}}$ marked by green arrow. Therefore, the fracture energy release rate $\mathcal{G}_I$ computed based on the first three-dimensional Tetrahedron crack pattern is defined as follows,
\begin{eqnarray}
\mathcal{G}_I = \frac{1}{2} \left( \hat{\bm{\sigma}}_{G_3} \cdot \hat{\bm{\delta}}_1 + \hat{\bm{\sigma}}_{G_4} \cdot \hat{\bm{\delta}}_2 \right) = \frac{1}{2} \left[ \left(\bm{\sigma}_{G_3} \cdot \bm{n} \right) \cdot \left( \bm{\delta}_1 \cdot \bm{n} \right) + \left( \bm{\sigma}_{G_4} \cdot \bm{n} \right) \cdot \left( \bm{\delta}_2 \cdot \bm{n} \right) \right]
\label{eq:G-I}
\end{eqnarray}

The second possible crack pattern in three-dimension (see Figure.\ref{fig35: Gc_compute_CSTet}(c)) also exhibits the formation of a fracture opening displacement along the crack front line, whose tensile stretches at $G_1$ and $G_2$ are $\bm{\delta}_1$ and $\bm{\delta}_2$, respectively. They are marked by blue arrows and computed by Eq.(\ref{eq:stretch-delta}). The maximum principal stress at opposite edge quadrature point $G_6$ is described as $\bm{\sigma}_{G_6}$, which is marked by yellow arrow. Also, the unit normal vector to the triangle plane $G_1G_2G_6$ is denoted by $\bm{n}$, marked by dashed purple line. The projected stretch $\hat{\bm{\delta}}$ and projected maximum principal stress $\hat{\bm{\sigma}}$ on the unit normal vector $\bm{n}$ follow the same definitions and graphical illustrations as crack pattern I. Therefore, the fracture energy release rate $\mathcal{G}_{II}$ computed based on the second three-dimensional Tetrahedron crack pattern is defined as follows,
\begin{eqnarray}
\mathcal{G}_{II} = \frac{1}{2} \left( \hat{\bm{\sigma}}_{G_6} \cdot \hat{\bm{\delta}}_1 + \hat{\bm{\sigma}}_{G_6} \cdot \hat{\bm{\delta}}_2 \right) = \frac{1}{2} \left[ \left(\bm{\sigma}_{G_6} \cdot \bm{n} \right) \cdot \left( \bm{\delta}_1 \cdot \bm{n} \right) + \left( \bm{\sigma}_{G_6} \cdot \bm{n} \right) \cdot \left( \bm{\delta}_2 \cdot \bm{n} \right) \right]
\label{eq:G-II}
\end{eqnarray}
when one of $\mathcal{G}_{I}$ and $\mathcal{G}_{II}$ exceeds the critical fracture energy release rate $\mathcal{G}_c$, the new fractured surface follows the corresponding three-dimensional crack pattern. 

\begin{figure}[htp]
	\centering
        \begin{minipage}{0.3\linewidth}
            \begin{center}
            \includegraphics[height=1.2in]{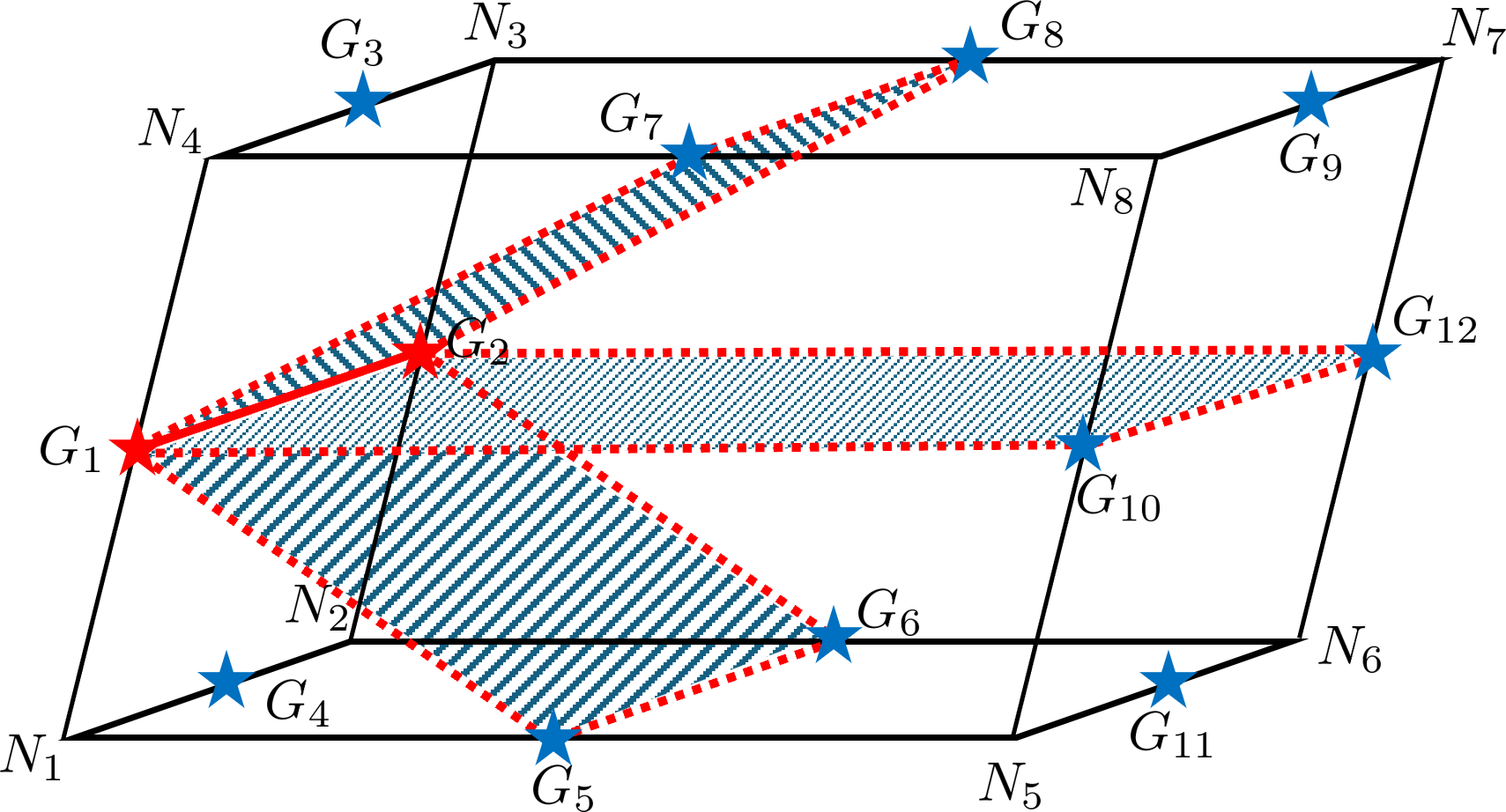}
            \end{center}
            \begin{center}
            (a)
            \end{center}
        \end{minipage}
        \hfill
        \begin{minipage}{0.3\linewidth}
            \begin{center}
            \includegraphics[height=1.2in]{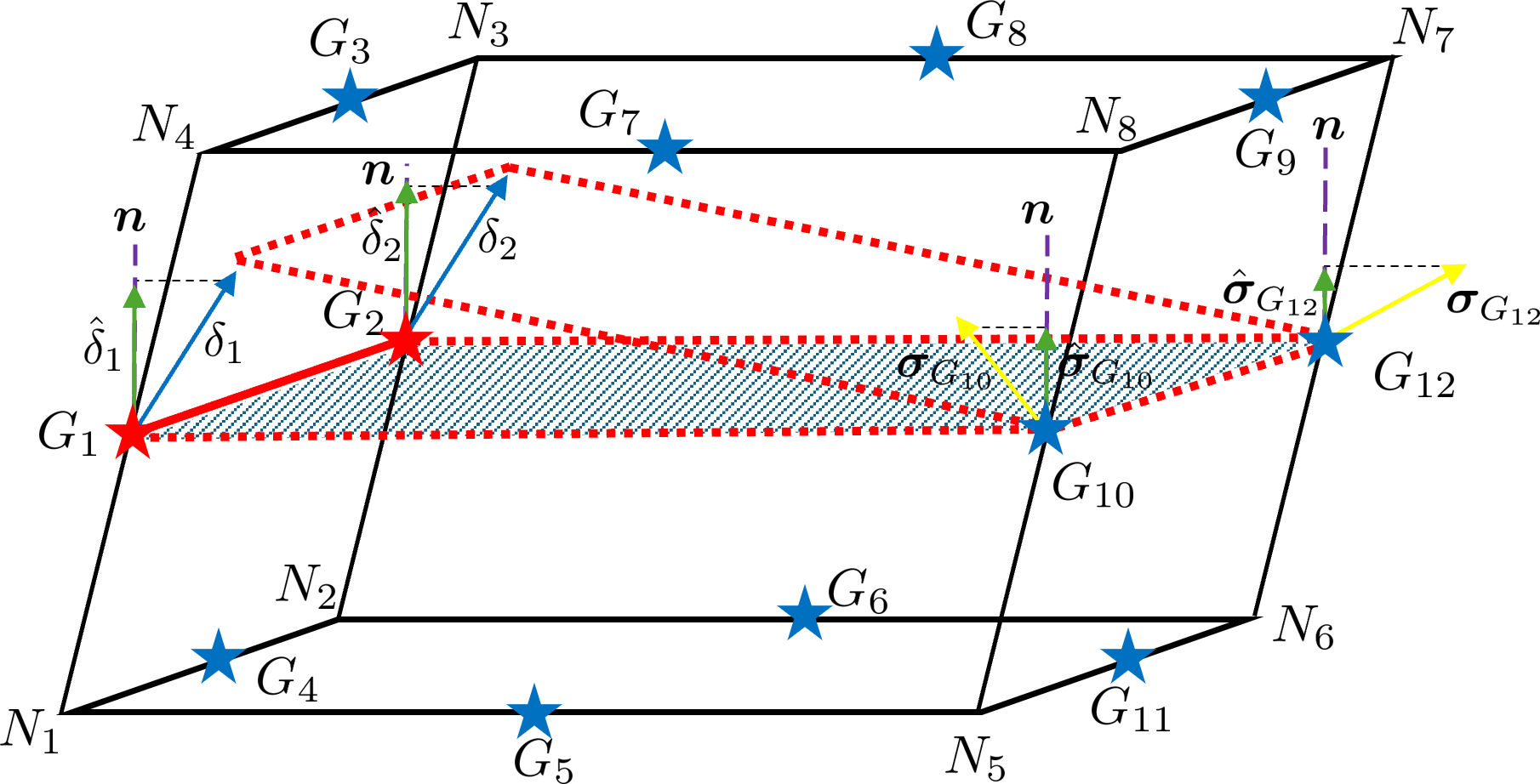}
            \end{center}
            \begin{center}
            (b)
            \end{center}
        \end{minipage}   
        \hfill
        \begin{minipage}{0.3\linewidth}
            \begin{center}
            \includegraphics[height=1.2in]{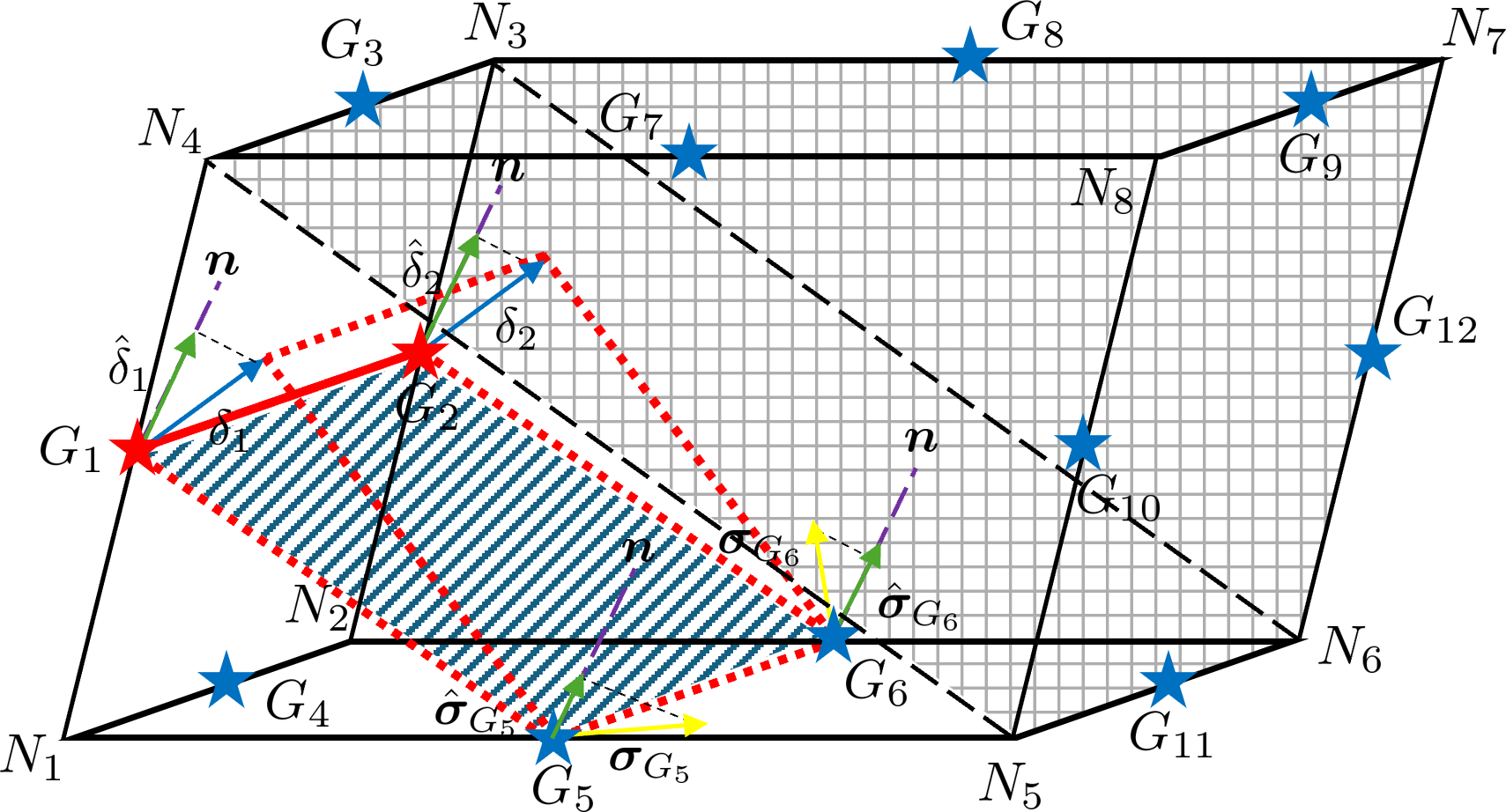}
            \end{center}
            \begin{center}
            (c)
            \end{center}
        \end{minipage}
        \hfill
        \begin{minipage}{0.3\linewidth}
            \begin{center}
            \includegraphics[height=1.2in]{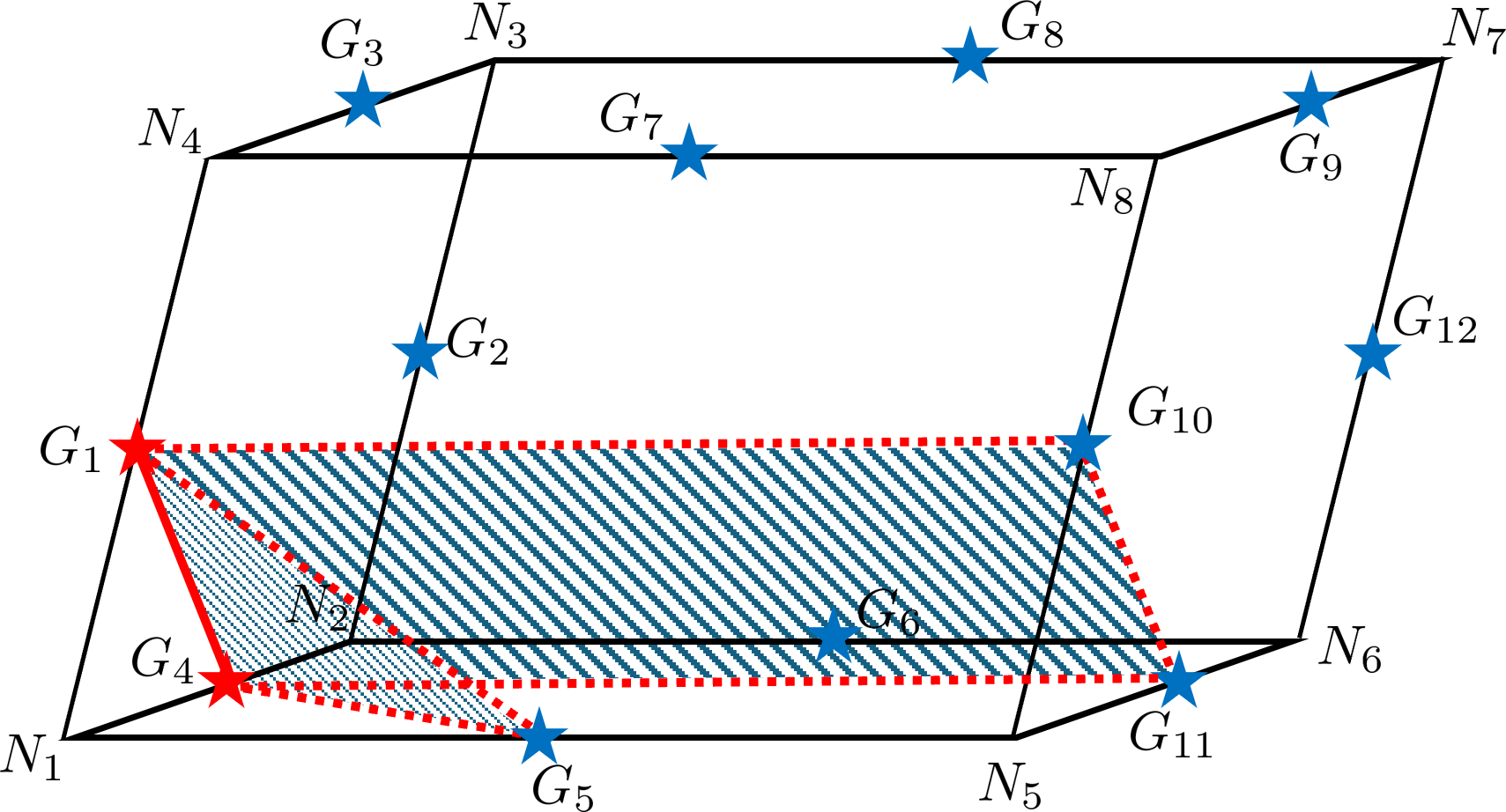}
            \end{center}
            \begin{center}
            (d)
            \end{center}
        \end{minipage}
        \hfill
        \begin{minipage}{0.3\linewidth}
            \begin{center}
            \includegraphics[height=1.25in]{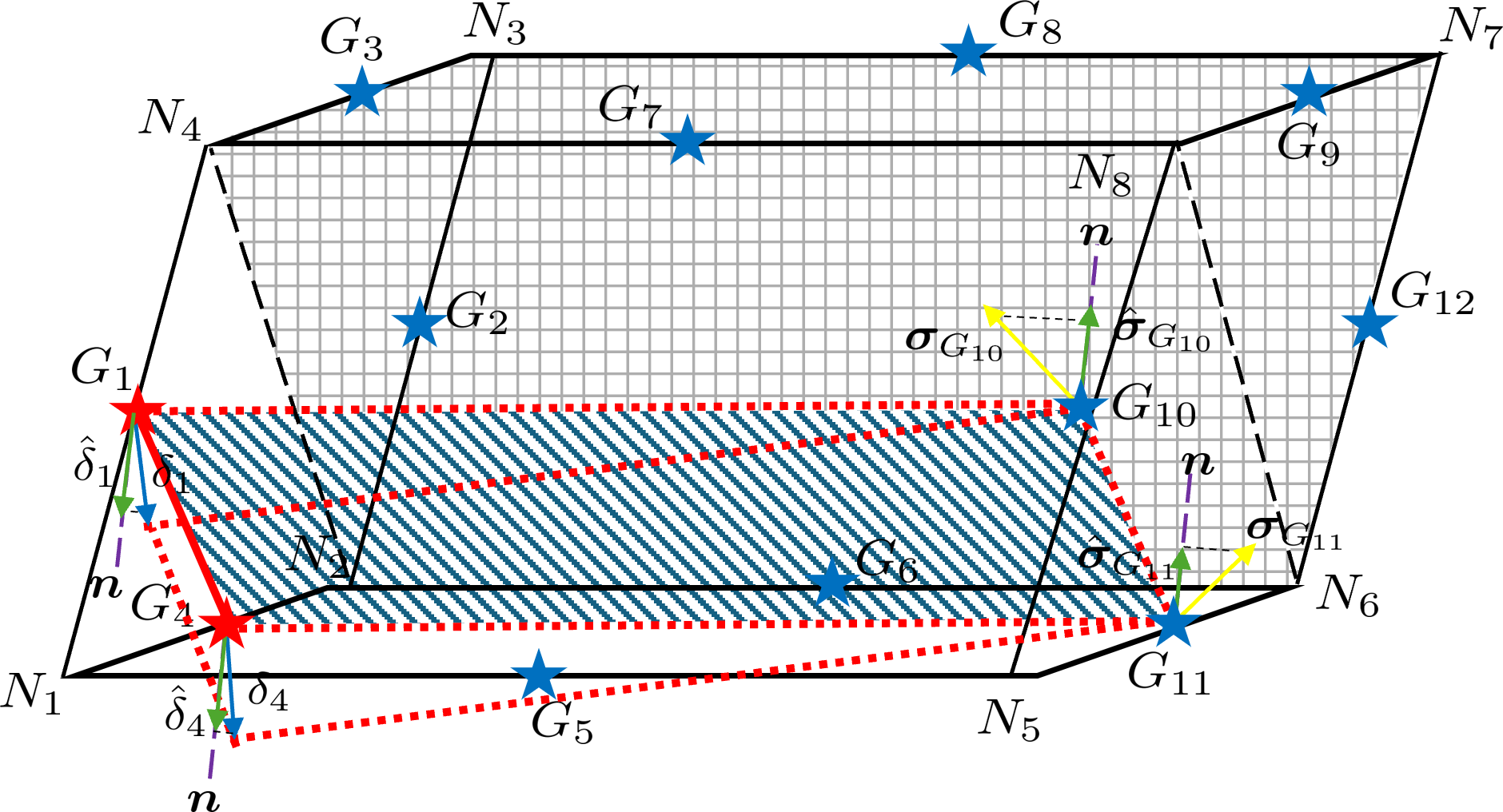}
            \end{center}
            \begin{center}
            (e)
            \end{center}
        \end{minipage}   
        \hfill
        \begin{minipage}{0.3\linewidth}
            \begin{center}
            \includegraphics[height=1.2in]{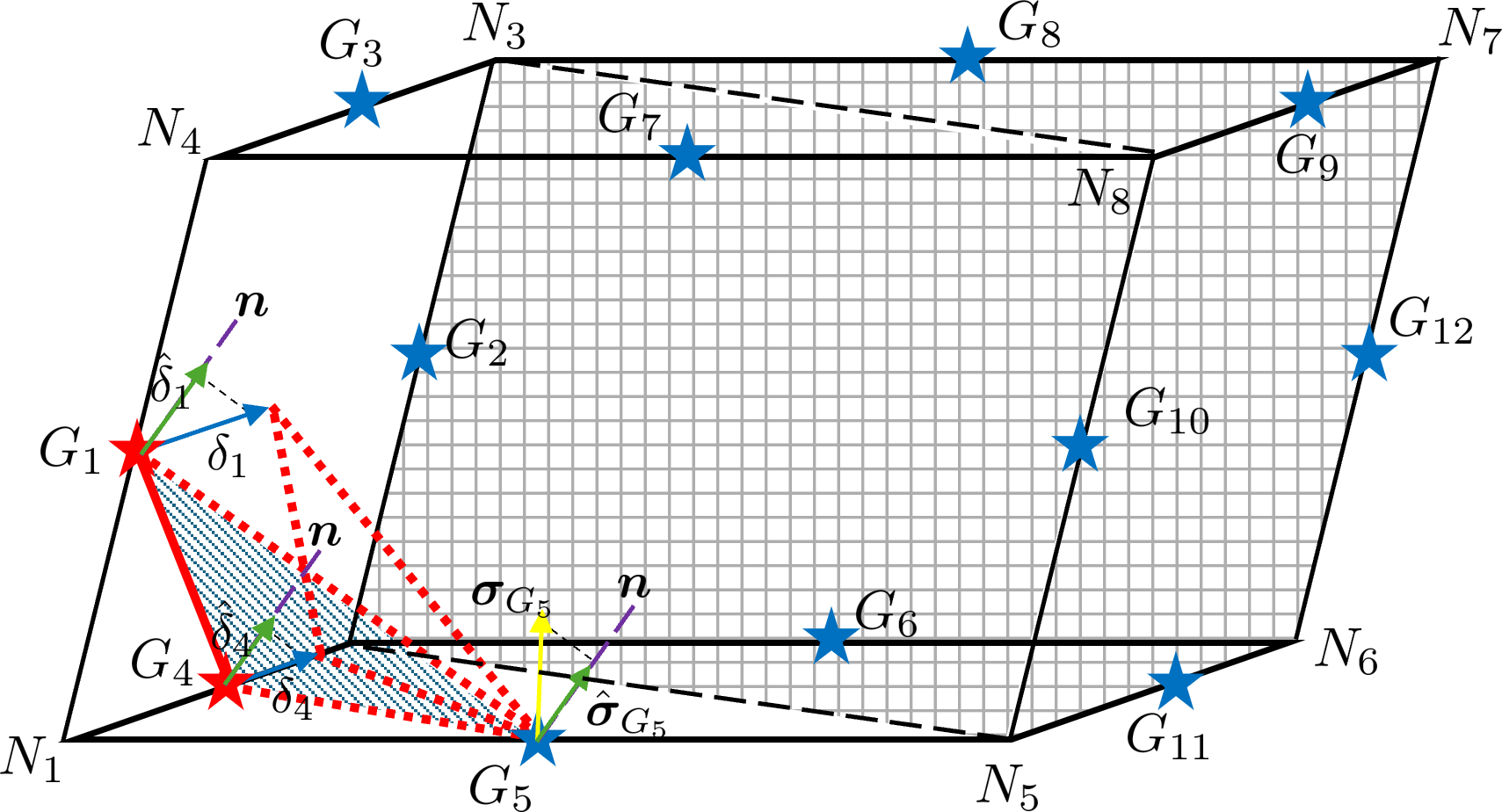}
            \end{center}
            \begin{center}
            (f)
            \end{center}
        \end{minipage}
        \caption{Three-dimensional crack types of trilinear Parallelepiped element, where $N_i$ denotes node ID, $G_j$ denotes edge quadrature point ID, $\delta_i$ denotes stretch of an edge, $\bm{n}$ denotes the normal vector of crack surface, $\hat{\bm{\delta}}_i = \left(\bm{\delta} \cdot \bm{n} \right) \bm{n}$ denotes the projection of $\delta_i$ on normal vector $\bm{n}$, $\bm{\sigma}_{G}$ denotes the maximum principal stress at edge quadrature $G$, $\hat{\bm{\sigma}}_G = \left(\bm{\sigma}_G \cdot \bm{n} \right)\bm{n}$ denotes the projection of $\bm{\sigma}_G$ on normal vector $\bm{n}$, blue grids demonstrate possible crack surface and grey grids demonstrate the remaining active portion after element fracturing: (a). possible three-dimensional crack patterns based on opposite quadrature points; (b). three-dimensional crack pattern $III$ and corresponding computation of fracture energy release rate $\mathcal{G}_{III}$; (c). three-dimensional crack pattern $IV$ and corresponding computation of fracture energy release rate $\mathcal{G}_{IV}$; (d). possible three-dimensional crack patterns based on neighboring quadrature points; (e). three-dimensional crack pattern $V$ and corresponding computation of fracture energy release rate $\mathcal{G}_{V}$; (f). three-dimensional crack pattern $VI$ and corresponding computation of fracture energy release rate $\mathcal{G}_{VI}$.}
        \label{fig36: Gc_compute_Hex}
\end{figure}

Contrary to the definitive coplanarity of fracture opening surfaces in Constant Strain Tetrahedron element, the coplanarity of fracture surfaces in trilinear Hexahedron element is NOT guaranteed. To be comprehensive, we consider a special Hexahedron in three-dimension, i.e., Parallelepiped, as shown in Figure.\ref{fig36: Gc_compute_Hex}. 

In a Parallelepiped element, the three-dimensional crack patterns can be categorized into four types, of which two types are formed by a crack front line with two quadrature points at opposite edges (see Figure.\ref{fig36: Gc_compute_Hex}(a-c)). The other two types are formed by a crack front line with two neighboring quadrature points (see Figure.\ref{fig36: Gc_compute_Hex}(d-f)). The definitions of notations and symbols in Figure.\ref{fig36: Gc_compute_Hex} follow Constant Strain Tetrahedron element in Figure.\ref{fig35: Gc_compute_CSTet}. The computational methods of $\mathcal{G}_{III}$, $\mathcal{G}_{IV}$ and $\mathcal{G}_{V}$ follow $\mathcal{G}_I$, i.e., Eq.(\ref{eq:G-I}) while the computation method of $\mathcal{G}_{VI}$ follows $\mathcal{G}_{II}$, i.e., Eq.(\ref{eq:G-II}). 

In addition, it is important to note that the remaining portion of the fractured parallelepiped element needs special handling. In Figure.\ref{fig36: Gc_compute_Hex}(c)(e)(f), the remaining portions (grey grids) continue participating after computation but their roles become that of a regular triangular prism, which can be divided into three tetrahedron elements. However, in Figure.\ref{fig36: Gc_compute_Hex}(b), the fractured Parallelepiped element is considered fully cracked and is deactivated in subsequent computations.

\section{Representative Numerical Examples}
Five classic three-dimensional transient dynamic crack propagation examples are considered in this section to demonstrate the effectiveness and efficiency of the proposed CEM. The first three examples, including the Kalthoff-Winkler experiment, compact compression test and pull-out anchorage example, show the effectiveness of the proposed CEM in the prediction of three-dimensional single crack propagation. The proposed three-dimensional CEM successfully captures crack branching in the final two examples, significantly exceeding our initial expectations, as it was previously applied only to single-crack propagation in three dimensions. It is significant to emphasize that the appearance of "branching" in three-dimensional solids is completely spontaneous without any locally/globally defined criteria of crack banching. To ensure consistency and efficiency, all benchmark examples are conducted using the Tetrahedron element formulation and the aid of GPU-acceleration.

\subsection{Dynamic shear loading experiment}
\begin{figure}[htp]
        \centering
        \begin{minipage}{0.45\linewidth}
            \begin{center}
            \includegraphics[height=3.0in]{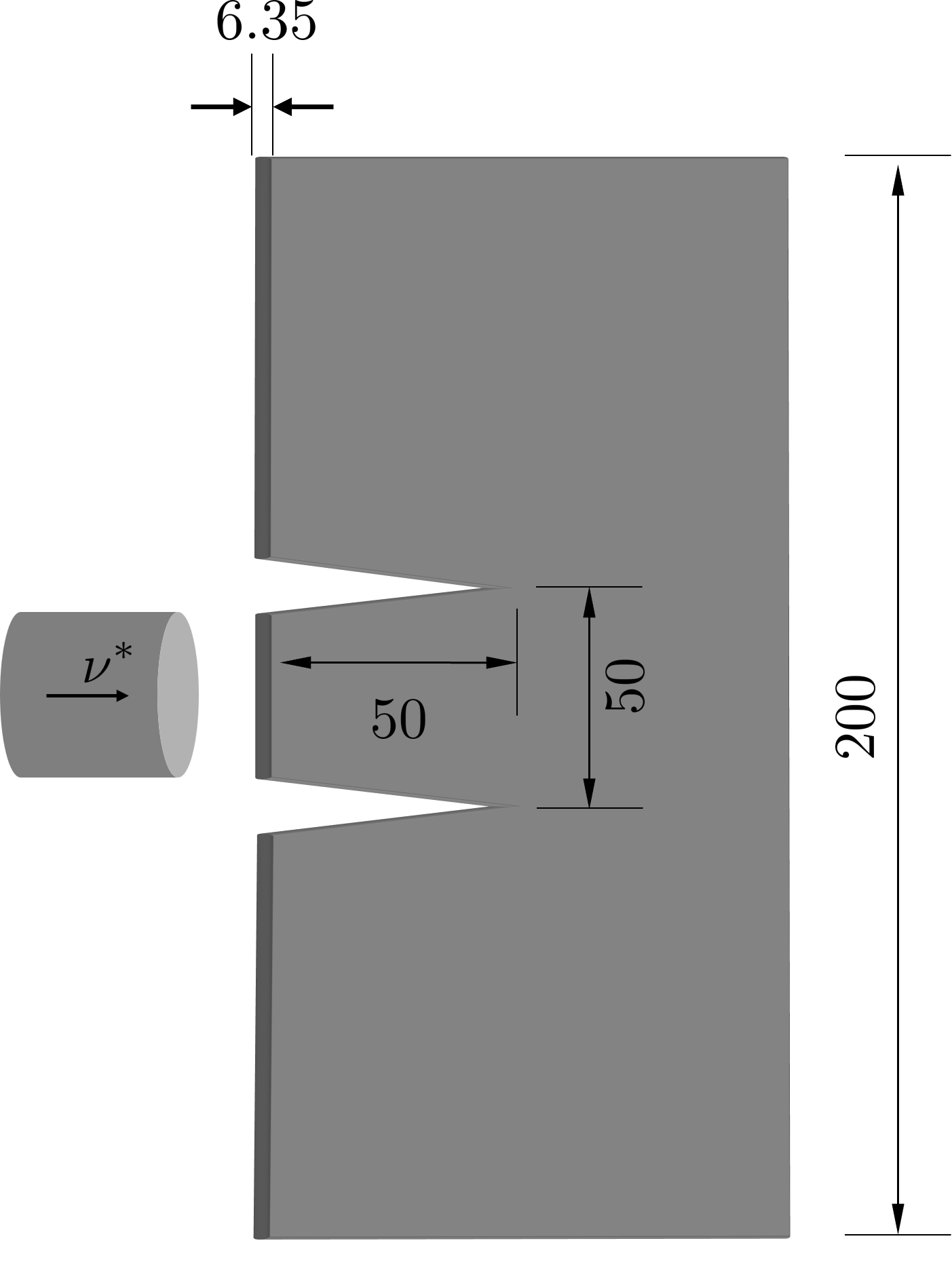}
            \end{center}
            \begin{center}
            (a)
            \end{center}
        \end{minipage}
        \begin{minipage}{0.45\linewidth}
            \begin{center}
            \includegraphics[height=2.4in]{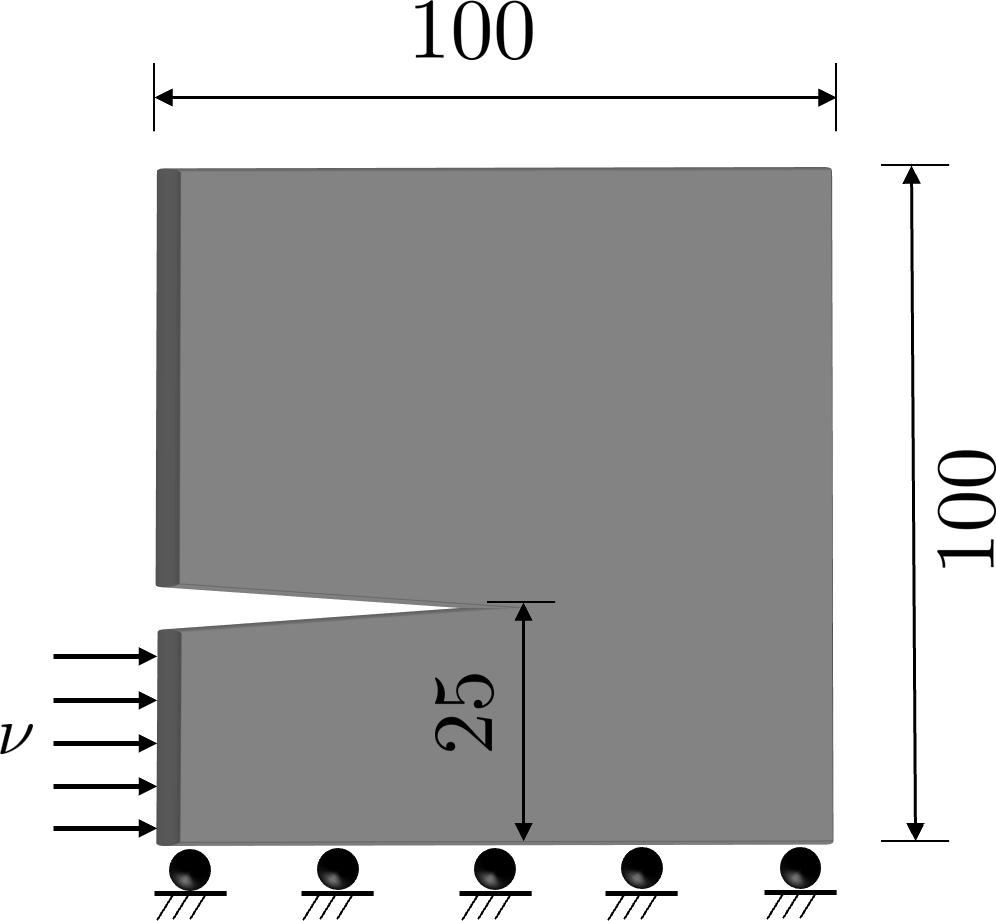}
            \end{center}
            \begin{center}
            (b)
            \end{center}
        \end{minipage}   
        \caption{(a). The whole geometry (unit: mm), boundary condition of three-dimensional Kalthoff-Winkler plate experiment; (b). Upper half of the three-dimensional Kalthoff-Winkler plate.}
        \label{fig1: kalthoff-plate-3D}
\end{figure}

In this three-dimensional benchmark example, we studied the transient-dynamic crack propagation induced by a projectile impact in an edge-notched steel plate. The experiment is completed by Kalthoff and Winkler (see \cite{kalthoff1988failure}, \cite{kalthoff2000modes}), known as the Kalthoff-Winkler experiment. The geometric sizes in three dimensions are illustrated in Figure.\ref{fig1: kalthoff-plate-3D}(a). It is recorded that the high-strength steel plate responds from brittlely to ductilely with escalating impact velocity. The single crack propagates brittlely in about $70^\circ$ from the existing notch if the impact velocity of the projectile is up to $33\ m/s$. The numerical simulation of the Kalthoff-Winkler plate has been extensively studied, including works by \cite{borden2012phase}, \cite{bui2022numerical}, \cite{rabczuk2007simplified}, \cite{song2009cracking}, \cite{park2012adaptive}, \cite{hirmand2019block}, \cite{geelen2019phase}. However, there are few studies on the Kalthoff-Winkler plate in three dimensions in academia and industry due to  the complexity of physical phenomena and more demanding numerical challenges. Therefore, the proposed CEM exhibits obvious advantages, including strong extensibility from 2D to 3D, a valid methodology, and high computational efficiency. 

The material properties utilized in simulation are listed below: Young's modulus $E=190\ GPa$, Poisson ratio $\nu = 0.3$, critical fracture energy release rate $G_f = 2.213 \times 10^4\ J/m^2$ and density $\rho = 8000\ kg/m^3$. Accordingly, the resulting Rayleigh wave velocity is $v_R = 2803\ m/s$. The Dirichlet and Neumann boundary conditions are: external impact is applied horizontally at the left center part with specified velocity and the bottom boundary of the upper half plate is constrained vertically, as shown in Figure.\ref{fig1: kalthoff-plate-3D}(b) due to model symmetry. The applied velocity of the projectile is firstly ramping and then stays constant, as shown in the equation below 
\begin{equation}
v =
\begin{cases} 
\frac{t}{t_0}v_0,  & \text{if } \ t \le t_0, \\
v_0, & \text{if } \ t > t_0.
\end{cases}
\end{equation}
in which, $v_0 = 16.5\ m/s$ and $t_0 = 1\ \mu s$.

\begin{figure}[htp]
        \centering
        \begin{minipage}{0.3\linewidth}
            \begin{center}
            \includegraphics[height=2.0in]{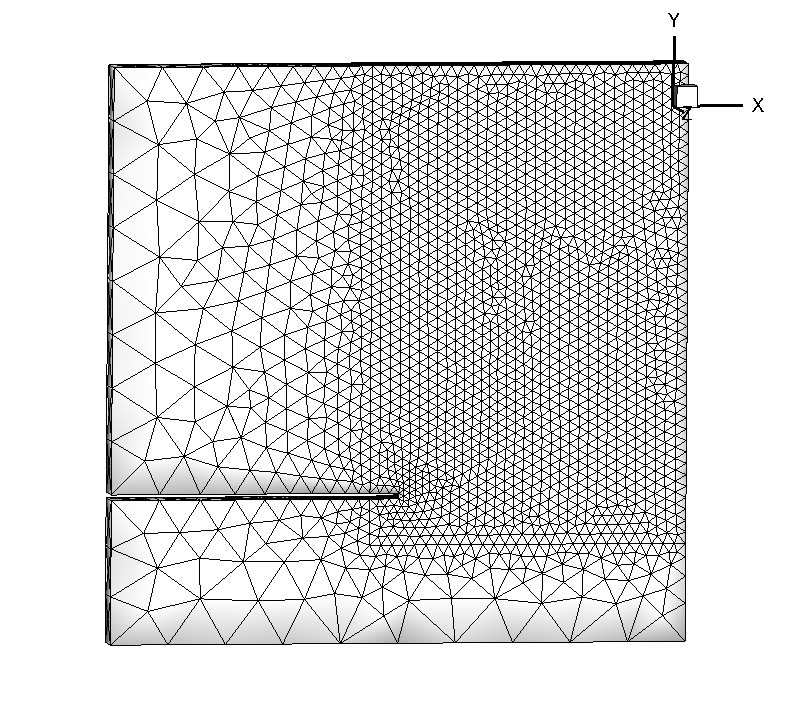}
            \end{center}
            \begin{center}
            (a)
            \end{center}
        \end{minipage}
        \begin{minipage}{0.3\linewidth}
            \begin{center}
            \includegraphics[height=2.0in]{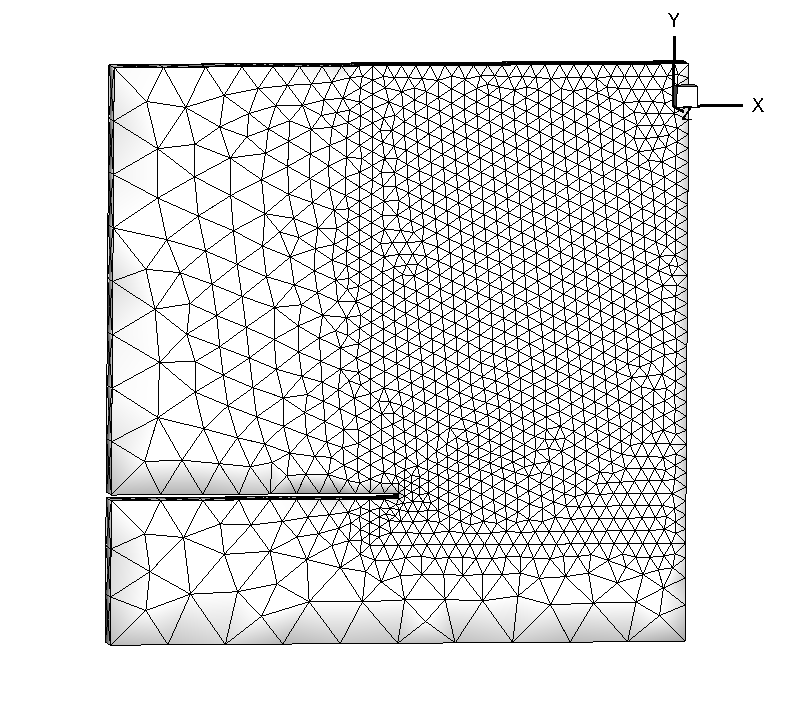}
            \end{center}
            \begin{center}
            (b)
            \end{center}
        \end{minipage}   
        \begin{minipage}{0.3\linewidth}
            \begin{center}
            \includegraphics[height=2.0in]{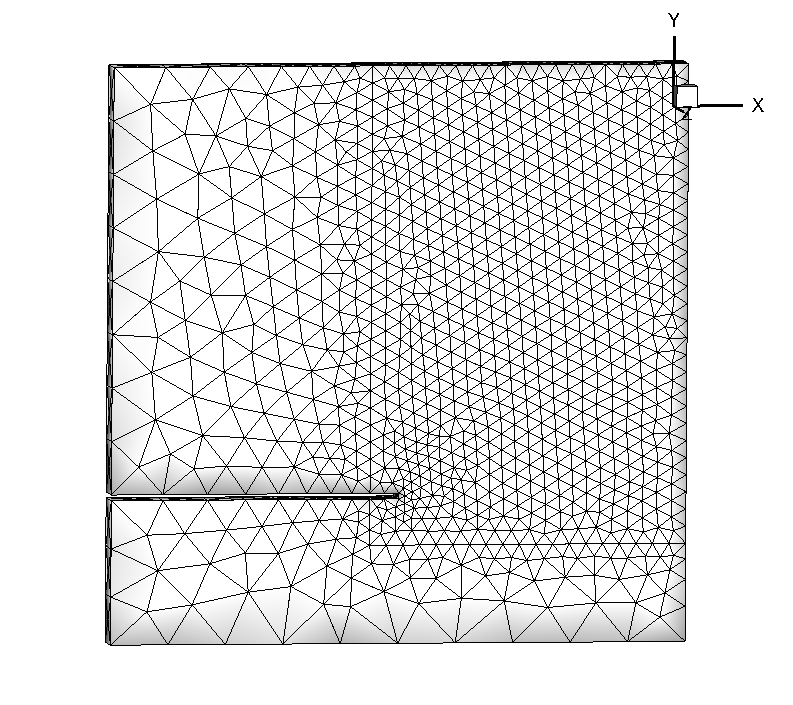}
            \end{center}
            \begin{center}
            (c)
            \end{center}
        \end{minipage}
        \caption{Three representative tetrahedron meshes are illustrated: (a). the finest mesh with $24137$ elements; (b). the medium mesh with $12294$ elements; (c). the coarsest mesh with $8396$ elements.}
        \label{fig2: kalthoff-meshes-3D}
\end{figure}

The three different tetrahedron meshes are considered: the first one is the finest mesh with $6051$ nodes and $24137$ elements, the second one is a medium mesh with $3494$ nodes and $12294$ elements, and the third one is the coarsest tetrahedron mesh with $2480$ nodes and $8396$ elements, as shown in Figure.\ref{fig2: kalthoff-meshes-3D}.

\begin{figure}[htp]
	\centering
        \begin{minipage}{0.3\linewidth}
            \begin{center}
            \includegraphics[height=2.0in]{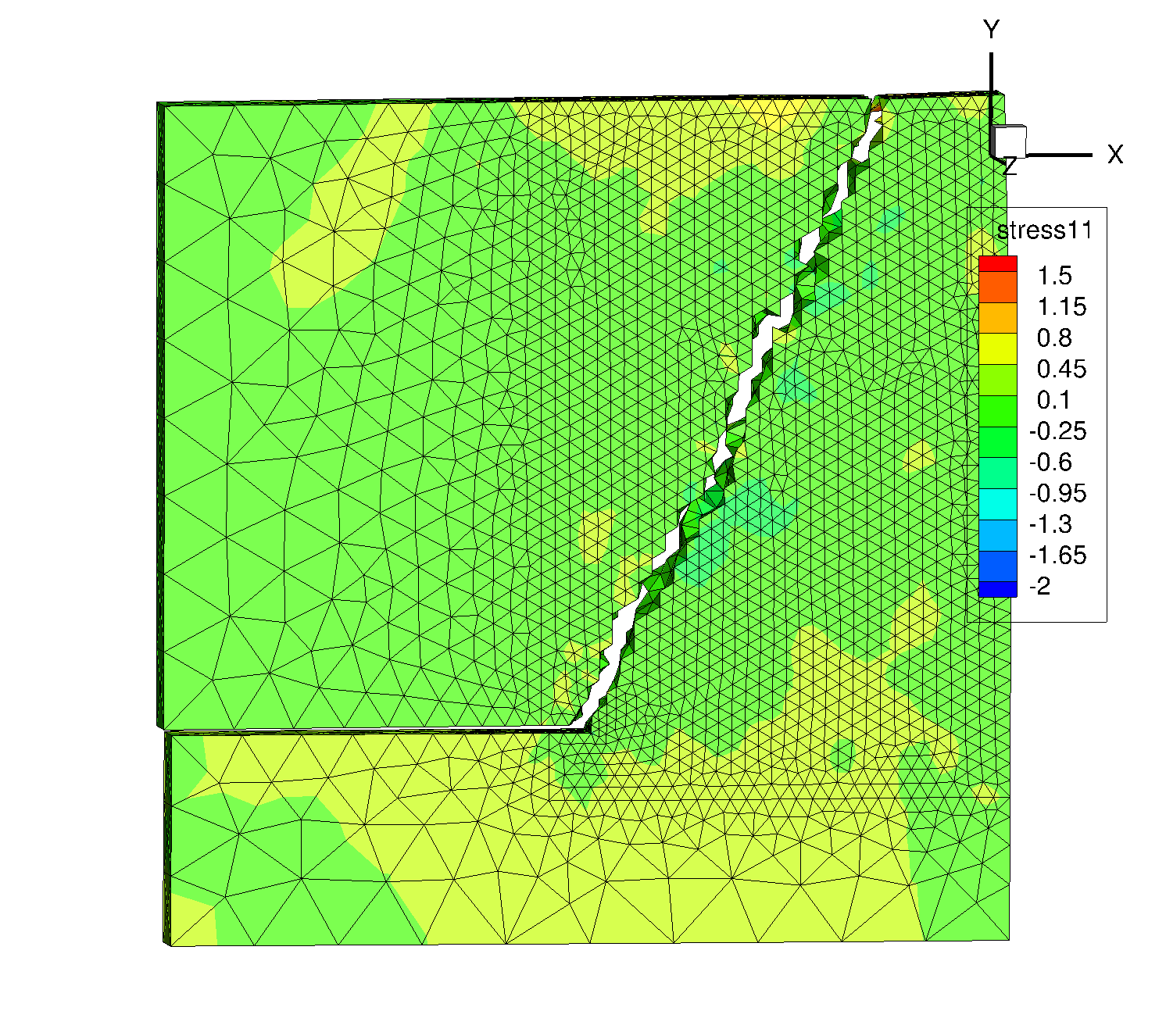}
            \end{center}
            \begin{center}
            (a)
            \end{center}
        \end{minipage}
        \hfill
        \begin{minipage}{0.3\linewidth}
            \begin{center}
            \includegraphics[height=2.0in]{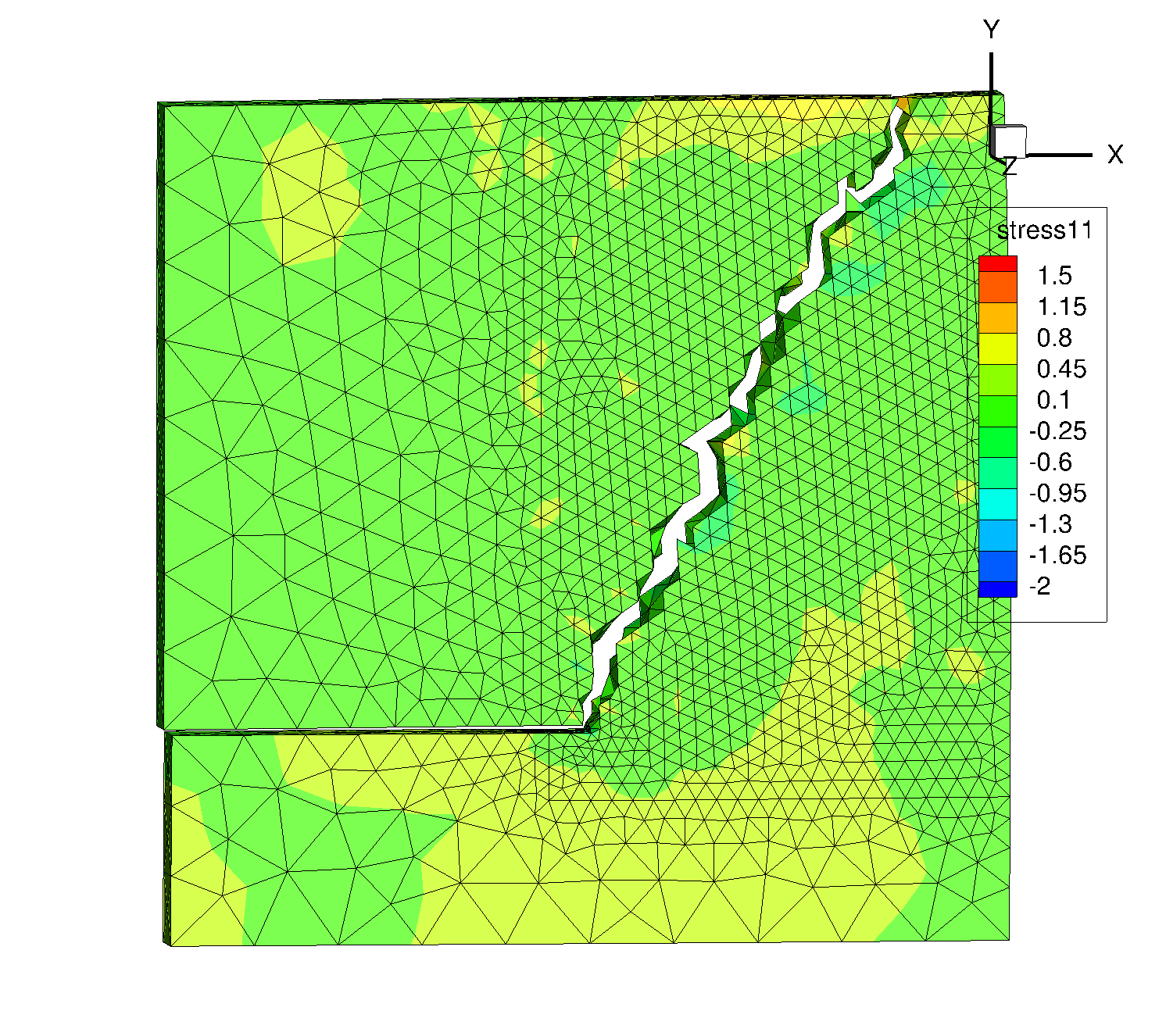}
            \end{center}
            \begin{center}
            (b)
            \end{center}
        \end{minipage}   
        \hfill
        \begin{minipage}{0.3\linewidth}
            \begin{center}
            \includegraphics[height=2.0in]{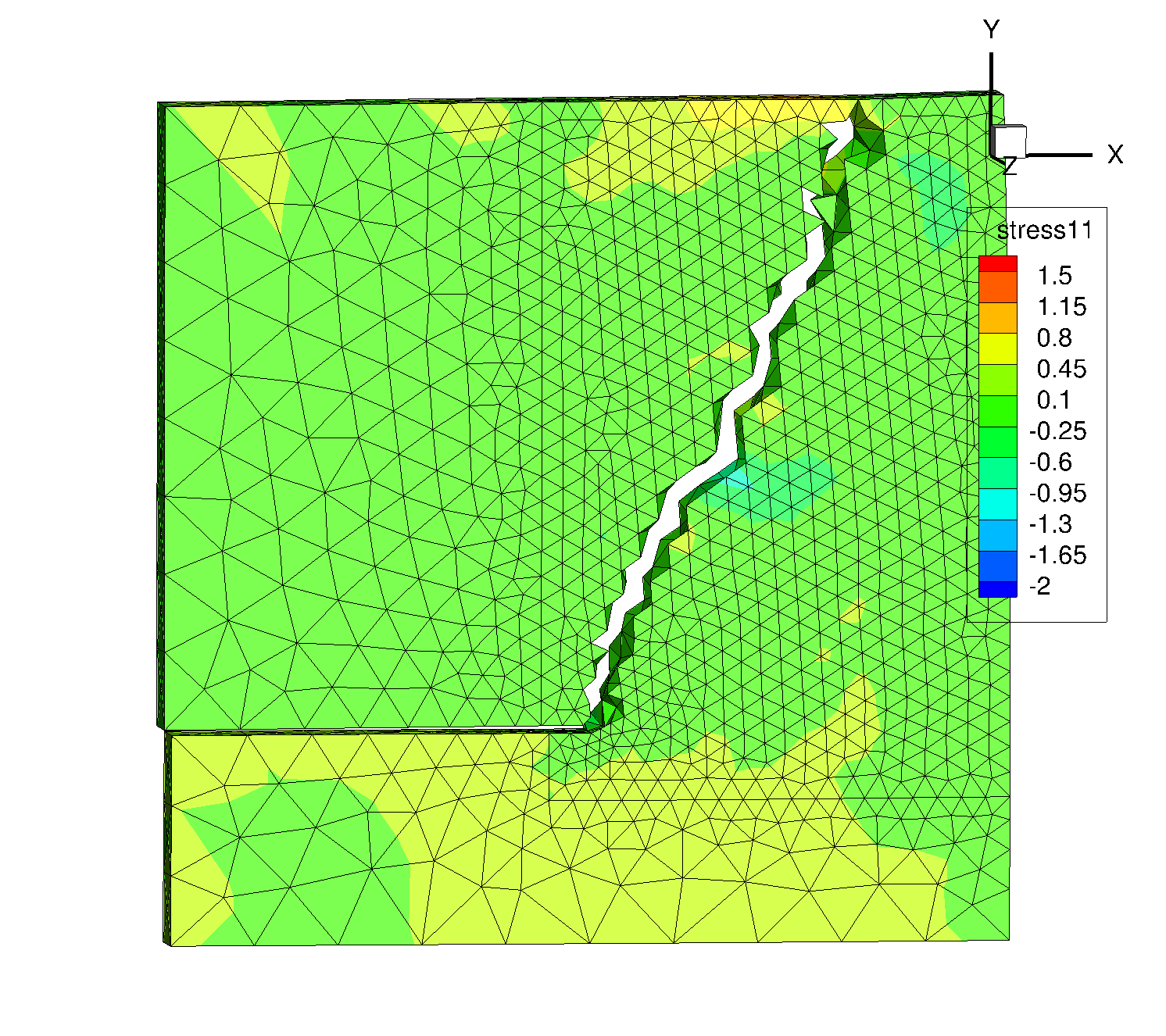}
            \end{center}
            \begin{center}
            (c)
            \end{center}
        \end{minipage}
        \caption{The final crack patterns of the three representative tetrahedron meshes are illustrated: (a). the fine mesh; (b). the medium mesh; (c). the coarse mesh.}
        \label{fig3: Kalthoff-crack-patterns-3D}
\end{figure}

\begin{figure}[htp]
	\centering
        \begin{minipage}{0.24\linewidth}
            \begin{center}
            \includegraphics[height=1.5in]{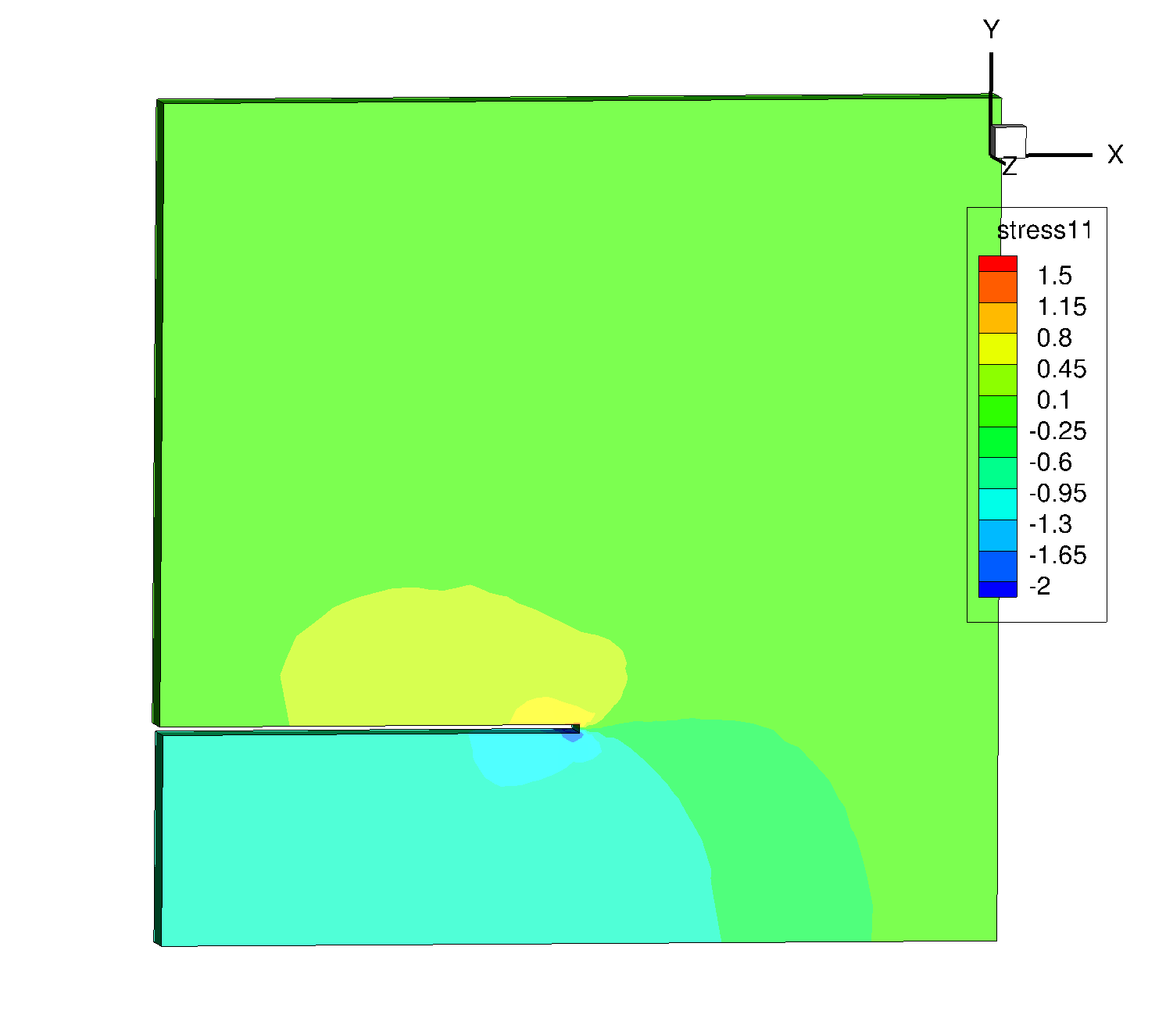}
            (a)
            \end{center}
        \end{minipage}
        \hfill
        \begin{minipage}{0.24\linewidth}
            \begin{center}
            \includegraphics[height=1.5in]{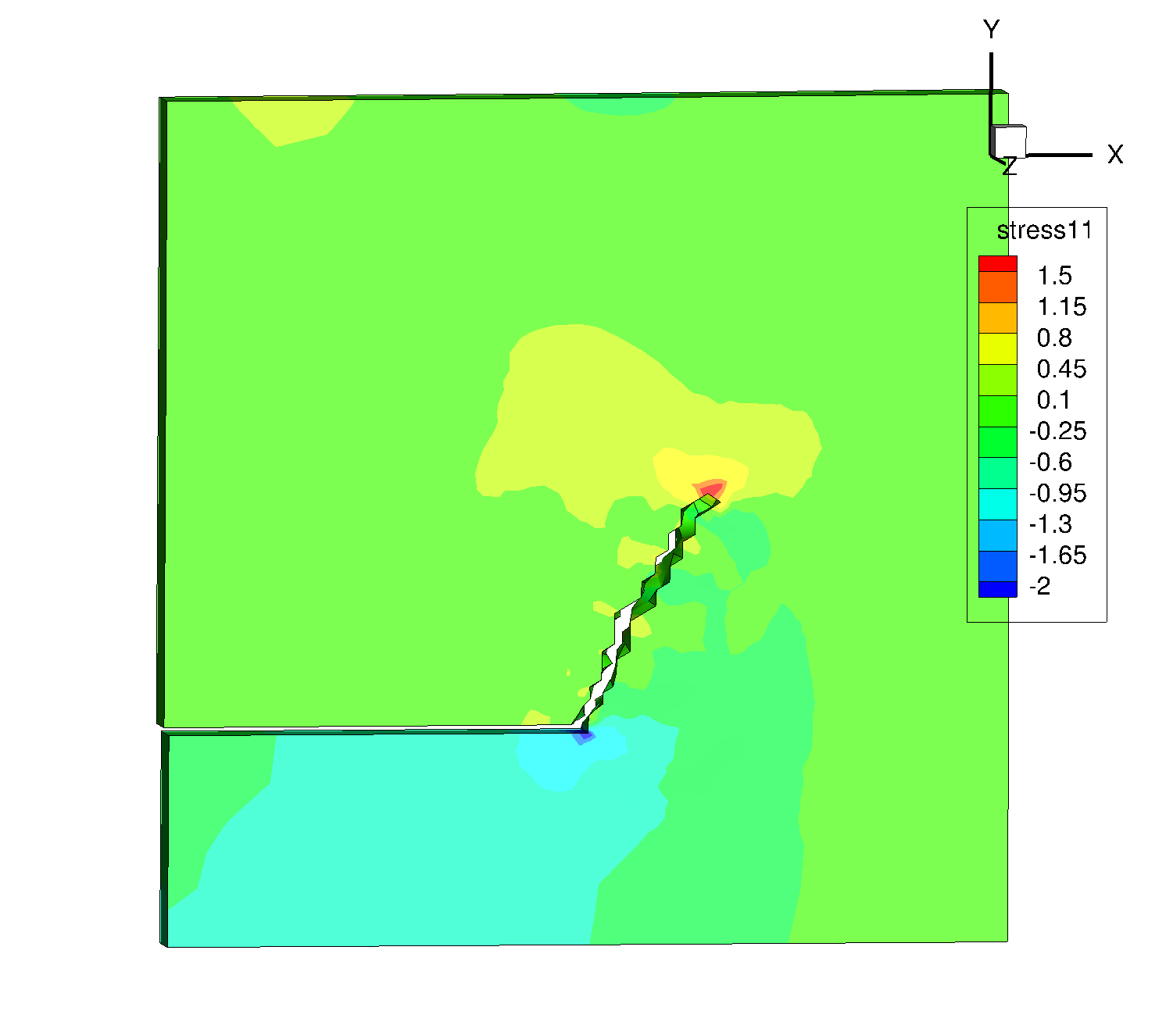}
            (b)
            \end{center}
        \end{minipage}   
        \hfill
        \begin{minipage}{0.24\linewidth}
            \begin{center}
            \includegraphics[height=1.5in]{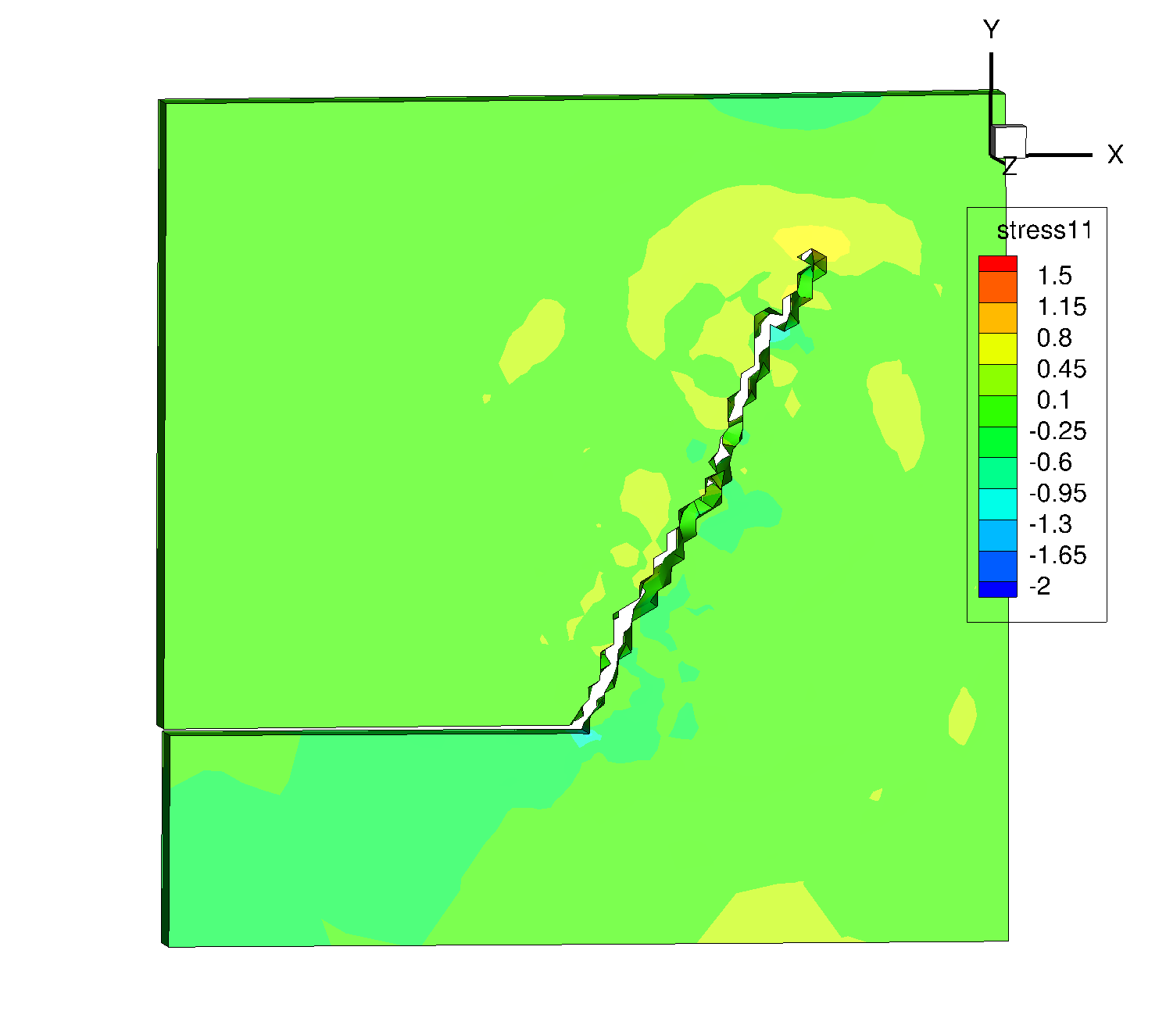}
            (c)
            \end{center}
        \end{minipage}
        \hfill
        \begin{minipage}{0.24\linewidth}
            \begin{center}
            \includegraphics[height=1.5in]{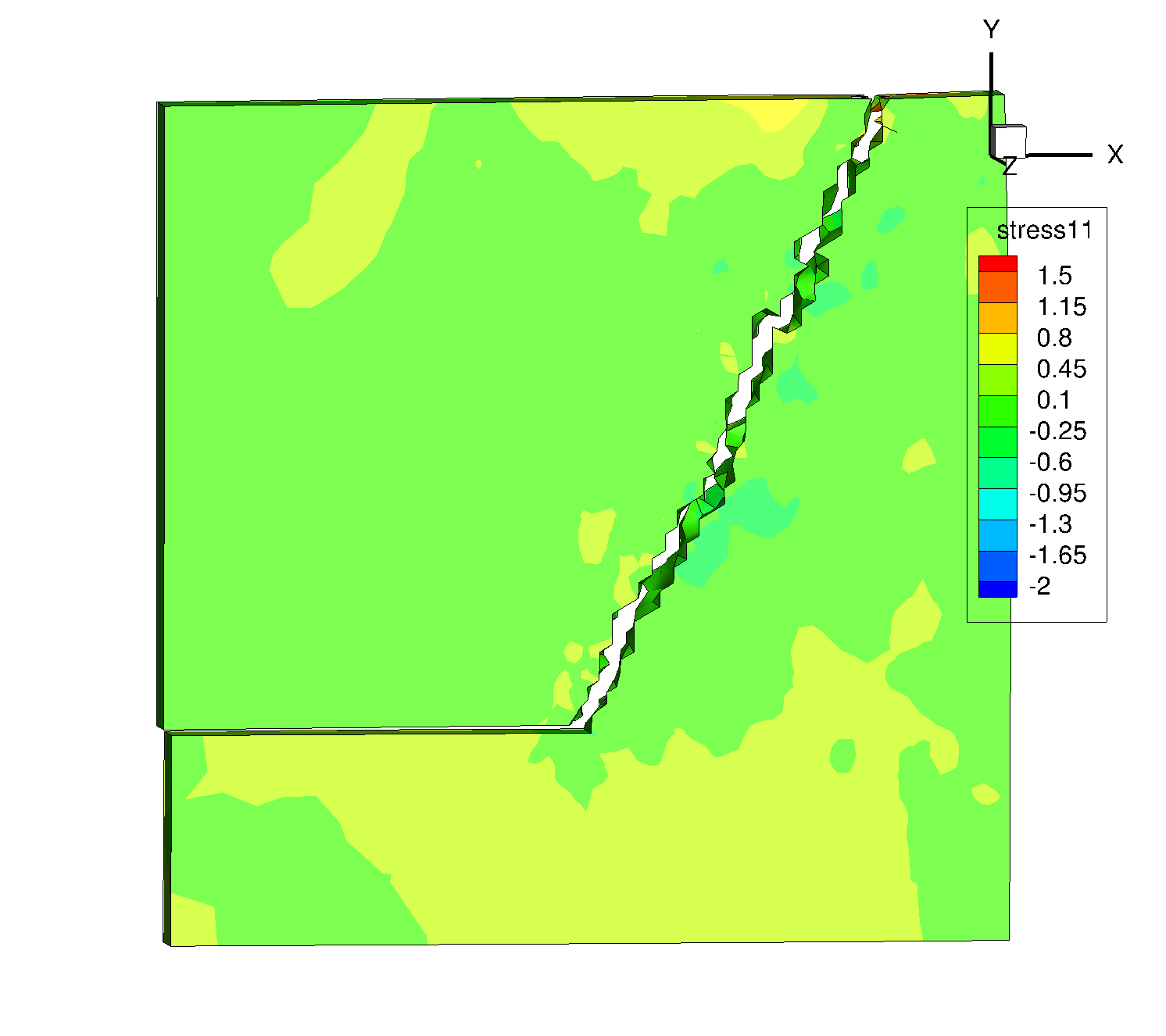}
            (d)
            \end{center}
        \end{minipage}
        \caption{Stress $\sigma_{xx}$ contour and crack patterns evolution for three-dimensional Kalthoff-Winkler plate fine mesh at times at (a). $t=23.49\ \mu s$, (b). $t=45\ \mu s$, (c). $t=64\ \mu s$, (d). $t=90\ \mu s$.}
        \label{fig4: Kalthoff-crack-evolution-1}
\end{figure}

\begin{figure}[htp]
	\centering
        \begin{minipage}{0.24\linewidth}
            \begin{center}
            \includegraphics[height=1.5in]{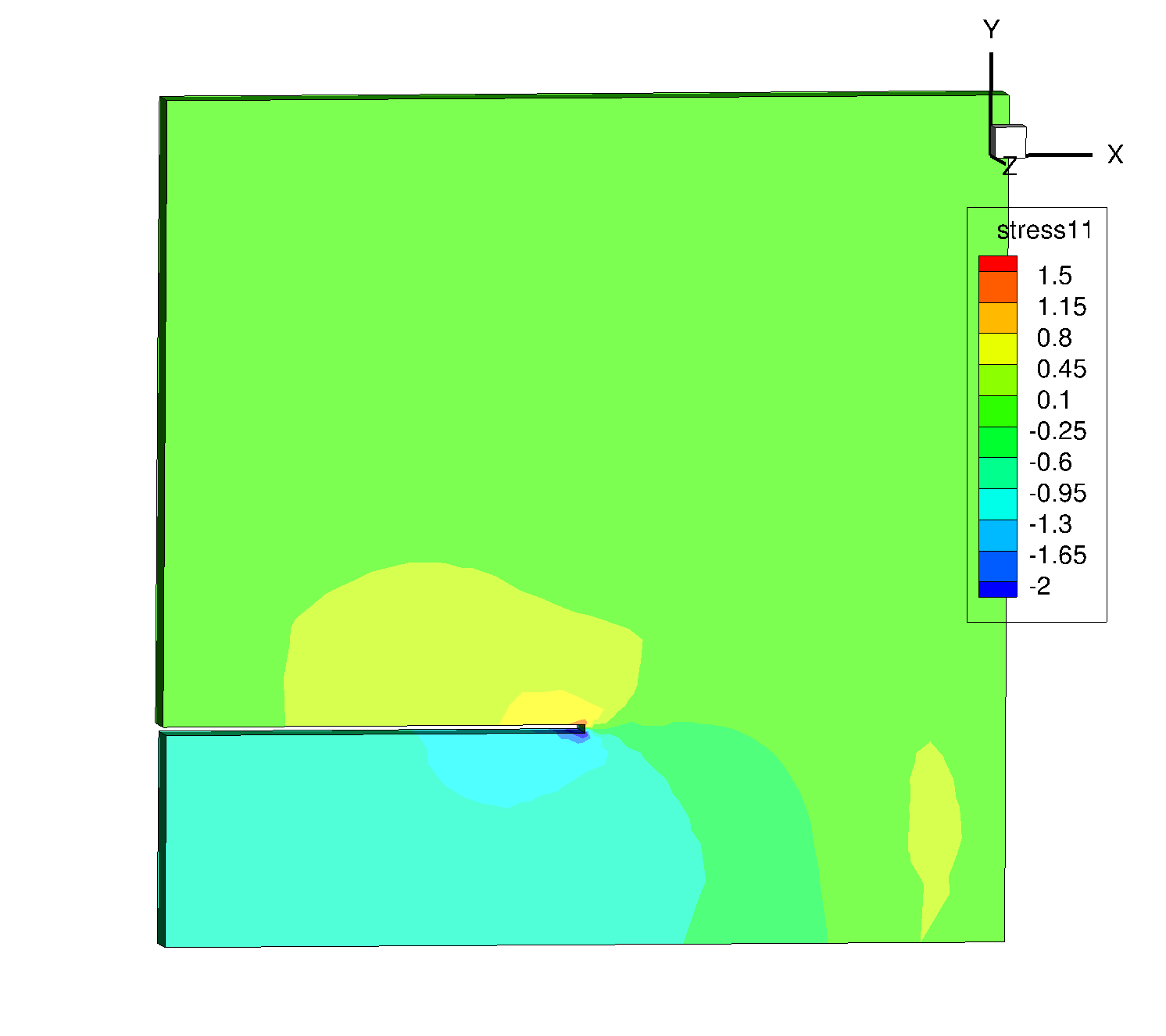}
            (a)
            \end{center}
        \end{minipage}
        \hfill
        \begin{minipage}{0.24\linewidth}
            \begin{center}
            \includegraphics[height=1.5in]{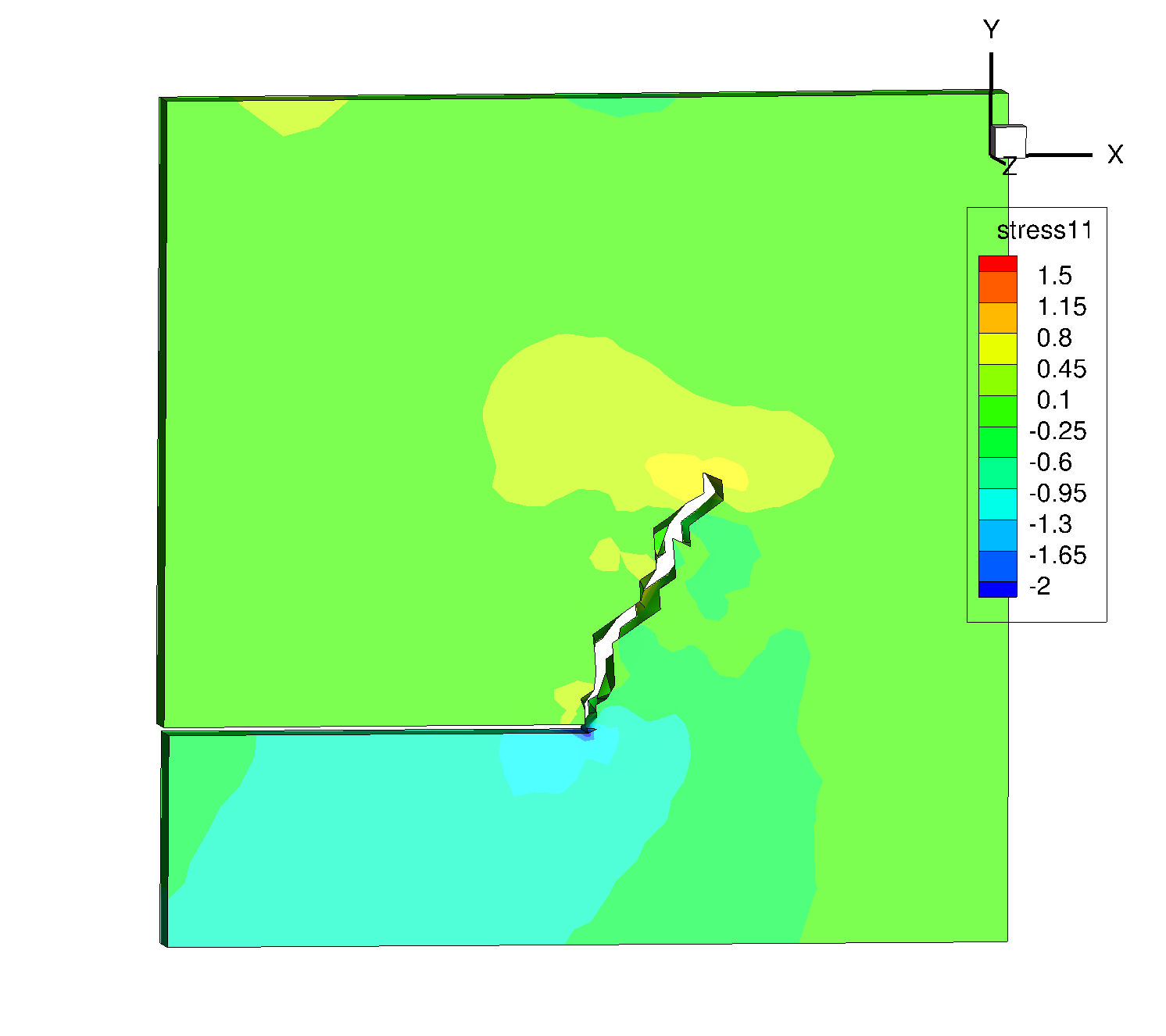}
            (b)
            \end{center}
        \end{minipage}   
        \hfill
        \begin{minipage}{0.24\linewidth}
            \begin{center}
            \includegraphics[height=1.5in]{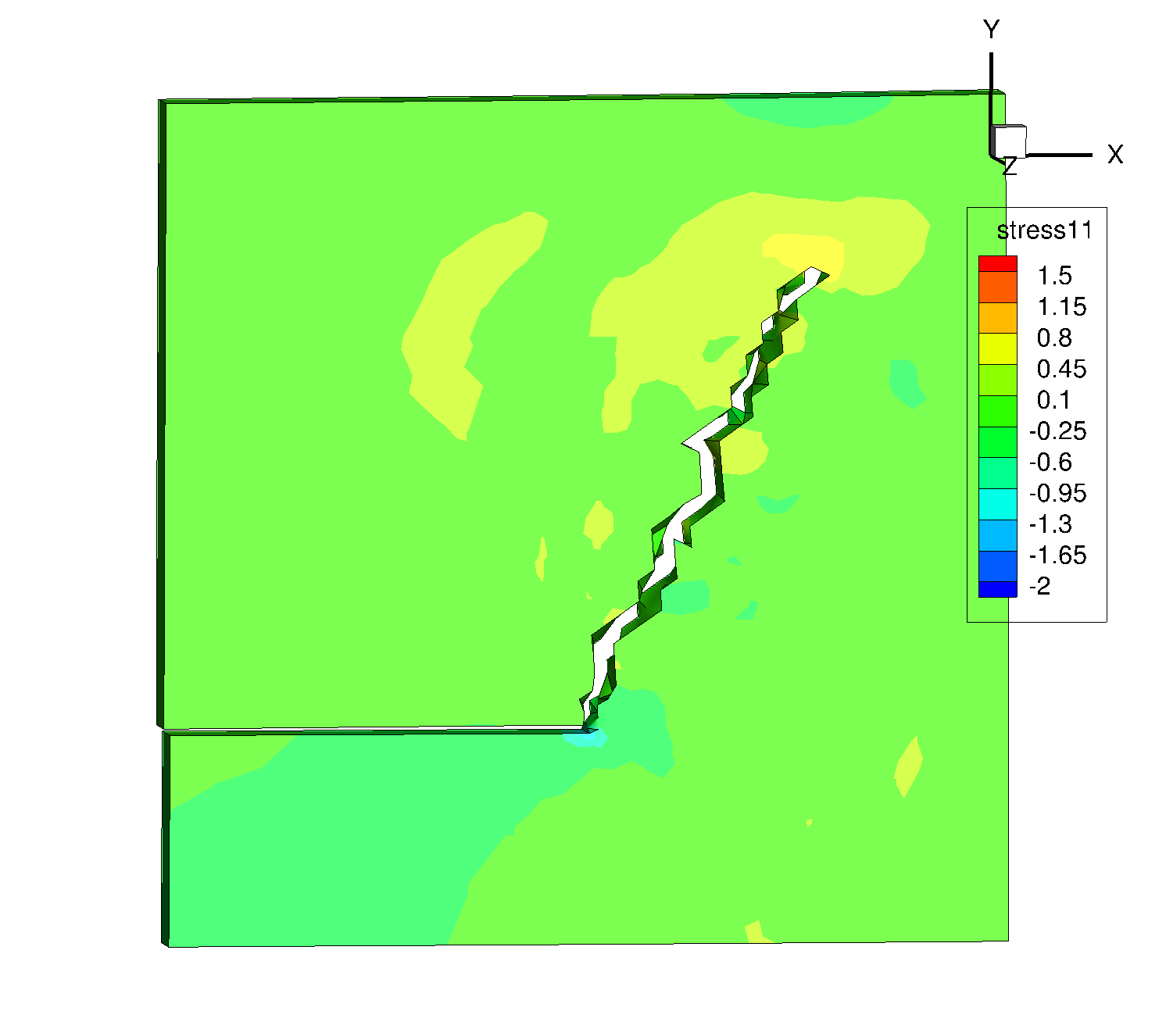}
            (c)
            \end{center}
        \end{minipage}
        \hfill
        \begin{minipage}{0.24\linewidth}
            \begin{center}
            \includegraphics[height=1.5in]{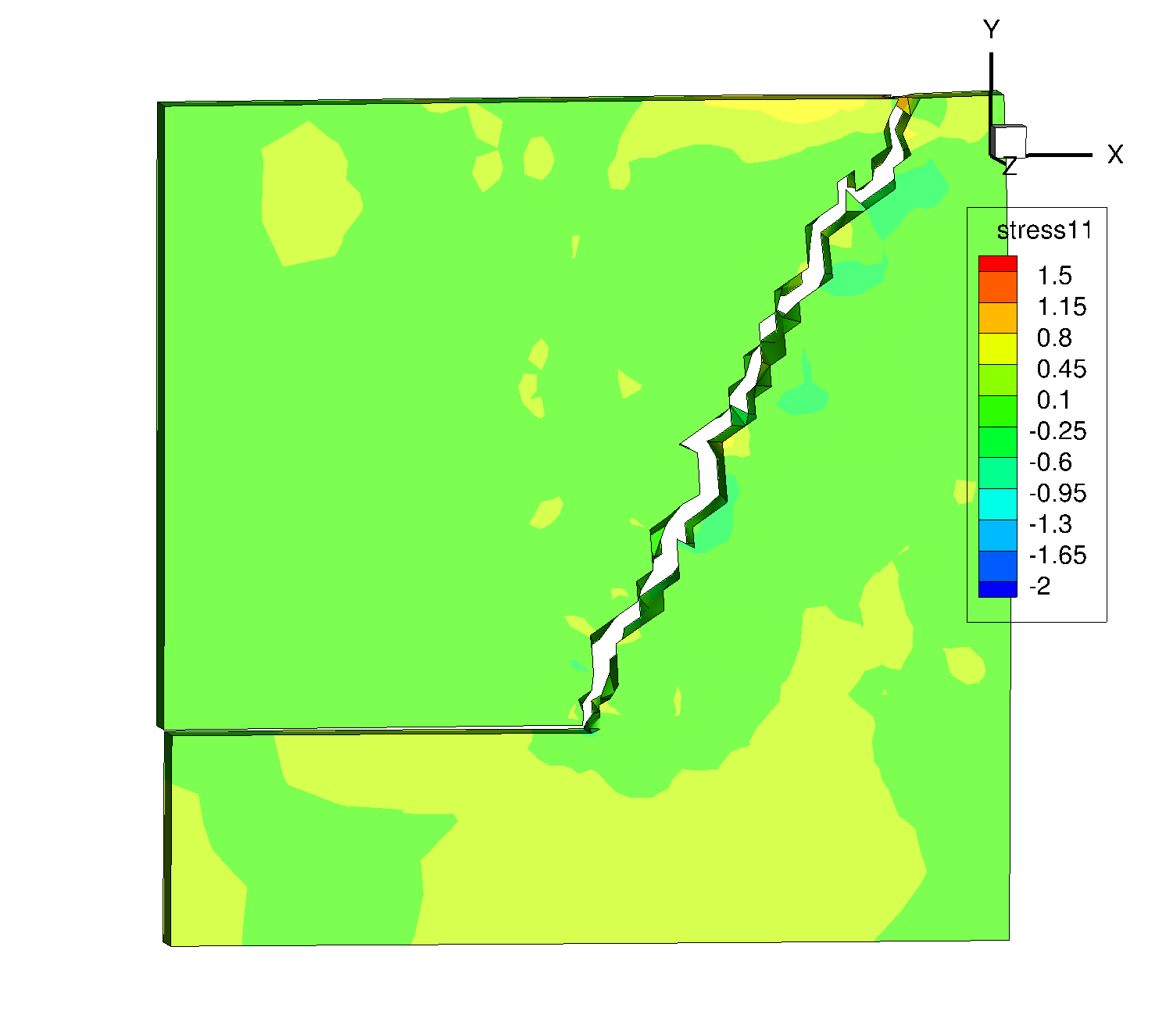}
            (d)
            \end{center}
        \end{minipage}
        \caption{Stress $\sigma_{xx}$ contour and crack patterns evolution for three-dimensional Kalthoff-Winkler plate medium mesh at times at (a). $t=25.92\ \mu s$, (b). $t=45\ \mu s$, (c). $t=64\ \mu s$, (d). $t=86\ \mu s$.}
        \label{fig5: Kalthoff-crack-evolution-2}
\end{figure}

\begin{figure}
	\centering
        \begin{minipage}{0.24\linewidth}
            \begin{center}
            \includegraphics[height=1.5in]{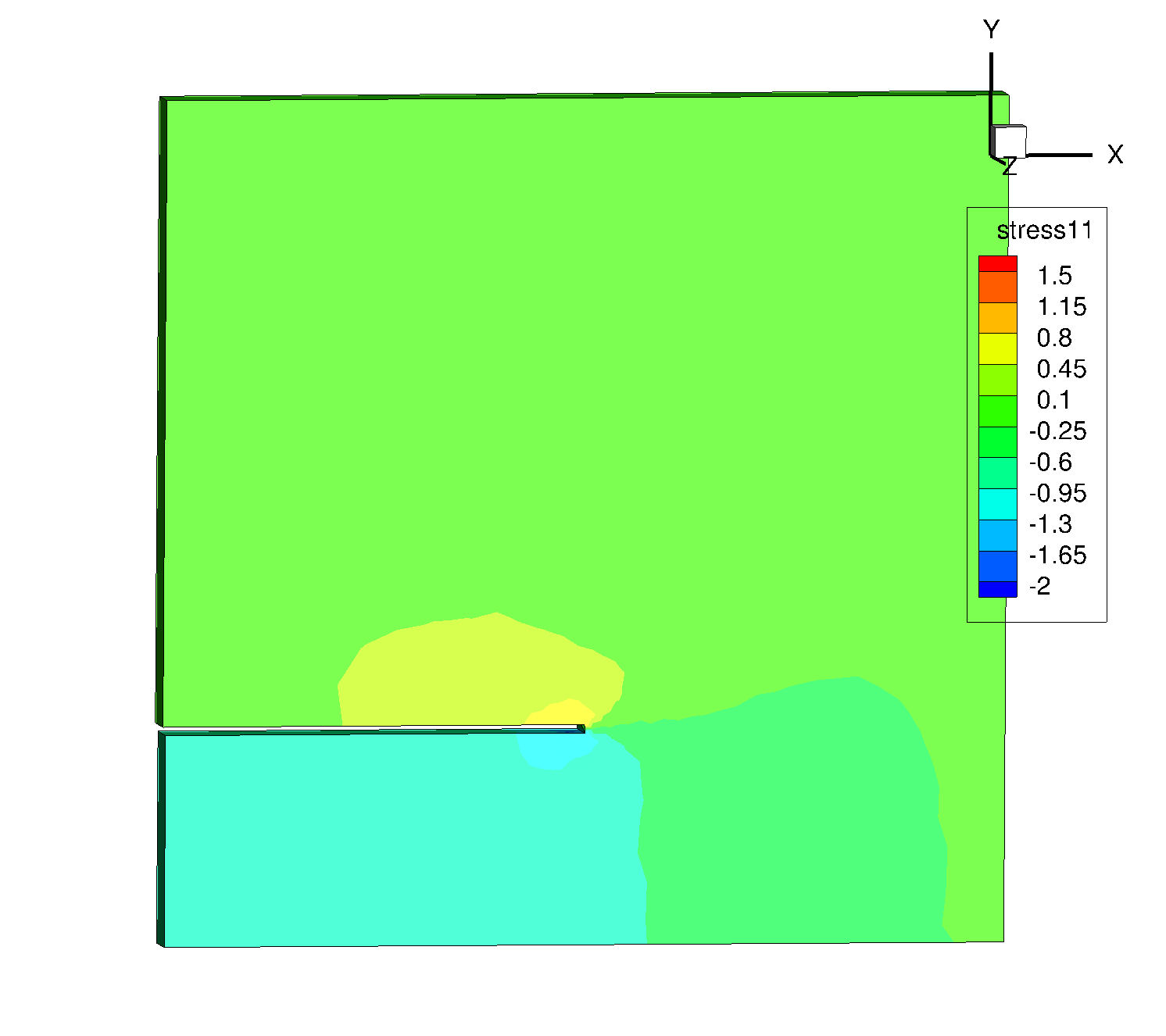}
            (a)
            \end{center}
        \end{minipage}
        \hfill
        \begin{minipage}{0.24\linewidth}
            \begin{center}
            \includegraphics[height=1.5in]{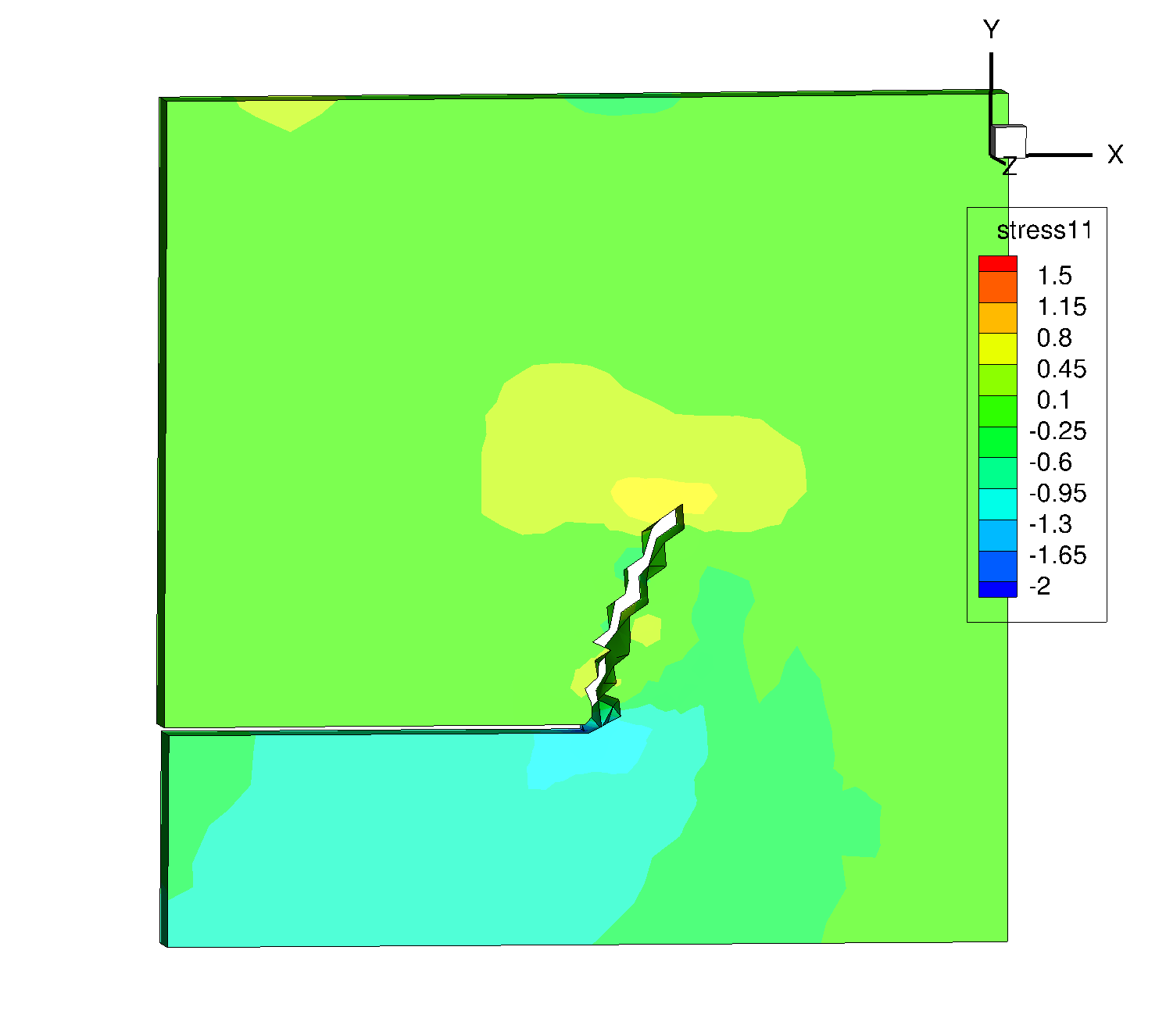}
            (b)
            \end{center}
        \end{minipage}   
        \hfill
        \begin{minipage}{0.24\linewidth}
            \begin{center}
            \includegraphics[height=1.5in]{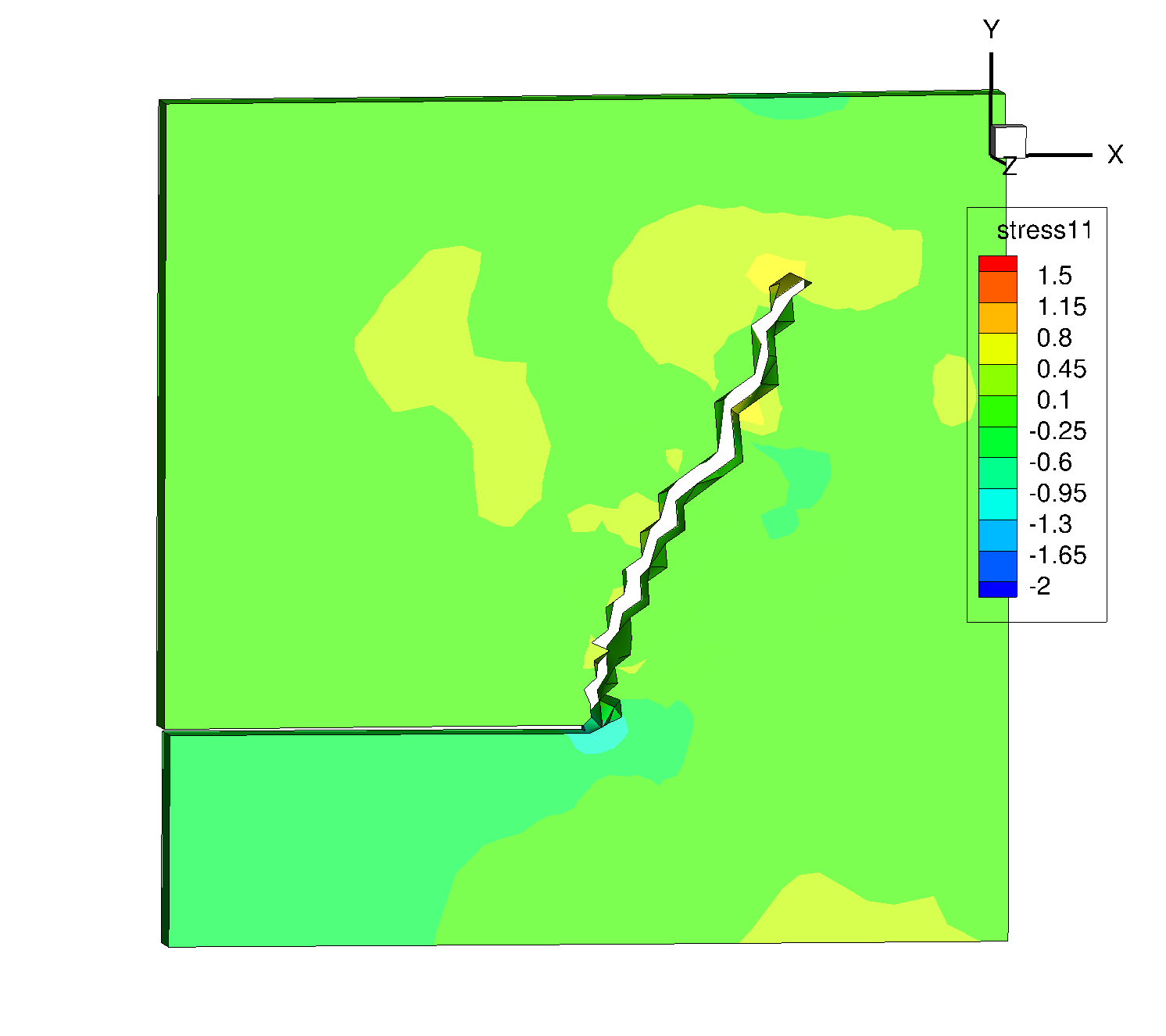}
            (c)
            \end{center}
        \end{minipage}
        \hfill
        \begin{minipage}{0.24\linewidth}
            \begin{center}
            \includegraphics[height=1.5in]{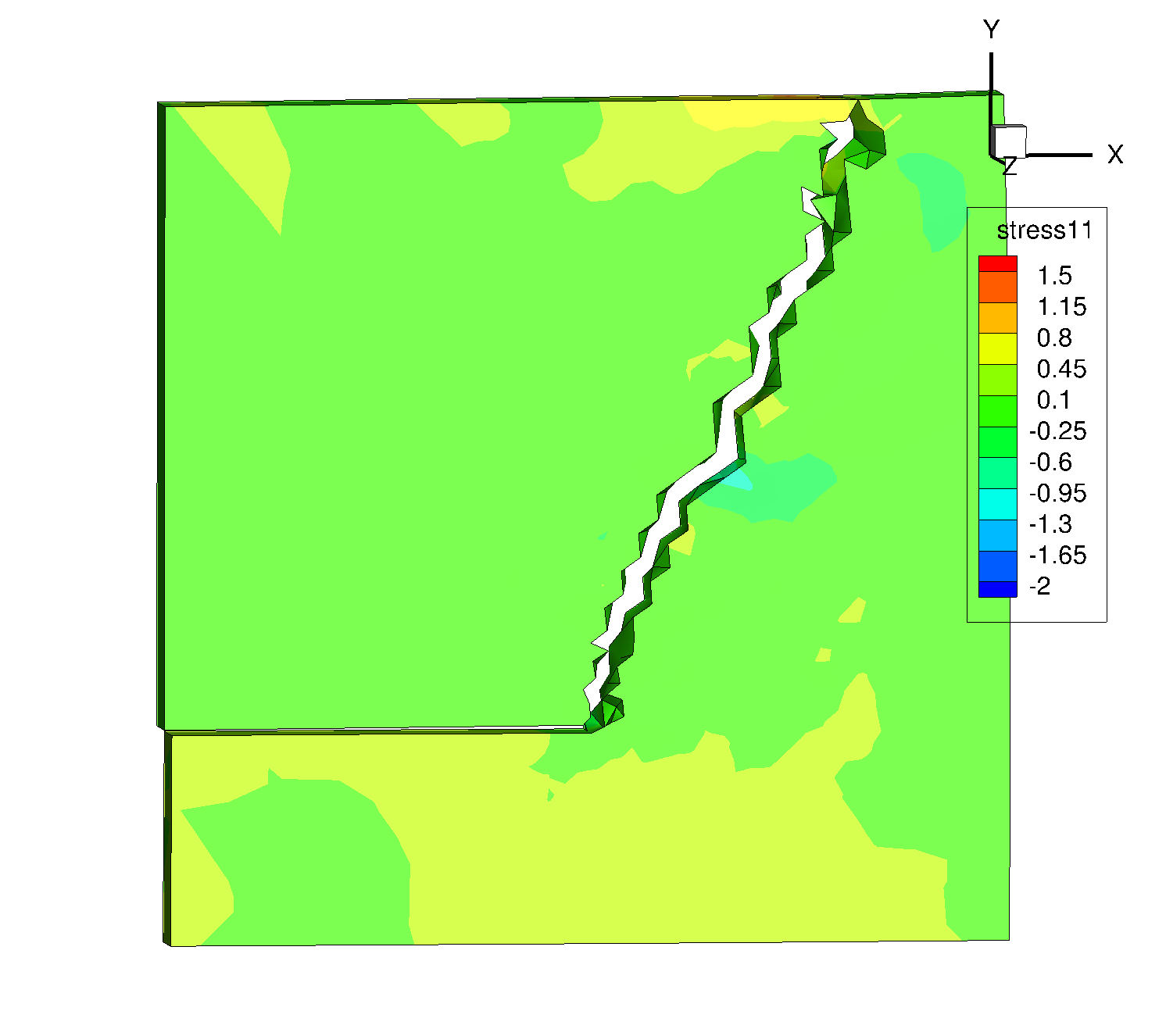}
            (d)
            \end{center}
        \end{minipage}
        \caption{Stress $\sigma_{xx}$ contour and crack patterns evolution for three-dimensional Kalthoff-Winkler plate coarse mesh at times at (a). $t=20.58\ \mu s$, (b). $t=45\ \mu s$, (c). $t=64\ \mu s$, (d). $t=90\ \mu s$.}
        \label{fig6: Kalthoff-crack-evolution-3}
\end{figure}

The three-dimensional crack patterns of the three different finite element meshes are illustrated in Figure.\ref{fig3: Kalthoff-crack-patterns-3D}. The angle of final crack patterns in the proposed CEM keeps in the range of $65^{\circ}\sim 70^{\circ}$, which aligns with the experimental result ($70^{\circ}$) and the other two-dimensional numerical results ($65^{\circ}$). It is worthy to emphasize that the direction of crack propagation is not sensitive to the element size and mesh. In other words, even though in coarse mesh, the crack pattern may be more tortuous, but it can still stay within the approximated range of $65^{\circ}\sim 70^{\circ}$. More importantly, the proposed CEM still accurately capturing crack path in the coarse mesh with only $8396$ tetrahedron elements, greatly outperforms other methods in efficiency, such as Bui et al. (\cite{tran2024nonlocal}), which uses $732756$ tetrahedron elements to reproduce crack patterns with similar accuracy.
And Figures.\ref{fig4: Kalthoff-crack-evolution-1} - \ref{fig6: Kalthoff-crack-evolution-3} show the crack pattern over time for the fine, medium, and coarse meshes, respectively.

\begin{figure}[htp]
	\centering
            \begin{center}
            \includegraphics[height=2.4in]{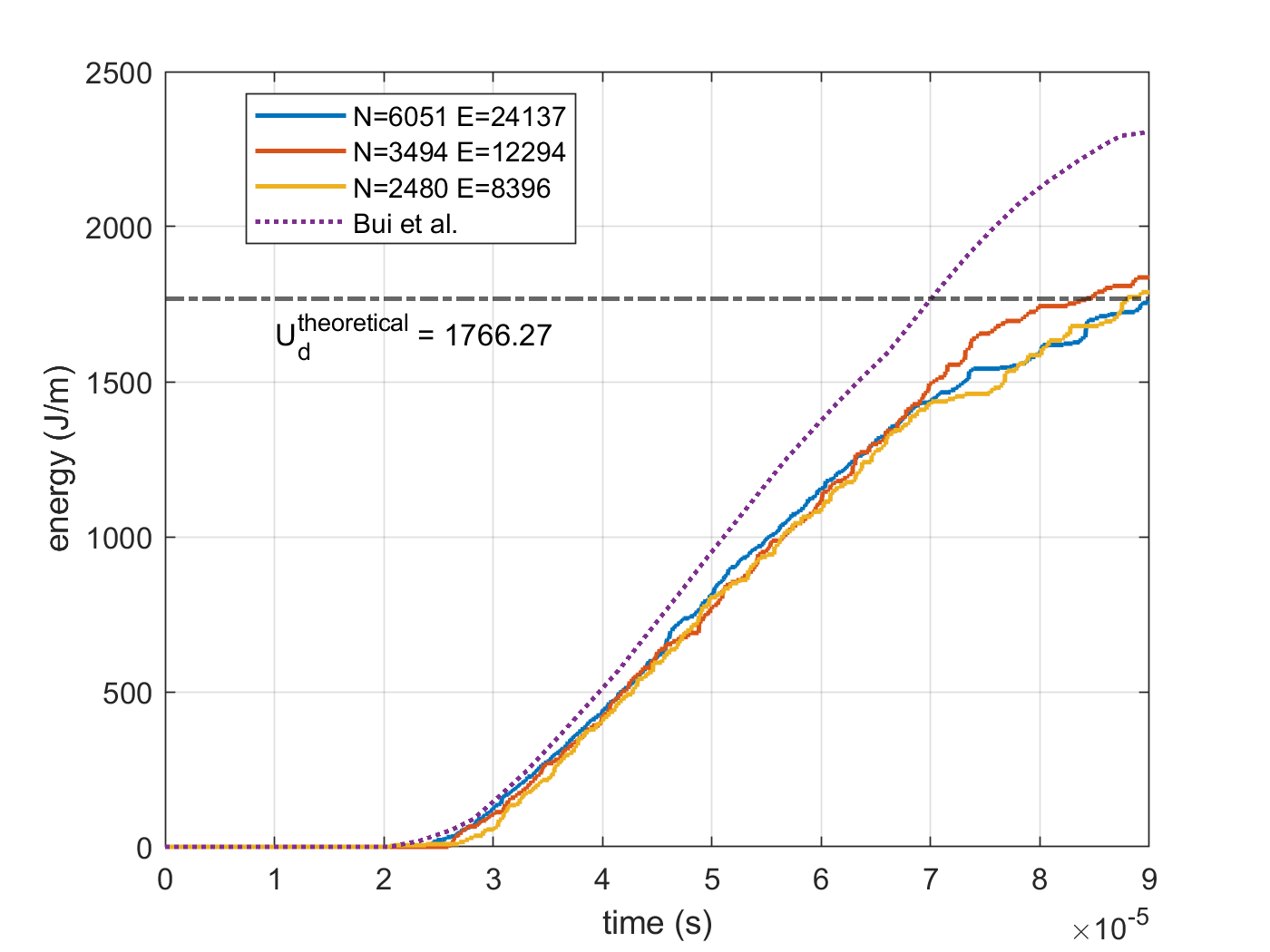}
            \end{center}
        \caption{Comparison of dissipated energy among three representative meshes and a reference mesh(\cite{tran2024nonlocal}).}
        \label{fig7: Kalthoff-Ud}
\end{figure}

The results of the dissipated energy in the three representative models are summarized in Figure.\ref{fig7: Kalthoff-Ud}. The comparison of dissipated energy shows that all three mesh types yield lower values than the numerical result from Bui et al. (\cite{tran2024nonlocal}), but they align well with the theoretical value. Besides, the dissipated energy evolution is consistent across the different meshes. These results support the effectiveness of the proposed CEM in the prediction of three-dimensional crack behavior. 

\subsection{Anchorage pull-out test}
\begin{figure}[htp]
	\centering
            \begin{center}
            \includegraphics[height=3.5in]{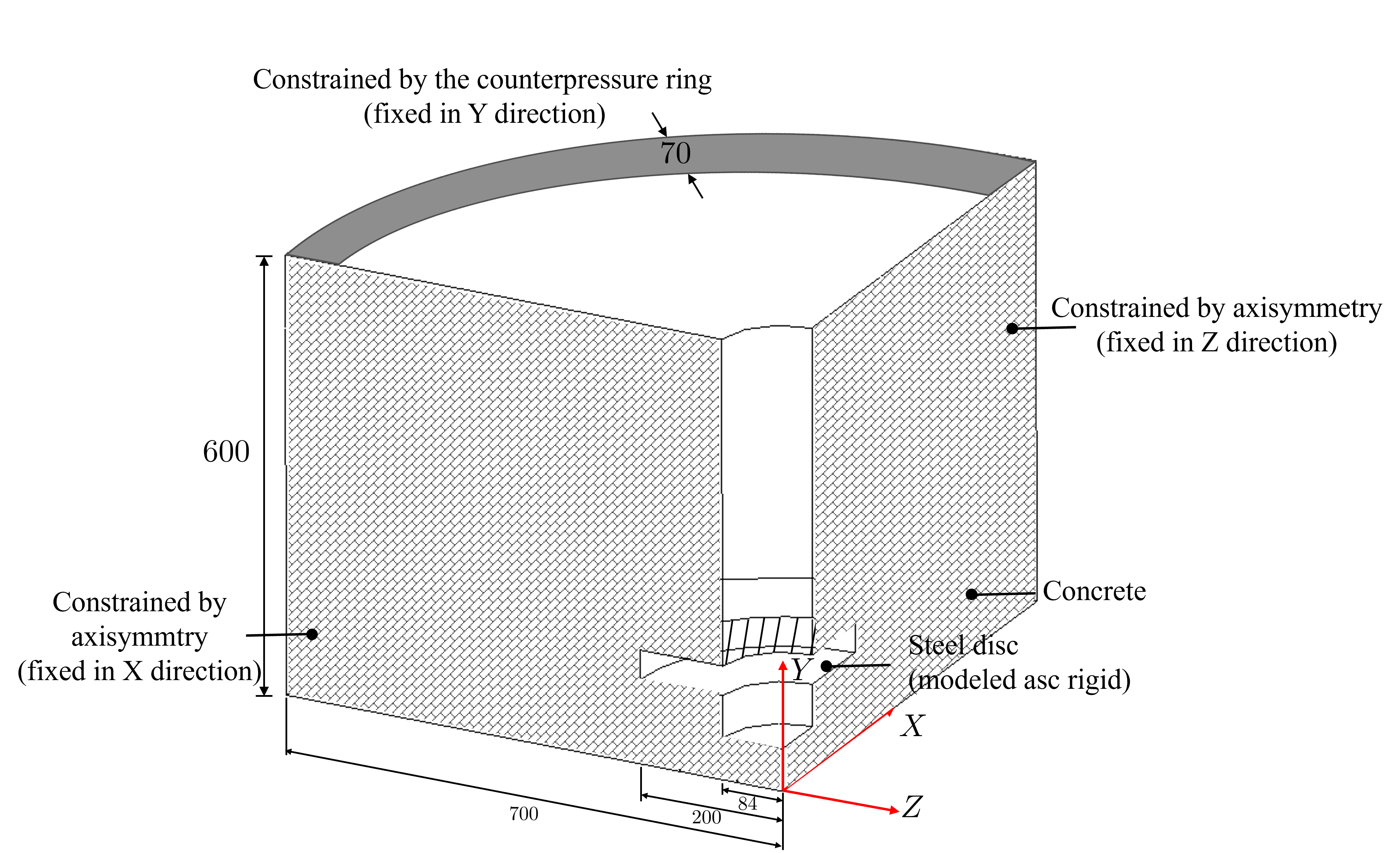}
            \end{center}
        \caption{Geometric dimensions of the anchorage structure (unit: mm)}
        \label{fig8: anchorage-geometry}
\end{figure}

This study simulates the extraction of a steel anchor embedded within a concrete body. Such steel anchors are commonly employed to create connections in concrete assemblies and serve as roof bolts in geotechnical applications. Specifically, the scenario involves a pull-out test of a circular steel plate measuring 400 mm in diameter and 40 mm in thickness, which is embedded in a large cylindrical concrete block. The concrete block has a diameter of 1400 mm and a height of 600 mm, replicating conditions representative of heavy structural anchoring systems. Figure.\ref{fig8: anchorage-geometry} provides comprehensive information on the geometry, boundary conditions, and applied loads, illustrating only one-quarter of the entire model for clarity. A vertical force is applied at the center of the steel disc, initiating the pull-out action. This load causes the embedded disc to be drawn upward against a concentric counteracting pressure located on the surface of the concrete block (illustrated by the shaded region at the top), which is restrained in the $Y$-direction. The two surfaces at $X=0$ and $Z=0$ are constrained in corresponding directions respectively due to four-fold symmtric structure. The process continues until the concrete material failes. The pull-out test serves as a dependable in-situ method for estimating the compressive strength of the targeted concrete. This approach correlates the observed pull-out force with the material’s strength through the use of a calibration curve. The test setup bears resemblance to the so-called Lok-Test, also known as the Danish punch-out test. Previous investigations into similar anchorage configurations can be found in \cite{de1986non} and \cite{rots1988computational}, where axisymmetric finite element models were employed. Notably, the study in \cite{rots1988computational} utilizes a smeared crack approach to represent the fracture behavior of concrete.

Belytschko et al. (\cite{areias2005analysis}) and Holzapfel et al. (\cite{gasser2005modeling}) studied this problem with quasi-static modeling. In our work, a transient-dynamic analysis is conducted on this problem to demonstrate validity of the proposed CEM in non-planar cracking problems. Therefore, the applied forces with steel plate anchor in quasi-static analysis are replaced by vertical velocities on the shaded upper surface of steel disc notch (see Figure.\ref{fig8: anchorage-geometry}) and the applied velocity is shown as below,
\begin{equation}
v =
\begin{cases} 
\frac{t}{t_0}v_0,  & \text{if } \ t \le t_0, \\
v_0, & \text{if } \ t > t_0.
\end{cases}
\end{equation}
in which, $v_0 = 0.01\ m/s$ and $t_0 = 5\ ms$.

\begin{figure}[htp]
	\centering
        \begin{minipage}{0.3\linewidth}
            \begin{center}
            \includegraphics[height=2.0in]{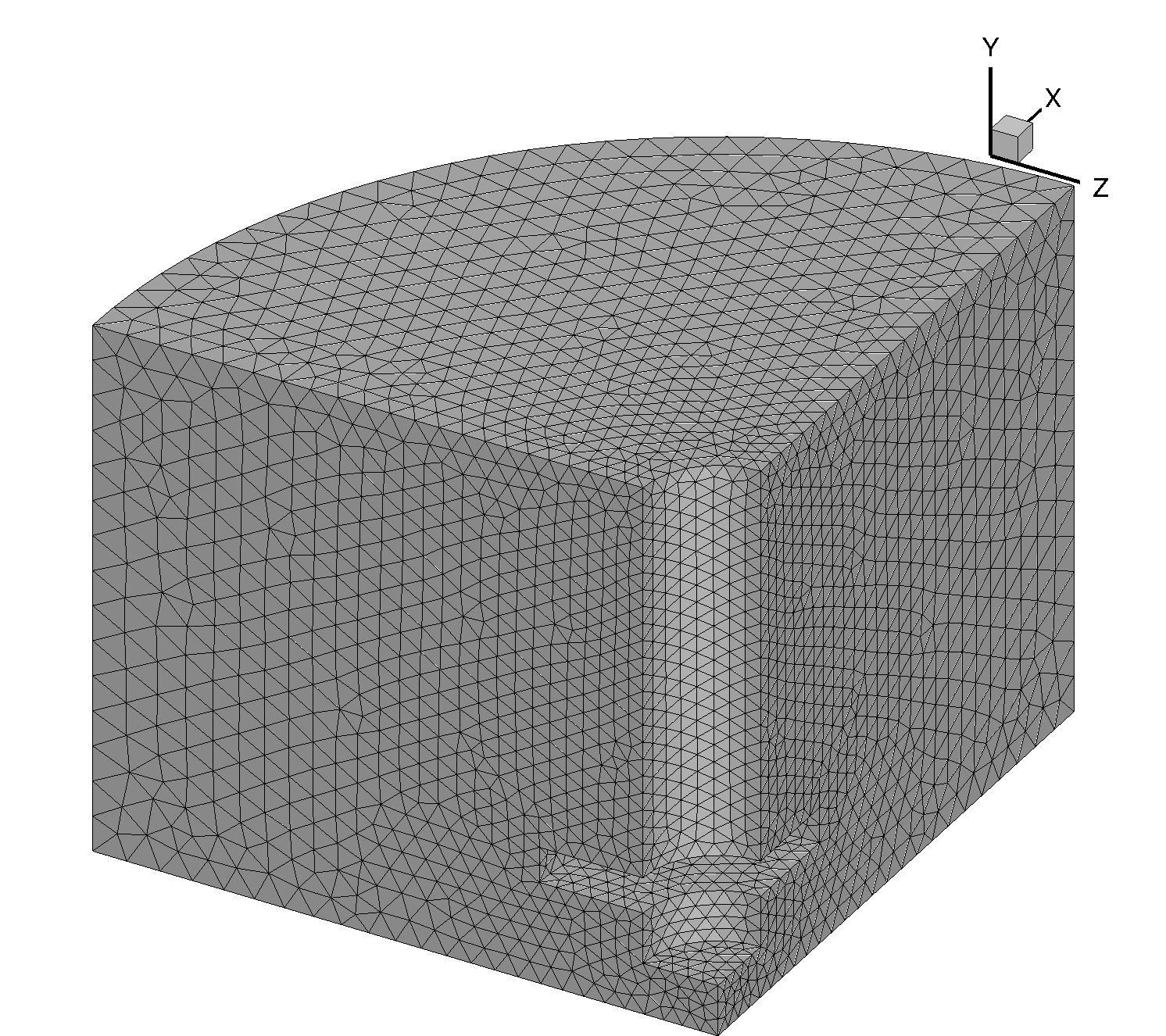}
            \end{center}
            \begin{center}
            (a)
            \end{center}
        \end{minipage}
        \hfill
        \begin{minipage}{0.3\linewidth}
            \begin{center}
            \includegraphics[height=2.0in]{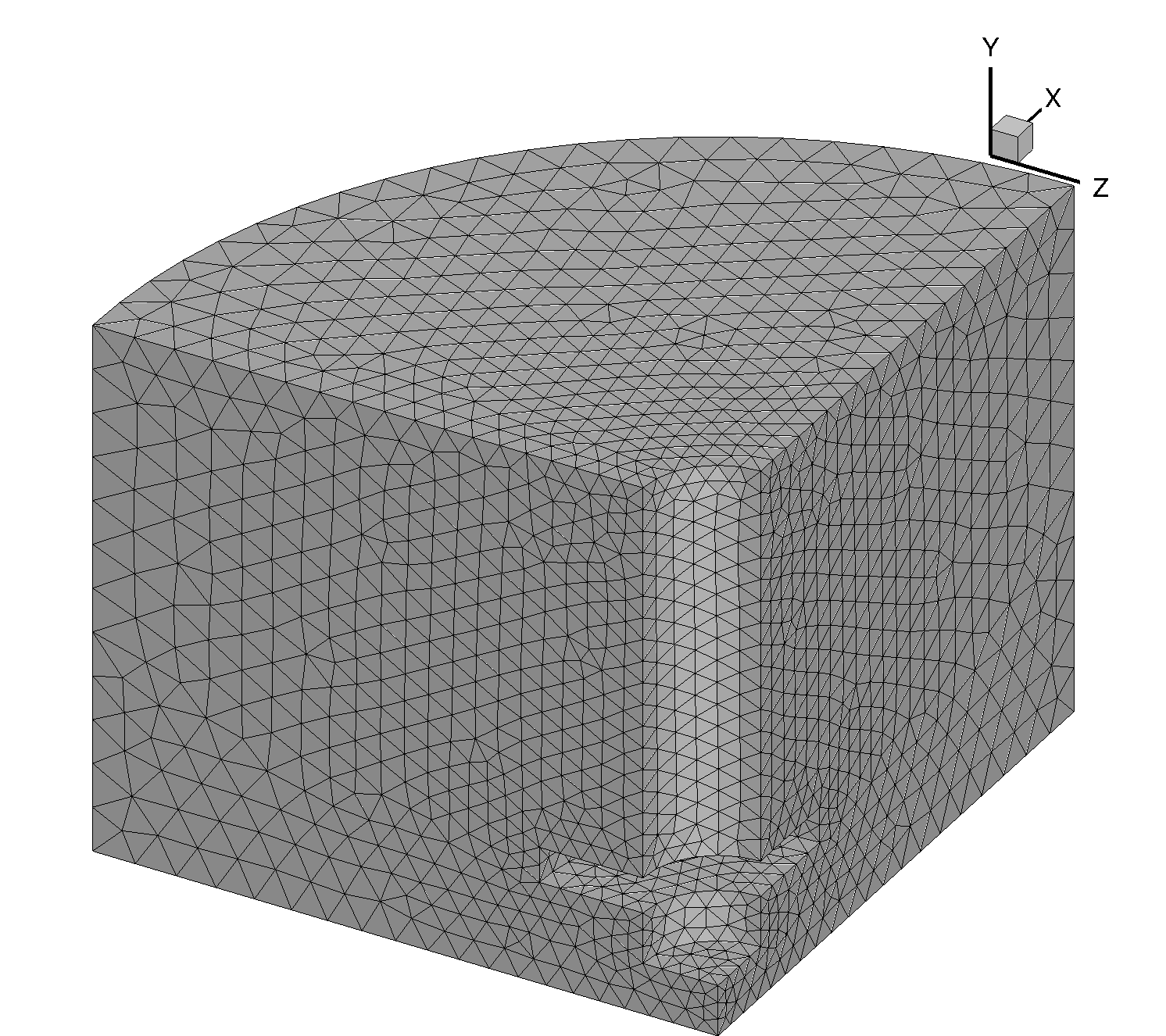}
            \end{center}
            \begin{center}
            (b)
            \end{center}
        \end{minipage}   
        \hfill
        \begin{minipage}{0.3\linewidth}
            \begin{center}
            \includegraphics[height=2.0in]{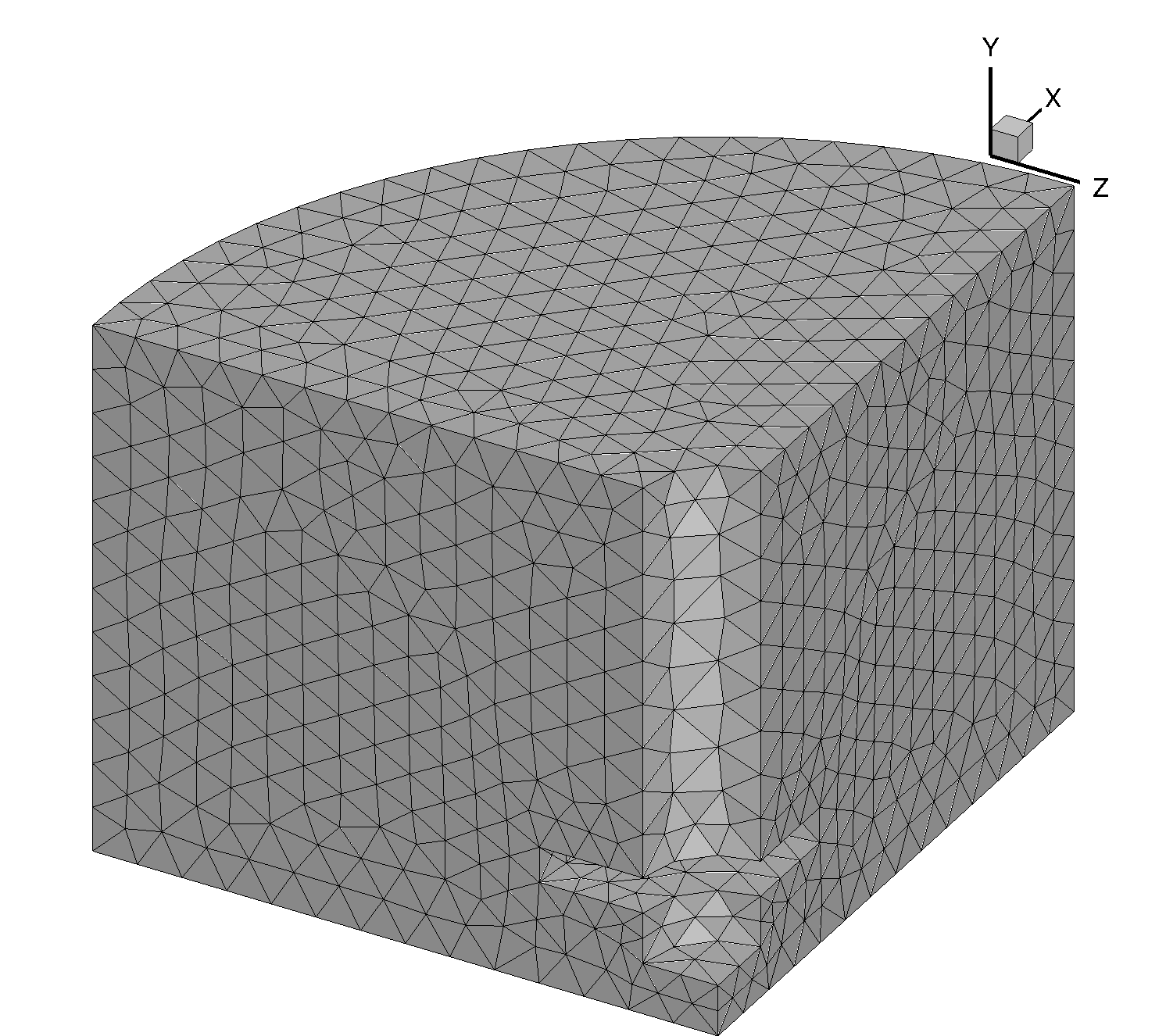}
            \end{center}
            \begin{center}
            (c)
            \end{center}
        \end{minipage}
        \caption{Three tetrahedron meshes of anchorage pull-out model are illustrated: (a). the fine mesh with $36866$ elements; (b). the medium mesh with $20523$ elements; (c). the coarse mesh with $9289$ elements.}
        \label{fig9: anchorage-mesh-3D}
\end{figure}

The material properties in the benchmark are given as follows: Young's modulus $E=30 \ GPa$, Poisson ratio $v=0.2$, density $\rho=2400 \ kg/m^3$ and critical fracture energy release rate $\mathcal{G}_c = 1.06 \times 10^2 \ J/m^2$. Three different discretizations with tetrahedron elements are utilized to discretize the three models, which have $36866$ elements with $7581$ nodes, $20523$ elements with $4401$ nodes, and $9289$ elements with $2150$ nodes, respectively. The finite element meshes of the three models are shown in Figure.\ref{fig9: anchorage-mesh-3D}.

\begin{figure}
	\centering
        \begin{minipage}{0.45\linewidth}
            \begin{center}
            \includegraphics[height=2.5in]{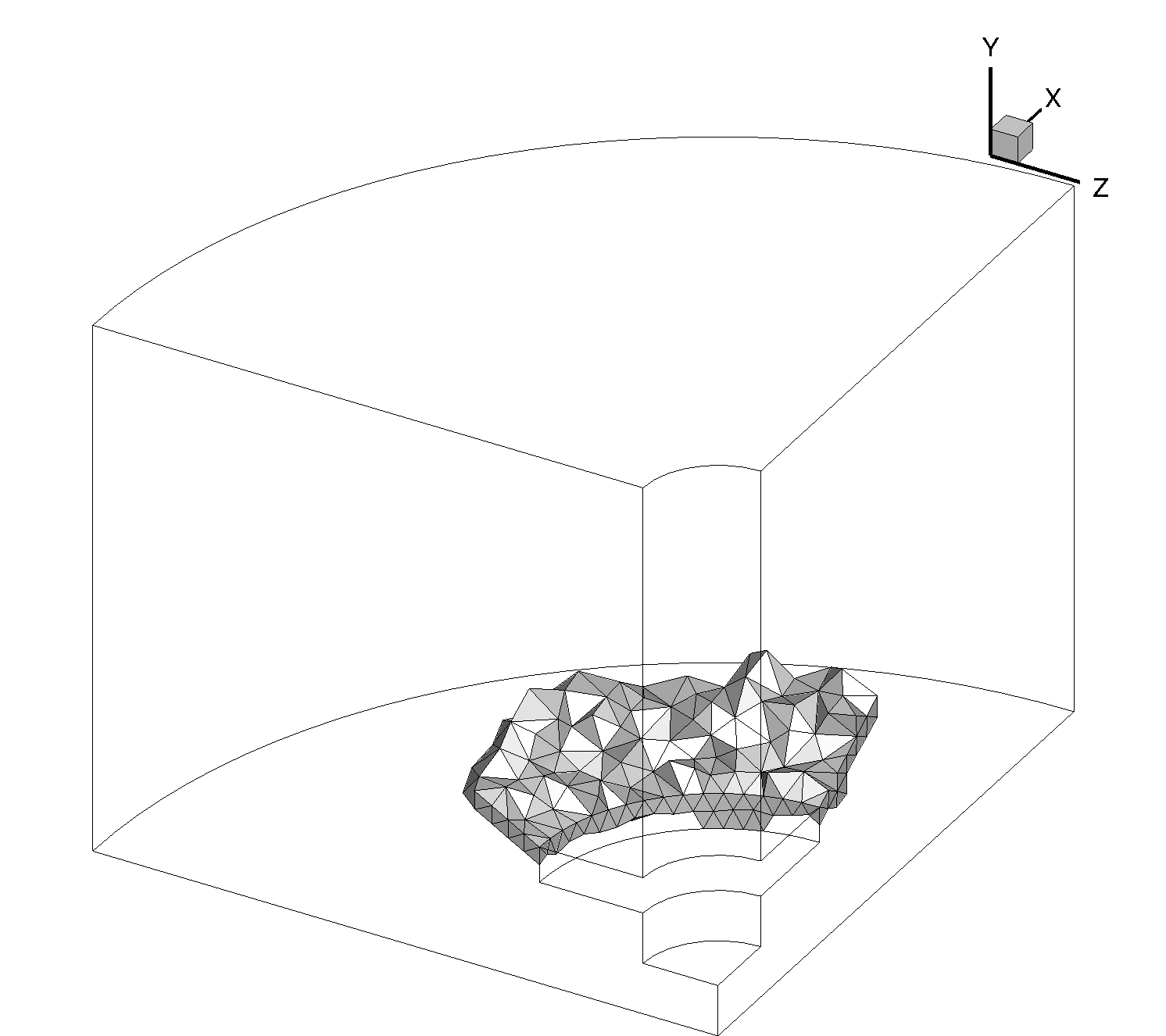}
            (a)
            \end{center}
        \end{minipage}
        \hfill
        \begin{minipage}{0.45\linewidth}
            \begin{center}
            \includegraphics[height=2.5in]{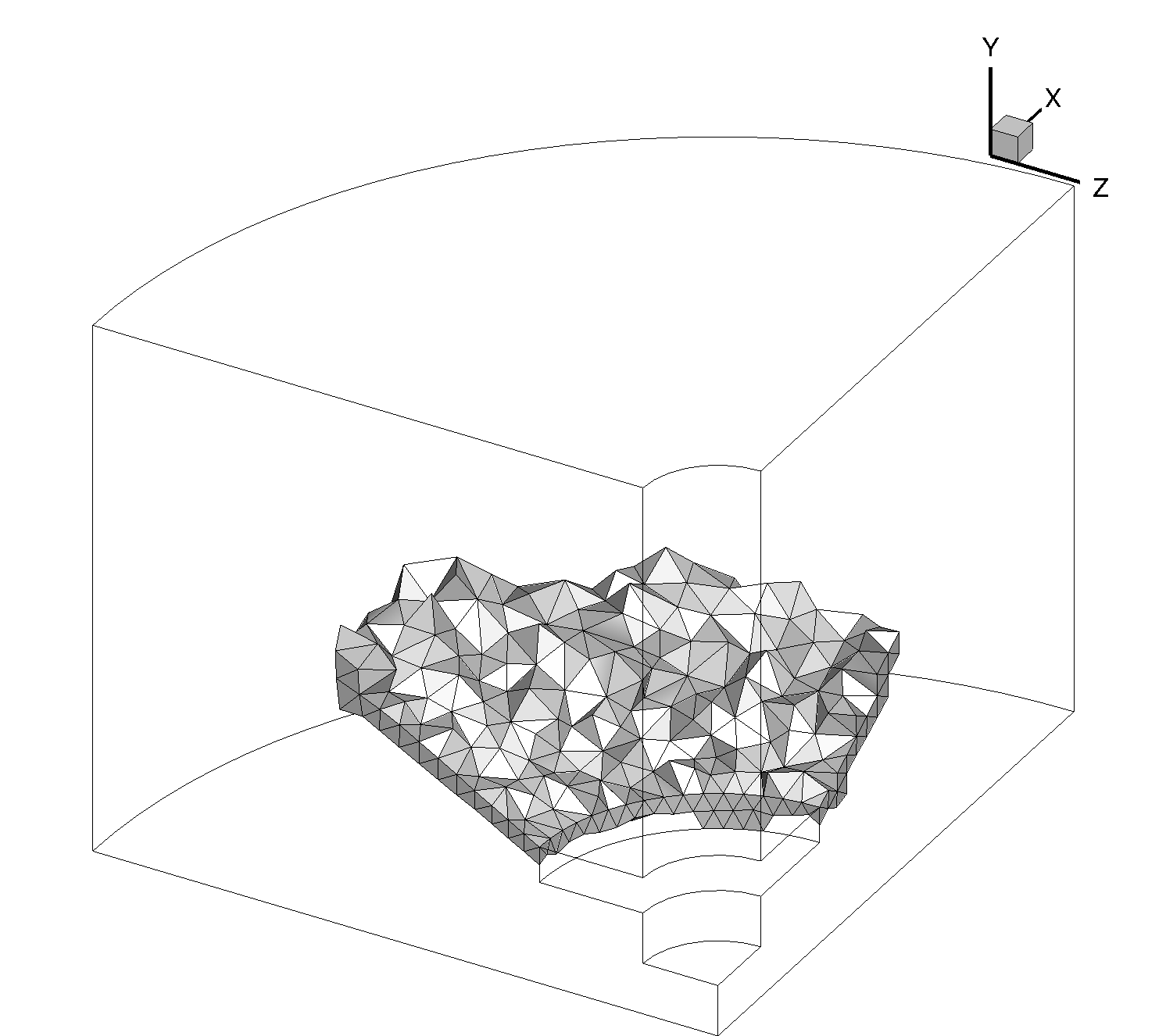}
            (b)
            \end{center}
        \end{minipage}   
        \hfill
        \begin{minipage}{0.45\linewidth}
            \begin{center}
            \includegraphics[height=2.5in]{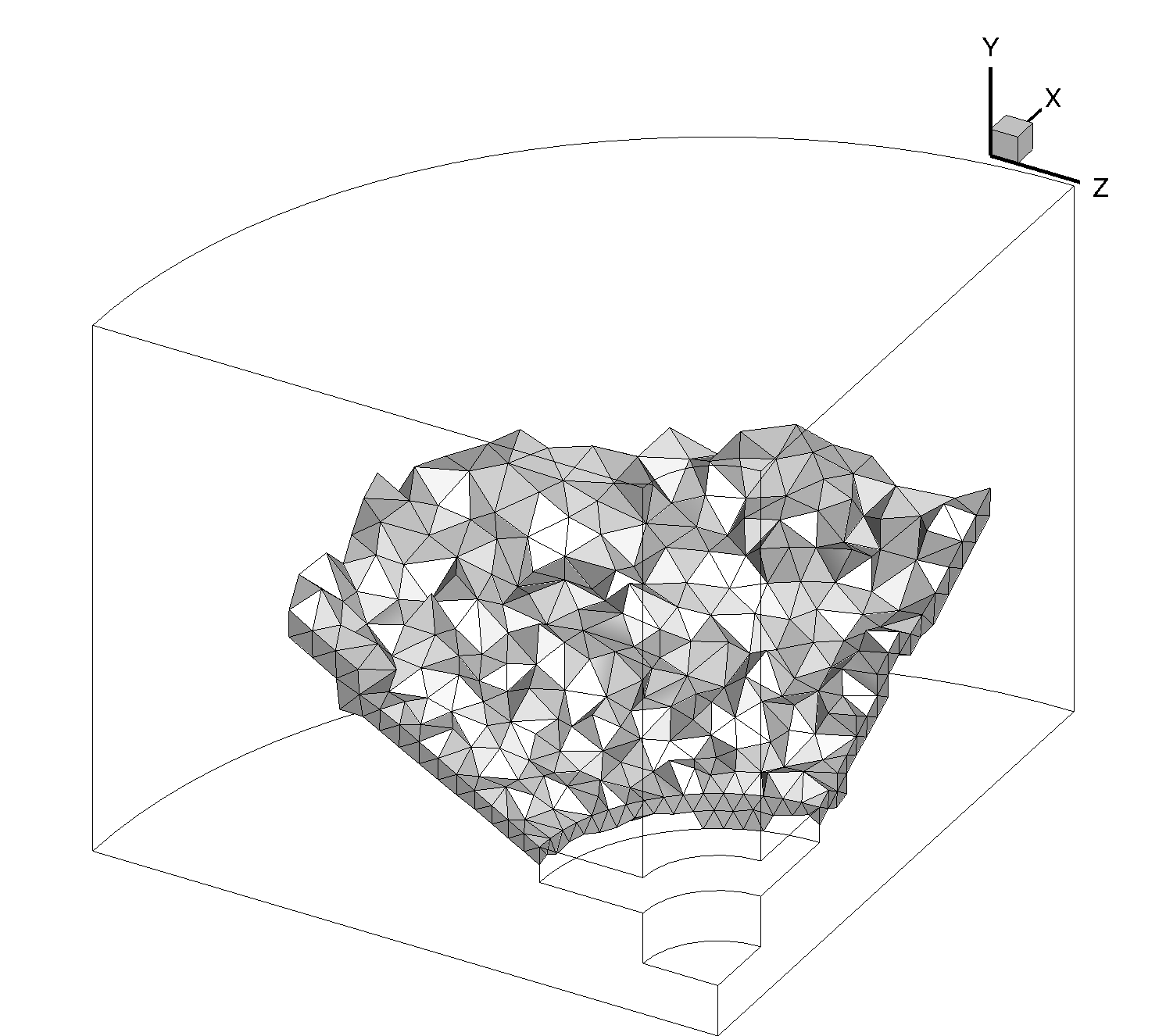}
            (c)
            \end{center}
        \end{minipage}
        \hfill
        \begin{minipage}{0.45\linewidth}
            \begin{center}
            \includegraphics[height=2.5in]{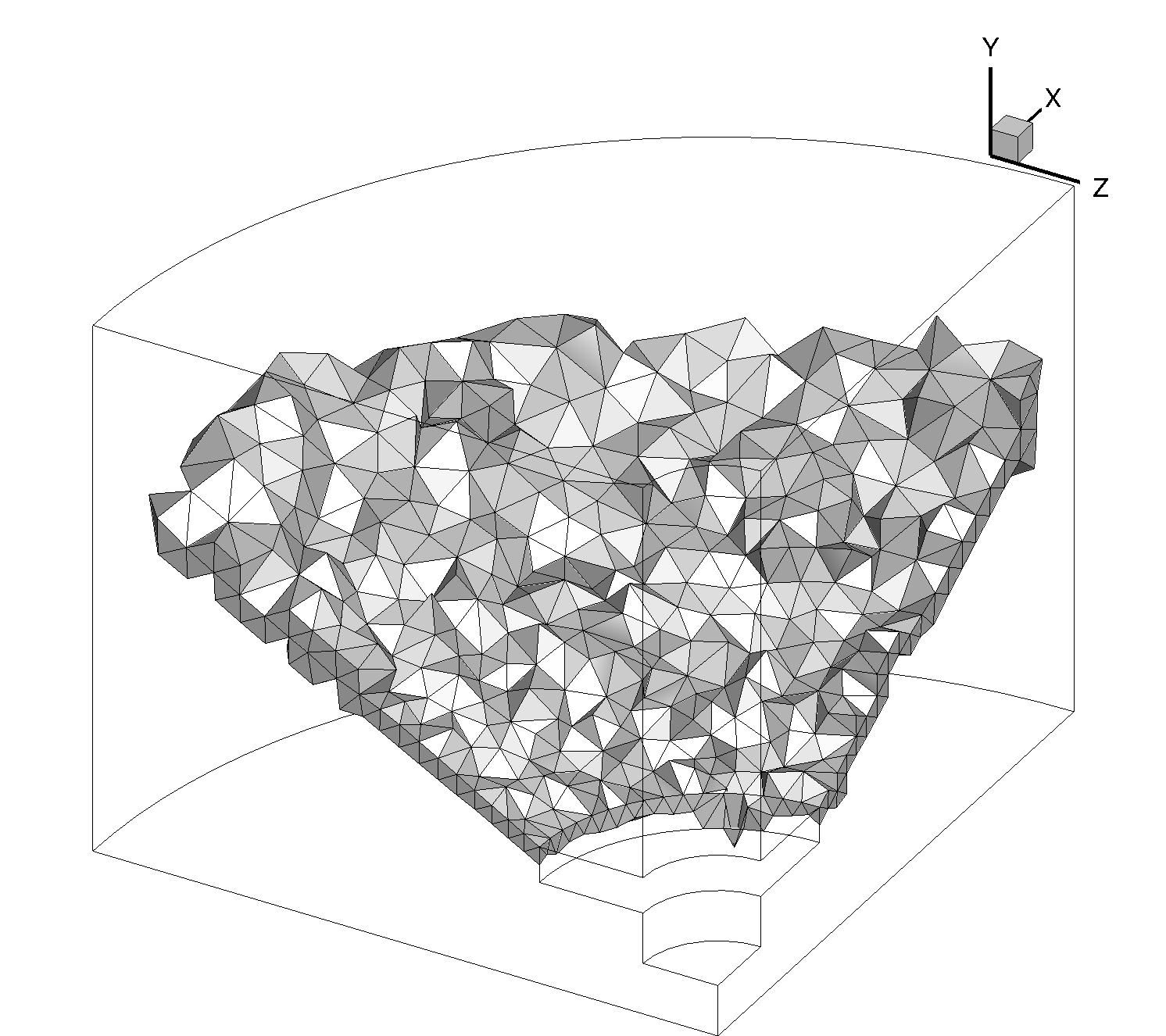}
            (d)
            \end{center}
        \end{minipage}
        \caption{The crack surface evolution for three-dimensional anchorage pull-out test with $36866$ elements at different times (a). $t=27.194\ ms$, (b). $t=27.541\ ms$, (c). $t=27.916\ ms$, (d). $t=42.5\ ms$.}
        \label{fig10: anchorage-crack-evolution-3}
\end{figure}

\begin{figure}
	\centering
        \begin{minipage}{0.45\linewidth}
            \begin{center}
            \includegraphics[height=2.5in]{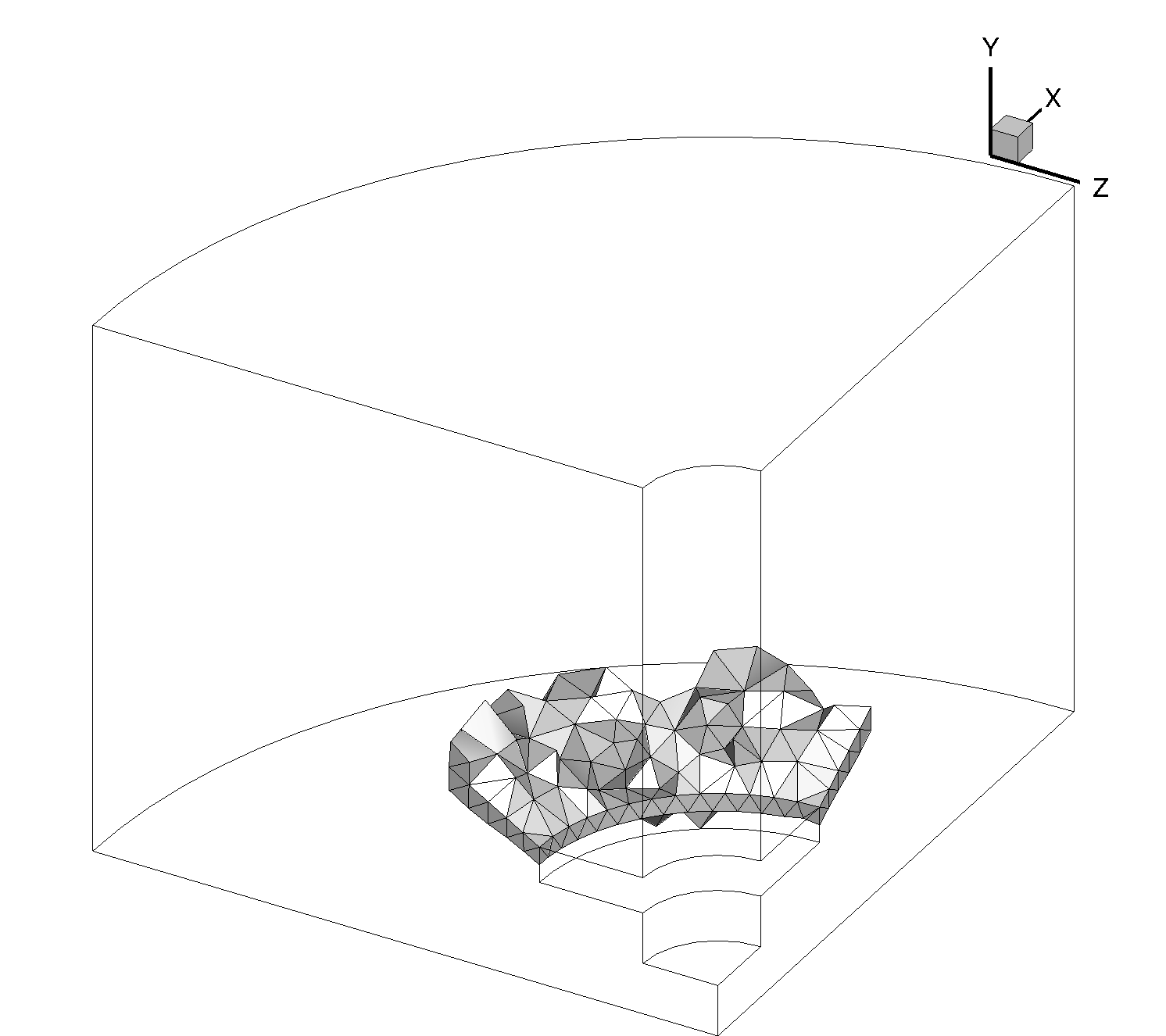}
            (a)
            \end{center}
        \end{minipage}
        \hfill
        \begin{minipage}{0.45\linewidth}
            \begin{center}
            \includegraphics[height=2.5in]{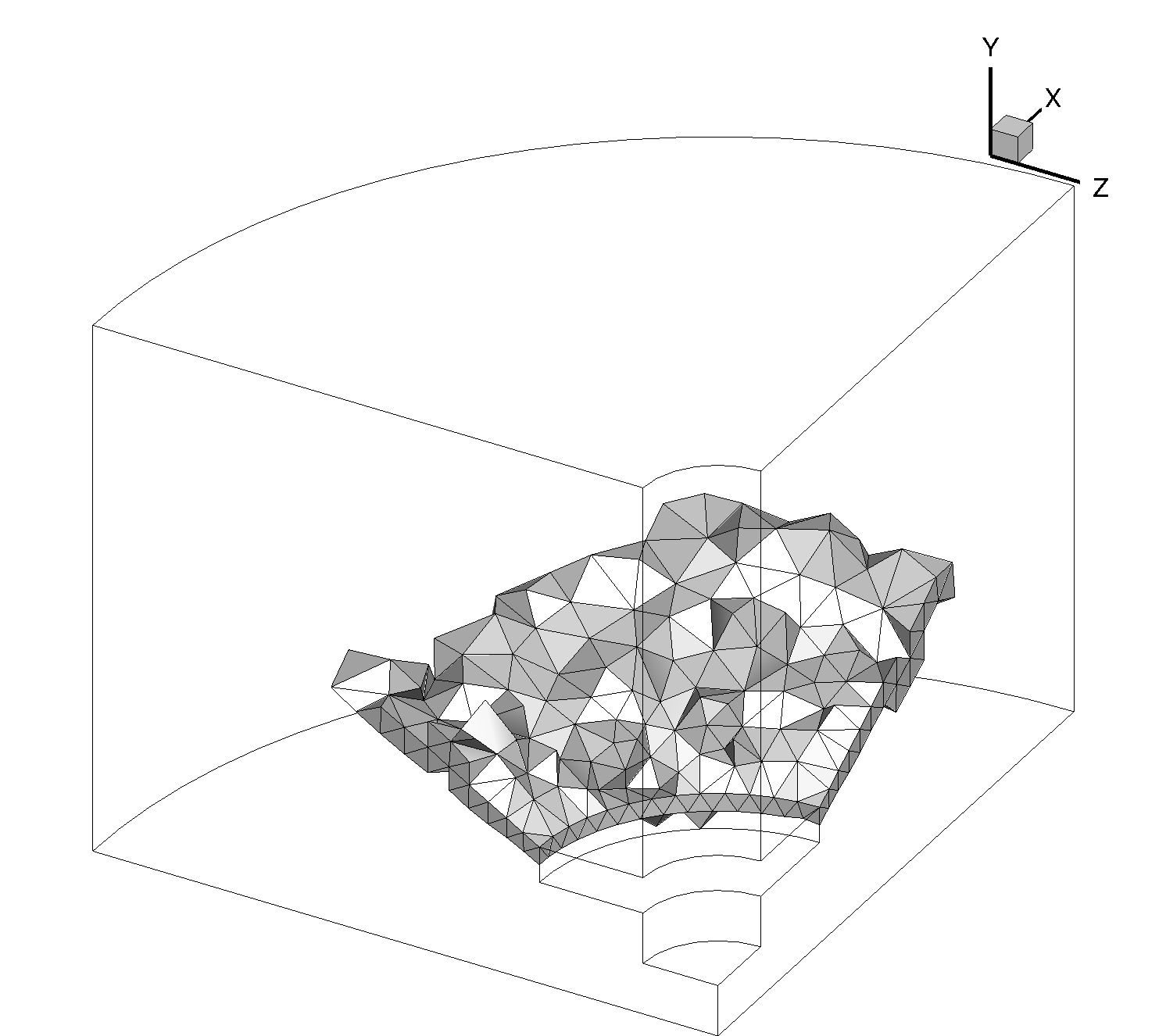}
            (b)
            \end{center}
        \end{minipage}   
        \hfill
        \begin{minipage}{0.45\linewidth}
            \begin{center}
            \includegraphics[height=2.5in]{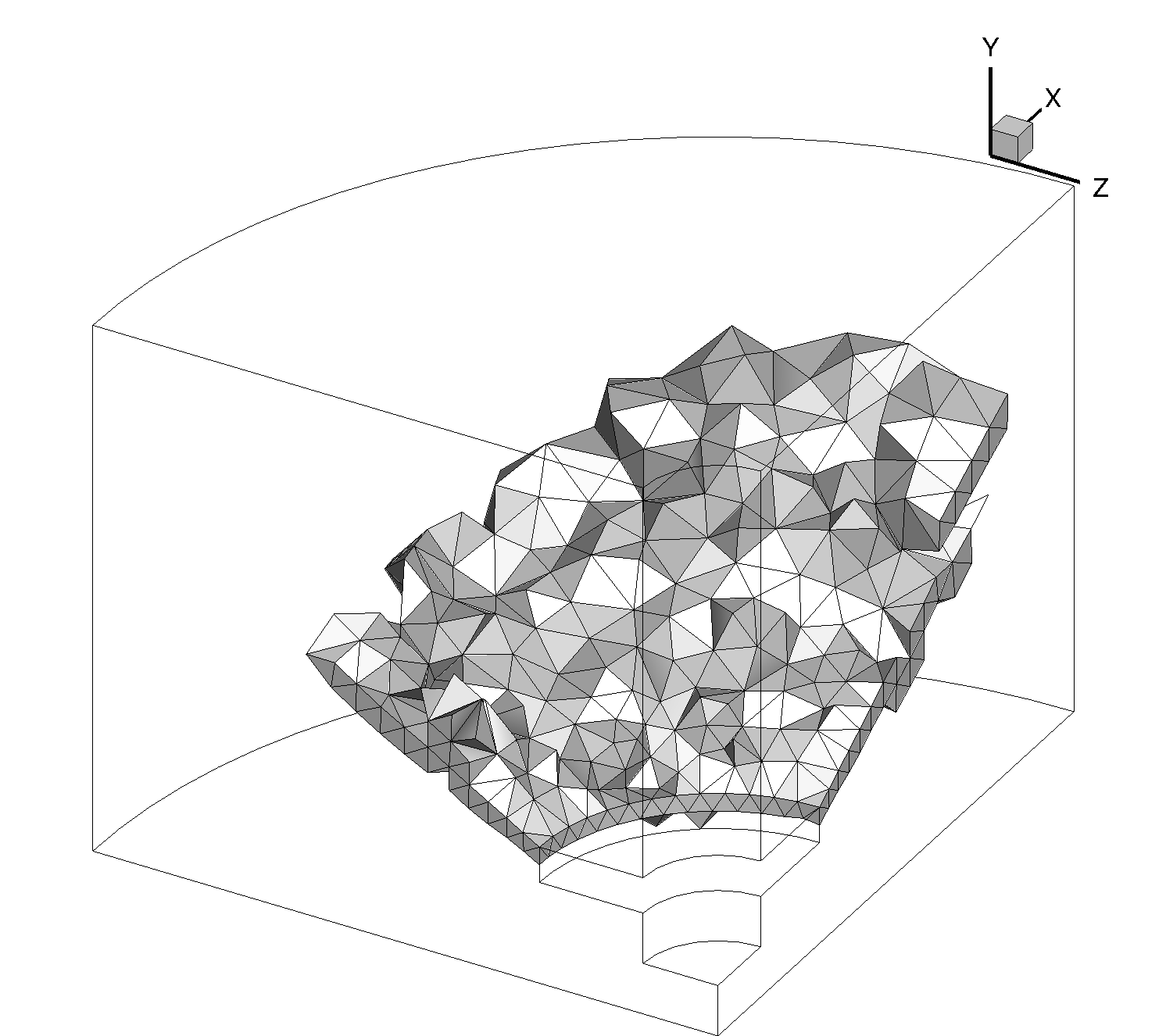}
            (c)
            \end{center}
        \end{minipage}
        \hfill
        \begin{minipage}{0.45\linewidth}
            \begin{center}
            \includegraphics[height=2.5in]{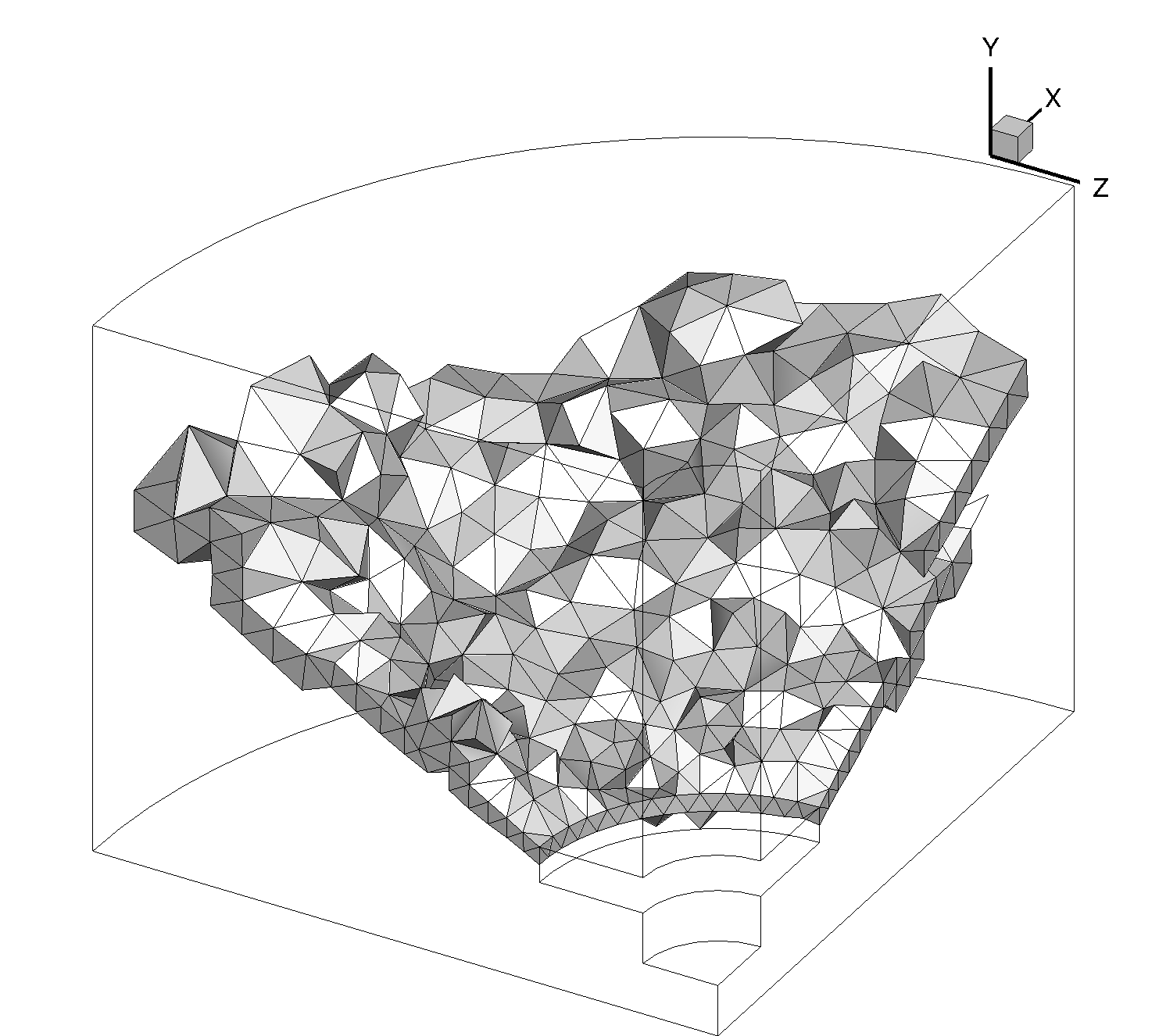}
            (d)
            \end{center}
        \end{minipage}
        \caption{The crack surface evolution for three-dimensional anchorage pull-out test with $20523$ elements at different times (a). $t=28.328\ ms$, (b). $t=28.683\ ms$, (c). $t=30.164\ ms$, (d). $t=42.5\ ms$.}
        \label{fig11: anchorage-crack-evolution-2}
\end{figure}

\begin{figure}
	\centering
        \begin{minipage}{0.45\linewidth}
            \begin{center}
            \includegraphics[height=2.5in]{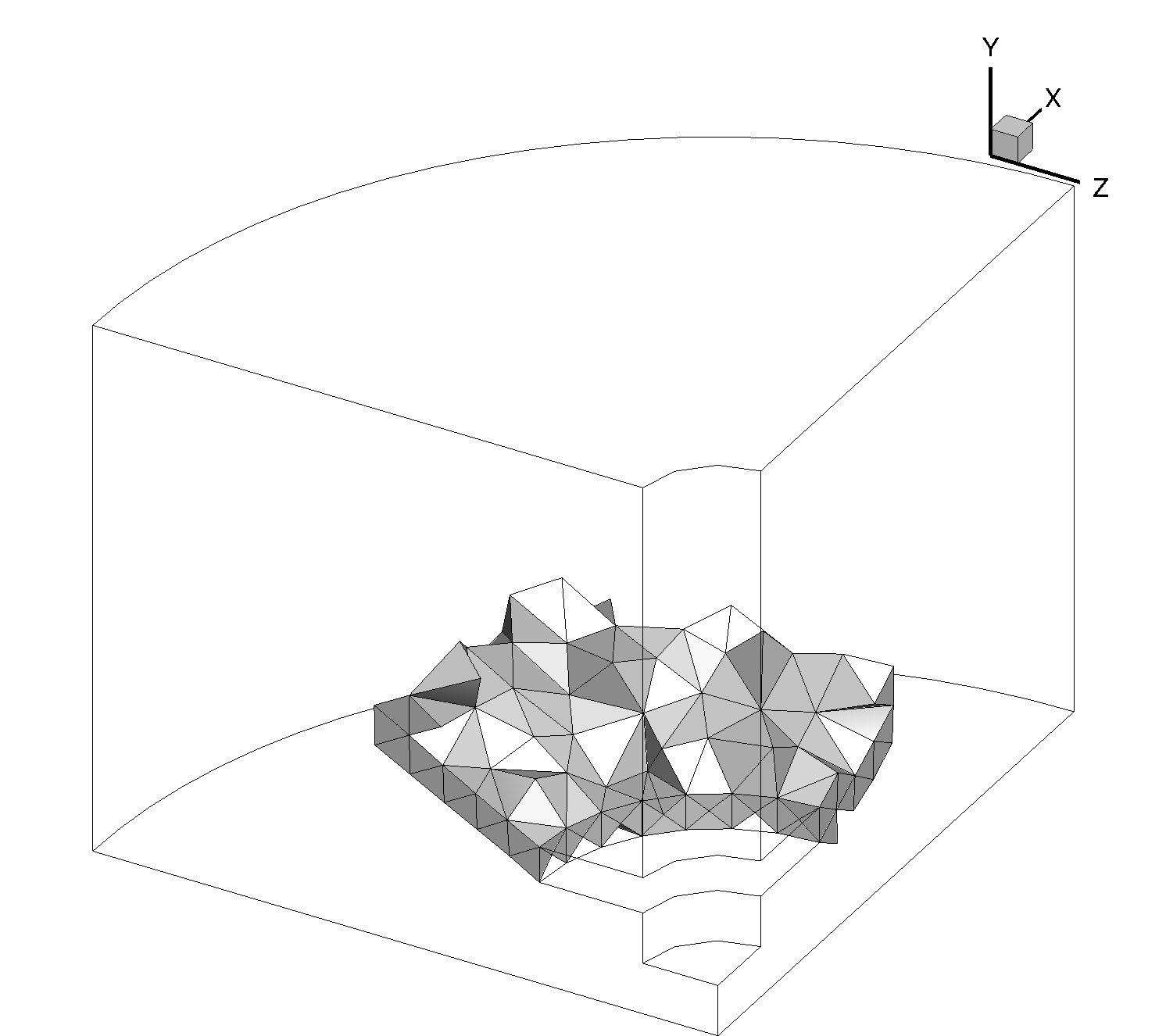}
            (a)
            \end{center}
        \end{minipage}
        \hfill
        \begin{minipage}{0.45\linewidth}
            \begin{center}
            \includegraphics[height=2.5in]{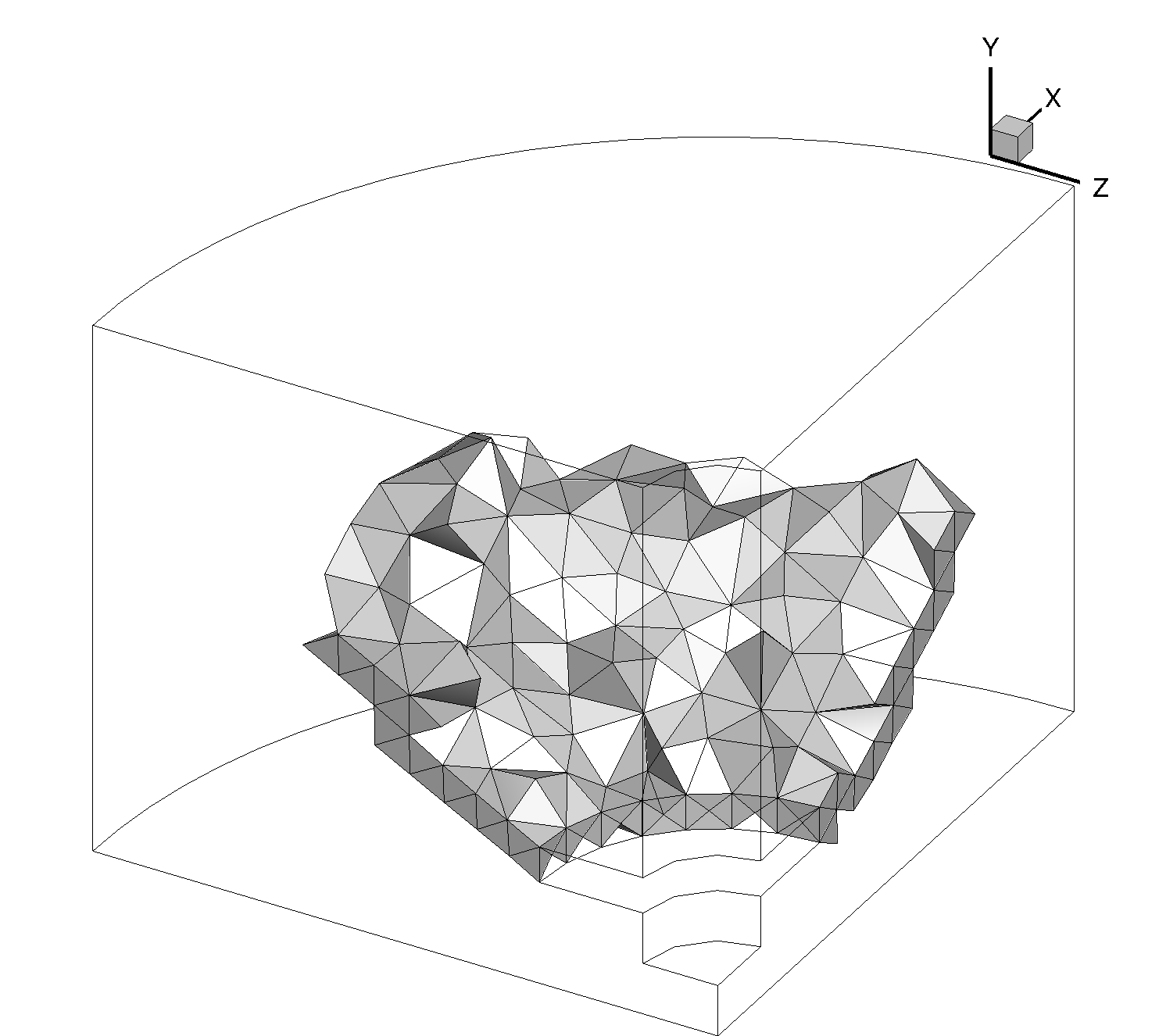}
            (b)
            \end{center}
        \end{minipage}   
        \hfill
        \begin{minipage}{0.45\linewidth}
            \begin{center}
            \includegraphics[height=2.5in]{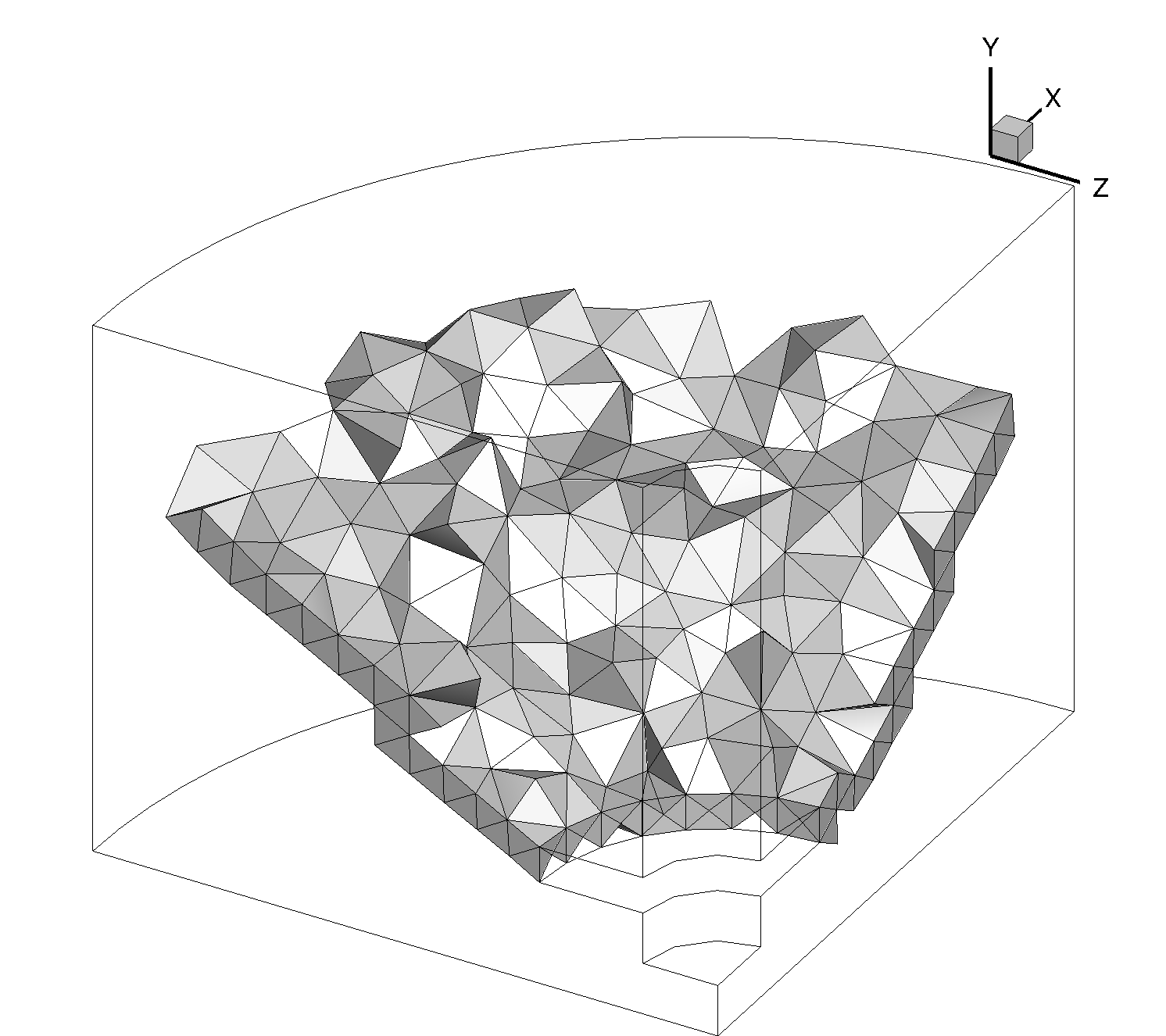}
            (c)
            \end{center}
        \end{minipage}
        \hfill
        \begin{minipage}{0.45\linewidth}
            \begin{center}
            \includegraphics[height=2.5in]{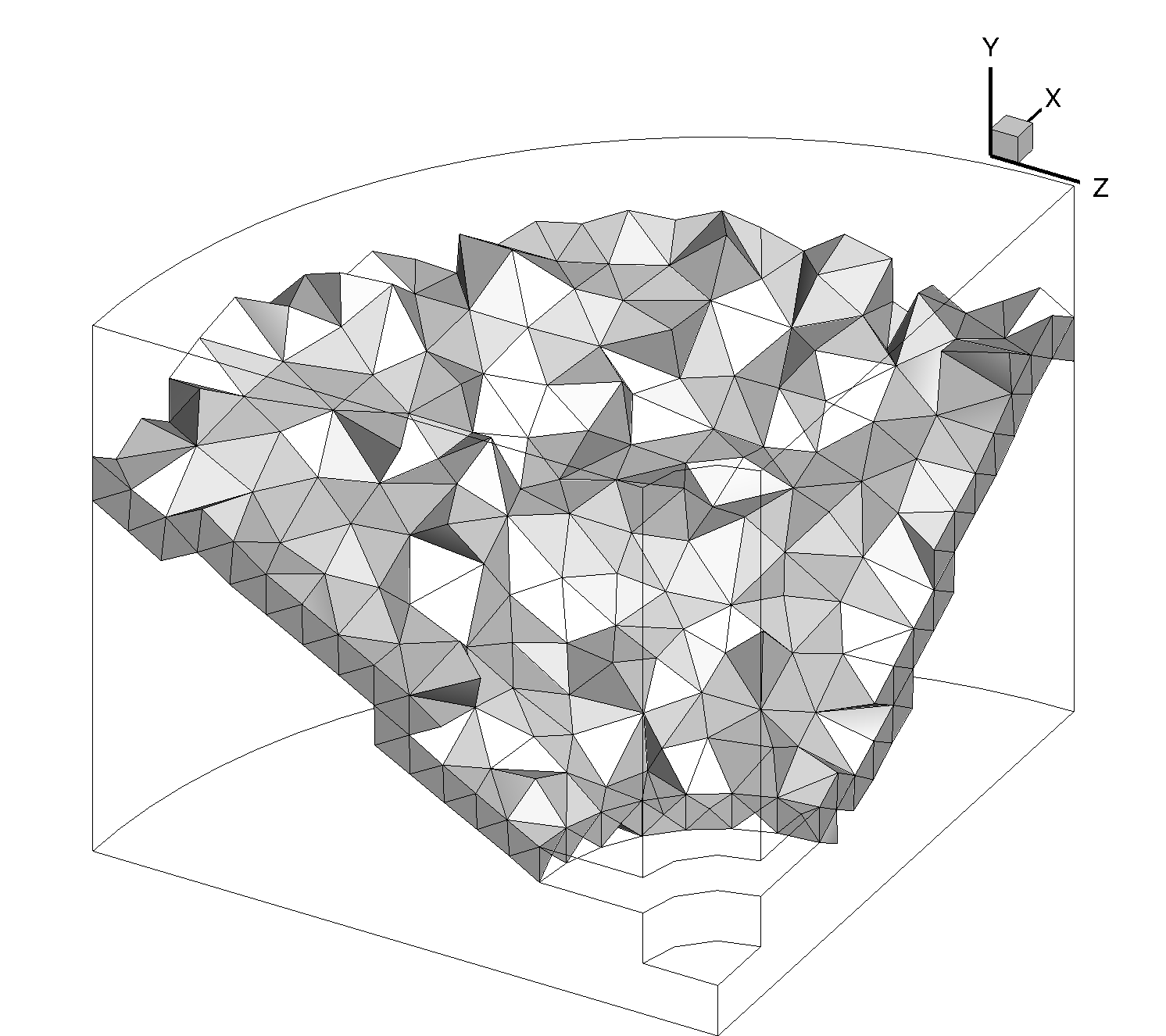}
            (d)
            \end{center}
        \end{minipage}
        \caption{The crack surface evolution for three-dimensional anchorage pull-out test with $9289$ elements at different times (a). $t=38.035\ ms$, (b). $t=38.244\ ms$, (c). $t=38.448\ ms$, (d). $t=42.5\ ms$.}
        \label{fig12: anchorage-crack-evolution-1}
\end{figure}

Figure.\ref{fig10: anchorage-crack-evolution-3} - \ref{fig12: anchorage-crack-evolution-1} show the crack surface evolution over time for models with $36866$, $20532$, and $9289$ elements, respectively.
As shown in Figure.\ref{fig10: anchorage-crack-evolution-3} - \ref{fig12: anchorage-crack-evolution-1}, the cracking process consistently initiates at the lower interface of the concrete block and propagates upward toward the top surface. Ultimately, the resulting failure surface takes on a conical shape. The crack patterns from all three mesh configurations exhibit similar geometries and show strong agreement with the results of the quasi-static analysis (\cite{areias2005analysis}, \cite{gasser2005modeling}). However, unlike the fine mesh and medium mesh, the coarse mesh with $9289$ elements delays crack initiation and generates a crack surface which penetrates the whole anchorage body.

\begin{figure}[htp]
	\centering
        \begin{minipage}{0.45\linewidth}
            \begin{center}
            \includegraphics[height=2.4in]{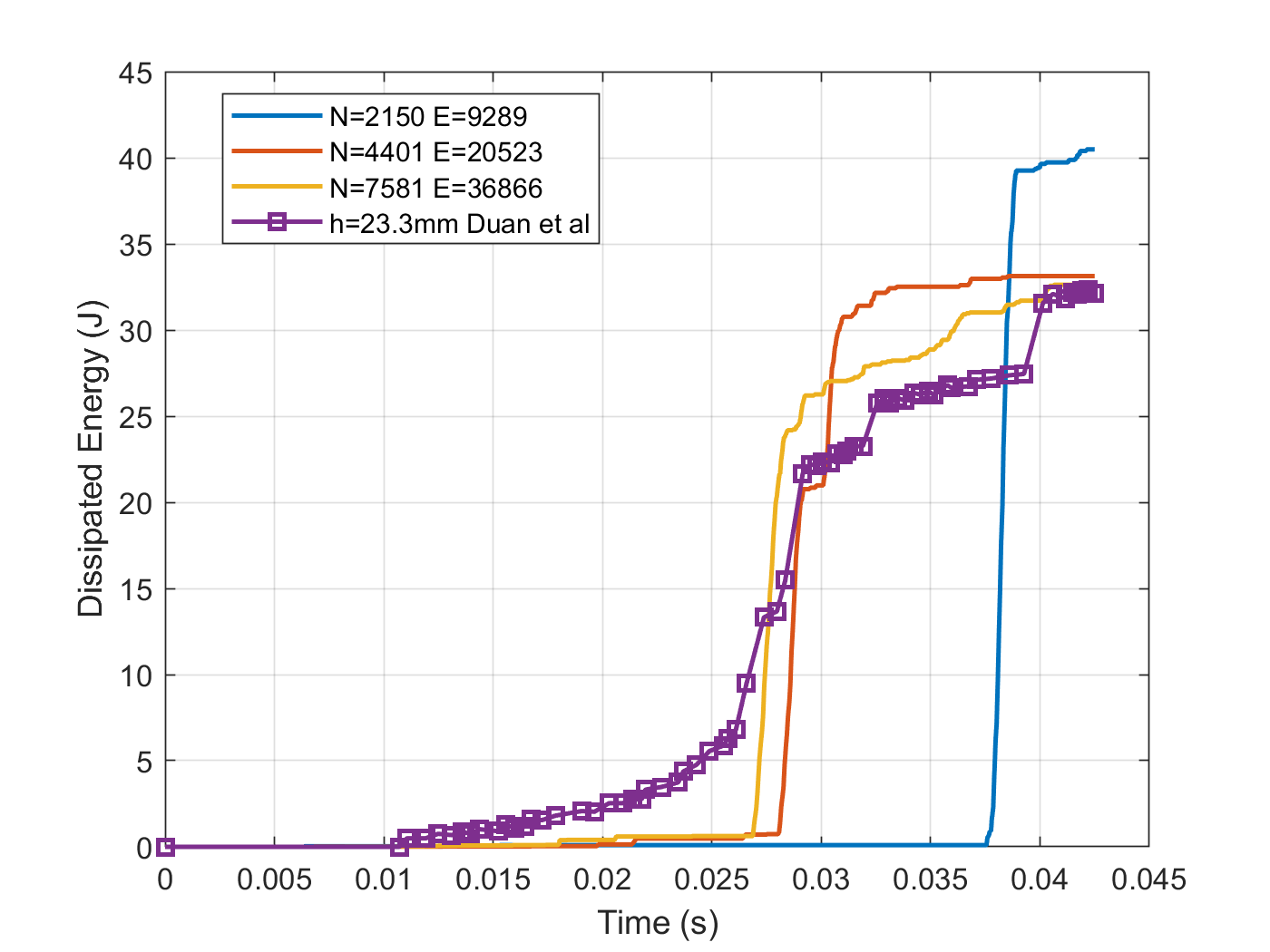}
            \end{center}
            \begin{center}
            (a)
            \end{center}
        \end{minipage}
        \hfill
        \begin{minipage}{0.45\linewidth}
        \begin{center}
        \includegraphics[height=2.4in]{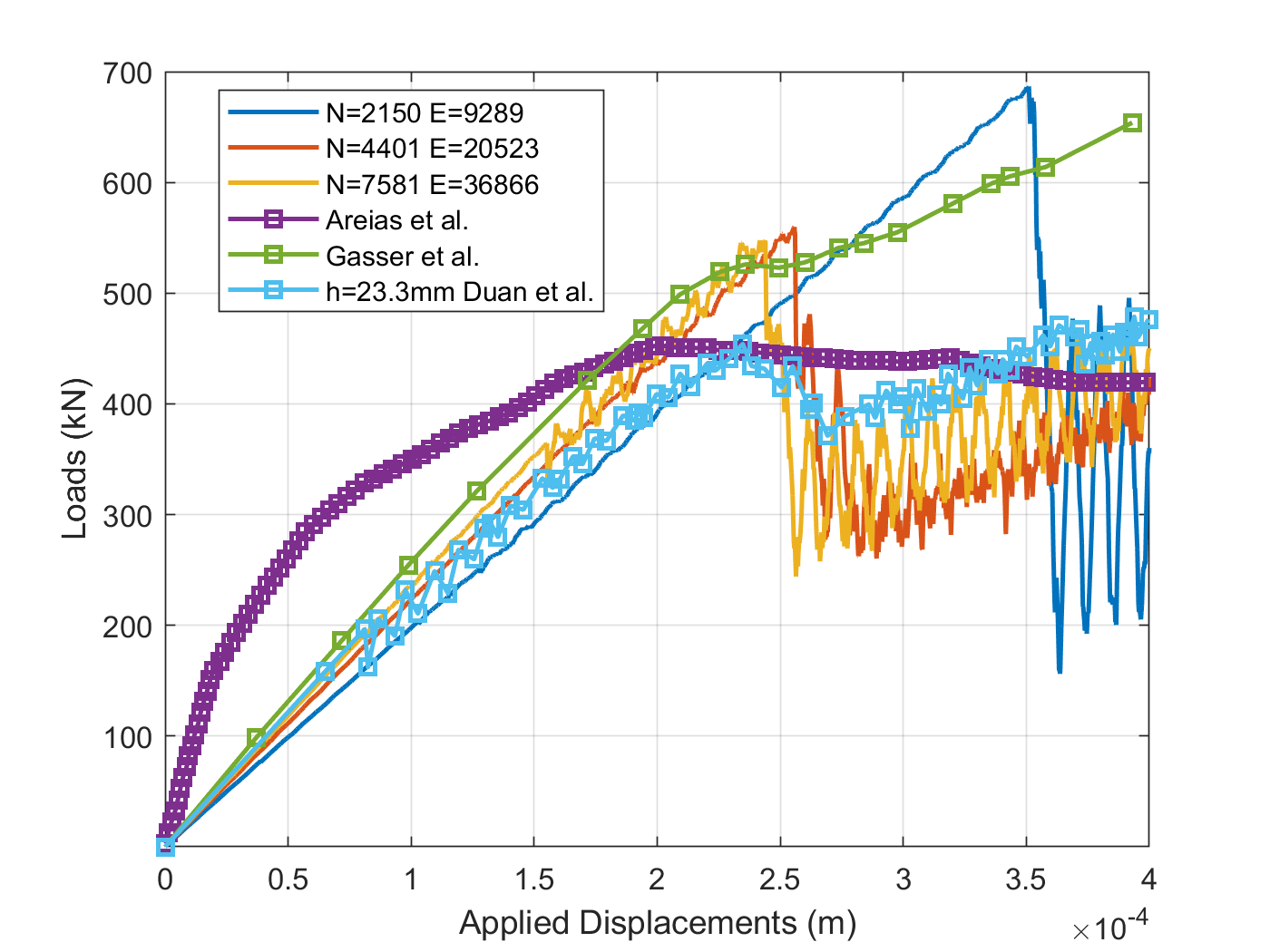}
        \end{center}
        \begin{center}
        (b)
        \end{center}
        \end{minipage}
        \caption{Comparison of (a). dissipated energy and (b). load-displacement of loading points among three representative meshes and reference results (\cite{areias2005analysis},\cite{duan2009element},\cite{gasser2005modeling}).}
        \label{fig13: anchorage-Ud-reaction}
\end{figure}

The dissipated energy evolution of three different meshes are shown in Figure.\ref{fig13: anchorage-Ud-reaction}(a). It is found that the crack surfaces in fine mesh and medium mesh start to initiate and quickly grow at time interval of $0.025$ seconds and $0.03$ seconds. The dissipated energy stabilizes at around $32.5$ J, which aligns with the results of Duan et al.(\cite{duan2009element}). In contrast, the coarse mesh shows delayed crack initiation and propagation at $0.035$ seconds and $0.04$ seconds, with a final dissipated energy of $40$ J, noticeably higher than the other two meshes.

In addition to the dissipated energy evolution, the load-displacement curves (upper surface of steel disc in Figure.\ref{fig8: anchorage-geometry}) are also studied, as shown in Figure.\ref{fig13: anchorage-Ud-reaction}(b). It is found that the fine mesh and medium mesh results initially coincide with results from the studies of Duan et al.  (\cite{duan2009element}) and Gasser et al. (\cite{gasser2005modeling}). At a displacement of $2.5 \times 10^{-4} m$, the loads suddenly drop due to accelerated crack surface evolution. Then the loads increase again at $3.0 \times 10^{-4} m$ following further crack propagation. Such "hardening" phenomenon is also observed in the results of Gasser and Duan, whereas Areias's result does not exhibit this behavior. It is noteworthy that Gasser et al. (\cite{gasser2005modeling}) considers elastic constitutive relation while Areias et al. (\cite{areias2005analysis}) includes plastic effects. On the other hand, the coarse mesh produces higher loads than any other cases, consistent with its fully penetrated crack surfaces shown in Figure.\ref{fig12: anchorage-crack-evolution-1}(d). 

\begin{figure}[htp]
	\centering
        \begin{minipage}{0.3\linewidth}
            \begin{center}
            \includegraphics[height=2.0in]{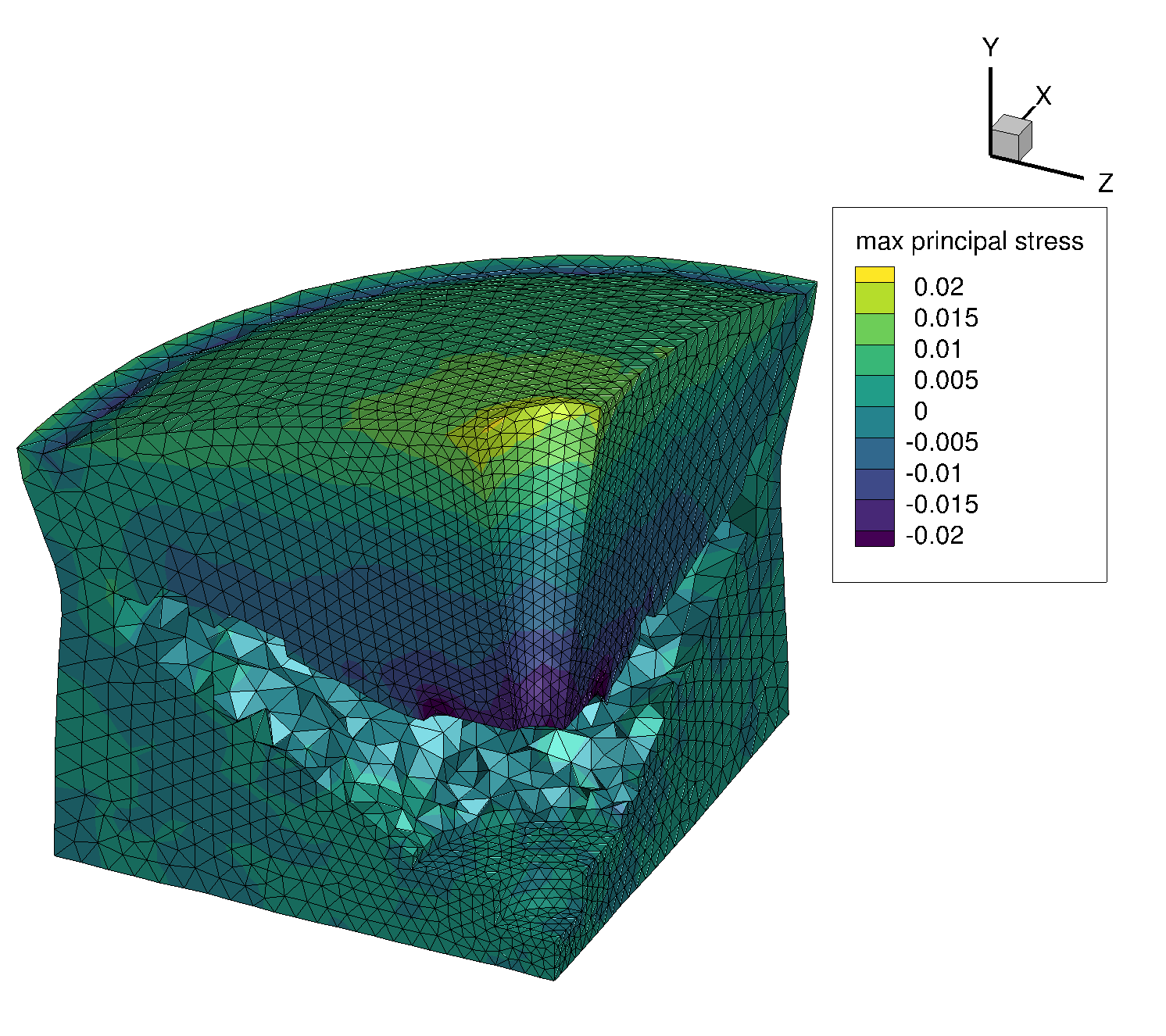}
            \end{center}
            \begin{center}
            (a)
            \end{center}
        \end{minipage}
        \hfill
        \begin{minipage}{0.3\linewidth}
            \begin{center}
            \includegraphics[height=2.0in]{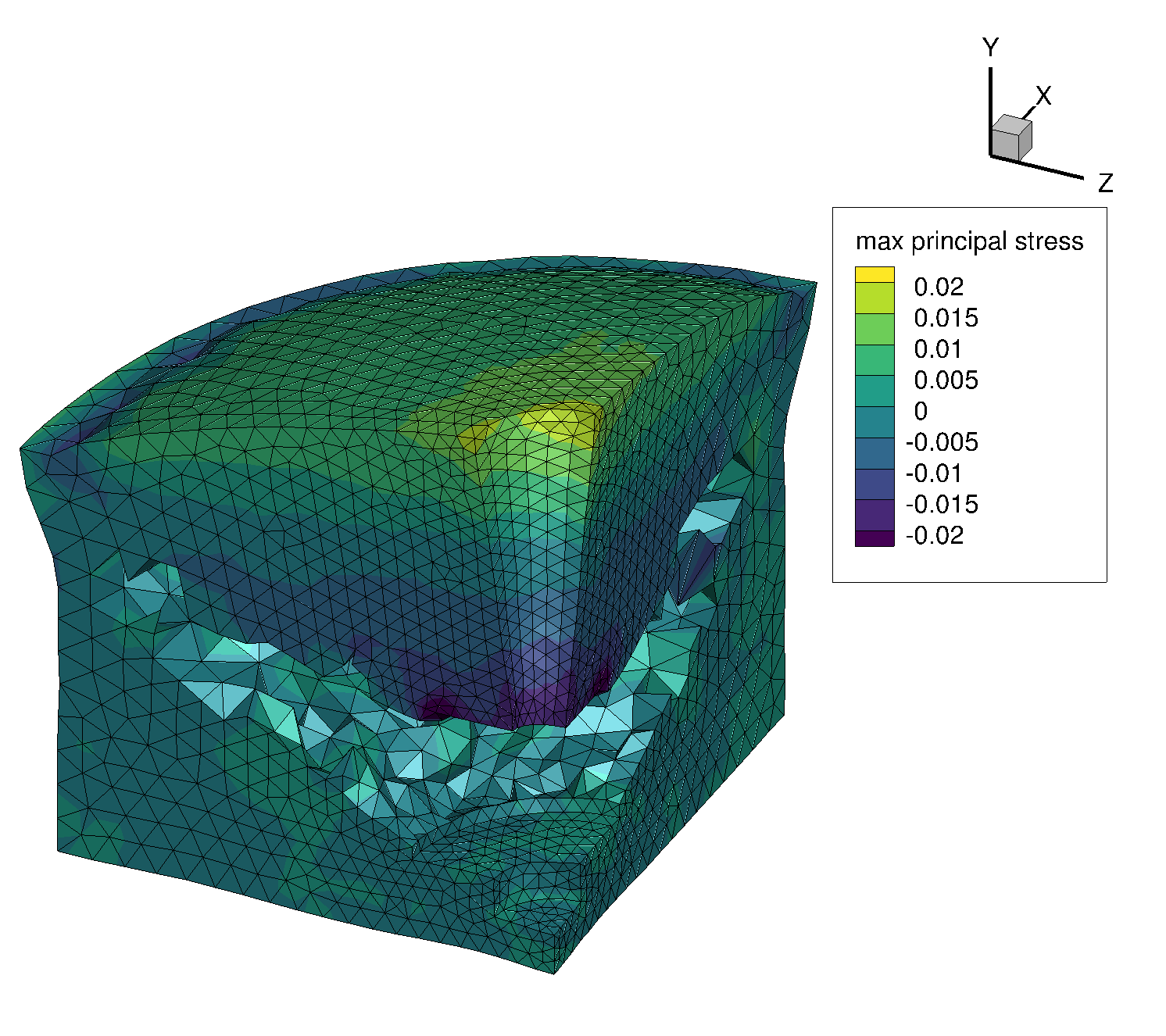}
            \end{center}
            \begin{center}
            (b)
            \end{center}
        \end{minipage}   
        \hfill
        \begin{minipage}{0.3\linewidth}
            \begin{center}
            \includegraphics[height=2.0in]{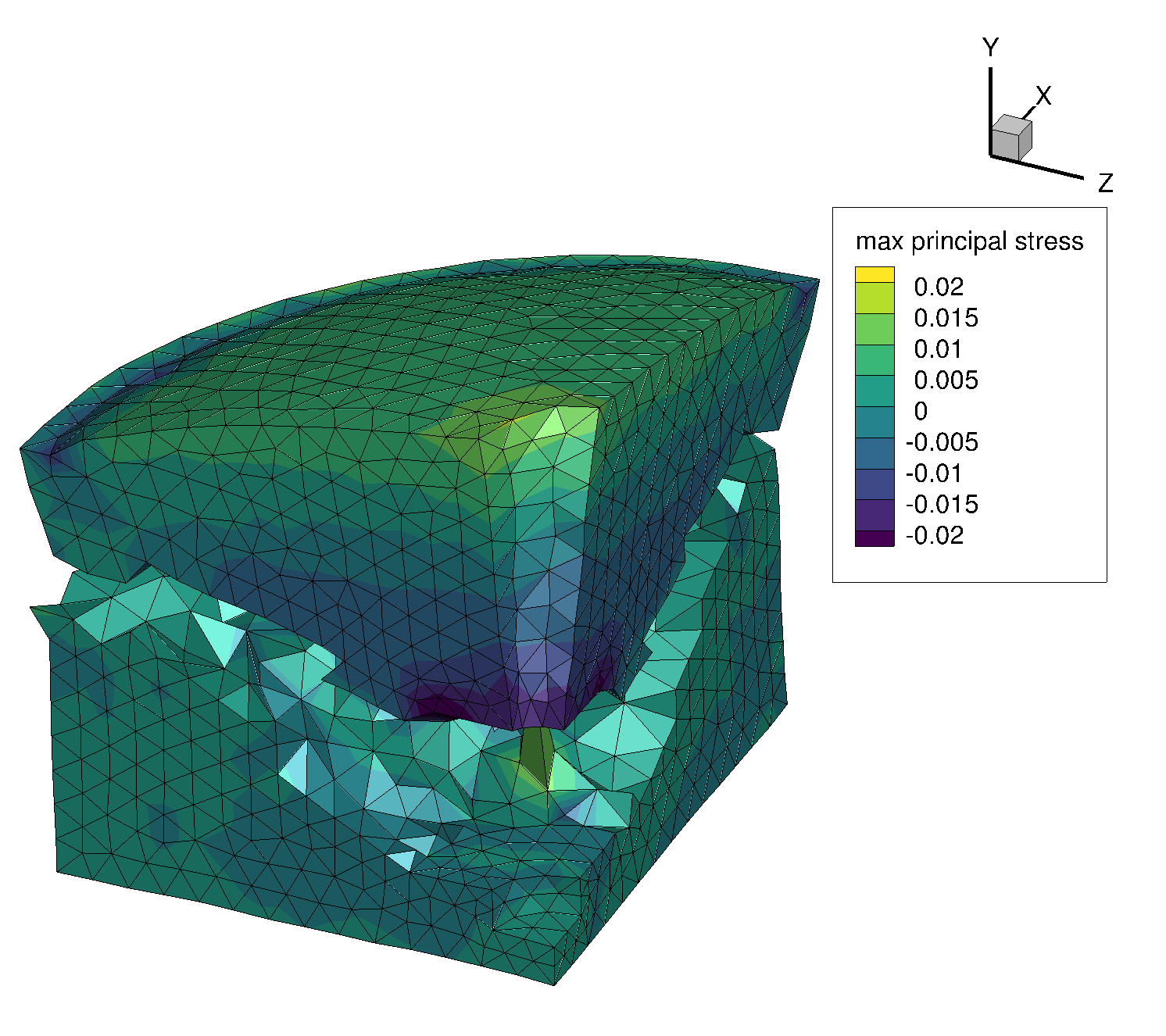}
            \end{center}
            \begin{center}
            (c)
            \end{center}
        \end{minipage}
        \caption{The deformed configuration (magnified by 500 times) of anchorage structure with three different meshes: (a). $36866$ elements; (b). $20523$ elements; (c). $9289$ elements.}
        \label{fig14: anchorage-defomed-magnified}
\end{figure}

Figure.\ref{fig14: anchorage-defomed-magnified} shows the final deformed shape of the concrete anchorage, with the displacement field magnified $500$ times, which aligns with the results of \cite{gasser2005modeling}. In the coarse mesh (see Figure.\ref{fig14: anchorage-defomed-magnified}(c)), the crack surface is found to have fully penetrated the anchorage structure.

In summary, when using relatively fine mesh (with over $20000$ elements), the proposed CEM can generate qualitatively accurate crack surfaces. Additionally, the dissipated energy and load-displacement curves show good quantitative agreement with the results of Duan et al., whose anchorage pull-out test was also conducted under transient-dynamic condition.

\subsection{Compact Compression Specimen}
This numerical example is to demonstrate the CEM's highly accurate and sensitive capability of capturing crack tip in dynamic simulation: even curved and arbitrary path can still be traced easily. The PMMA (Polymethyl methacrylate) specimen was experimented by Rittel et al. (\cite{rittel1996investigation}). Followed by numerical simulation, a lot of researchers validated or studied this experiment by their proposed modeling methods, including cohesive models (\cite{paulino2010adaptive}), Extedned FEM (\cite{menouillard2008mass}) and adaptive splitting Finite Elements (\cite{leon2014reduction}). However, to authors' knowledge, very few studies on the three-dimensional case due to more difficult to follow the curved crack surface in three-dimension than the curved crack line in two-dimension.

\begin{figure}[htp]
	\centering
        \begin{minipage}{0.45\linewidth}
            \centering
            \includegraphics[height=2.8in]{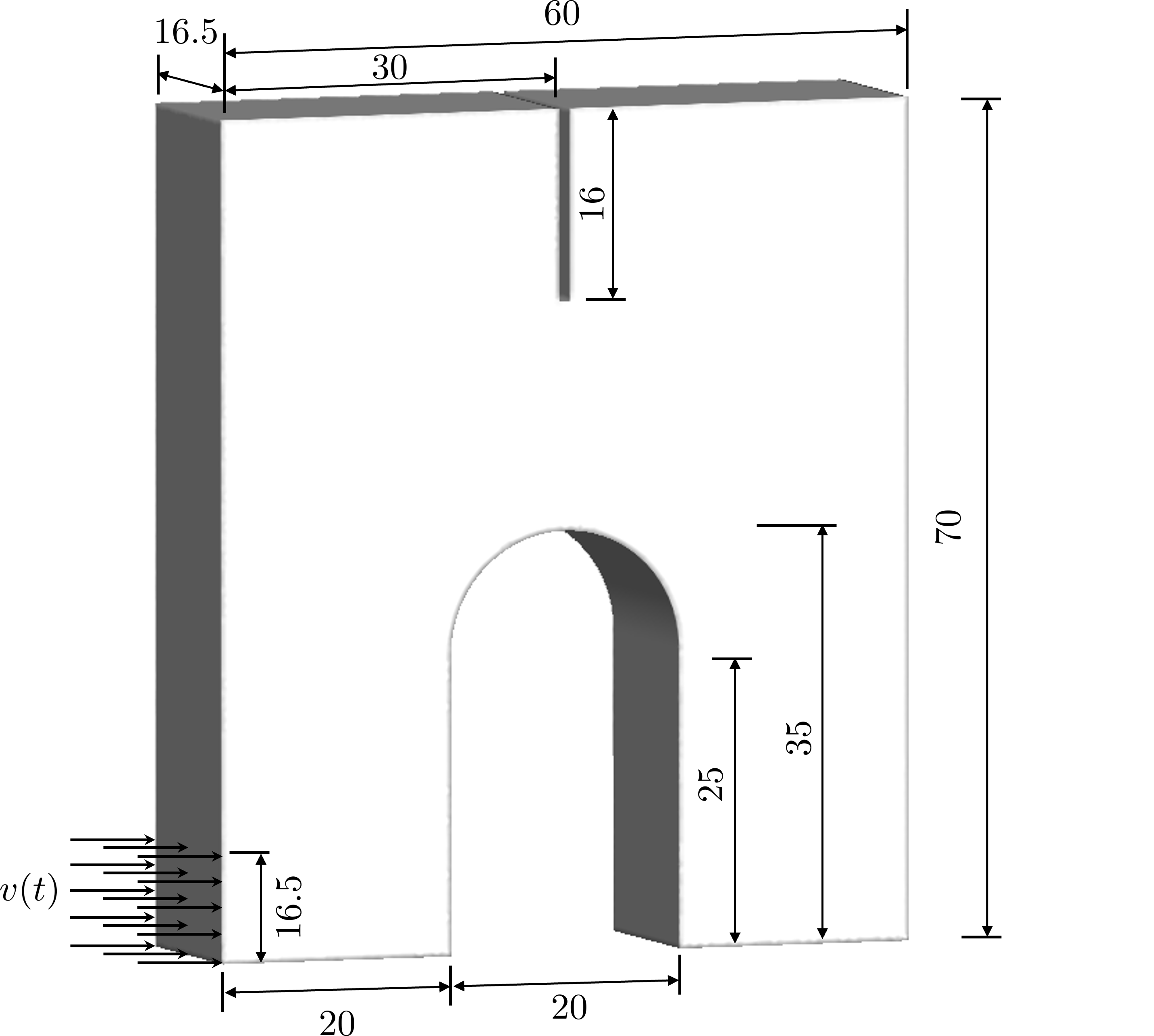}
            \begin{center}
            (a)
            \end{center}
        \end{minipage}
        \hfill
        \begin{minipage}{0.45\linewidth}
            \centering
            \includegraphics[height=2.8in]{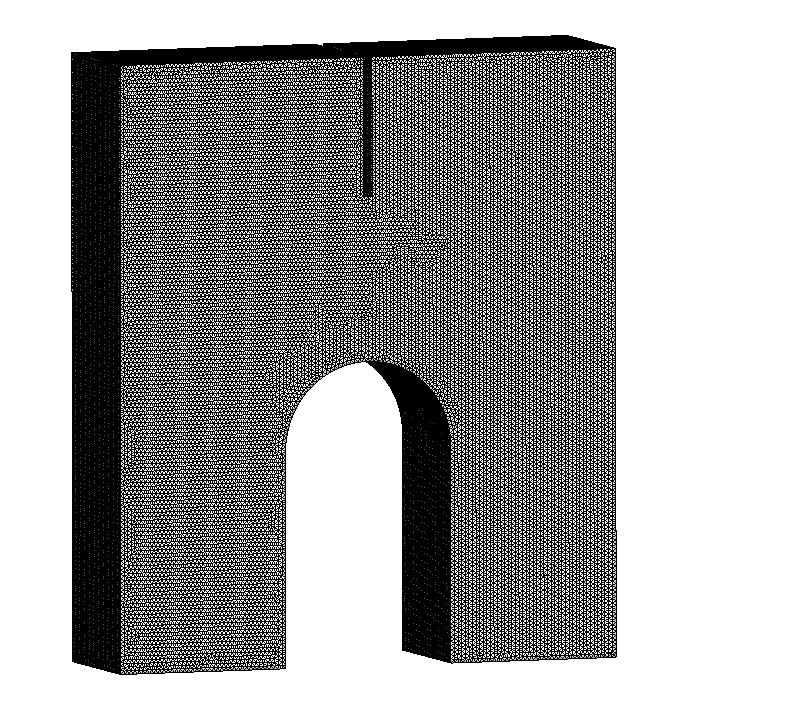}
            \begin{center}
            (b)
            \end{center}
        \end{minipage}   
        \hfill
        \begin{minipage}{0.45\linewidth}
            \centering
            \includegraphics[height=2.8in]{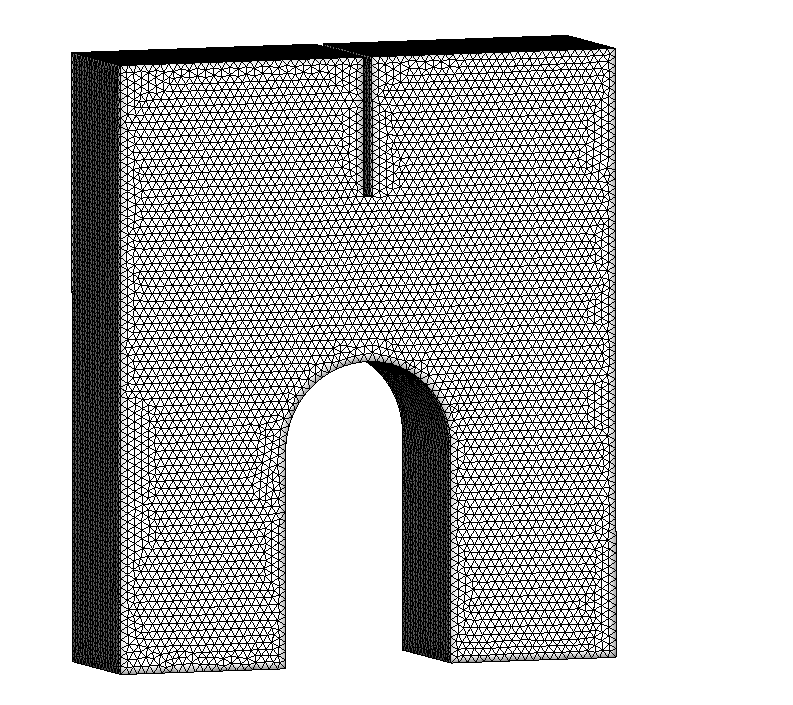}
            \begin{center}
            (c)
            \end{center}
        \end{minipage}  
        \hfill
        \begin{minipage}{0.45\linewidth}
            \centering
            \includegraphics[height=2.8in]{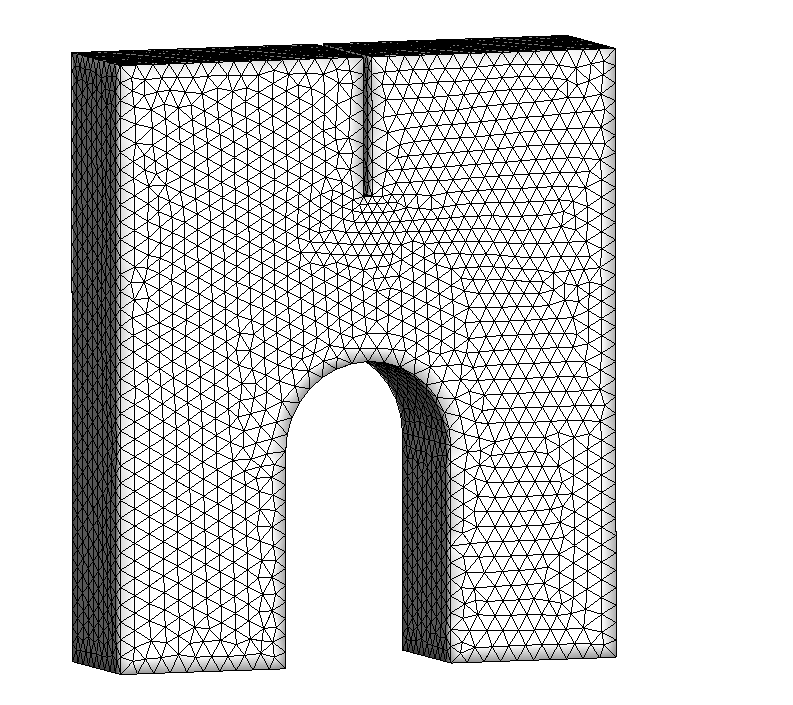}
            \begin{center}
            (d)
            \end{center}
        \end{minipage}  
        \caption{(a). The model geometric dimensions (unit: mm) and boundary conditions of compact compression experiment. (b). The mesh with $2069141$ elements; (c). The mesh with $267486$ elements; (d). The mesh with $39509$ elements.}
        \label{fig15: ccompression-model-grids}
\end{figure}

The model geometric dimensions are provided in Figure.\ref{fig15: ccompression-model-grids}(a). The impact by Hopkinson bar in experiment is located at left bottom and is modeled by prescribed velocity as 
\begin{equation}
v =
\begin{cases} 
\frac{t}{t_0}v_0,  & \text{if }\ t \le t_0, \\
v_0, & \text{if } \ t > t_0.
\end{cases}
\end{equation}
in which, $v_0 = 20\ m/s$ and $t_0 = \ 40 \mu s$. Other boundary conditions of the model are unspecified. 

The material properties utilized in simulation are as follows (see \cite{asareh2020general}): Young's modulus $E = 5.76 \ GPa$, Poisson ration $\nu = 0.42$, critical fracture energy release rate $\mathcal{G}_c = 352.3\ J/m^2$, density $\rho_0 = 1180\ kg/m^3$, tensile strength $\sigma_c = 129.6 \ MPa$ and Rayleigh wave speed based on material constants given above is $v_R = 1237.5\ m/s$. Total simulation time of the experiment process is $T = 140\times 10^{-6} s$ and a fixed time step $\Delta t = 1.0 \times 10^{-8} s$. Three different meshes, including 360078 nodes with 2069141 elements, 50661 nodes with 267486 elements and 8612 nodes with 39509 elements (see Figure.\ref{fig15: ccompression-model-grids}(b-d)), are generated to validate robustness of the proposed algorithm. 

\begin{figure}[htp]
	\centering
        \begin{minipage}{0.45\linewidth}
            \begin{center}
            \includegraphics[height=2.8in]{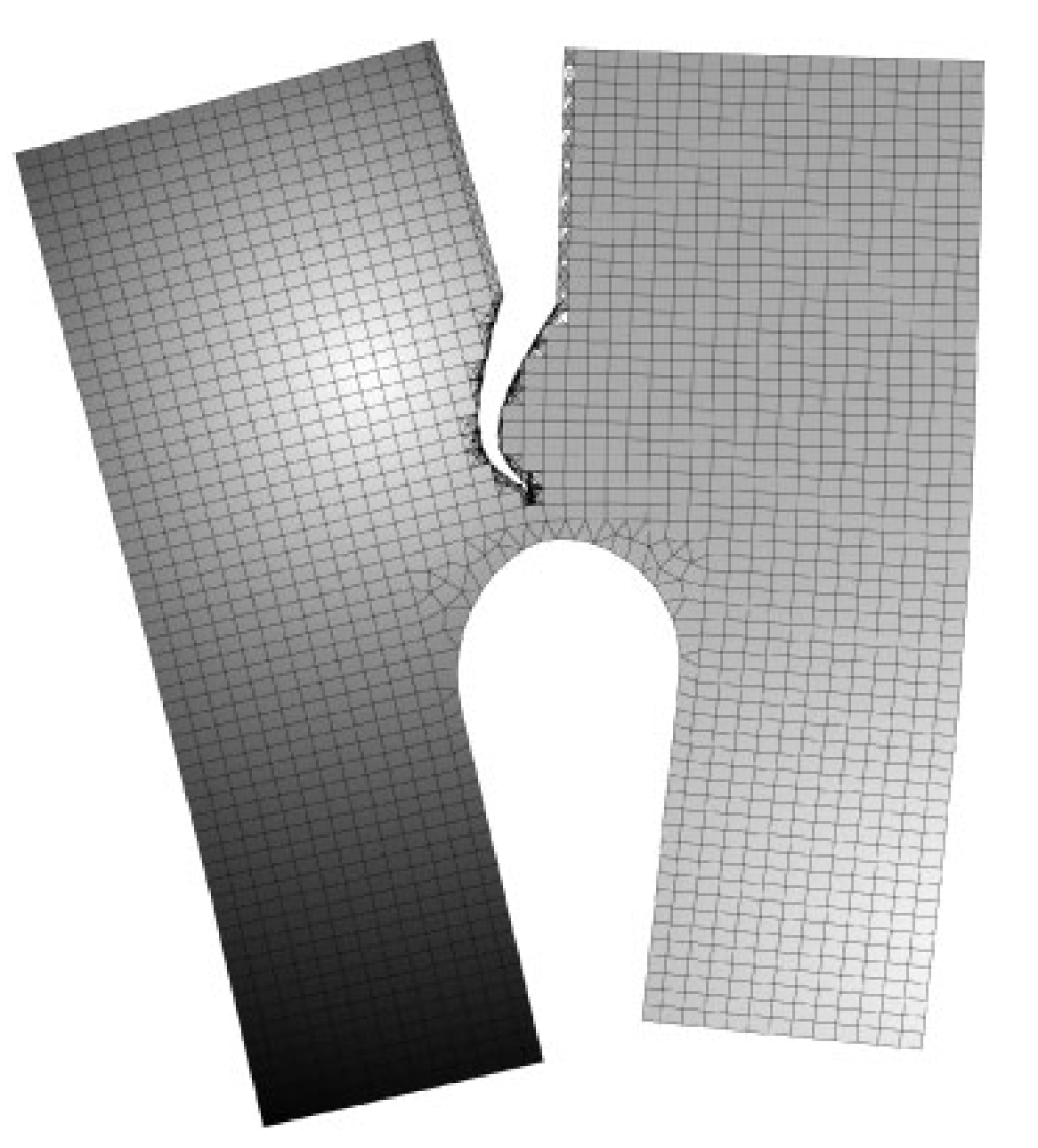}
            \end{center}
            \begin{center}
            (a)
            \end{center}
        \end{minipage}   
        \hfill
        \begin{minipage}{0.45\linewidth}
            \begin{center}
            \includegraphics[height=2.8in]{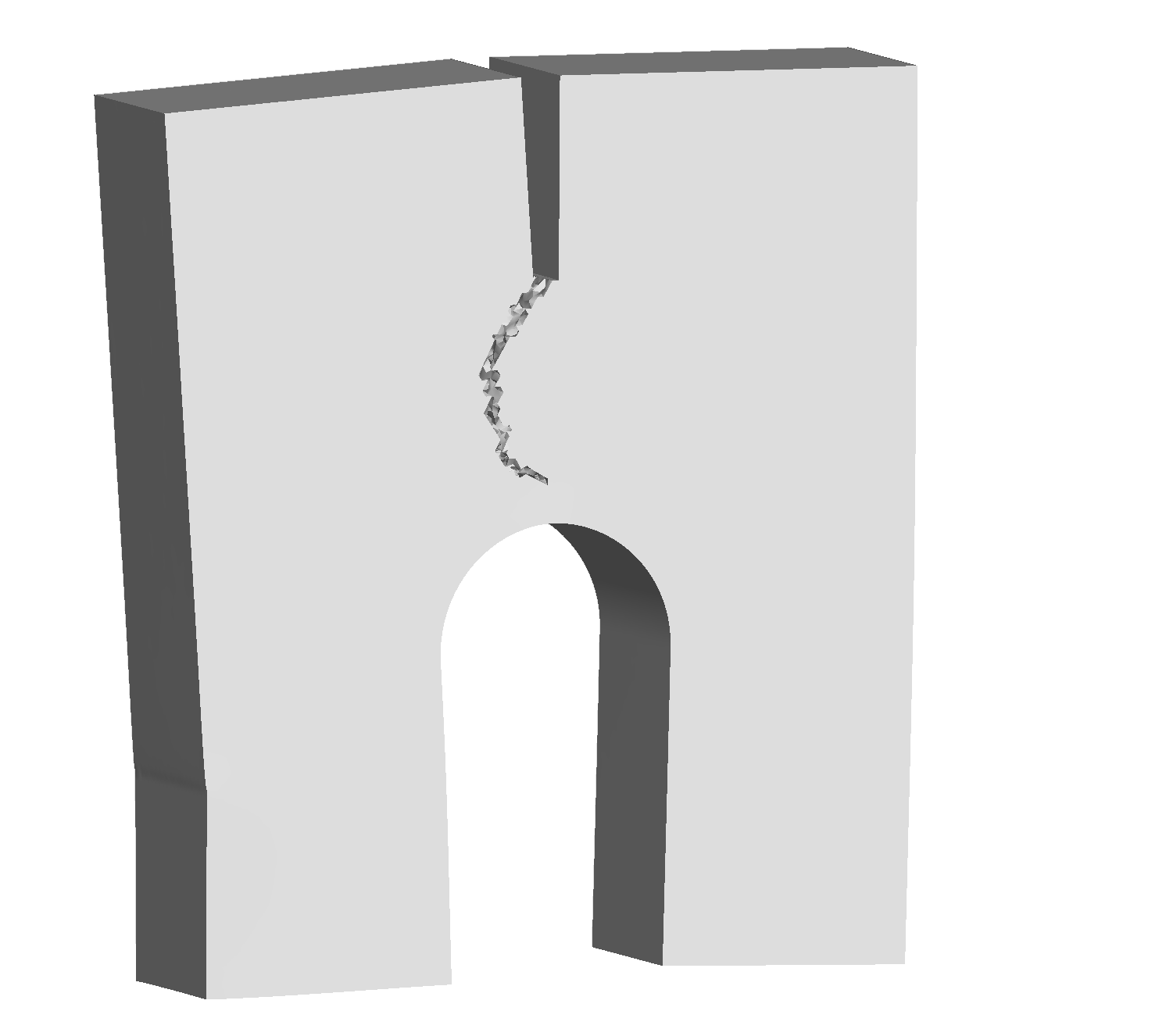}
            \end{center}
            \begin{center}
            (b)
            \end{center}
        \end{minipage}
        \hfill
        \begin{minipage}{0.45\linewidth}
            \begin{center}
            \includegraphics[height=2.8in]{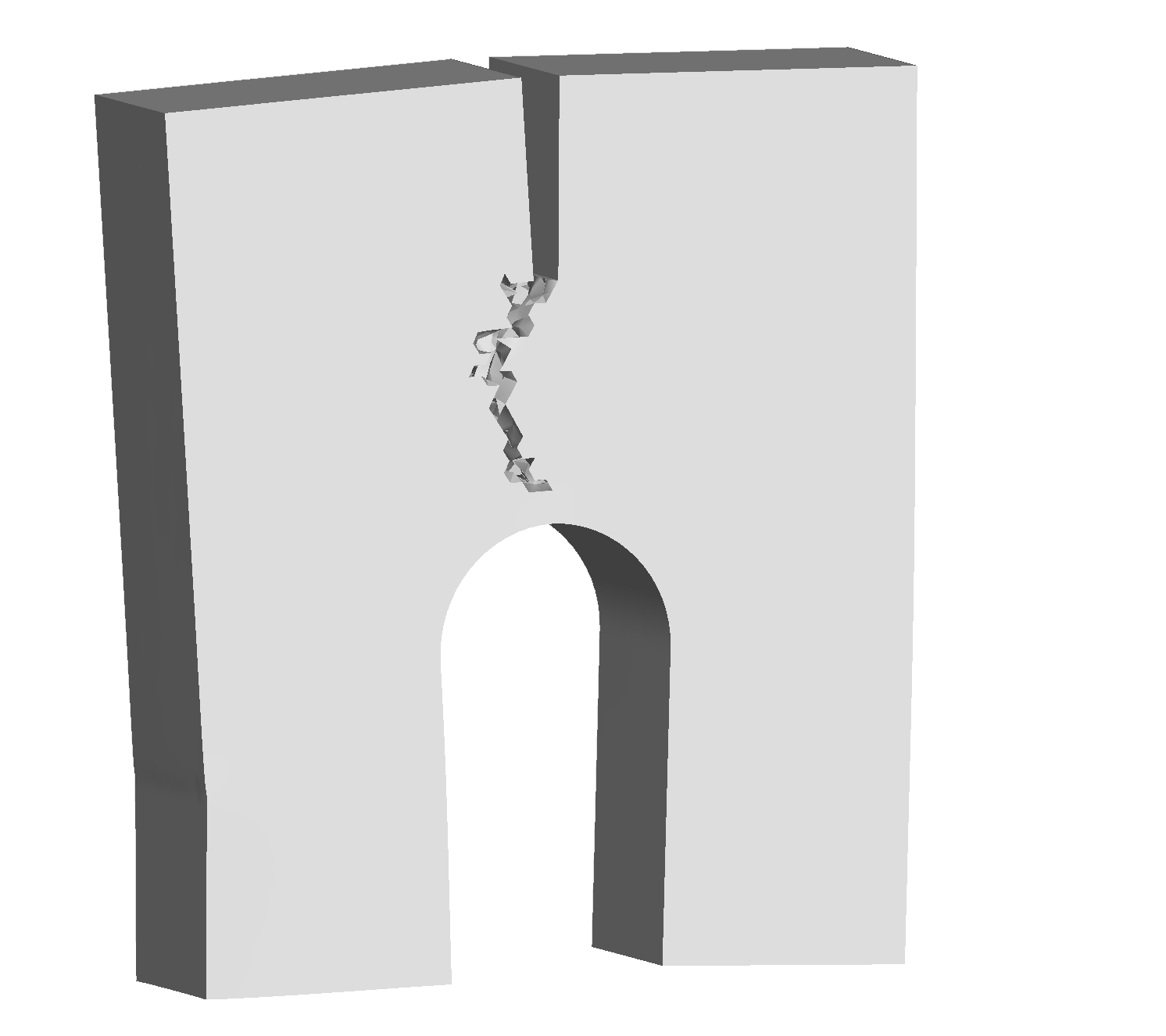}
            \end{center}
            \begin{center}
            (c)
            \end{center}
        \end{minipage}   
        \hfill
        \begin{minipage}{0.45\linewidth}
            \begin{center}
            \includegraphics[height=2.8in]{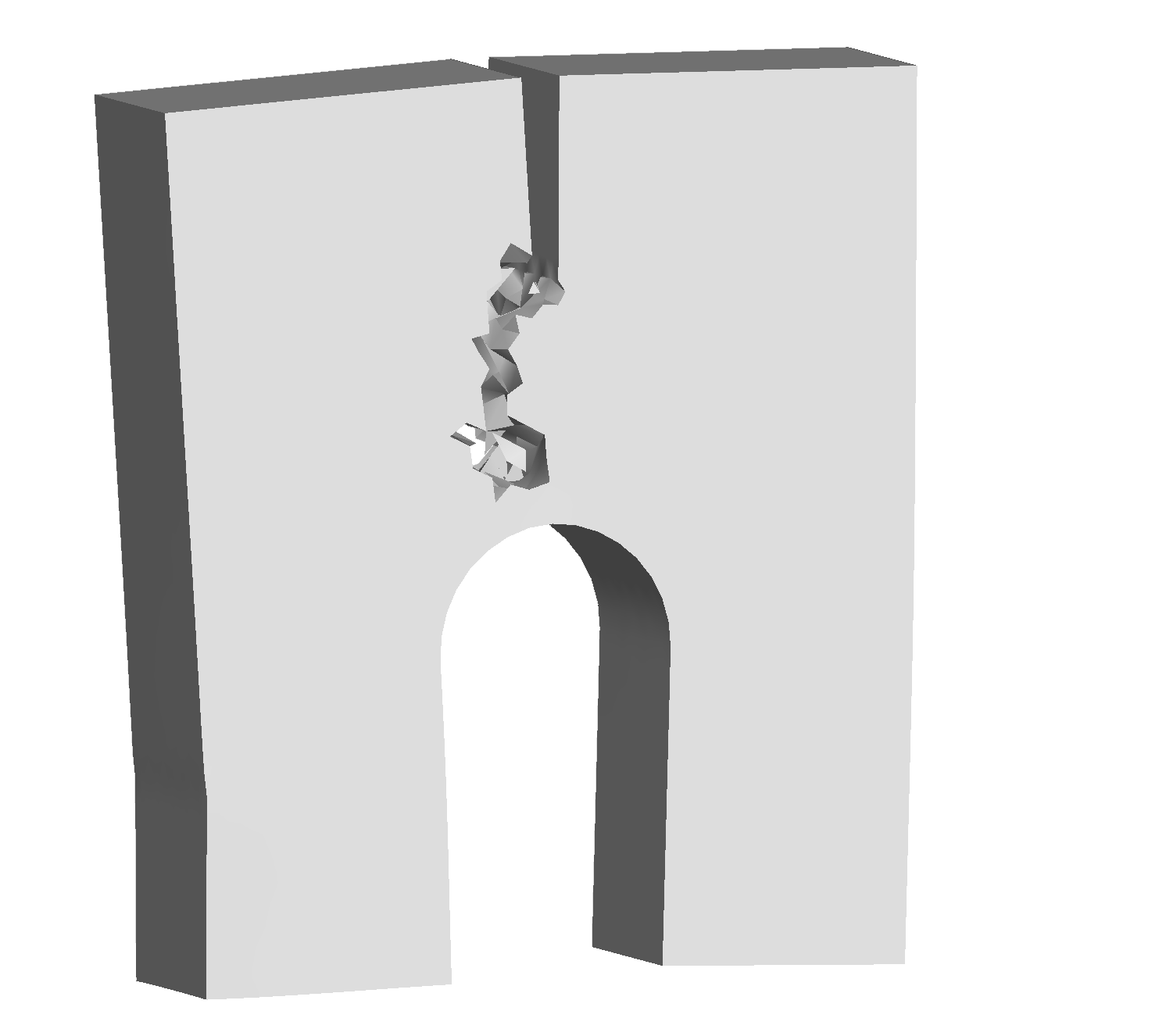}
            \end{center}
            \begin{center}
            (d)
            \end{center}
        \end{minipage}  
        \caption{(a). Reference crack path (see \cite{menouillard2006efficient}); (b). The crack path with $2069141$ elements; (c). The crack path with $267486$ elements; (d). The crack path with $39509$ elements.}
        \label{fig16: ccompression-crack-paths}
\end{figure}

Figure.\ref{fig16: ccompression-crack-paths} presents the crack propagation paths predicts by the three different meshes using the proposed method. Compared to crack patterns in the reference result (\cite{menouillard2006efficient}), it is shown that the curved, arc-shaped crack path is accurately captured by the CEM. It is concluded that the proposed crack tracking algorithm can reliably capture cracks even under very complicated three-dimensional stress fields. 

\begin{figure}[htp]
	\centering
            \begin{center}
            \includegraphics[height=2.4in]{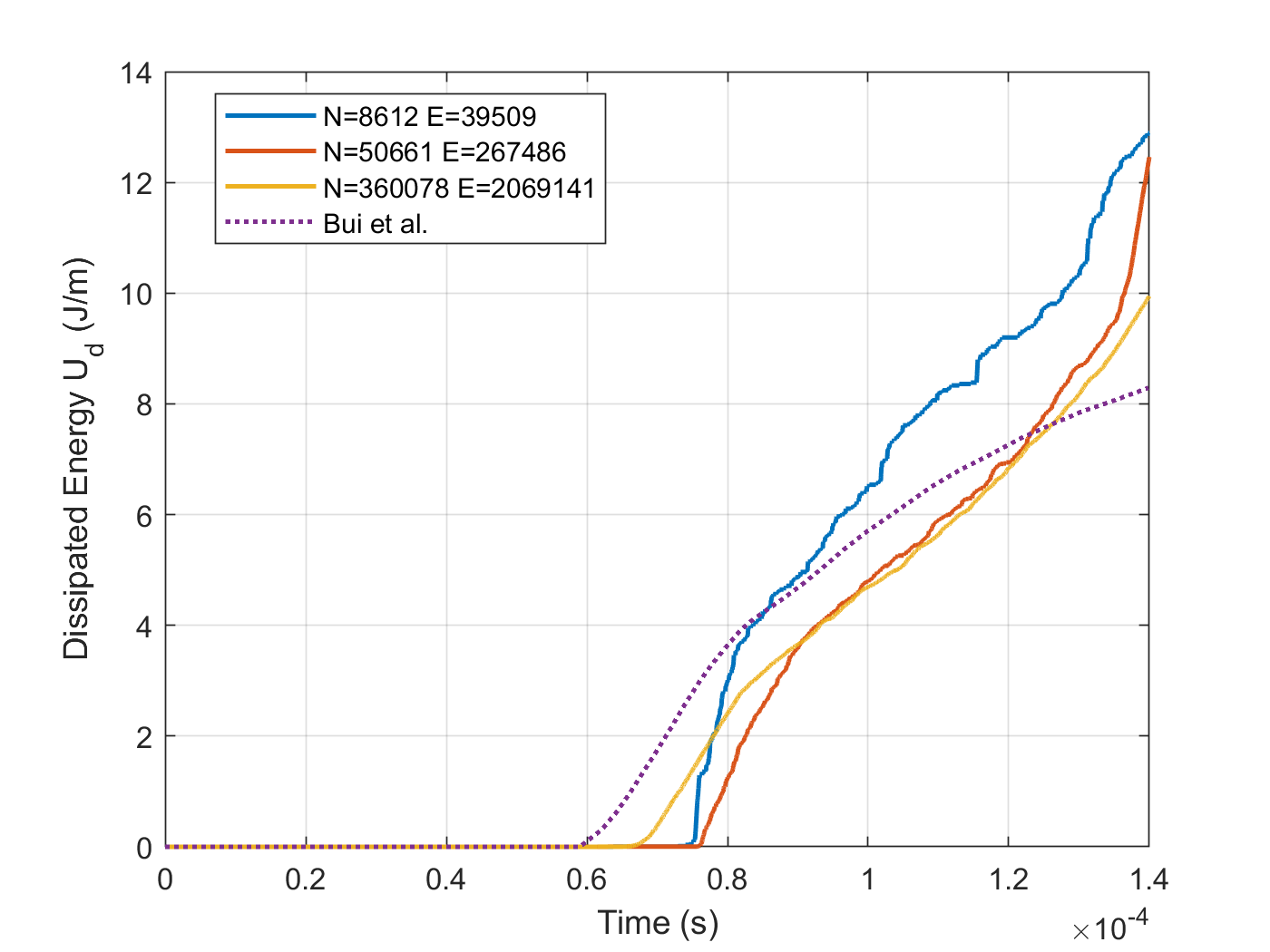}
            \end{center}
        \caption{Comparison of dissipated energy among three representative meshes and a reference result(\cite{tran2024nonlocal}).}
        \label{fig17: ccompression-energy}
\end{figure}

Since very few studies focus on three-dimensional compact compression test, a two-dimensional dissipated energy result (\cite{tran2024nonlocal}) is chosen for comparison. It is to note that three-dimensional results data from the proposed CEM need to be post-processed: three-dimensional dissipated energy $U_d$ is divided by $16.5\ mm$ thickness of the model so that three-dimensional dissipated energy unit $J$ is converted into two-dimensional unit $J/m$. Therefore a quantitative dissipated energy comparison result is provided in Figure.\ref{fig17: ccompression-energy}. It is found that the results from CEM using coarse mesh are higher than the result from Bui et al. (\cite{tran2024nonlocal}). As the finer meshes are used (from $39509$ elements to $2069141$ elements), the dissipated energy $U_d$ at the final stage decreases and converges toward numerical result from Bui et al.

\begin{figure}[htp]
	\centering
        \begin{minipage}{0.45\linewidth}
            \begin{center}
            \includegraphics[height=2.8in]{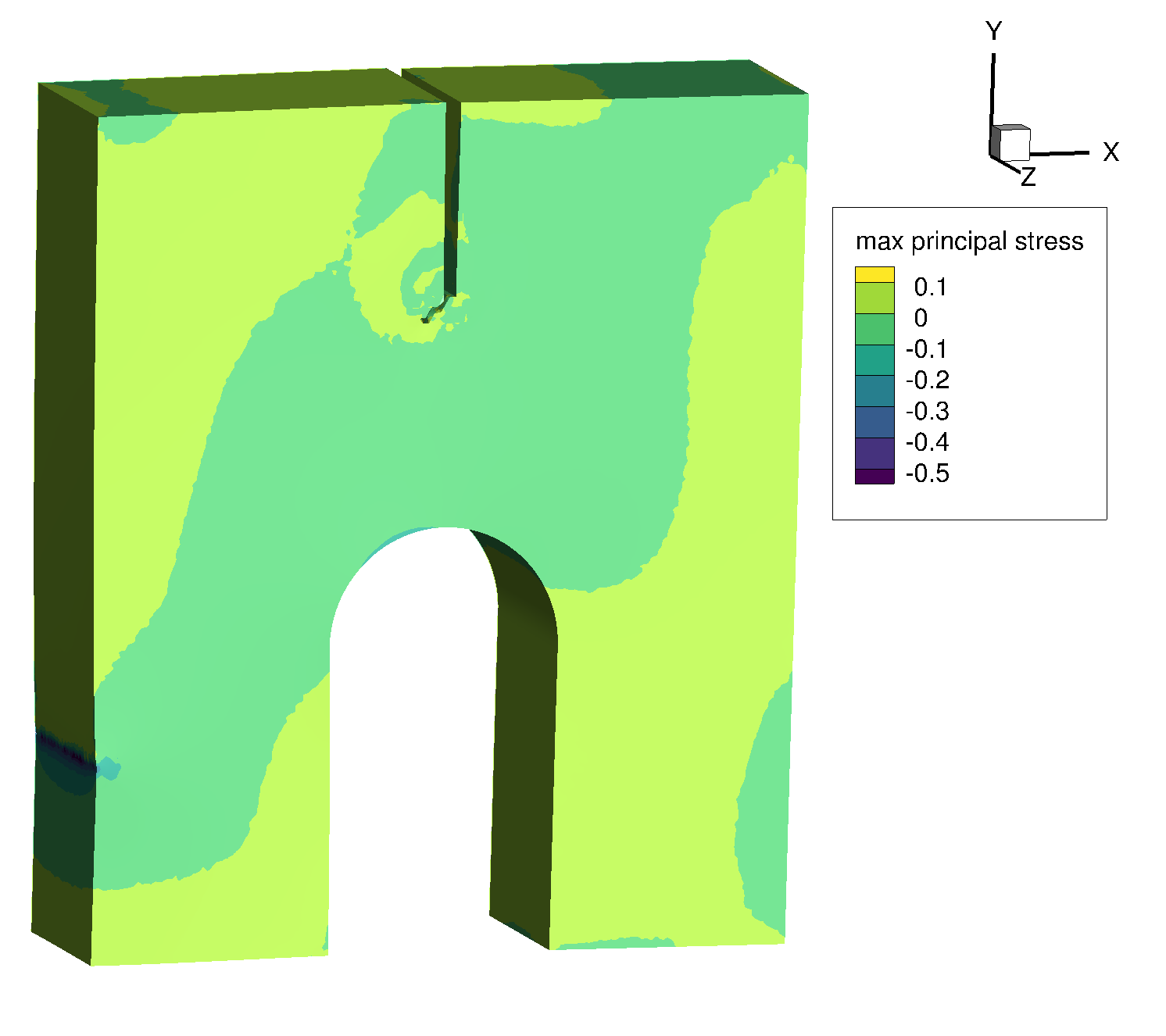}
            \end{center}
            \begin{center}
            (a)
            \end{center}
        \end{minipage}
        \hfill
        \begin{minipage}{0.45\linewidth}
            \begin{center}
            \includegraphics[height=2.8in]{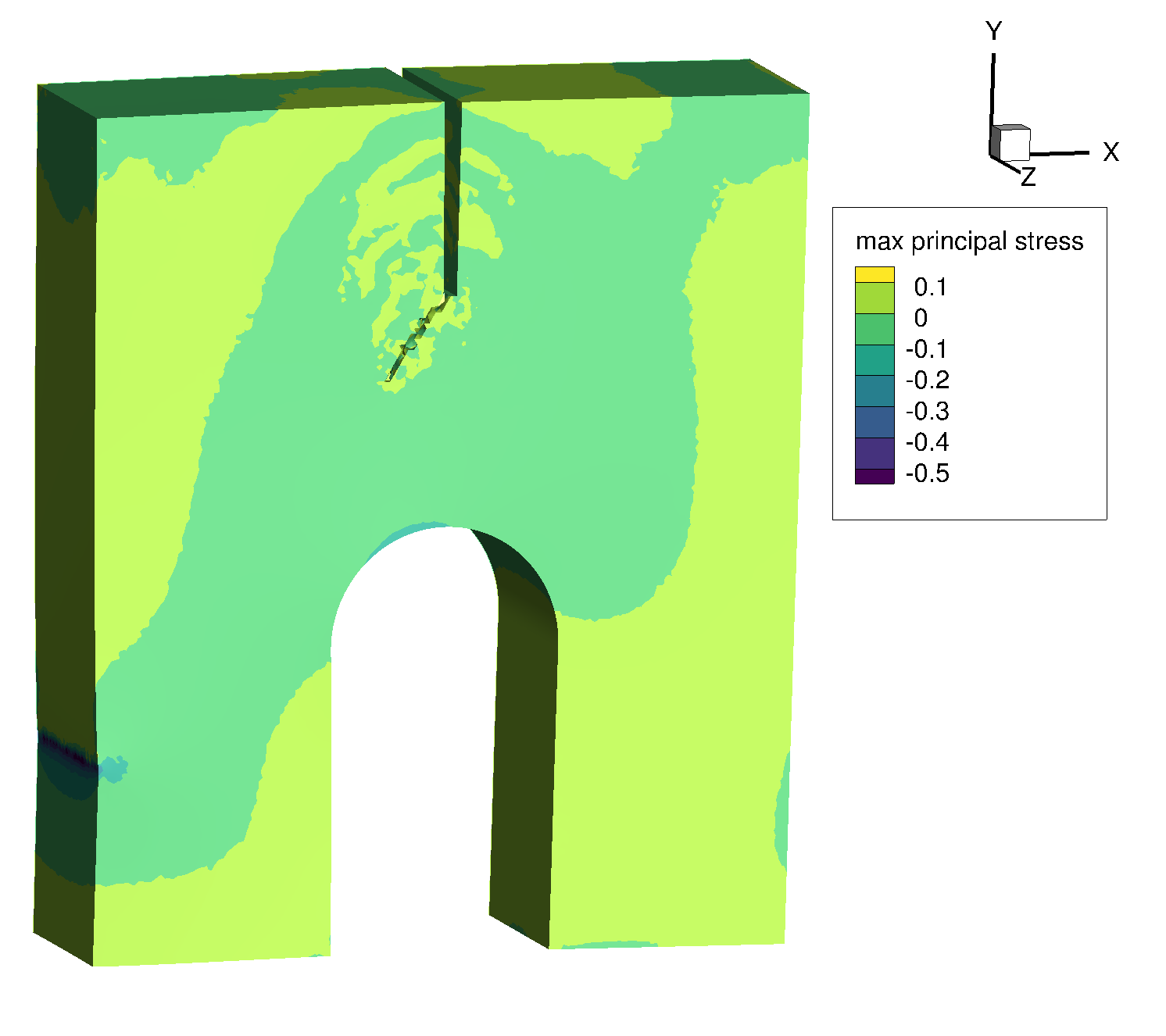}
            \end{center}
            \begin{center}
            (b)
            \end{center}
        \end{minipage}     
        \hfill
        \begin{minipage}{0.45\linewidth}
            \begin{center}
            \includegraphics[height=2.8in]{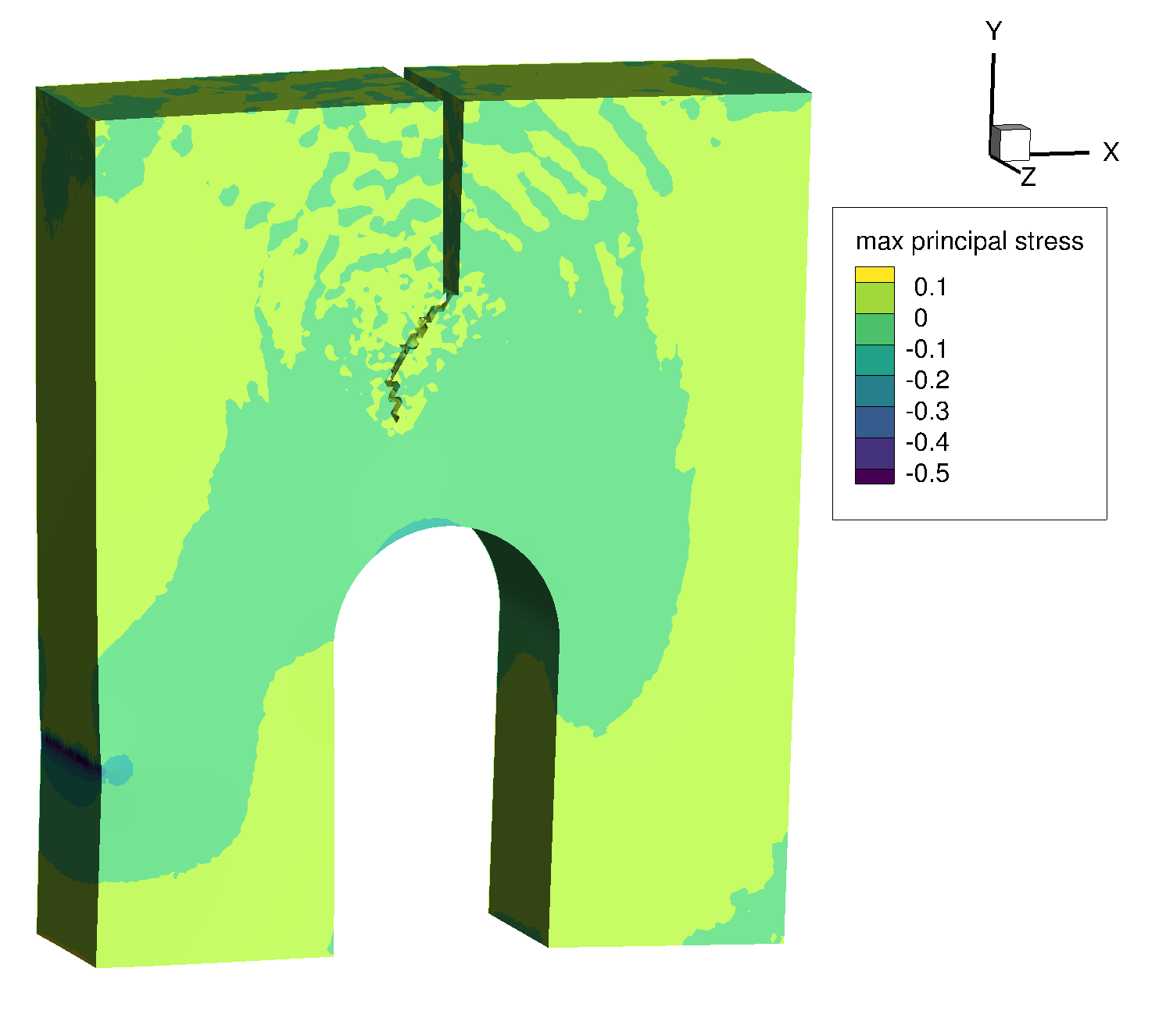}
            \end{center}
            \begin{center}
            (c)
            \end{center}
        \end{minipage}      
        \hfill
        \begin{minipage}{0.45\linewidth}
            \begin{center}
            \includegraphics[height=2.8in]{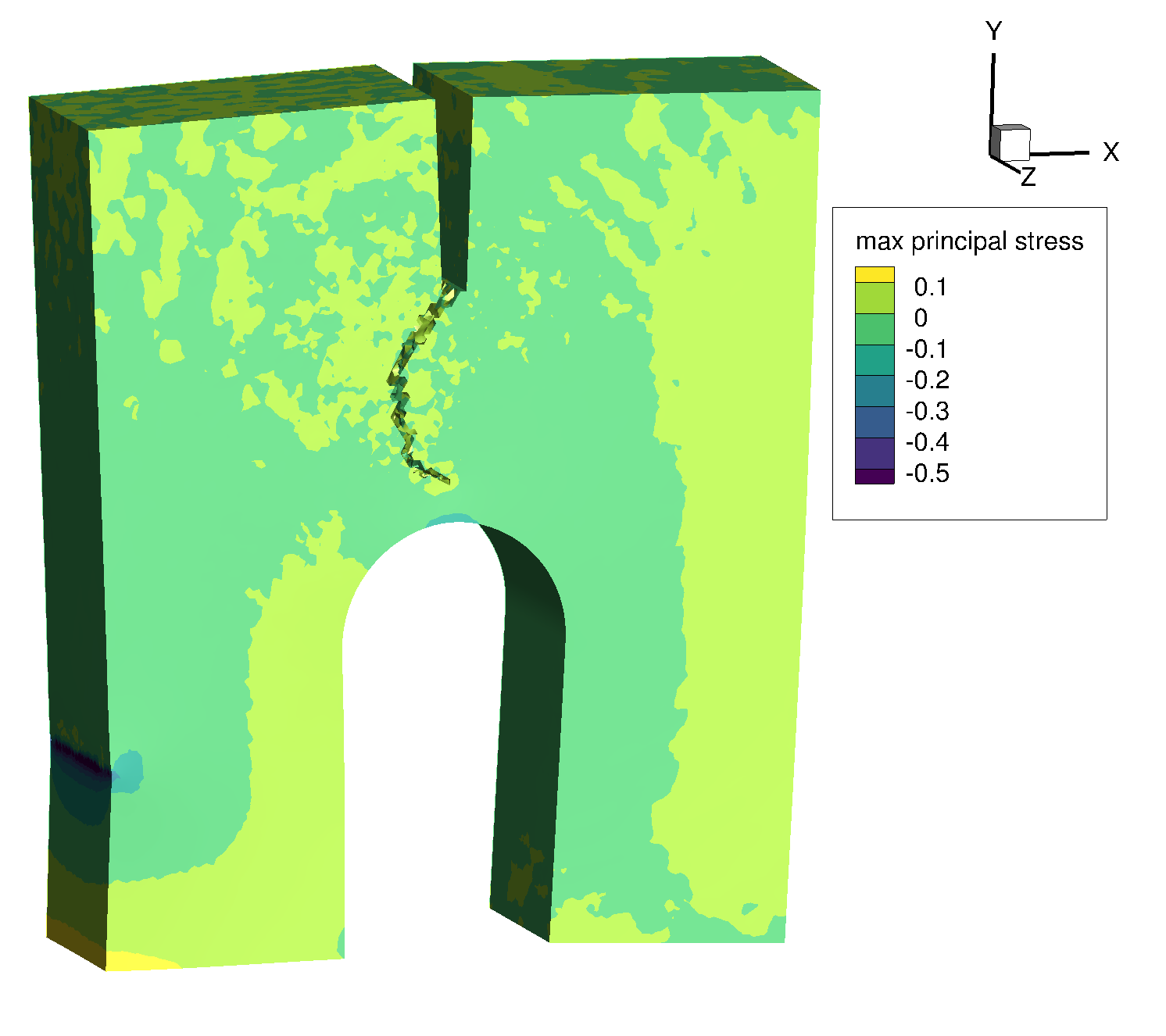}
            \end{center}
            \begin{center}
            (d)
            \end{center}
        \end{minipage} 
        \caption{Crack path evolution and maximum principal stress contour with $2069141$ elements model: (a). time $t=61.6\ \mu s$; (b). time $t=75.6\ \mu s$; (c). time $t=98\ \mu s$; (d). time $t=140\ \mu s$.}
        \label{fig18: ccompression-crack-evolution}
\end{figure}

Furthermore, the crack propagation with $2069141$ elements model is illustrated in Figure.\ref{fig18: ccompression-crack-evolution}. The crack initiates at the notch tip and curves as it extends toward the left side of the specimen. Ultimately, the crack follows an arc-like path. Despite the specimen’s shape is symmetrical, the uneven loading causes asymmetric deformation. This, in turn, leads to a three-dimensional irregular damage distribution that the CEM successfully captures.

\subsection{Crack branching with Neumann boundary condition}
\begin{figure}[htp]
	\centering
            \begin{center}
            \includegraphics[height=3.5in]{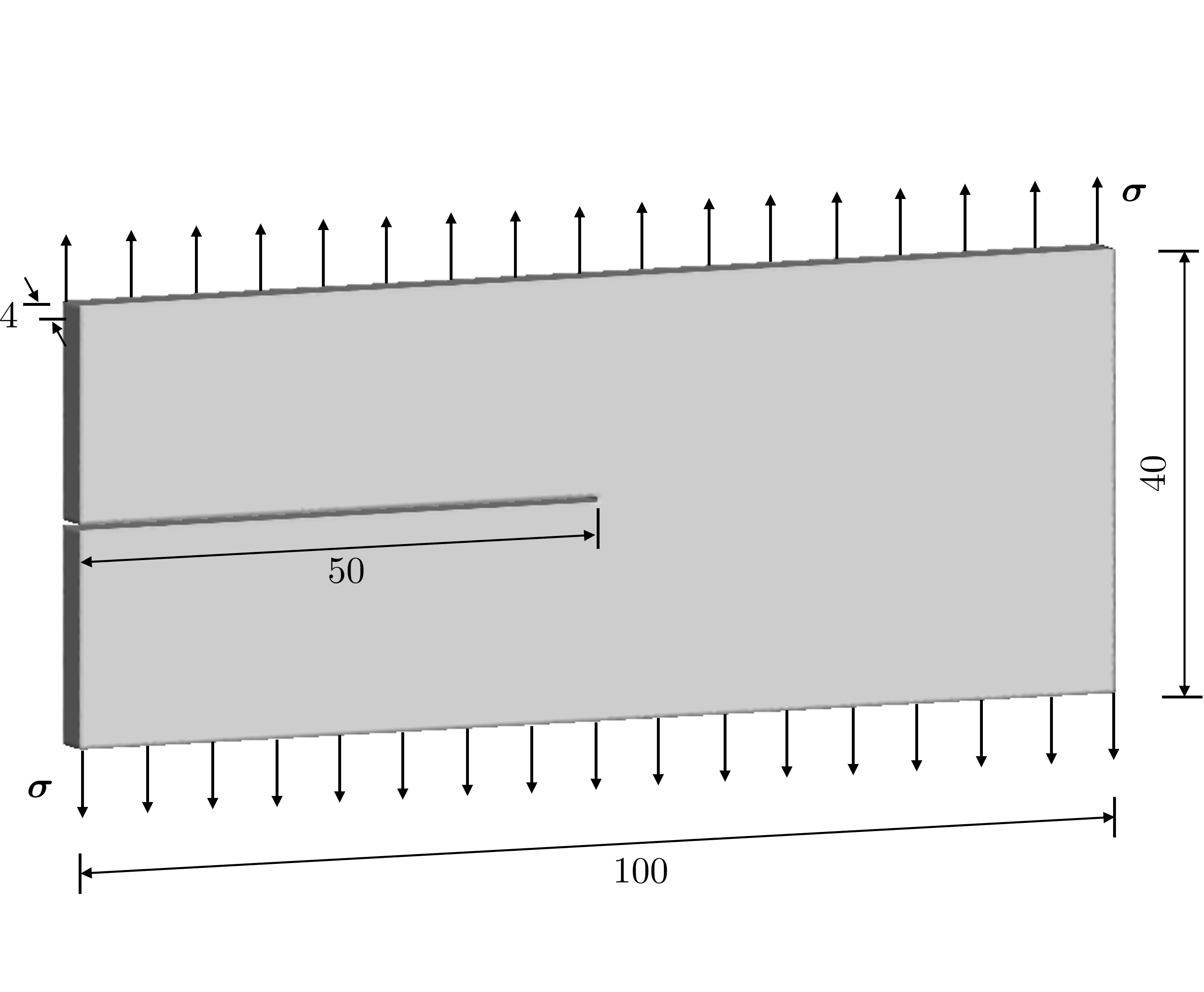}
            \end{center}
        \caption{Geometric dimensions of the three-dimensional pre-notched plate under traction-applied boundary condition. (unit: mm)}
        \label{fig19: cbranching-1-geometry}
\end{figure}

Starting with this benchmark example, we present our investigation into the application of the CEM to a three-dimensional transient-dynamic crack branching problem. In this example, a classic pre-notched three-dimensional plate under two-sided traction-driven tension is studied. The geometric dimensions and boundary conditions are shown in Figure.\ref{fig19: cbranching-1-geometry}. The plate specimen has length of $0.1\ m$, width of $0.04\ m$ and thickness of $0.004\ m$. It is important to note that the Neumann boundary conditions with traction of stress $\bm{\sigma} = \left(0, 1, 0 \right)\ MPa$ are applied on the top and bottom sides of the plate (see Figure.\ref{fig19: cbranching-1-geometry}). The tractions are applied at the beginning and remain constant until the simulation ends. To assist with crack initiation, there exits a pre-notch with a length of $0.05\ m$. 

Many researchers have studied this crack branching problem. For example, Rabczuk et al. investigated it with a cracking particle method (\cite{rabczuk2004cracking}), both Xu et al. (\cite{xu1993void}) and Falk et al. (\cite{falk2001critical}) applied interelement methods to it, Borden et al. studied it with phase field theory (\cite{borden2012phase}), Bui et al. analyzed it with a local damage model (\cite{bui2022simulation}), and Hirmand et al. used cohesive fracture energy theory on it (\cite{hirmand2019block}). To the authors' knowledge, very limited studies address this problem in three-dimension, such as \cite{rabczuk2007three} and \cite{bordas2008three}, reflecting the complexity and difficulty of capturing crack branching in three-dimension.

\begin{figure}[htp]
	\centering
        \begin{minipage}{0.9\linewidth}
            \begin{center}
            \includegraphics[height=2.0in]{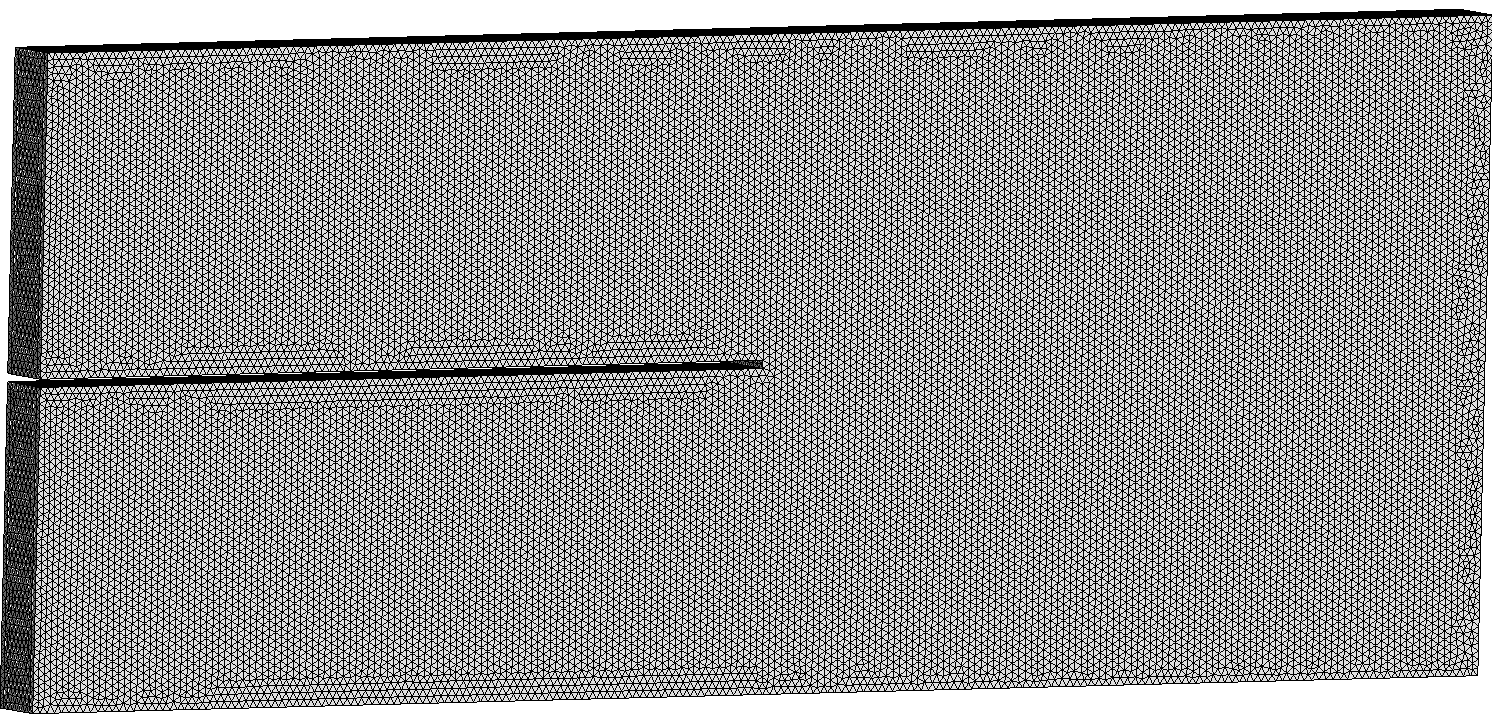}
            \end{center}
            \begin{center}
            (a)
            \end{center}
        \end{minipage}
        \hfill
        \begin{minipage}{0.9\linewidth}
            \begin{center}
            \includegraphics[height=2.0in]{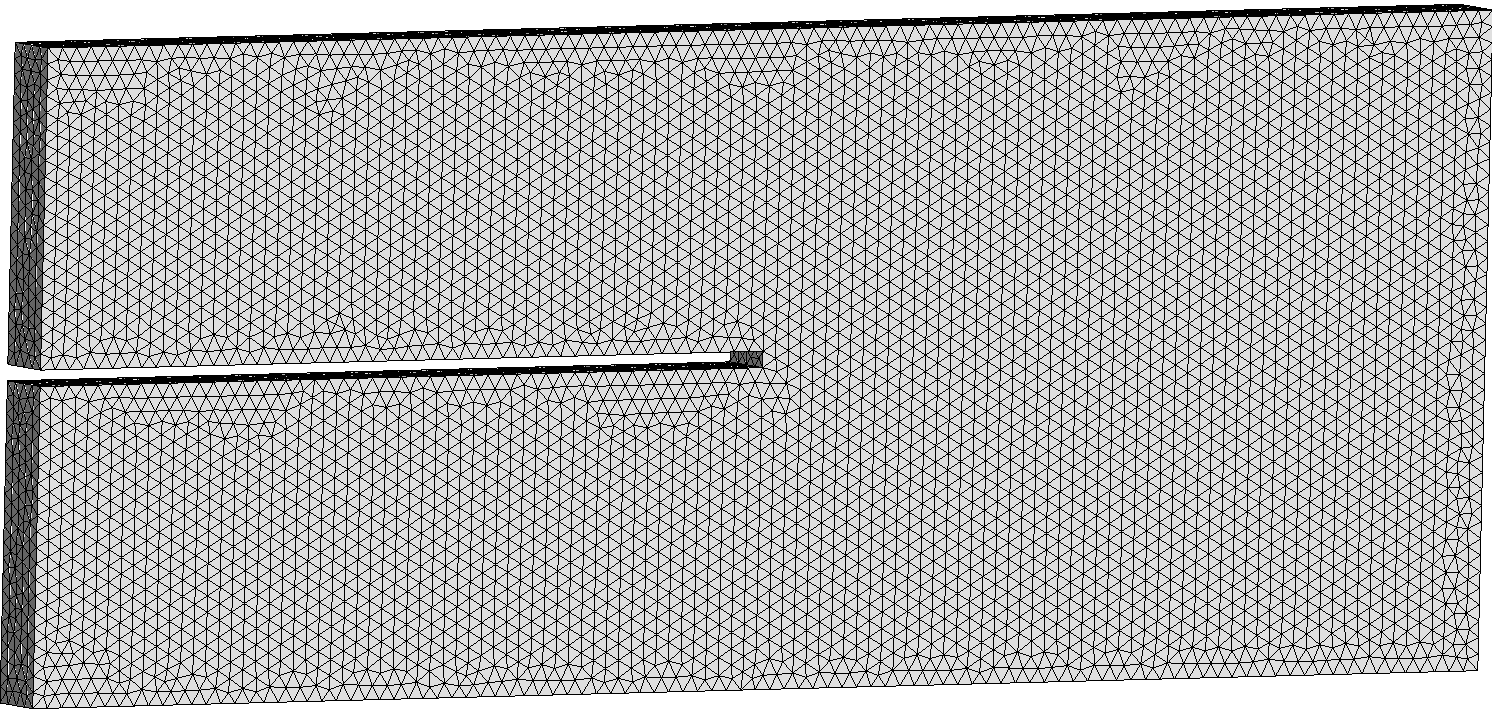}
            \end{center}
            \begin{center}
            (b)
            \end{center}
        \end{minipage}   
        \hfill
        \begin{minipage}{0.9\linewidth}
            \begin{center}
            \includegraphics[height=2.0in]{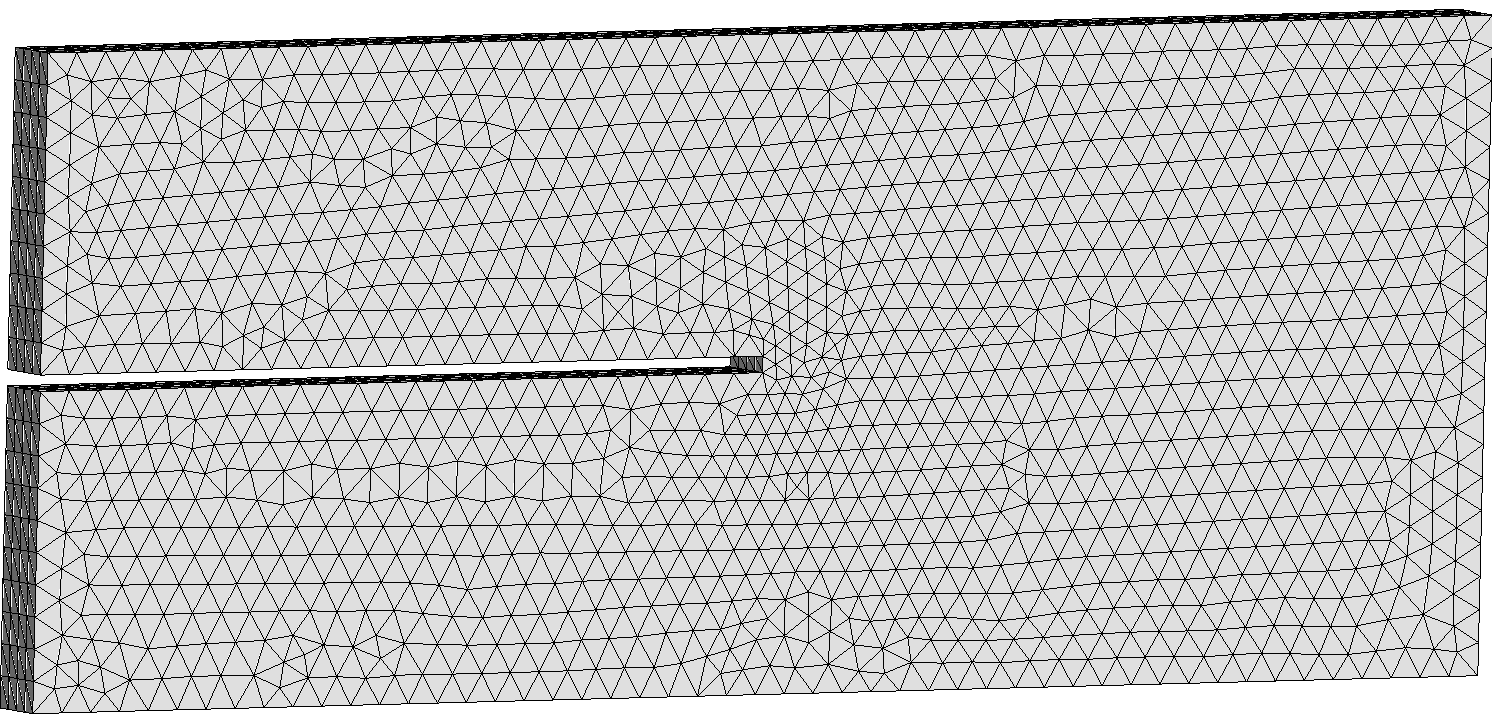}
            \end{center}
            \begin{center}
            (c)
            \end{center}
        \end{minipage}
        \caption{Three different three-dimensional meshes of the plate with tetrahedron element are illustrated: (a). $586624$ elements; (b). $75115$ elements; (c). $30996$ elements.}
        \label{fig20: cbranching-1-meshes}
\end{figure}

In this study, the three-dimensional model is considered. The material properties of the specimen are given below: Young's modulus $E=32\ GPa$, Poisson ratio $v=0.2$, critical fracture energy release rate $\mathcal{G}_c = 3\ J/m^2$, and density $\rho = 2450\ kg/m^3$. A total of $80 \ \mu s$ simulation time is applied.  Three different meshes of the three-dimensional plate are considered as shown in Figure.\ref{fig20: cbranching-1-meshes}.

\begin{figure}[htp]
	\centering
        \begin{minipage}{0.9\linewidth}
            \begin{center}
            \includegraphics[height=2.0in]{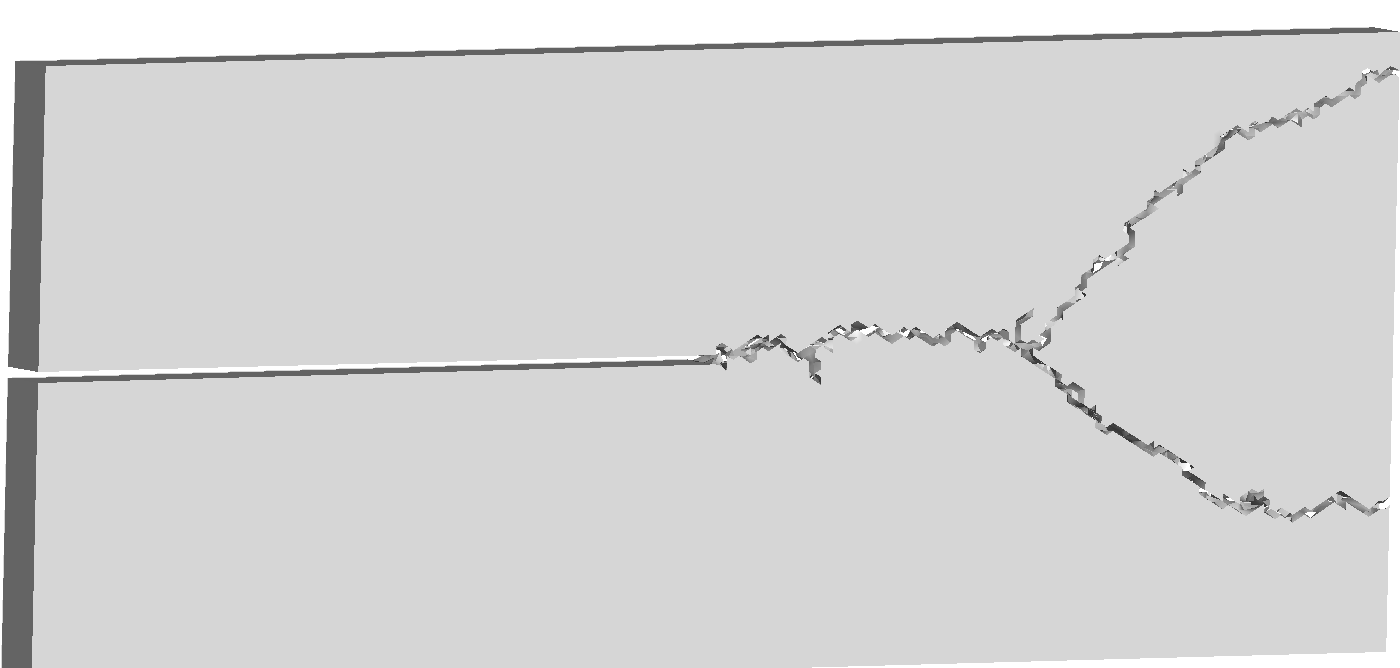}
            \end{center}
            \begin{center}
            (a)
            \end{center}
        \end{minipage}
        \hfill
        \begin{minipage}{0.9\linewidth}
            \begin{center}
            \includegraphics[height=2.0in]{cbranching-1-N18036_E75115-crack-pattern-crop.png}
            \end{center}
            \begin{center}
            (b)
            \end{center}
        \end{minipage}   
        \hfill
        \begin{minipage}{0.9\linewidth}
            \begin{center}
            \includegraphics[height=2.0in]{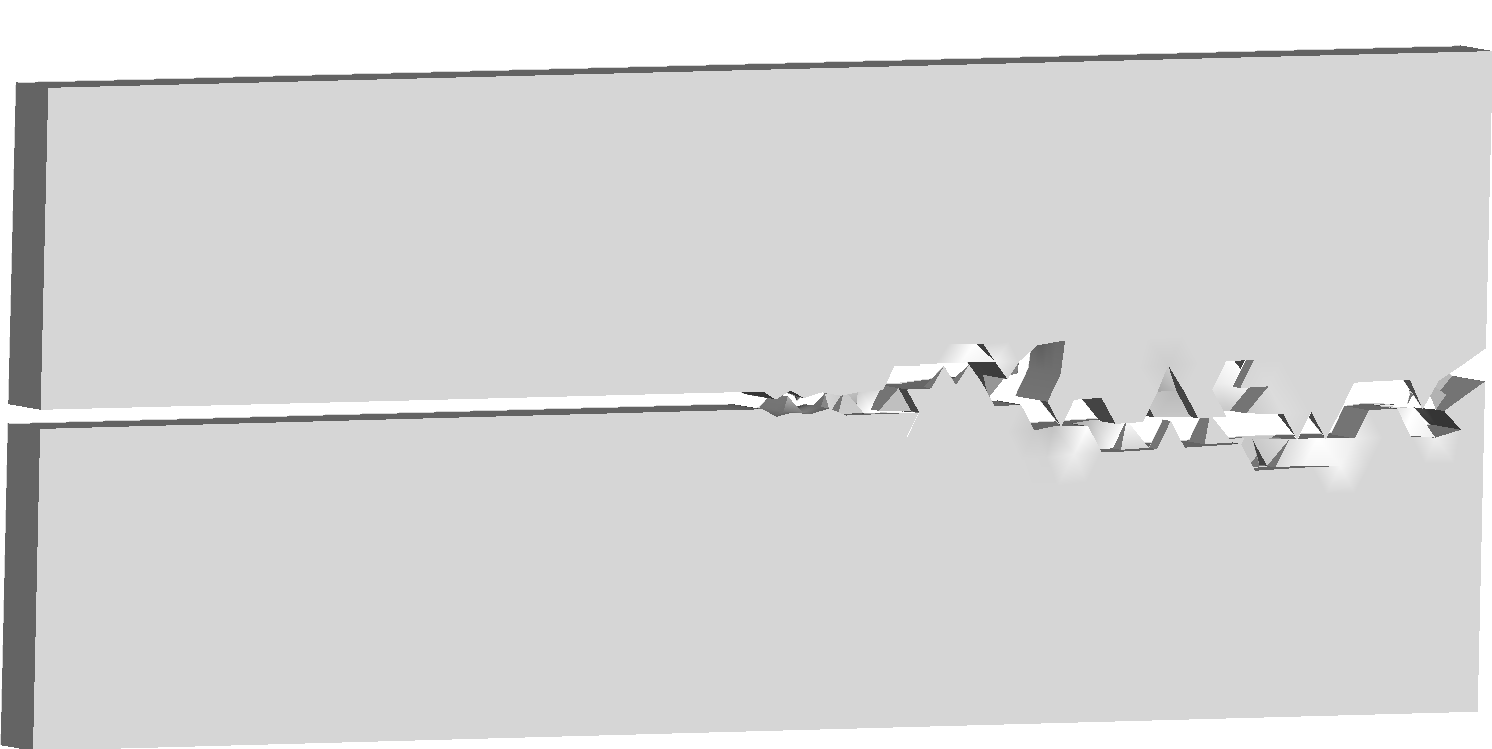}
            \end{center}
            \begin{center}
            (c)
            \end{center}
        \end{minipage}
        \caption{The final crack patterns of three representative tetrahedron meshes are illustrated: (a). $586624$ elements; (b). $75115$ elements; (c). $30996$ elements.}
        \label{fig21: cbranching-1-crack-pattern}
\end{figure}

The final crack patterns of the three different meshes are shown in Figure.\ref{fig21: cbranching-1-crack-pattern}. It is observed that the meshes of $e=586624, n=116362$, and $e=75115, n=18036$ are capable of capturing branching crack patterns, while $e=30996, n=6940$ misses the branching point. 
%

\begin{figure}[htp]
	\centering
        \begin{minipage}{0.45\linewidth}
            \begin{center}
            \includegraphics[height=2.8in]{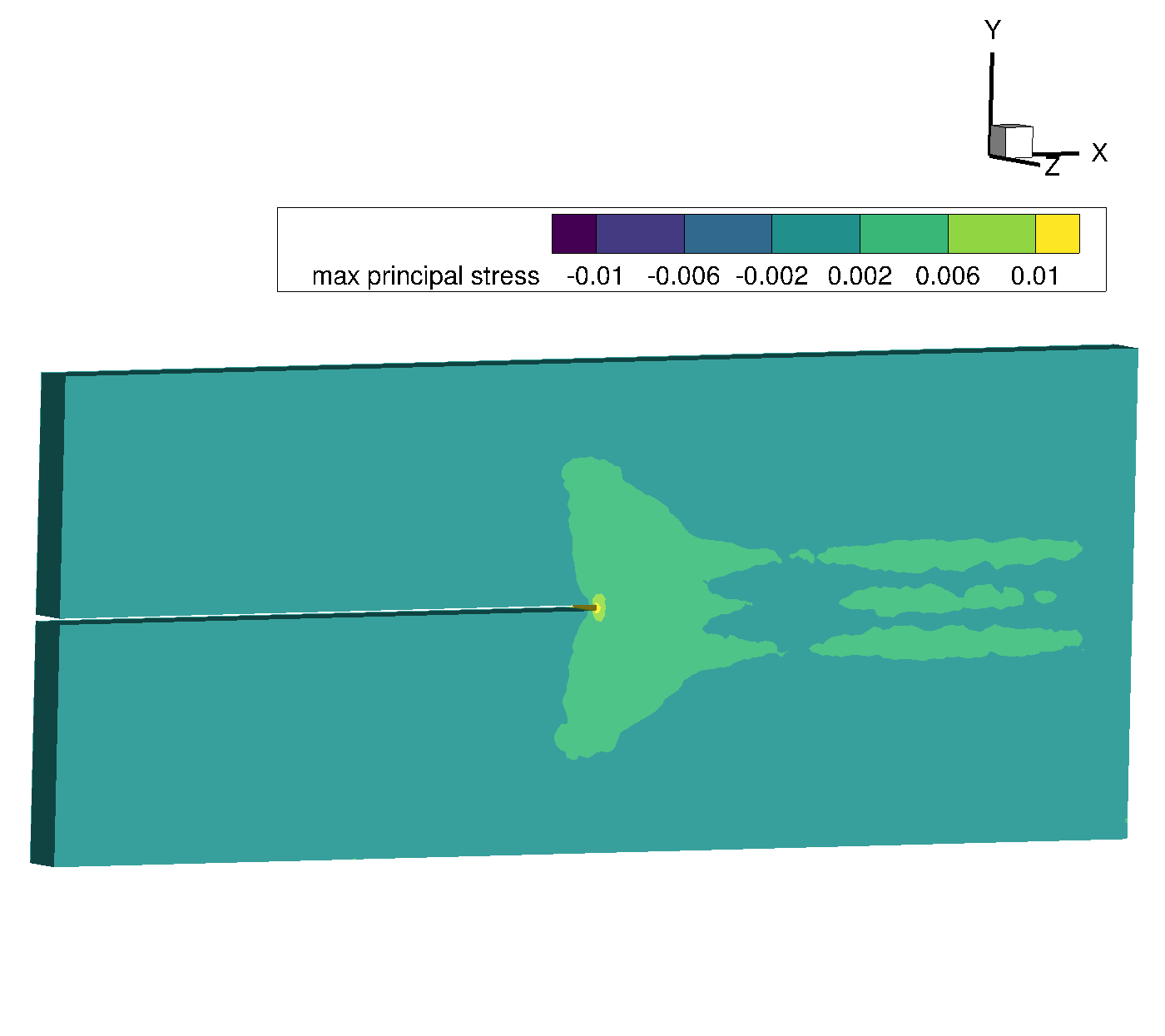}
            \end{center}
            \begin{center}
            (a)
            \end{center}
        \end{minipage}
        \hfill
        \begin{minipage}{0.45\linewidth}
            \begin{center}
            \includegraphics[height=2.8in]{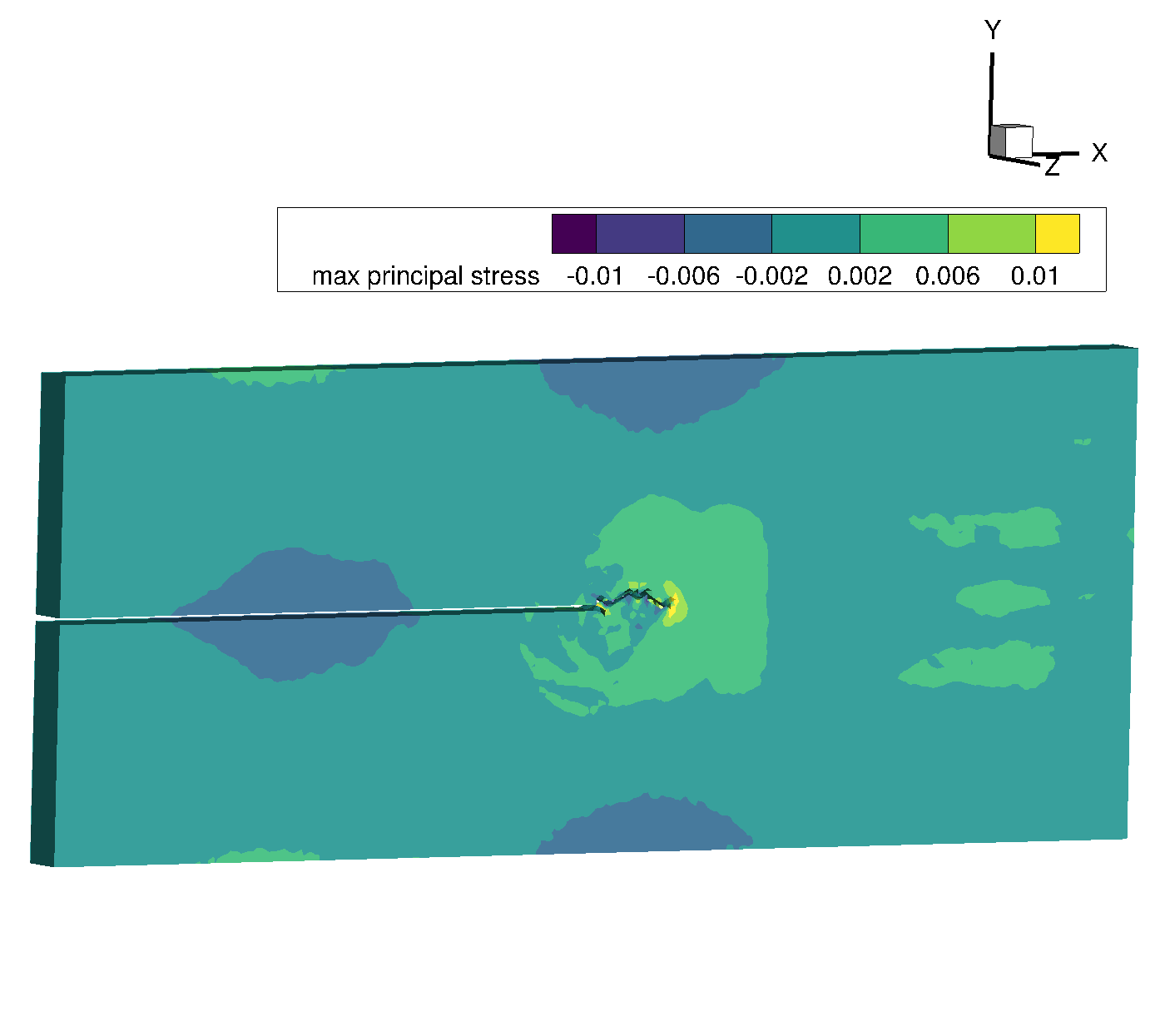}
            \end{center}
            \begin{center}
            (b)
            \end{center}
        \end{minipage}   
        \hfill
        \begin{minipage}{0.45\linewidth}
            \begin{center}
            \includegraphics[height=2.8in]{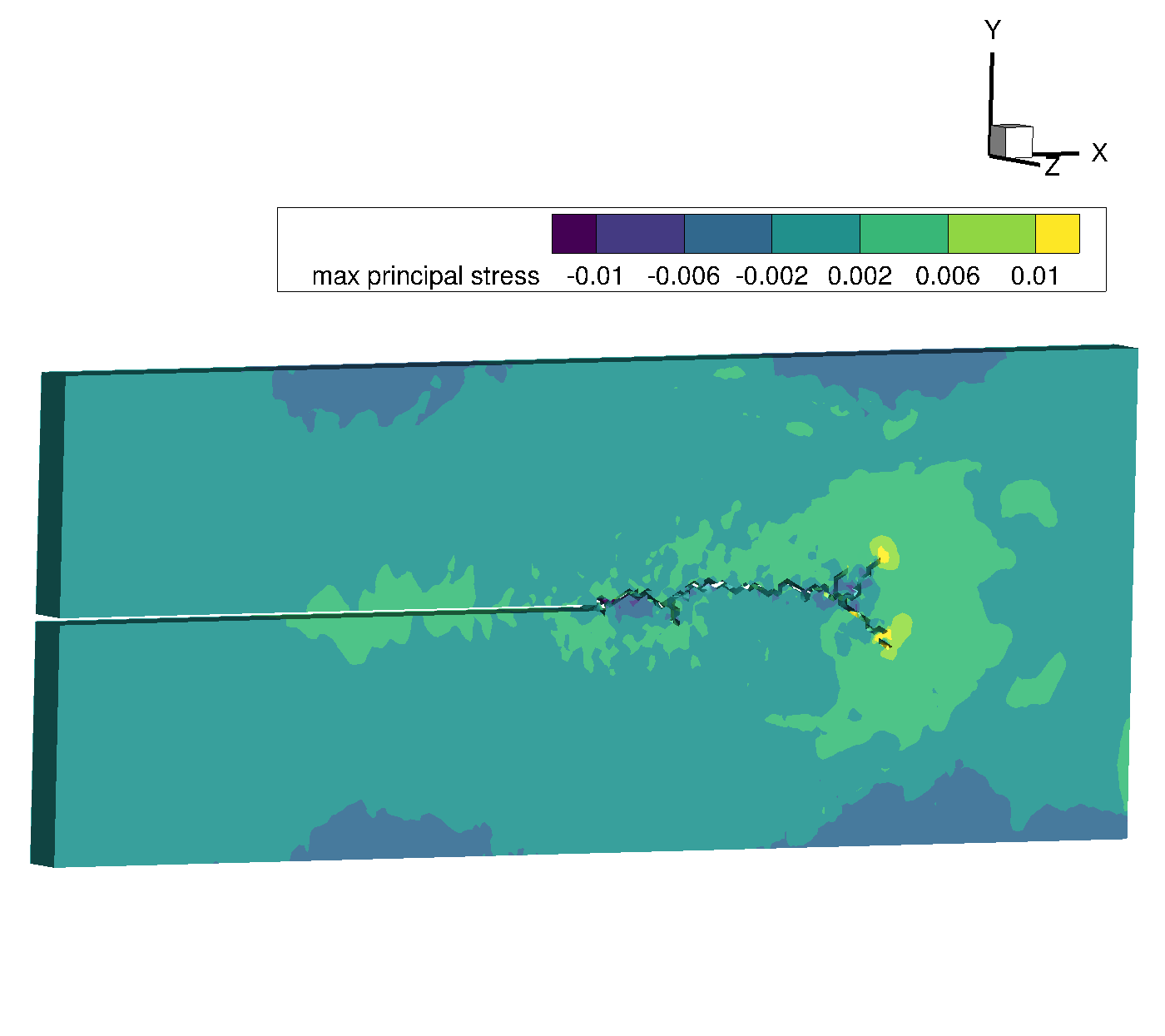}
            \end{center}
            \begin{center}
            (c)
            \end{center}
        \end{minipage}
        \hfill
        \begin{minipage}{0.45\linewidth}
            \begin{center}
            \includegraphics[height=2.8in]{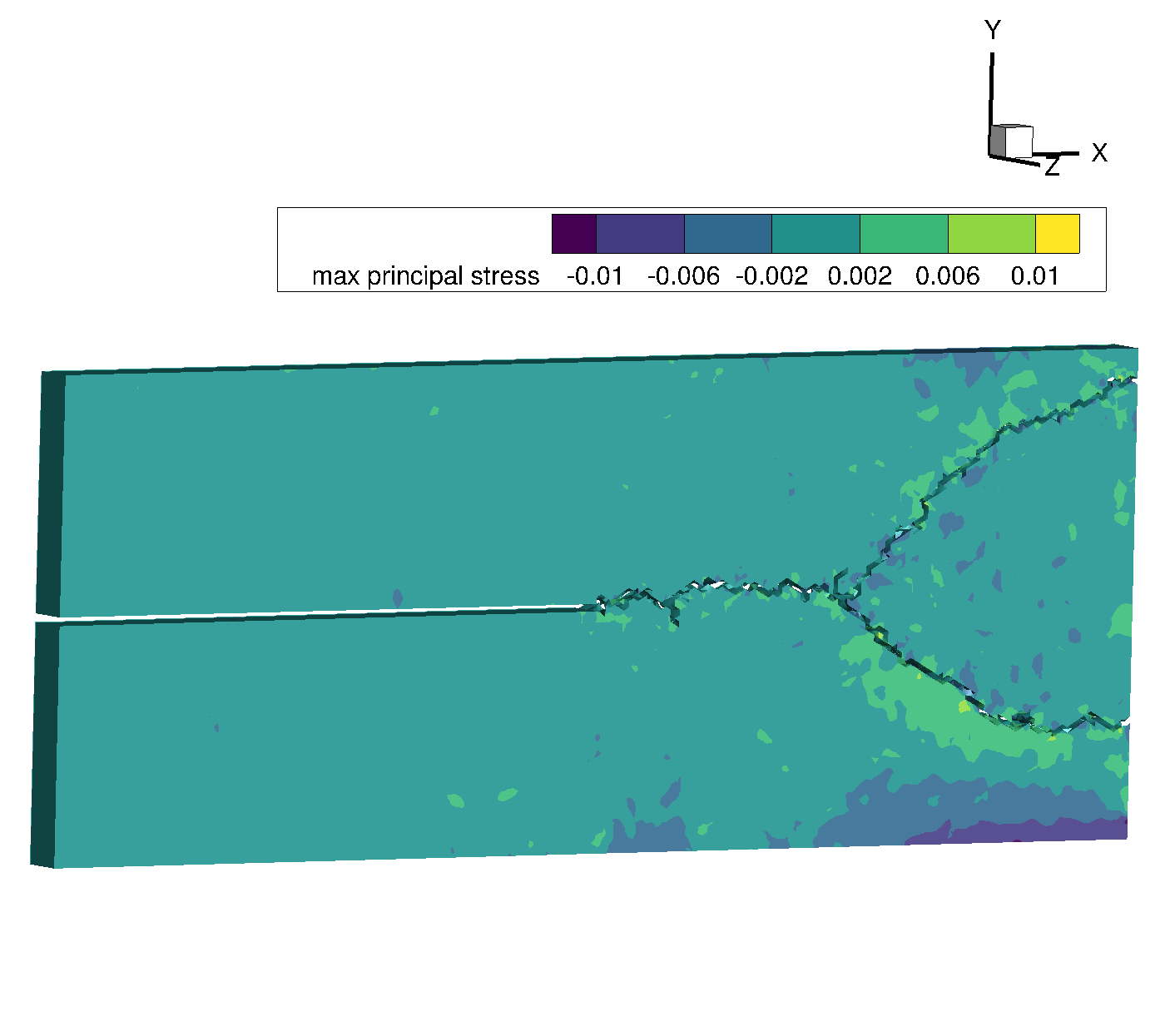}
            \end{center}
            \begin{center}
            (d)
            \end{center}
        \end{minipage}
        \caption{The crack pattern evolution of fine mesh with $586624$ elements are provided: (a). time $t=14.4\ \mu s$; (b). time $t=33.6\ \mu s$; (c). time $t=56\ \mu s$; (d). time $t=96\ \mu s$.}
        \label{fig22: cbranching-1-mesh3-crack-evolution}
\end{figure}

\begin{figure}[htp]
	\centering
        \begin{minipage}{0.45\linewidth}
            \begin{center}
            \includegraphics[height=2.8in]{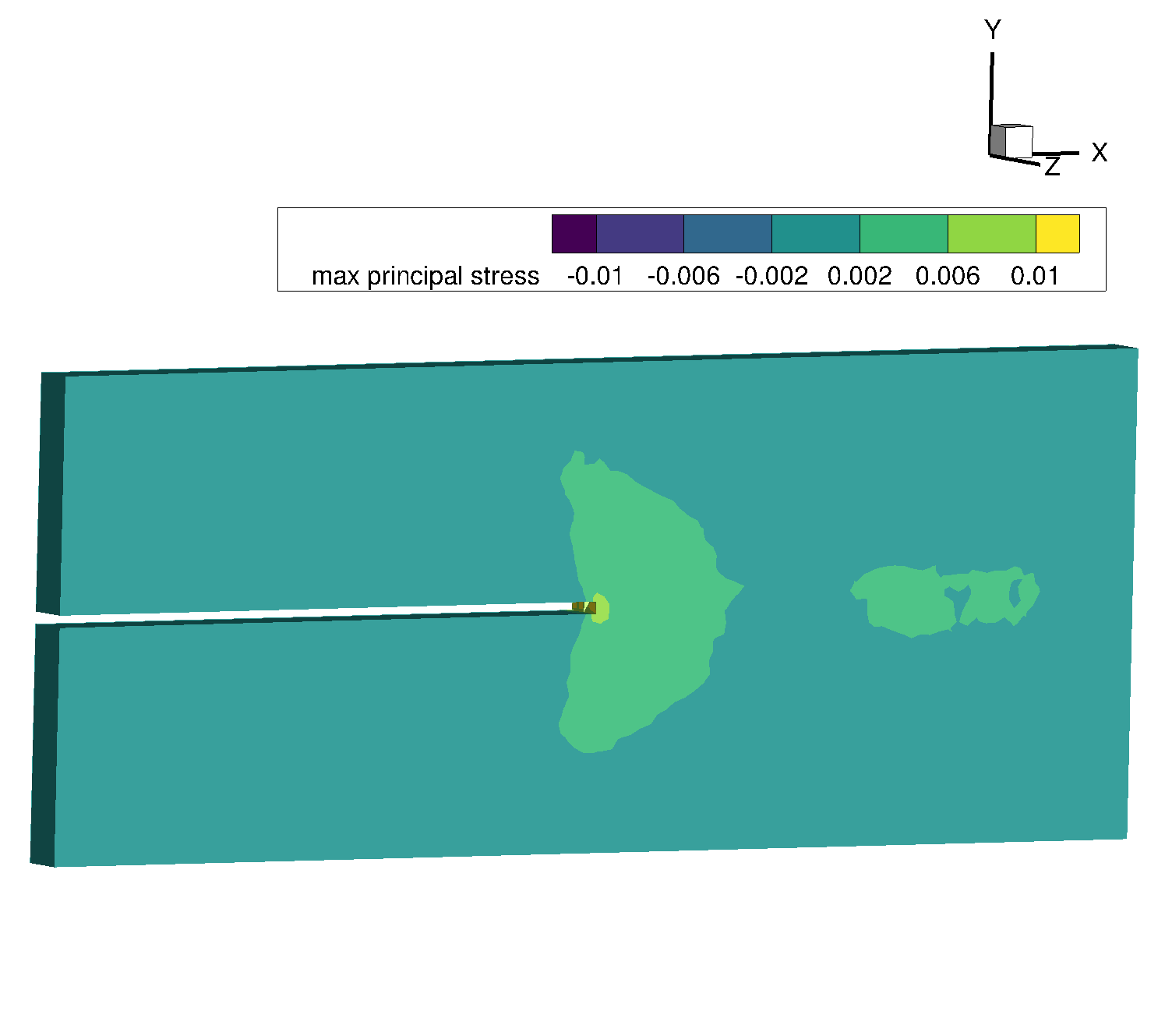}
            \end{center}
            \begin{center}
            (a)
            \end{center}
        \end{minipage}
        \hfill
        \begin{minipage}{0.45\linewidth}
            \begin{center}
            \includegraphics[height=2.8in]{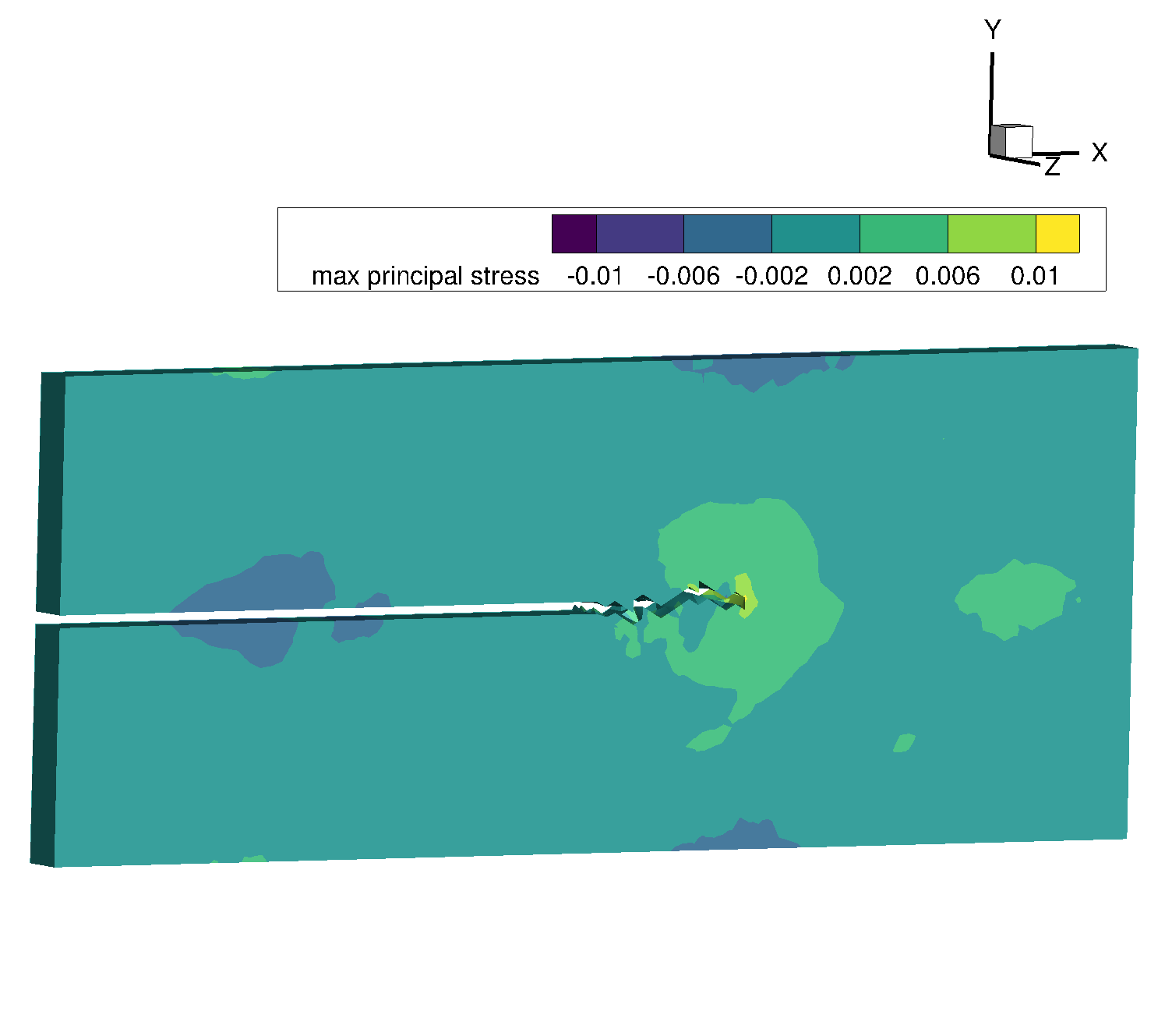}
            \end{center}
            \begin{center}
            (b)
            \end{center}
        \end{minipage}   
        \hfill
        \begin{minipage}{0.45\linewidth}
            \begin{center}
            \includegraphics[height=2.8in]{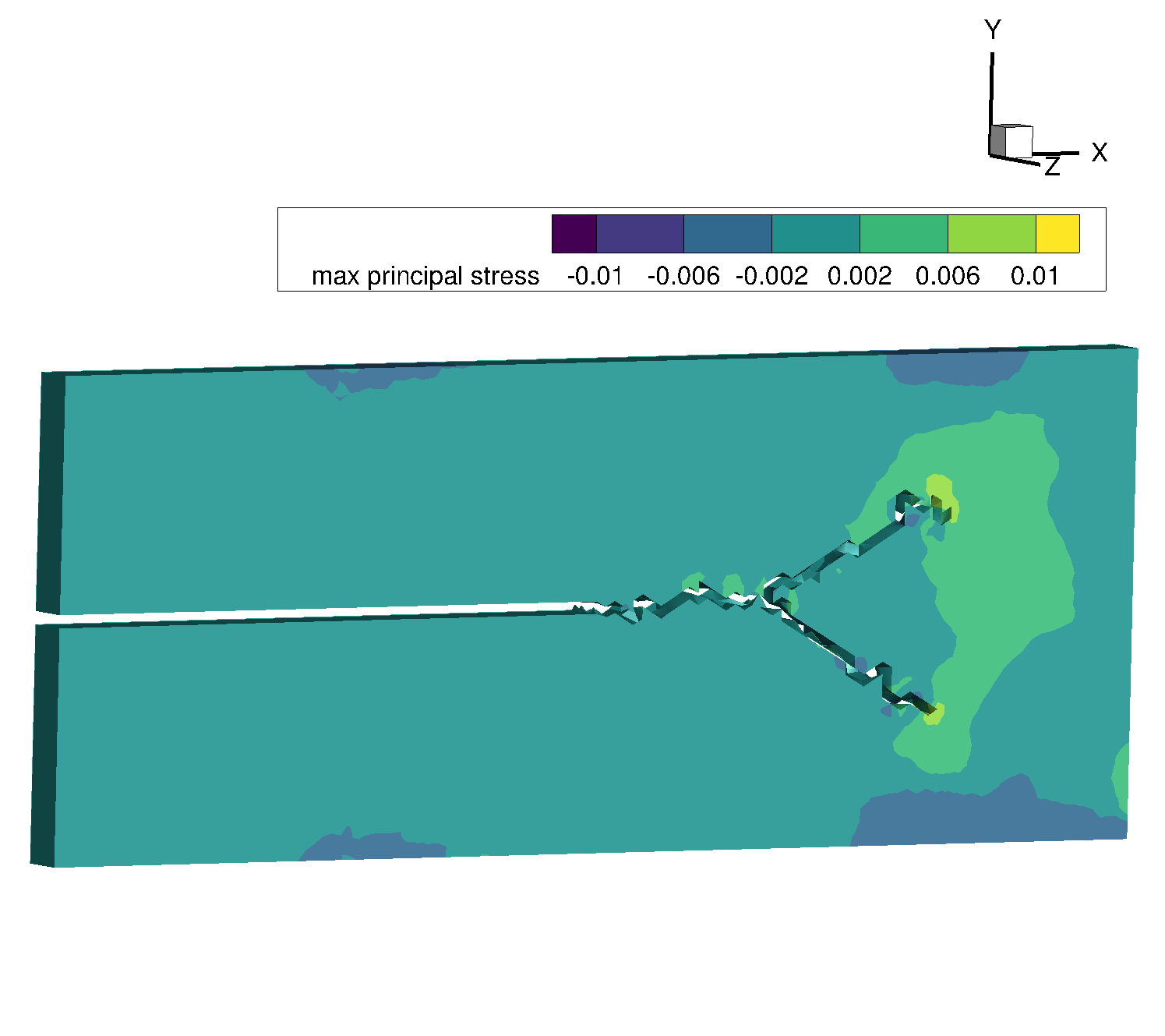}
            \end{center}
            \begin{center}
            (c)
            \end{center}
        \end{minipage}
        \hfill
        \begin{minipage}{0.45\linewidth}
            \begin{center}
            \includegraphics[height=2.8in]{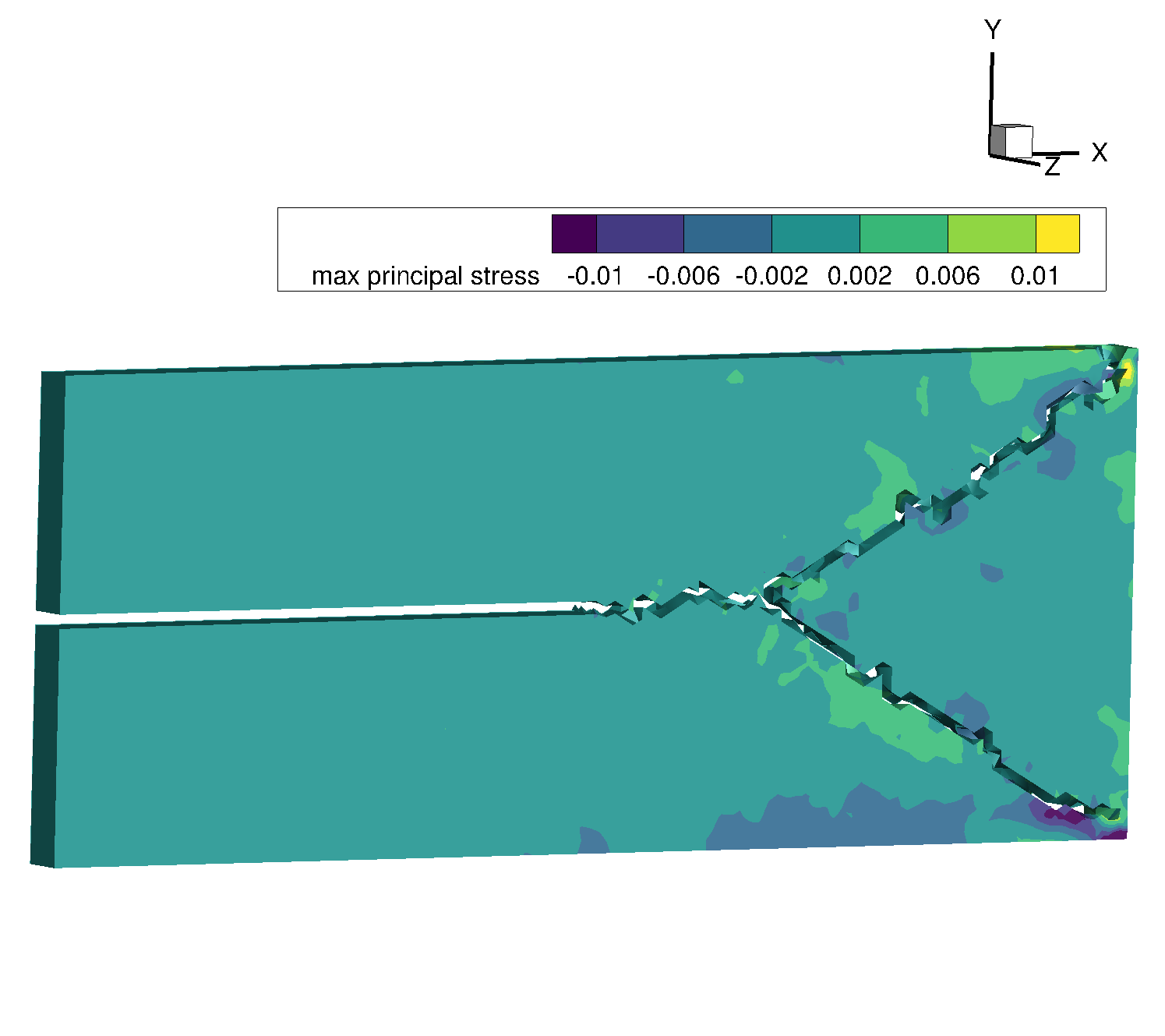}
            \end{center}
            \begin{center}
            (d)
            \end{center}
        \end{minipage}
        \caption{The crack pattern evolution of medium mesh with $75115$ elements are provided: (a). time $t=14.4\ \mu s$; (b). time $t=33.6\ \mu s$; (c). time $t=56\ \mu s$; (d). time $t=80\ \mu s$.}
        \label{fig23: cbranching-1-mesh2-crack-evolution}
\end{figure}

\begin{figure}[htp]
	\centering
        \begin{minipage}{0.45\linewidth}
            \begin{center}
            \includegraphics[height=2.8in]{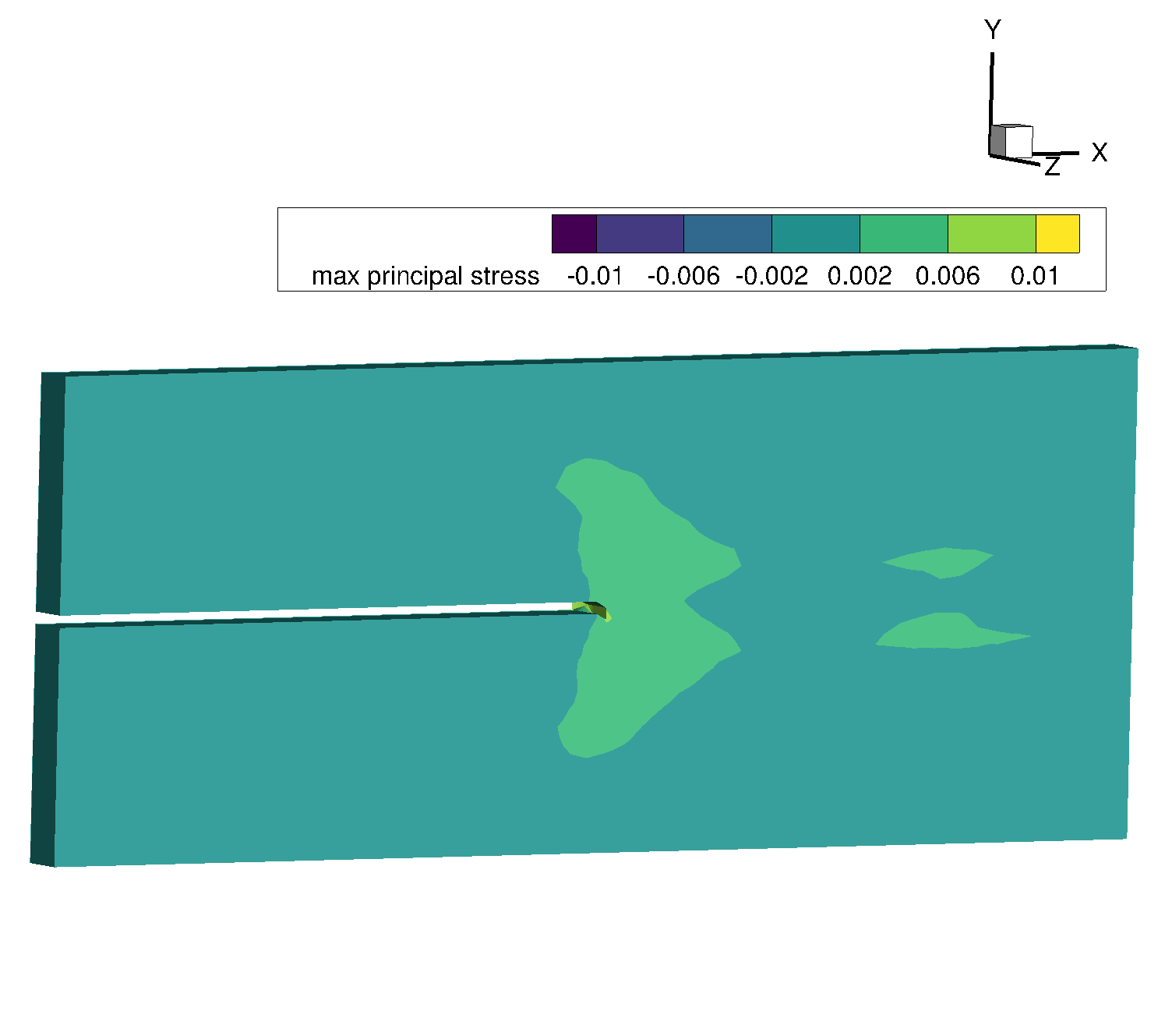}
            \end{center}
            \begin{center}
            (a)
            \end{center}
        \end{minipage}
        \hfill
        \begin{minipage}{0.45\linewidth}
            \begin{center}
            \includegraphics[height=2.8in]{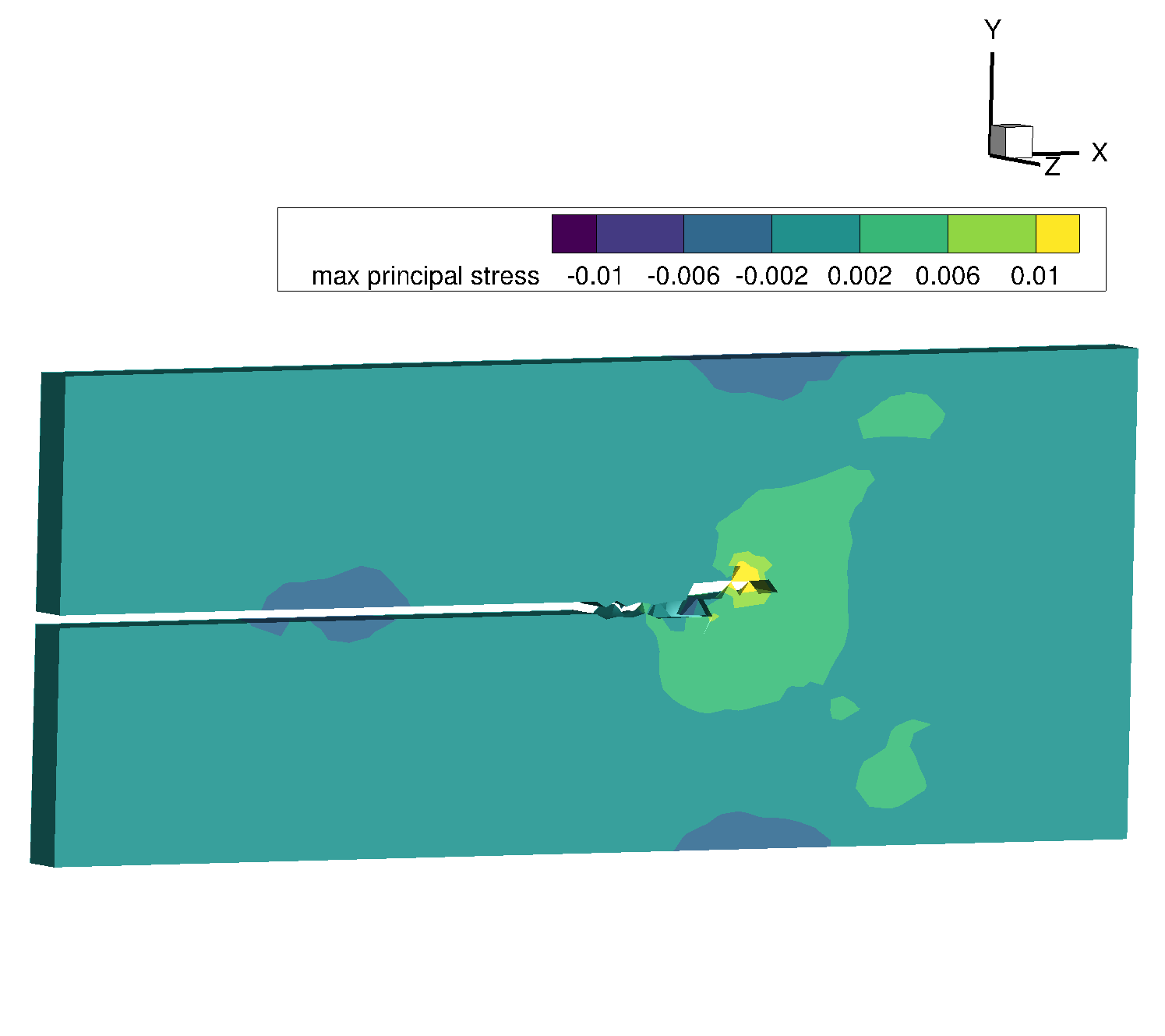}
            \end{center}
            \begin{center}
            (b)
            \end{center}
        \end{minipage}   
        \hfill
        \begin{minipage}{0.45\linewidth}
            \begin{center}
            \includegraphics[height=2.8in]{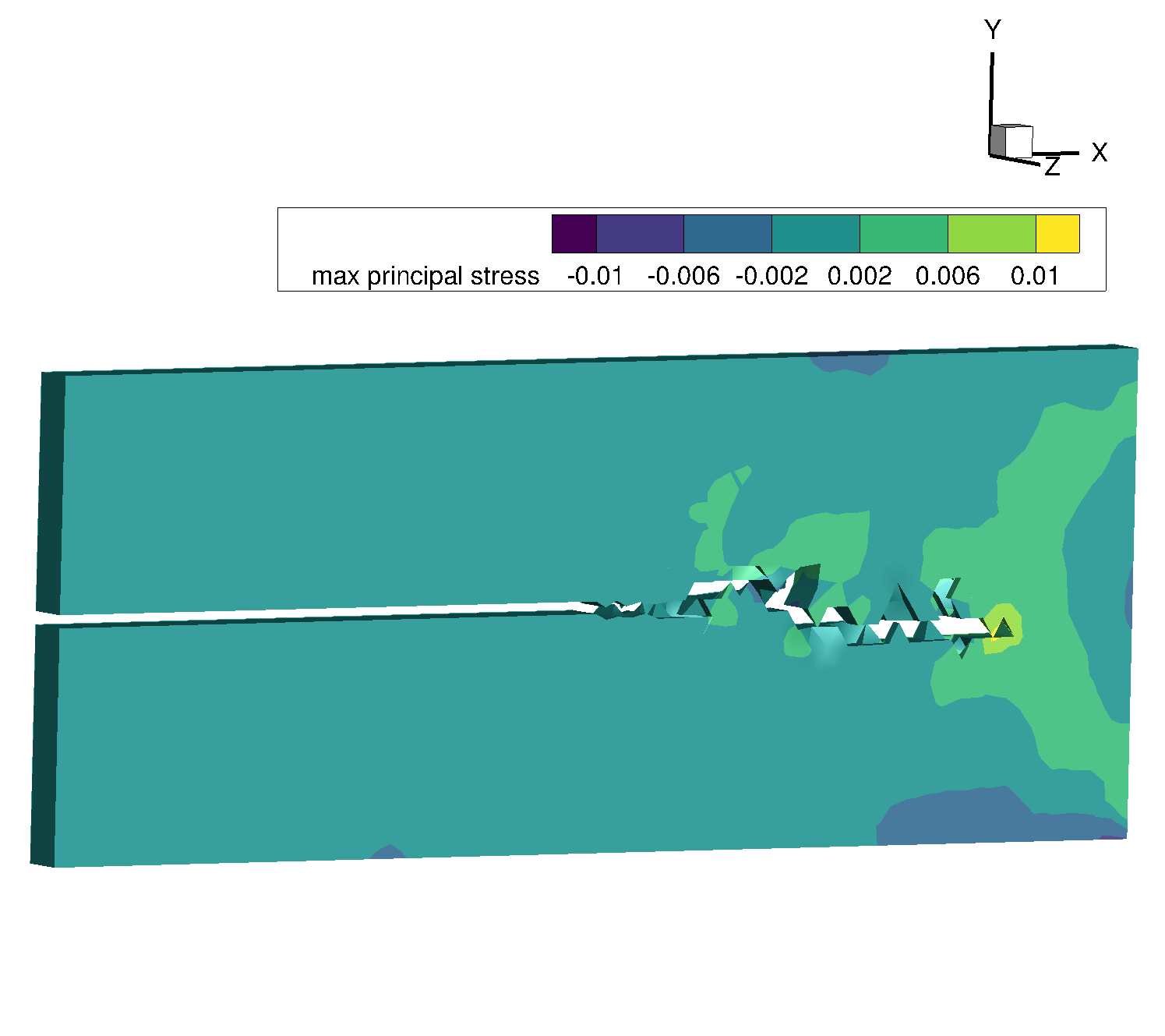}
            \end{center}
            \begin{center}
            (c)
            \end{center}
        \end{minipage}
        \hfill
        \begin{minipage}{0.45\linewidth}
            \begin{center}
            \includegraphics[height=2.8in]{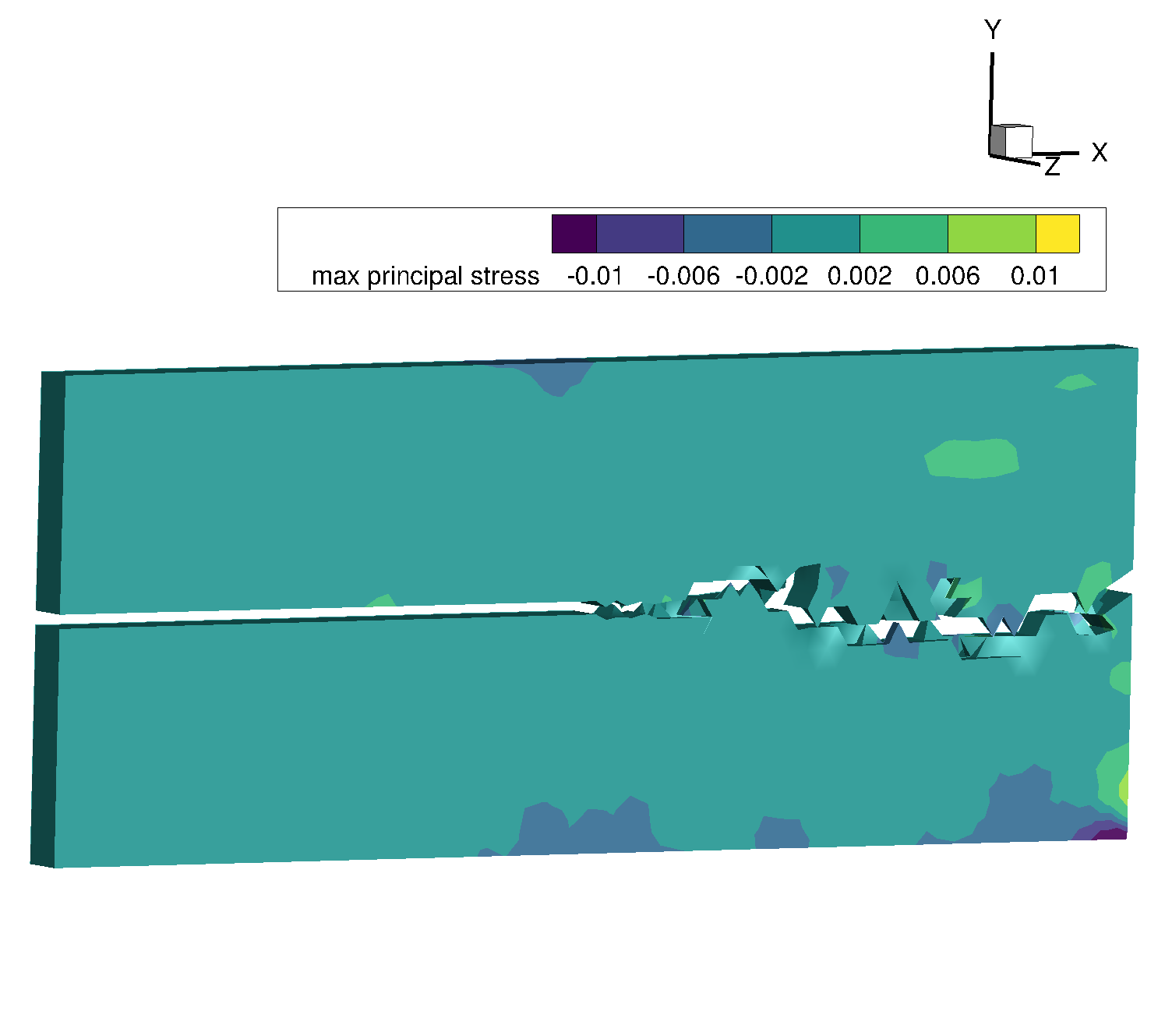}
            \end{center}
            \begin{center}
            (d)
            \end{center}
        \end{minipage}
        \caption{The crack pattern evolution of coarse mesh with $30996$ elements are provided: (a). time $t=14.4\ \mu s$; (b). time $t=33.6\ \mu s$; (c). time $t=56\ \mu s$; (d). time $t=80\ \mu s$.}
        \label{fig24: cbranching-1-mesh1-crack-evolution}
\end{figure}

The crack pattern propagation of the fine mesh ($586624$ elements) at different time steps are shown in Figure.\ref{fig22: cbranching-1-mesh3-crack-evolution},
the crack pattern propagation of the medium mesh ($75115$ elements) at different time steps are shown in Figure.\ref{fig23: cbranching-1-mesh2-crack-evolution},
and the crack pattern propagation of the coarse mesh ($30996$ elements) at different time steps are shown in Figure.\ref{fig24: cbranching-1-mesh1-crack-evolution}. By comparing the crack pattern evolution of three types of meshes, it is found that at time $t=33.6\ \mu s$, the crack branching start to initiate. However, only fine mesh ($586624$ elements) and medium mesh ($75115$ elements) successfully branch, while coarse mesh ($30996$ elements) fails and only forms a single crack (see Figure.\ref{fig24: cbranching-1-mesh1-crack-evolution}(c-d)). 

\begin{figure}[htp]
	\centering
            \begin{center}
            \includegraphics[height=2.4in]{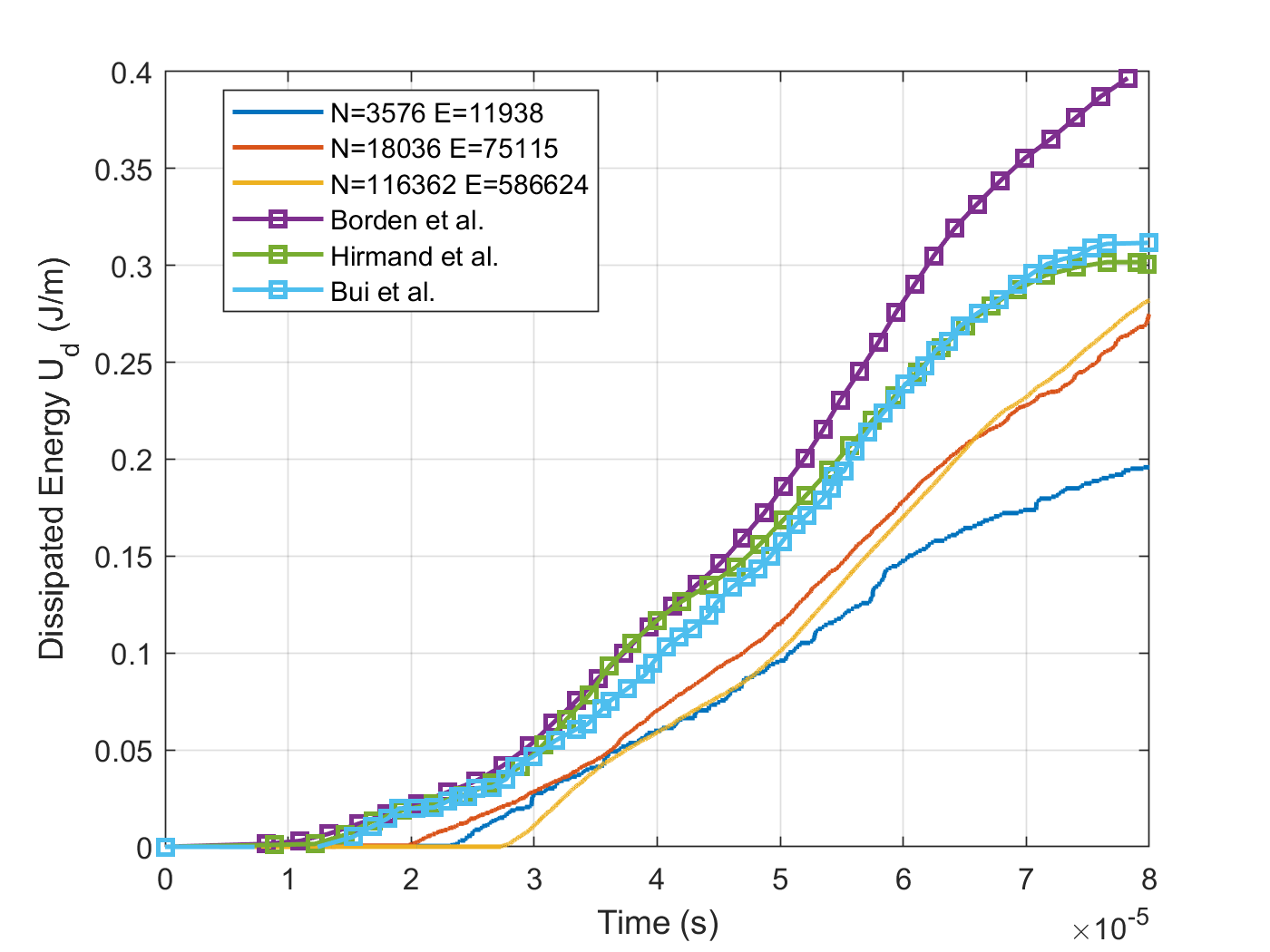}
            \end{center}
        \caption{Comparison of dissipated energy among three representative meshes and three reference results (\cite{bui2022simulation}, \cite{borden2012phase}, \cite{hirmand2019block}).}
        \label{fig25: cbranching-1-Ud}
\end{figure}

Figure.\ref{fig25: cbranching-1-Ud} illustrates the energy dissipation associated with crack formation. For comparison, reference data from the phase-field model by Borden et al. (\cite{borden2012phase}), the local damage model by Bui et al. (\cite{bui2022simulation}) and the cohesive fracture approach by Hirmand and Papoulia (\cite{hirmand2019block}) are also included. The energy dissipation curves from the fine and the medium meshes using the proposed CEM closely align with the results of the cohesive fracture and local damage model, whereas the phase-field model predicts higher dissipated energy. Only the coarse CEM mesh shows a clear discrepancy, attributing to its inability to capture crack branching.

In summary, the proposed CEM can capture three-dimensional crack branching under transient-dynamic loading when a sufficiently fine mesh (typically over $70000$ elements is used). Notably, this is achieved without employing any explicit crack branching criteria.

\subsection{Crack branching with Dirichlet boundary condition}
The previous numerical example demonstrates the effectiveness of the proposed CEM in capturing three-dimensional crack branching under Neumann boundary conditions. This benchmark example examines the validity of the proposed CEM in capturing three-dimensional crack branching with Dirichlet boundary conditions. Therefore, a transient-dynamic compact tension problem in concrete is selected, based on experiment data conducted by Ozbolt et al. (\cite{ovzbolt2013dynamic}). The impact of different loading rates on the failure modes and crack trajectories of plain concrete specimens is investigated. According to findings by Ožbolt et al. (\cite{ovzbolt2013dynamic}), an increase in strain rate causes a transition in the failure behavior of plain concrete from mode-I fracture to a mixed-mode fracture pattern. To examine the influence of loading rates under dynamic compact tension conditions, three specific cases out of twelve tested velocities are highlighted: Case 1 with a velocity of $1.375\ m/s$, Case 2 at $3.318\ m/s$, and Case 3 at $3.993\ m/s$. 

\begin{figure}[htp]
	\centering
        \begin{minipage}{0.45\linewidth}
            \centering
            \includegraphics[height=2.8in]{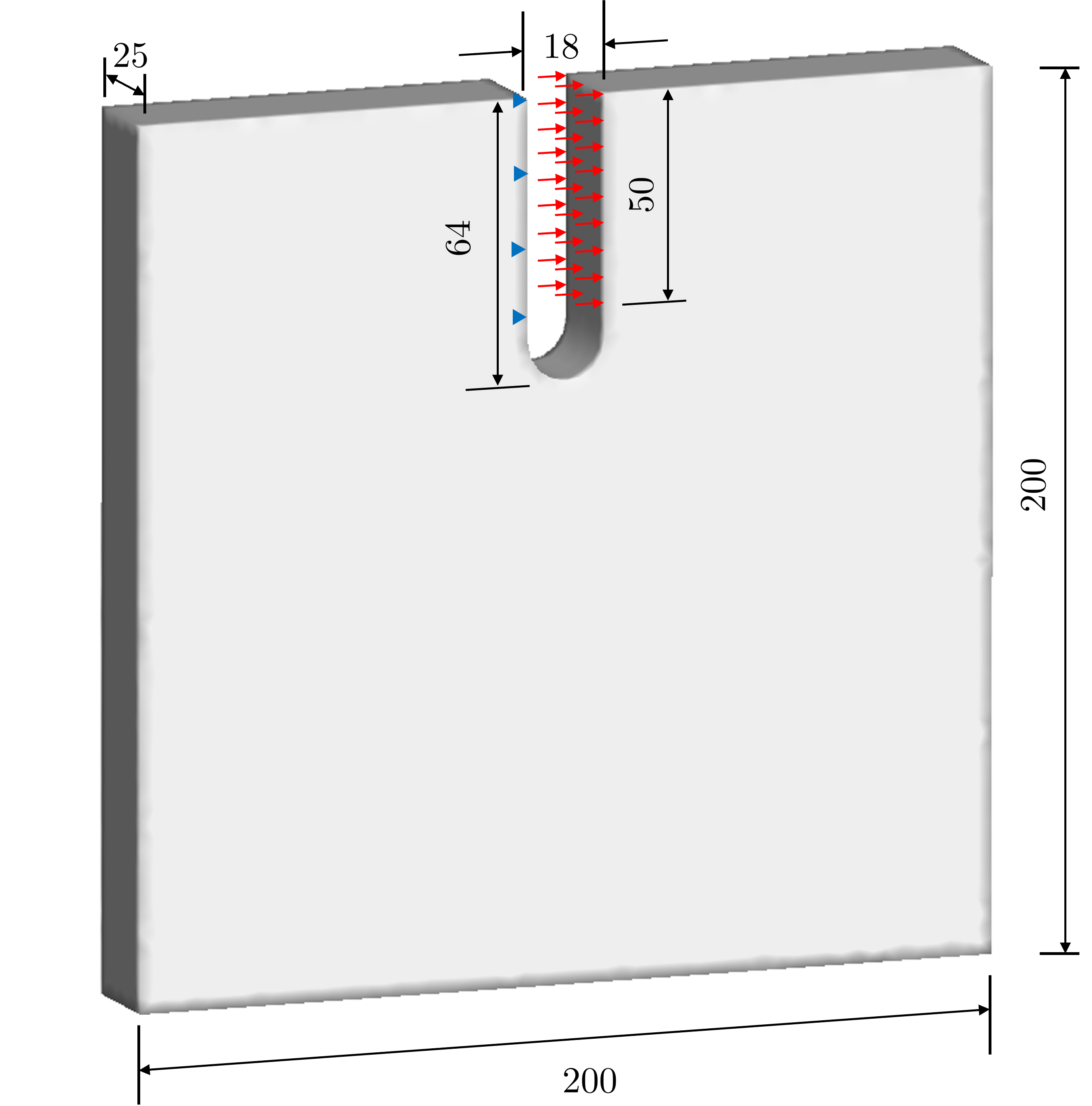}
            \begin{center}
            (a)
            \end{center}
        \end{minipage}
        \hfill
        \begin{minipage}{0.45\linewidth}
            \centering
            \includegraphics[height=2.8in]{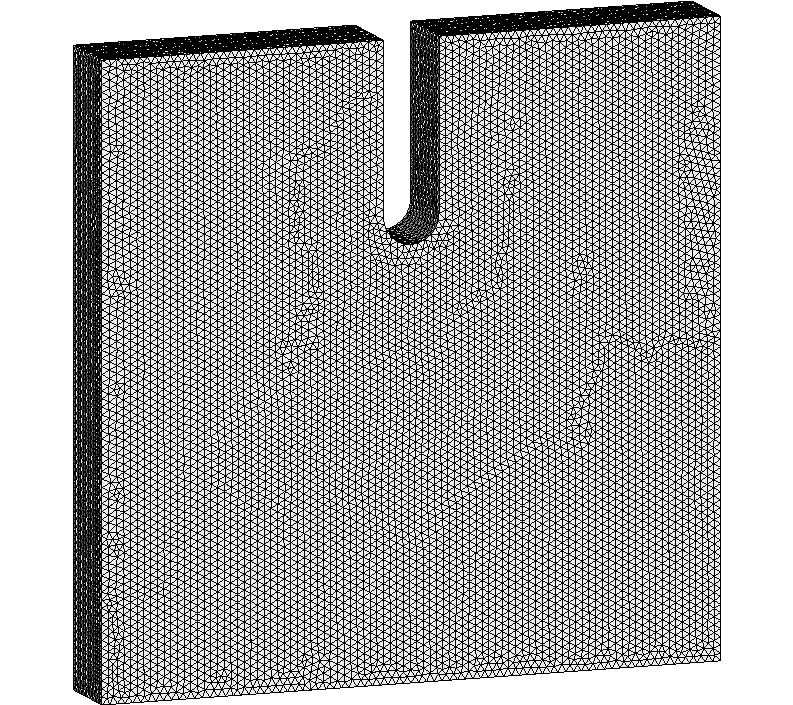}
            \begin{center}
            (b)
            \end{center}
        \end{minipage}   
        \hfill
        \begin{minipage}{0.45\linewidth}
            \centering
            \includegraphics[height=2.8in]{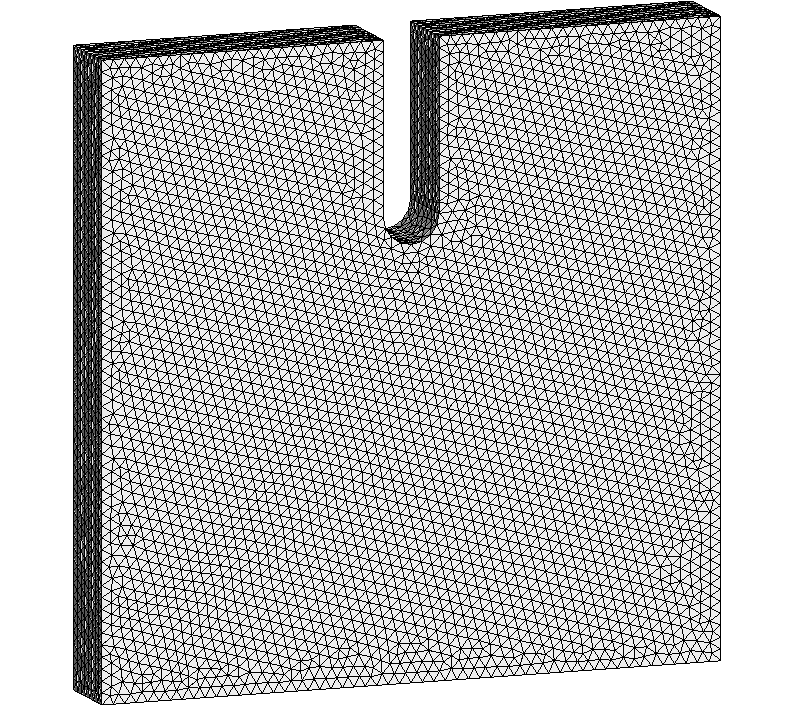}
            \begin{center}
            (c)
            \end{center}
        \end{minipage}  
        \hfill
        \begin{minipage}{0.45\linewidth}
            \centering
            \includegraphics[height=2.8in]{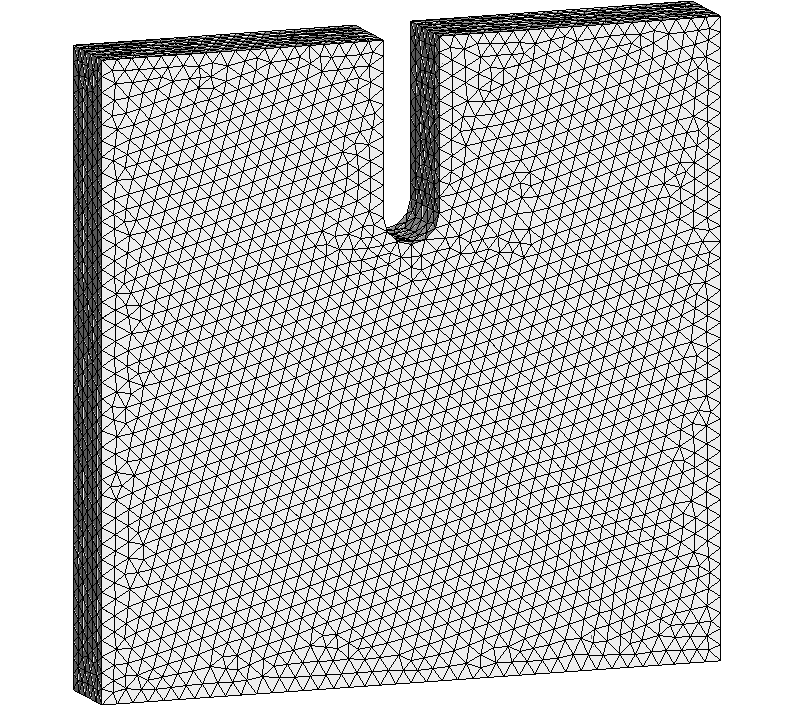}
            \begin{center}
            (d)
            \end{center}
        \end{minipage}  
        \caption{(a). The model geometric dimensions (unit: mm) and boundary conditions of compact tension experiment. (b). The mesh with $288693$ elements; (c). The mesh with $91939$ elements; (d). The mesh $38960$ elements.}
        \label{fig26: cbranching-2-model-grids}
\end{figure}

The geometric setup and loading conditions for the numerical simulations are depicted in Figure.\ref{fig26: cbranching-2-model-grids}(a). A prescribed velocity in the $x_1$-direction (from left to right) is applied along the right face of the notch, indicated by the red arrows, while the left face, marked by the blue triangles, is constrained in the $x_1$-direction. The prescribed velocity is formulated as following equation,
\begin{equation}
v =
\begin{cases} 
\frac{t}{t_0}v_0,  & \text{if }\ t \le t_0, \\
v_0, & \text{if } \ t > t_0.
\end{cases}
\end{equation}
in which, $t_0 = \ 100 \mu s$ and $v_0 = 1.375\ m/s, \  3.318\ m/s, \  3.993 \ m/s$ for three cases, respectively. 

The material properties of the concrete specimen are as following: Young's modulus $E=36\ GPa$, Poisson ratio $v=0.18$, critical fracture energy release rate $\mathcal{G}_c=65\ N/m$ and density $\rho = 2400\ kg/m^3$ (\cite{ovzbolt2013dynamic}). Three different three-dimensional finite element meshes with tetrahedron elements, i.e., fine mesh with $56048$ nodes and $288693$ elements, medium mesh with $19324$ nods and $91939$ elements, and coarse mesh with $8819$ nodes and $38960$ elements, are used for numerical simulation, as shown in Figure.\ref{fig26: cbranching-2-model-grids}(b-d). Total simulation lasts $300\ \mu s$. 

\begin{figure}[htp]
	\centering
        \begin{minipage}{0.3\linewidth}
            \begin{center}
            \includegraphics[height=1.5in]{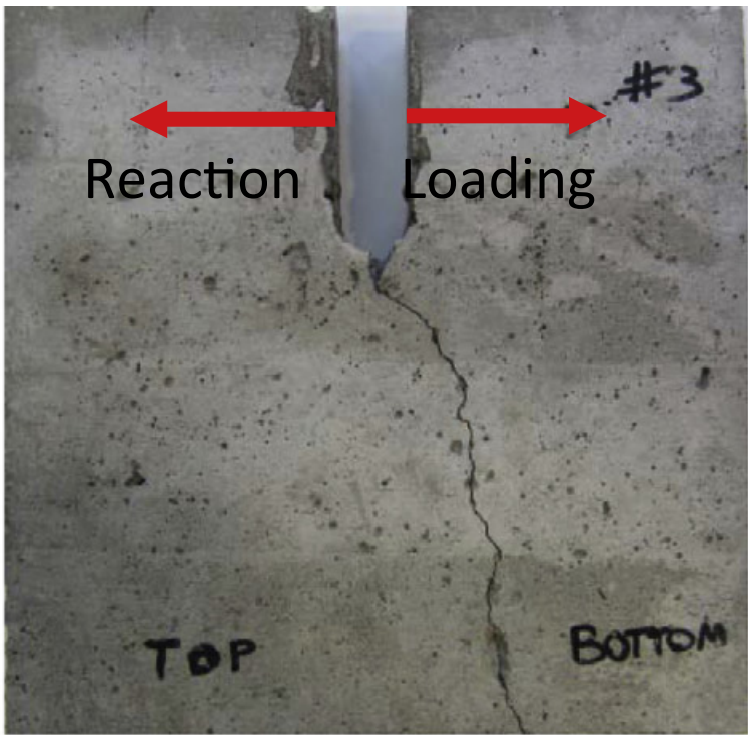}
            \end{center}
            \begin{center}
            (a)
            \end{center}
        \end{minipage}
        \hfill
        \begin{minipage}{0.3\linewidth}
            \begin{center}
            \includegraphics[height=1.5in]{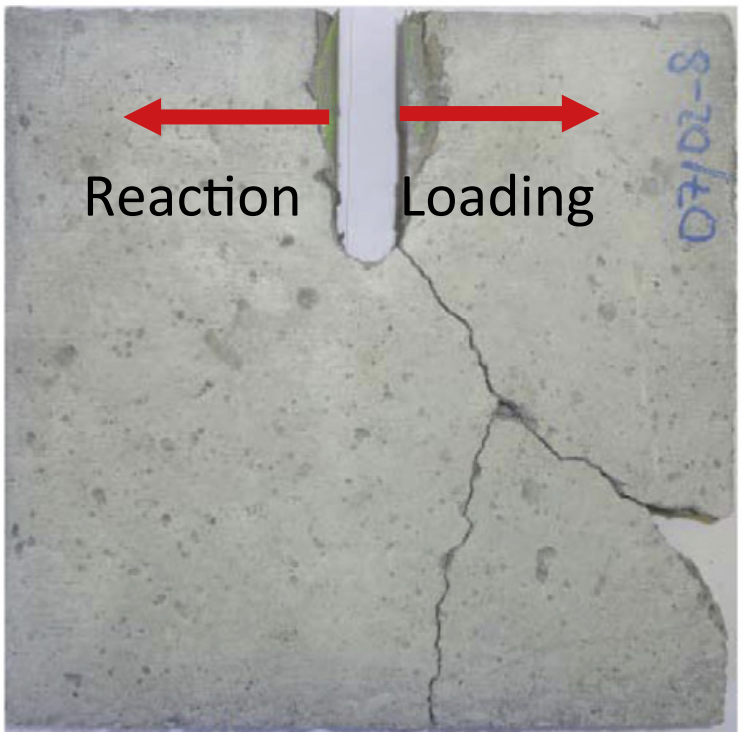}
            \end{center}
            \begin{center}
            (b)
            \end{center}
        \end{minipage}   
        \hfill
        \begin{minipage}{0.3\linewidth}
            \begin{center}
            \includegraphics[height=1.5in]{cbranching-2-crack-pattern-exp-v3.993.png}
            \end{center}
            \begin{center}
            (c)
            \end{center}
        \end{minipage}
        \hfill
        \begin{minipage}{0.3\linewidth}
            \begin{center}
            \includegraphics[height=1.5in]{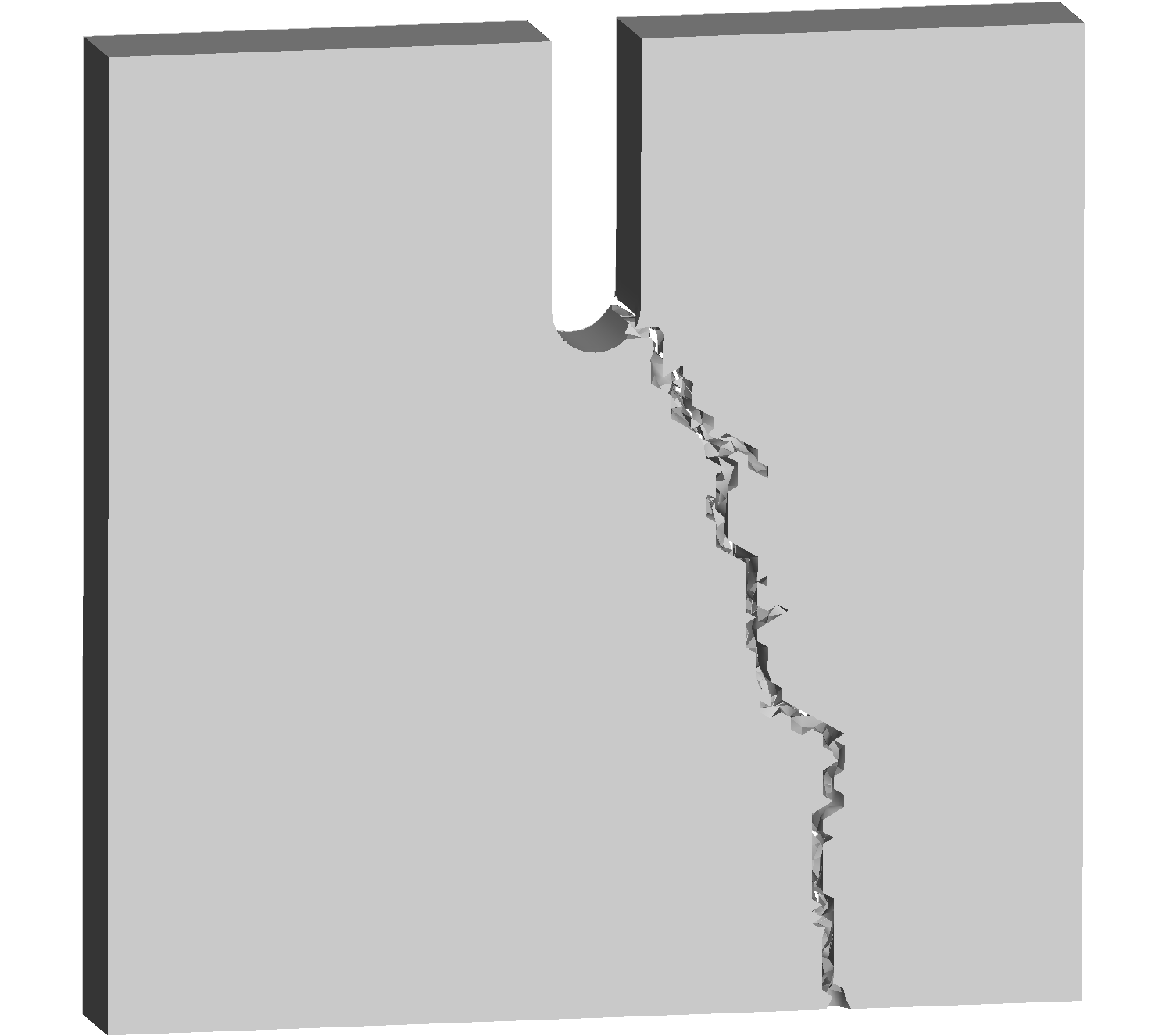}
            \end{center}
            \begin{center}
            (d)
            \end{center}
        \end{minipage}
        \hfill
        \begin{minipage}{0.3\linewidth}
            \begin{center}
            \includegraphics[height=1.5in]{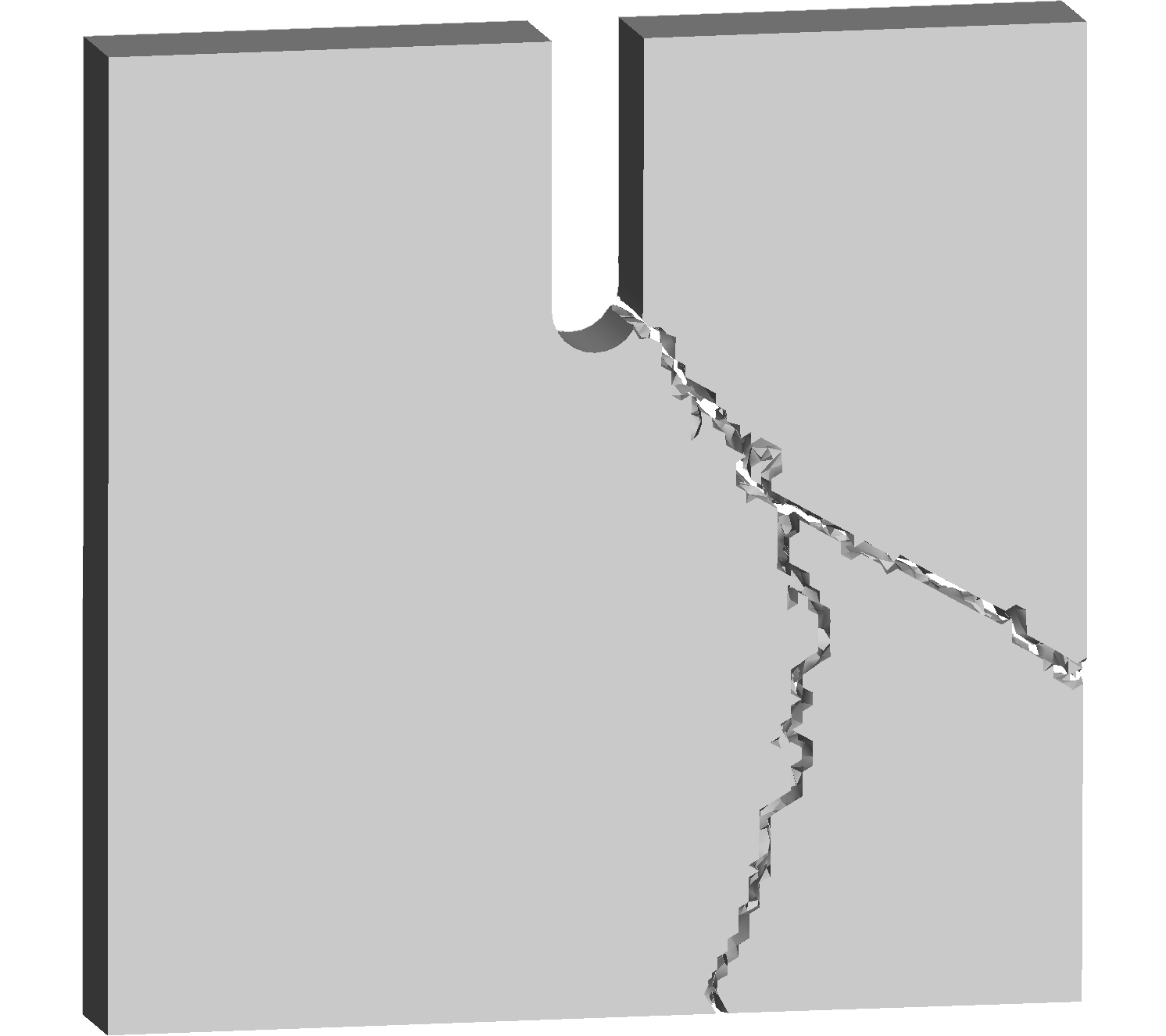}
            \end{center}
            \begin{center}
            (e)
            \end{center}
        \end{minipage}   
        \hfill
        \begin{minipage}{0.3\linewidth}
            \begin{center}
            \includegraphics[height=1.5in]{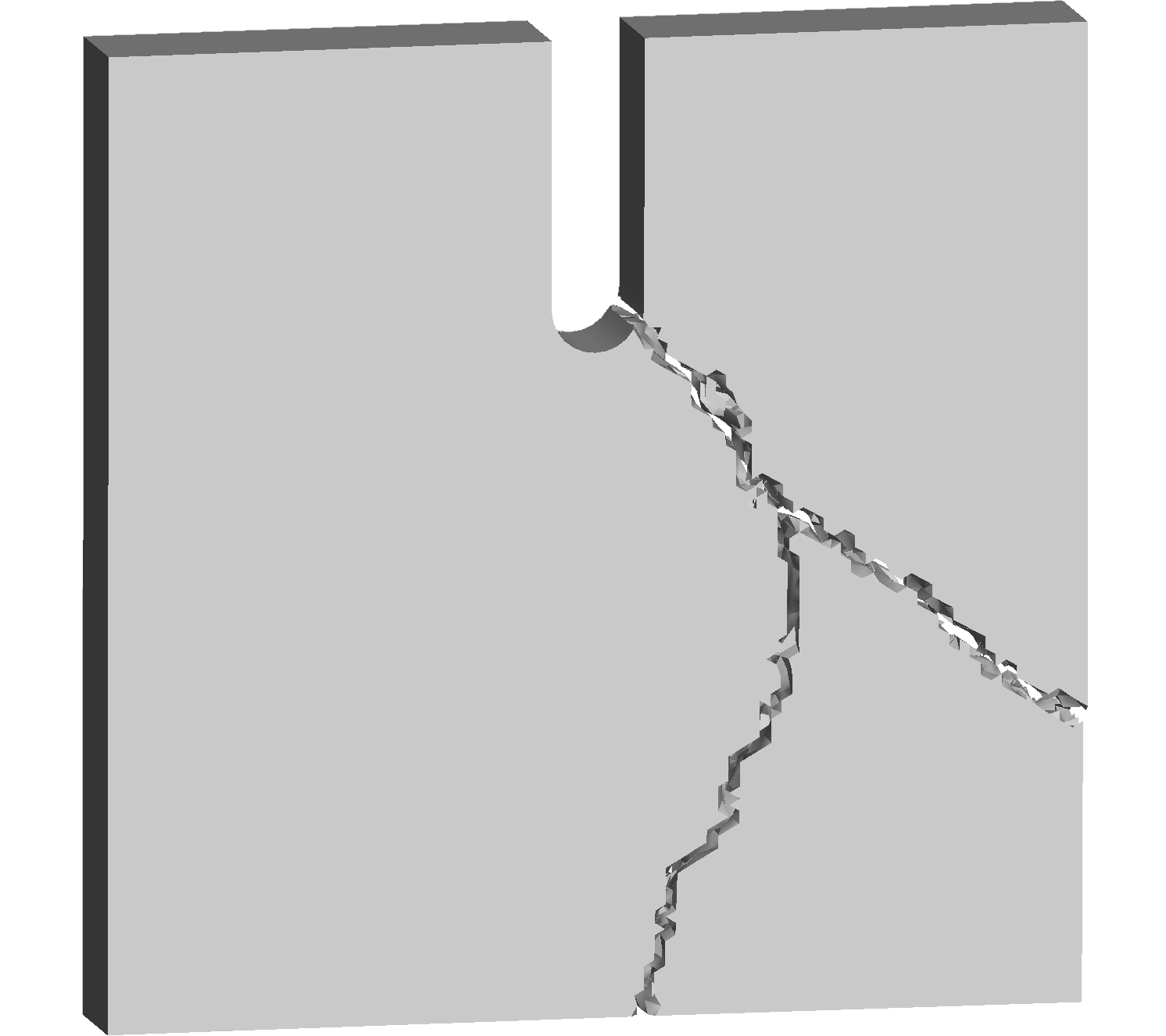}
            \end{center}
            \begin{center}
            (f)
            \end{center}
        \end{minipage}
        \hfill
        \begin{minipage}{0.3\linewidth}
            \begin{center}
            \includegraphics[height=1.5in]{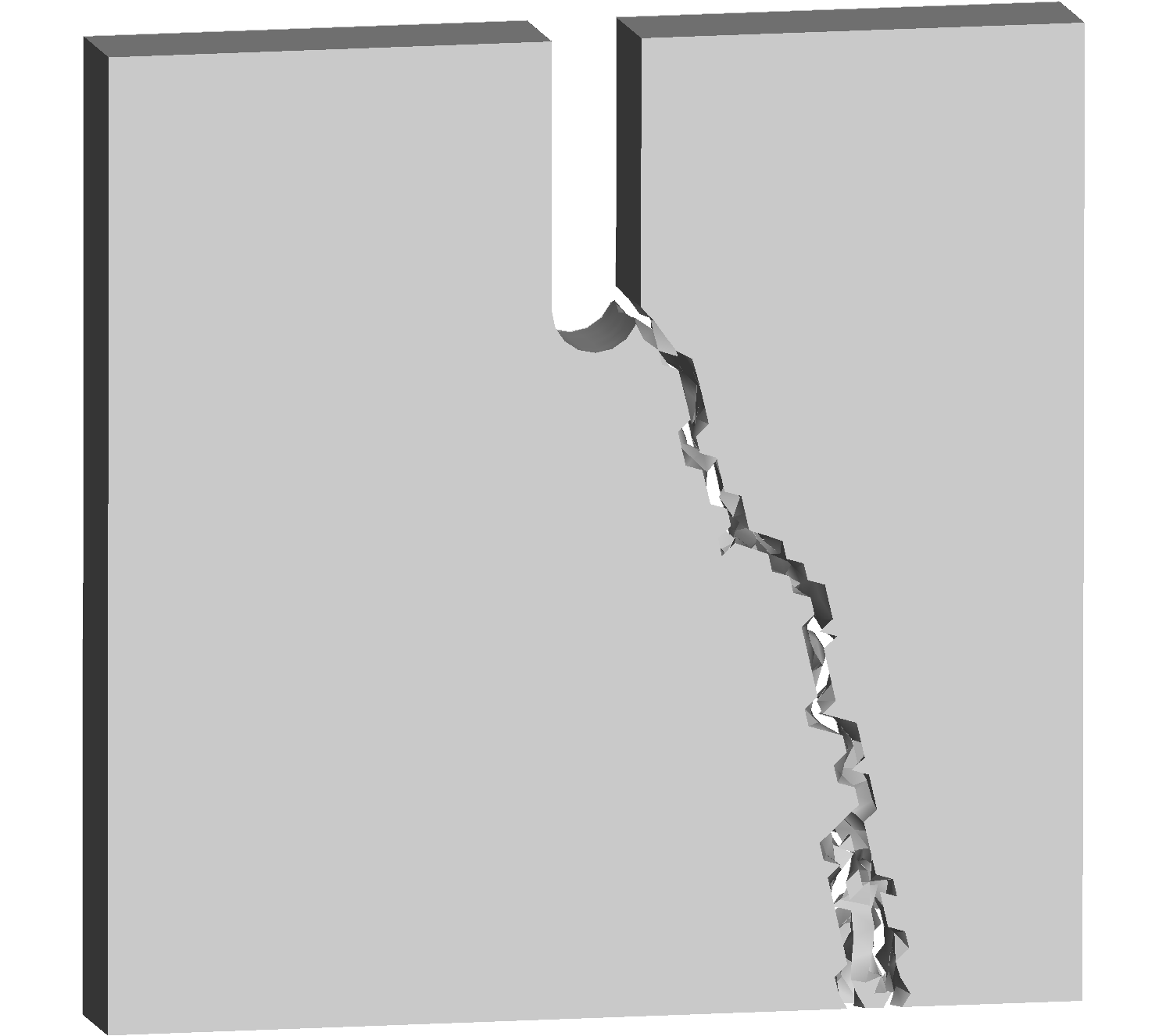}
            \end{center}
            \begin{center}
            (g)
            \end{center}
        \end{minipage}
        \hfill
        \begin{minipage}{0.3\linewidth}
            \begin{center}
            \includegraphics[height=1.5in]{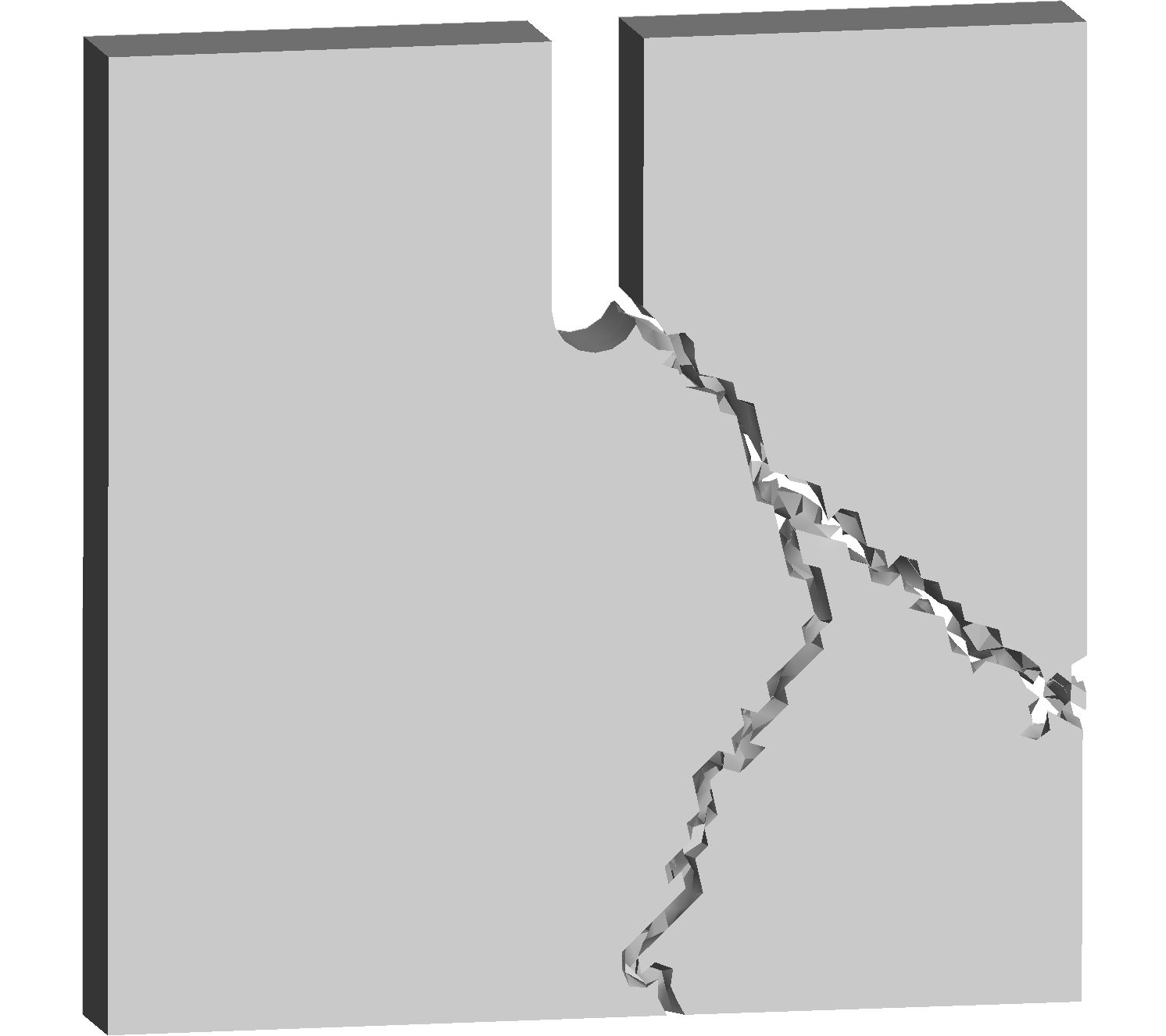}
            \end{center}
            \begin{center}
            (h)
            \end{center}
        \end{minipage}   
        \hfill
        \begin{minipage}{0.3\linewidth}
            \begin{center}
            \includegraphics[height=1.5in]{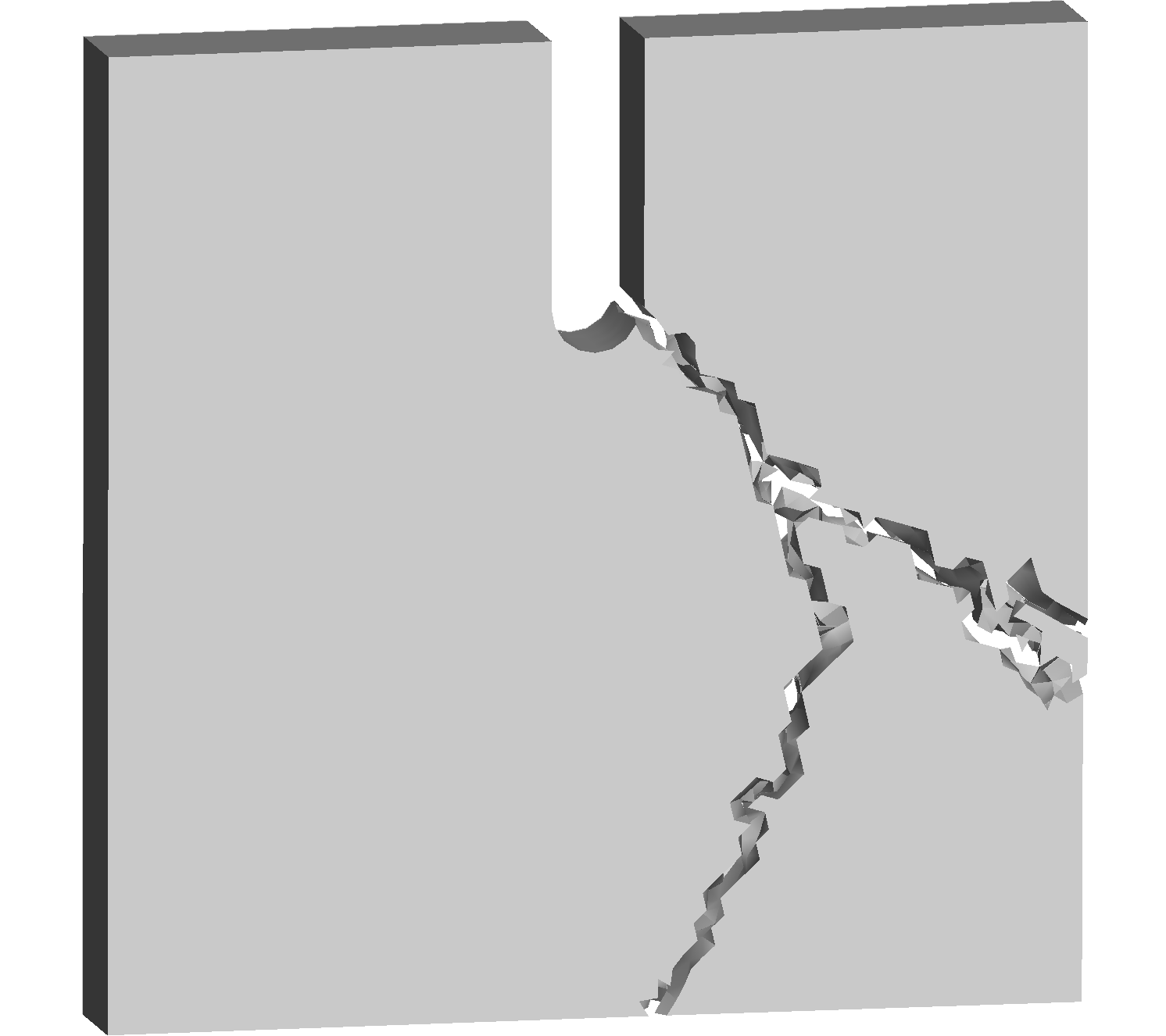}
            \end{center}
            \begin{center}
            (i)
            \end{center}
        \end{minipage}
        \hfill
        \begin{minipage}{0.3\linewidth}
            \begin{center}
            \includegraphics[height=1.5in]{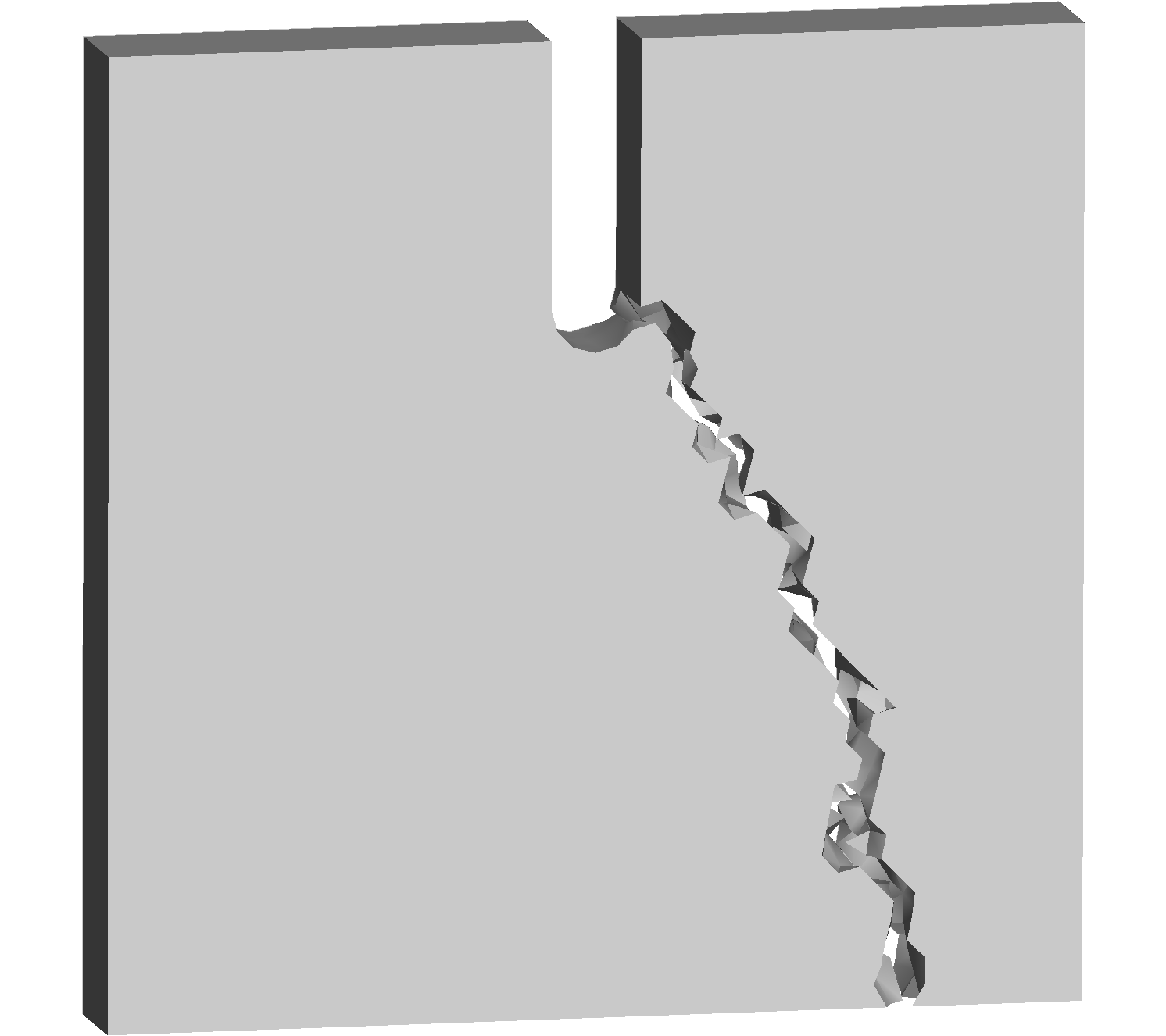}
            \end{center}
            \begin{center}
            (j)
            \end{center}
        \end{minipage}
        \hfill
        \begin{minipage}{0.3\linewidth}
            \begin{center}
            \includegraphics[height=1.5in]{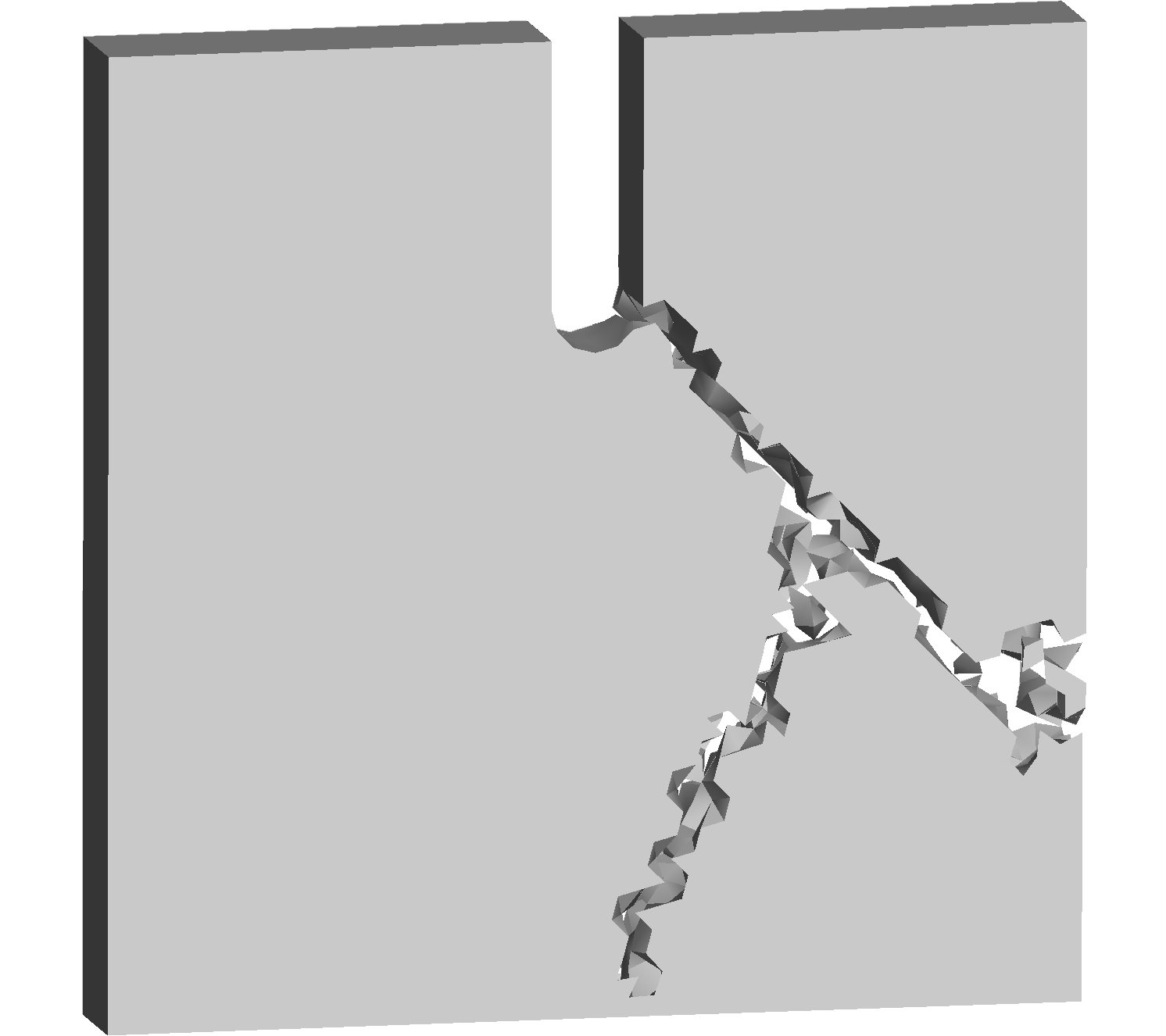}
            \end{center}
            \begin{center}
            (k)
            \end{center}
        \end{minipage}   
        \hfill
        \begin{minipage}{0.3\linewidth}
            \begin{center}
            \includegraphics[height=1.5in]{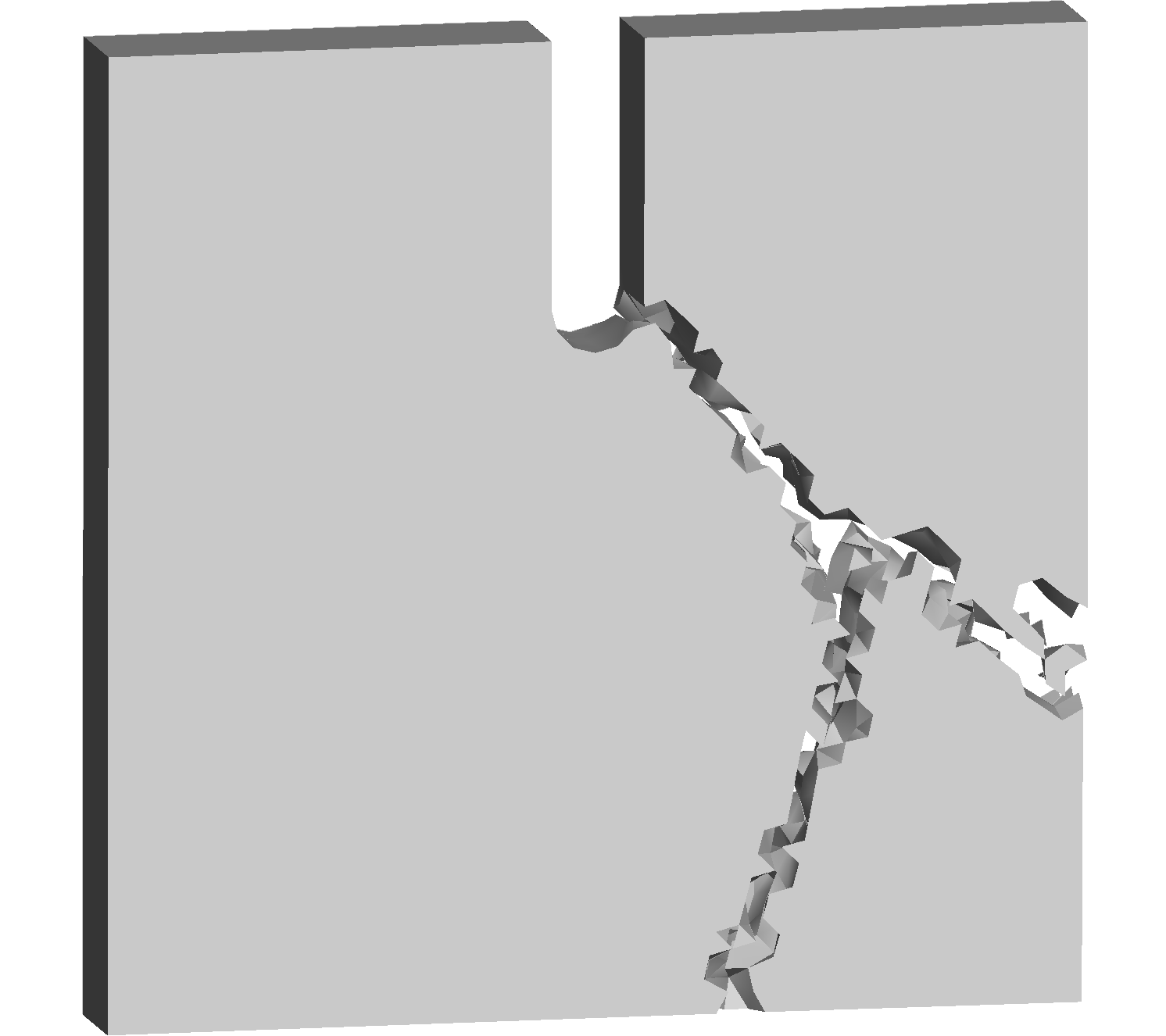}
            \end{center}
            \begin{center}
            (l)
            \end{center}
        \end{minipage}
        \caption{(a-c). experimental crack patterns under applied velocity of $1.375\ m/s$, $3.318\ m/s$ and $3.993\ m/s$ (\cite{ovzbolt2013dynamic}); (d-f). numerical crack patterns of $288693$ elements model under applied velocity of $1.375\ m/s$, $3.318\ m/s$ and $3.993\ m/s$; (g-i). numerical crack patterns of $91939$ elements model under applied velocity of $1.375\ m/s$, $3.318\ m/s$ and $3.993\ m/s$; (j-l). numerical crack patterns of $38960$ elements model under applied velocity of $1.375\ m/s$, $3.318\ m/s$ and $3.993\ m/s$. }
        \label{fig27: cbranching-2-crack-patterns}
\end{figure}

Figure.\ref{fig27: cbranching-2-crack-patterns} presents the final crack patterns for the three different meshes under three applied velocity cases, alongside corresponding experimental results, demonstrating the effectiveness of the proposed CEM in capturing three-dimensional crack branching under Dirichlet boundary conditions.

The proposed CEM successfully captures branching crack patterns, consistent with experimental observations, when the applied velocity exceeds $3.0$ m/s, even when using a coarse mesh with only $38960$ elements. Additionally, at an applied velocity of $1.375$ m/s, the CEM accurately reproduces a single crack pattern, also in agreement with experimental results. Therefore, it is concluded that the proposed CEM has the capability in accurately tracking the three-dimensional transient-dynamic crack branching under Dirichlet boundary condition.

\begin{figure}[htp]
	\centering
        \begin{minipage}{0.24\linewidth}
            \begin{center}
            \includegraphics[height=1.5in]{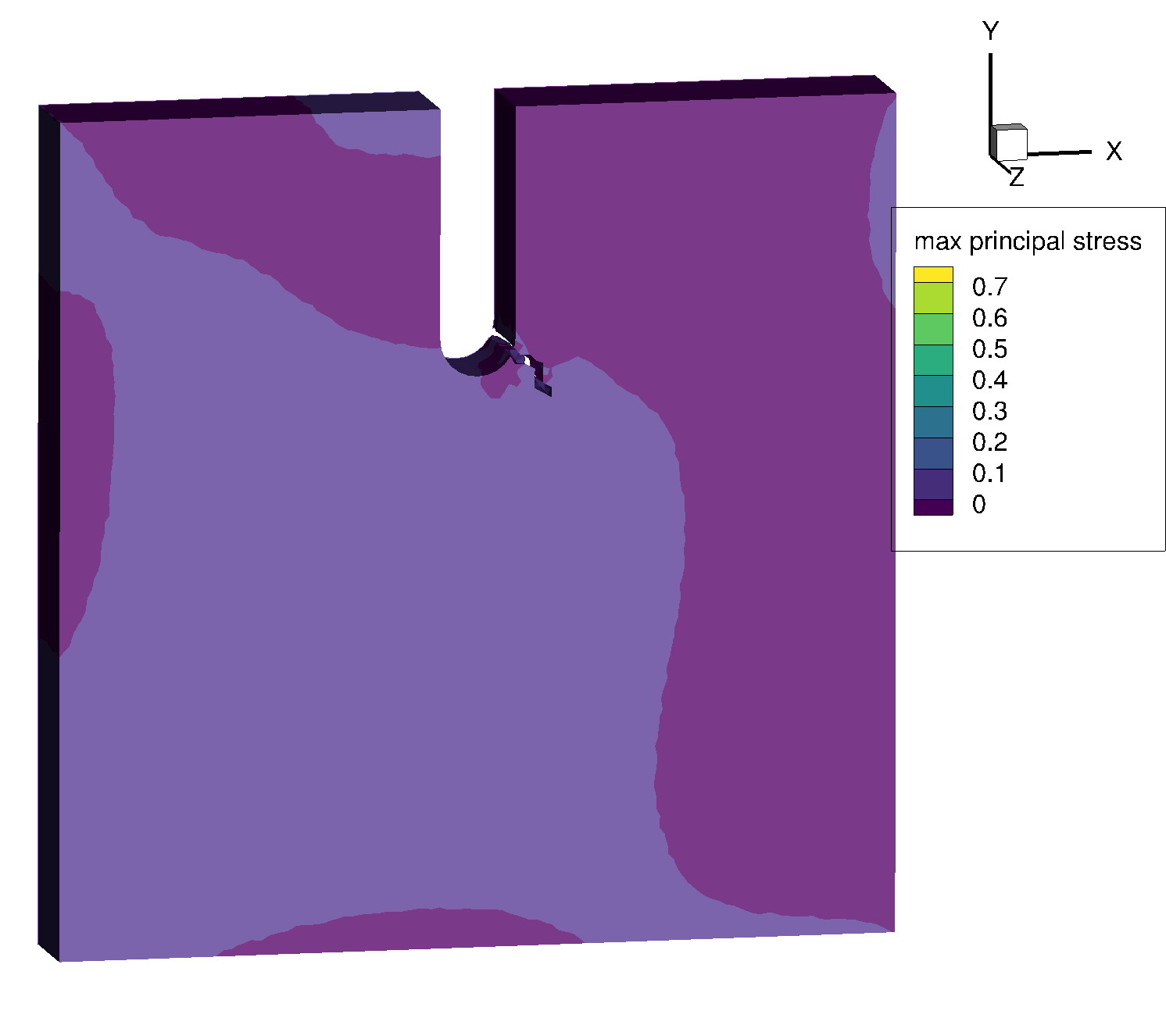}
            (a)
            \end{center}
        \end{minipage}
        \hfill
        \begin{minipage}{0.24\linewidth}
            \begin{center}
            \includegraphics[height=1.5in]{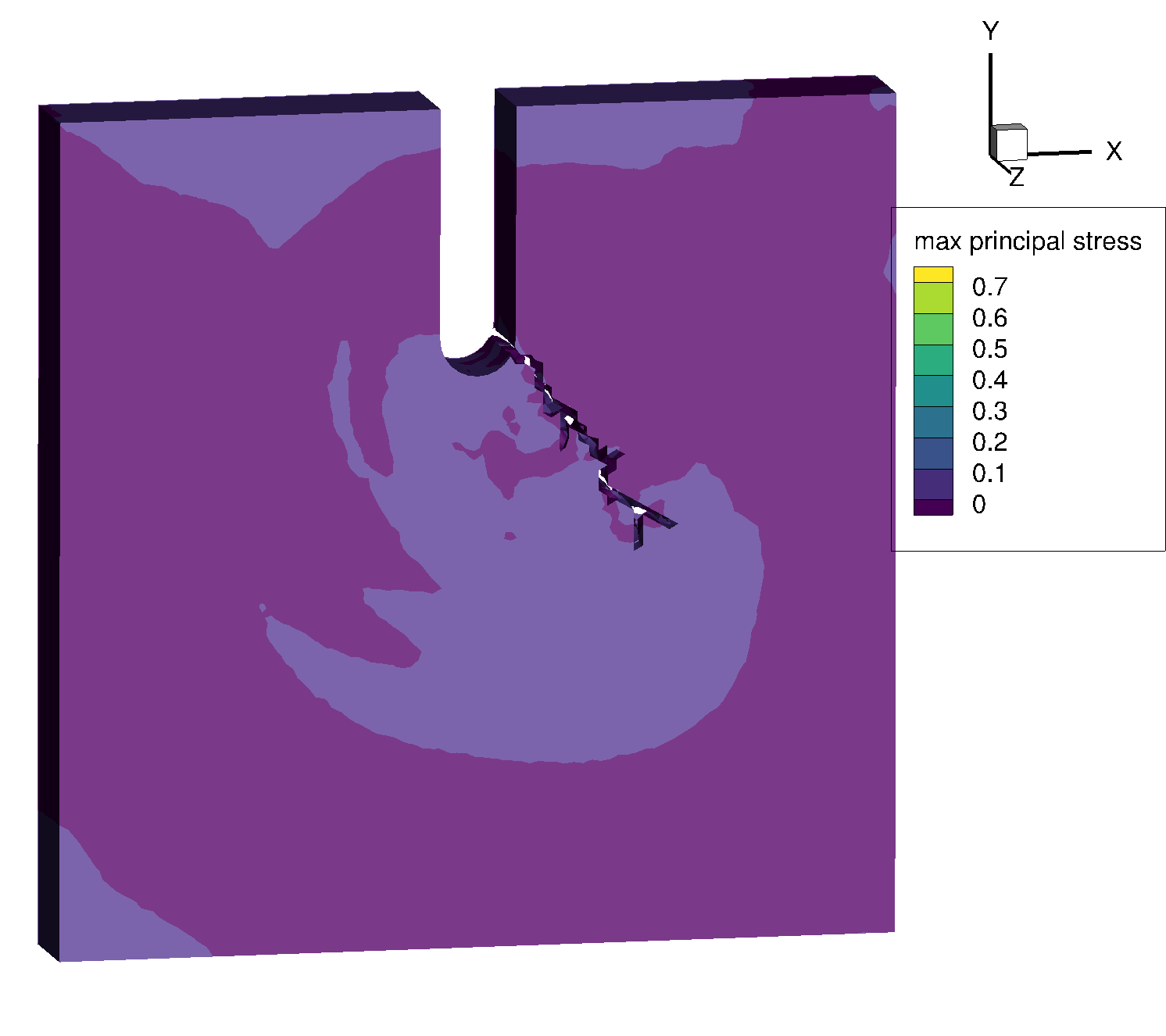}
            (b)
            \end{center}
        \end{minipage}   
        \hfill
        \begin{minipage}{0.24\linewidth}
            \begin{center}
            \includegraphics[height=1.5in]{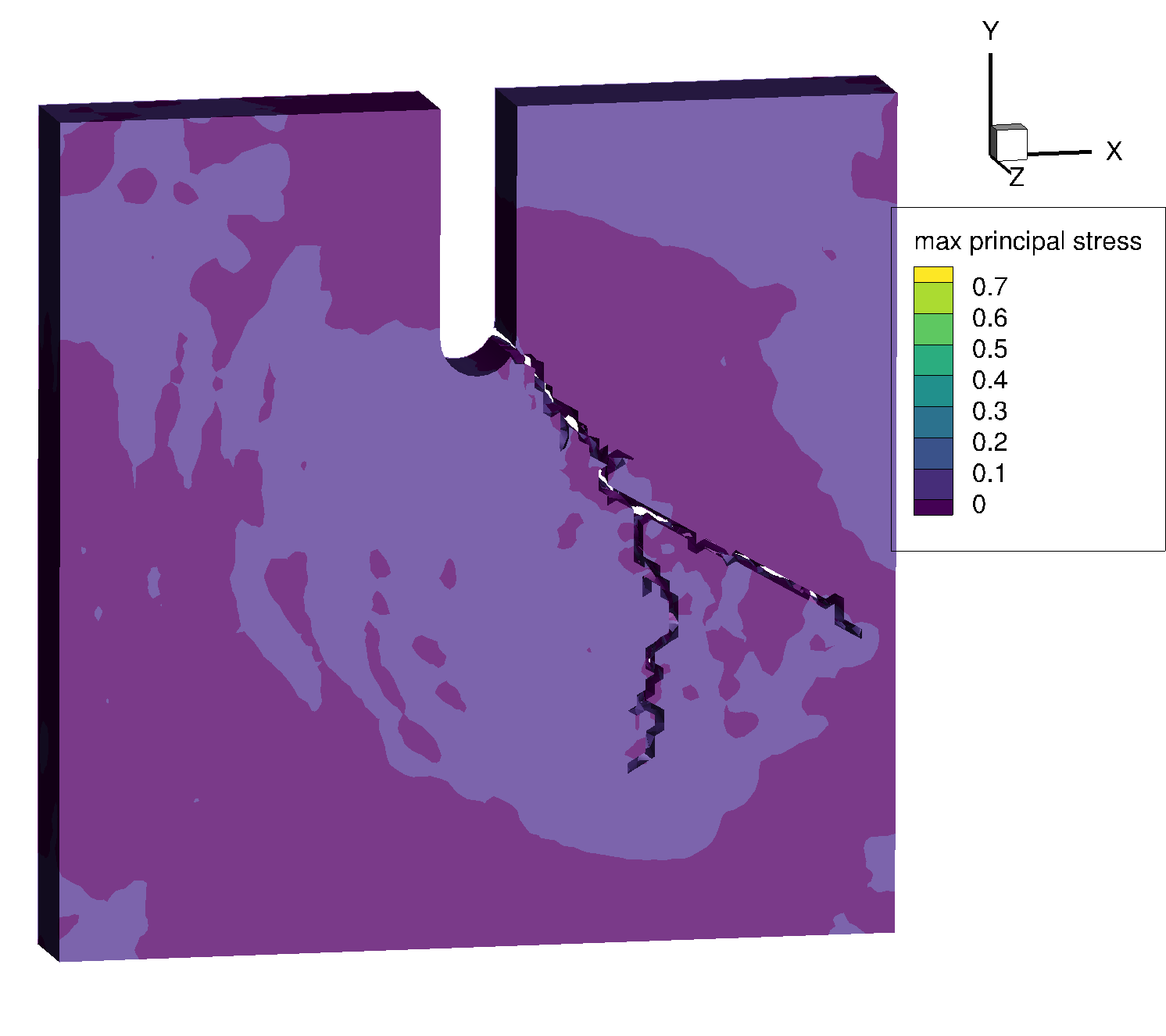}
            (c)
            \end{center}
        \end{minipage}
        \hfill
        \begin{minipage}{0.24\linewidth}
            \begin{center}
            \includegraphics[height=1.5in]{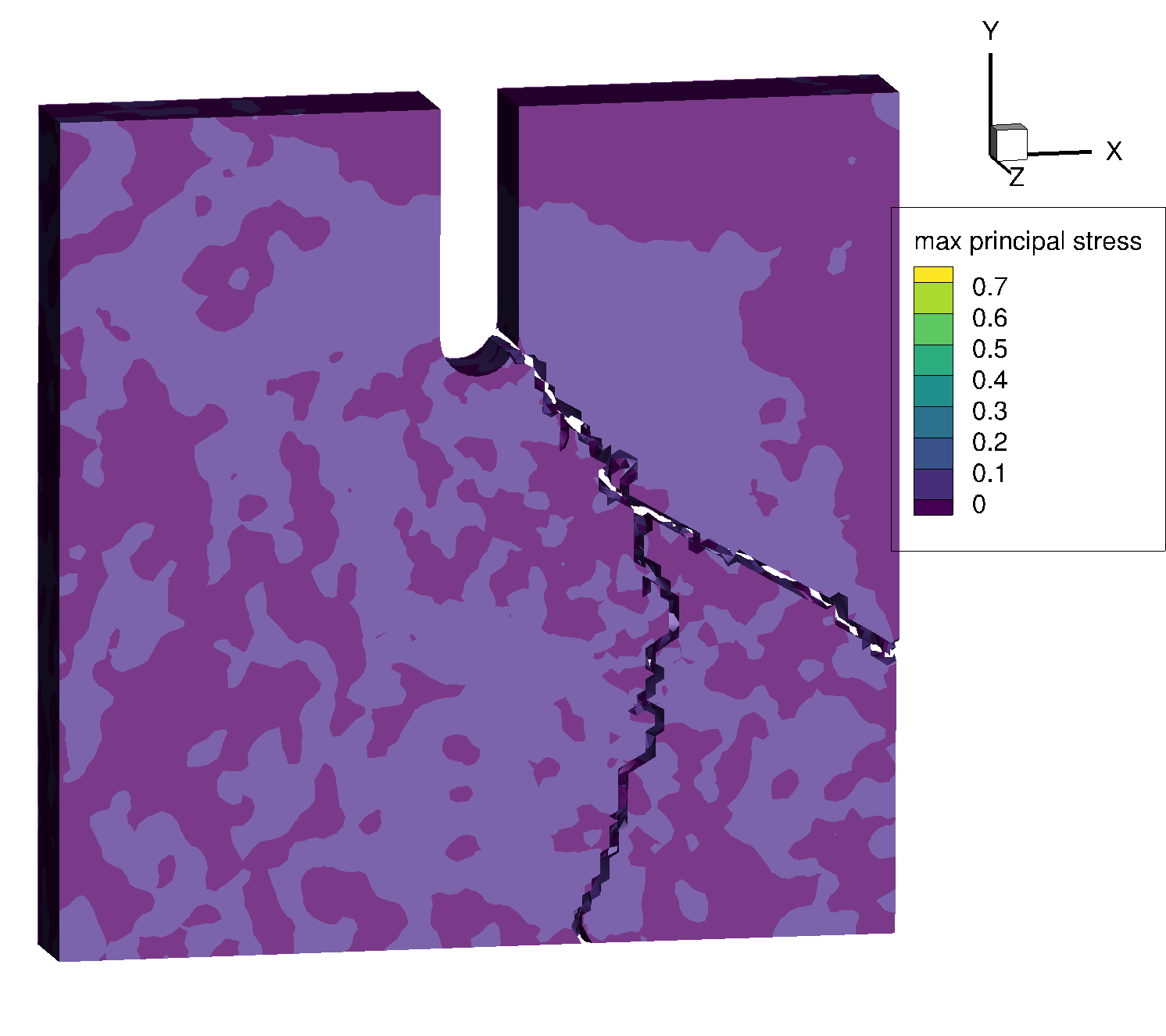}
            (d)
            \end{center}
        \end{minipage}
        \caption{Max. principle stress contour and crack patterns evolution of fine mesh $288693$ elements under the velocity of $v_0=3.318\ m/s$ at times (a). $t=60\ \mu s$, (b). $t=84\ \mu s$, (c). $t=144\ \mu s$, (d). $t=300\ \mu s$.}
        \label{fig28: cbranching-2-crack-evolution-1}
\end{figure}

\begin{figure}[htp]
	\centering
        \begin{minipage}{0.24\linewidth}
            \begin{center}
            \includegraphics[height=1.5in]{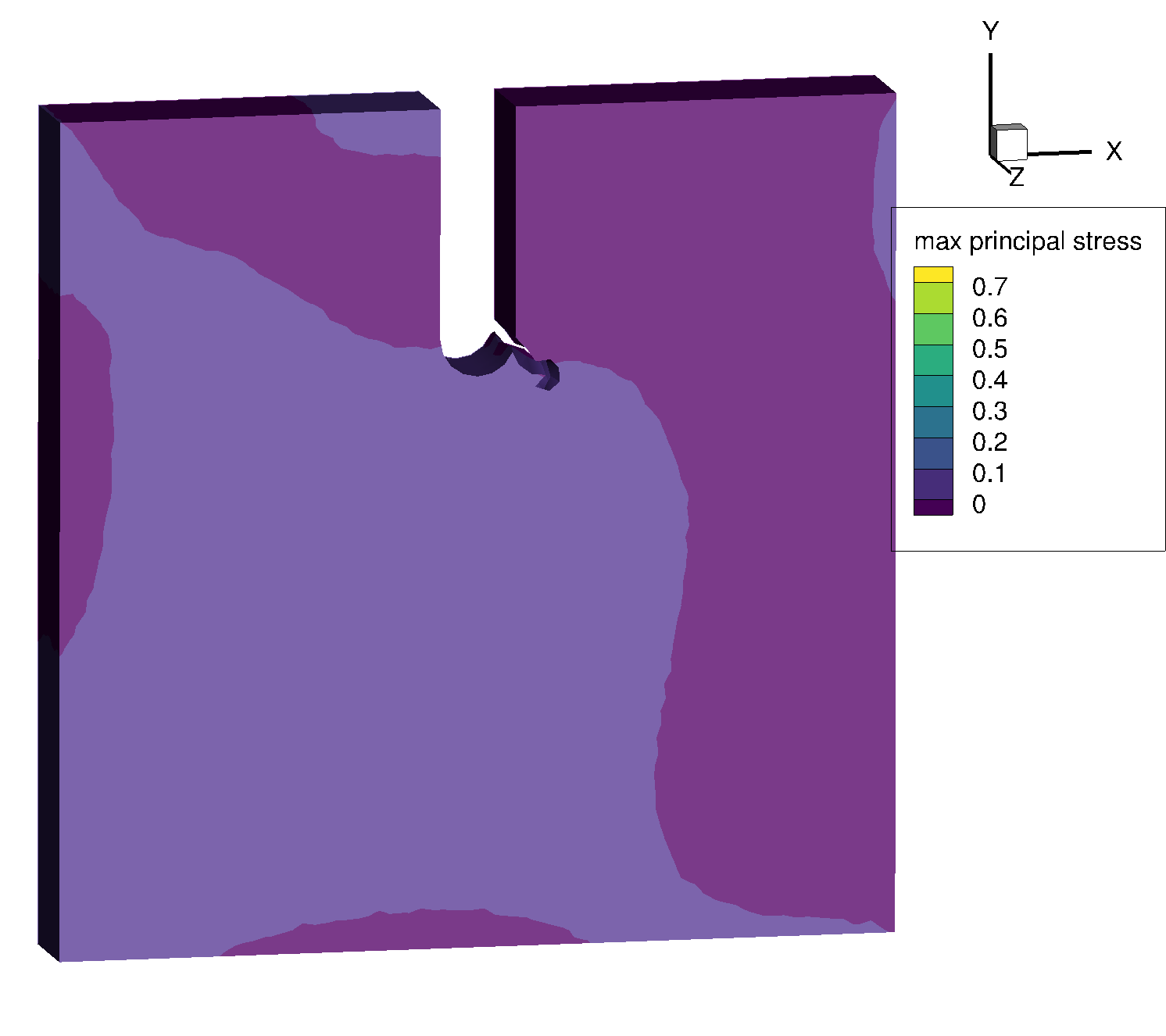}
            (a)
            \end{center}
        \end{minipage}
        \hfill
        \begin{minipage}{0.24\linewidth}
            \begin{center}
            \includegraphics[height=1.5in]{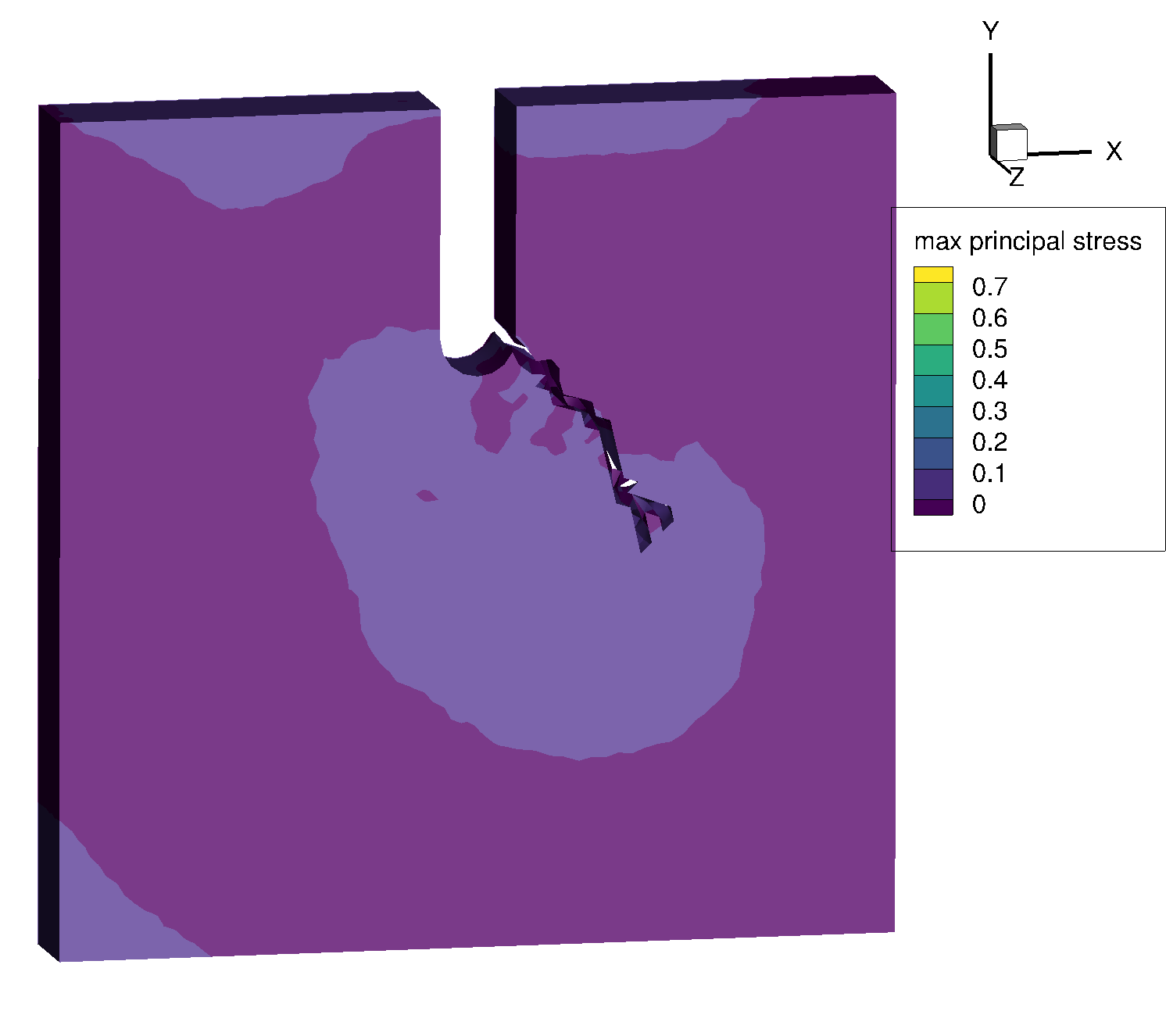}
            (b)
            \end{center}
        \end{minipage}   
        \hfill
        \begin{minipage}{0.24\linewidth}
            \begin{center}
            \includegraphics[height=1.5in]{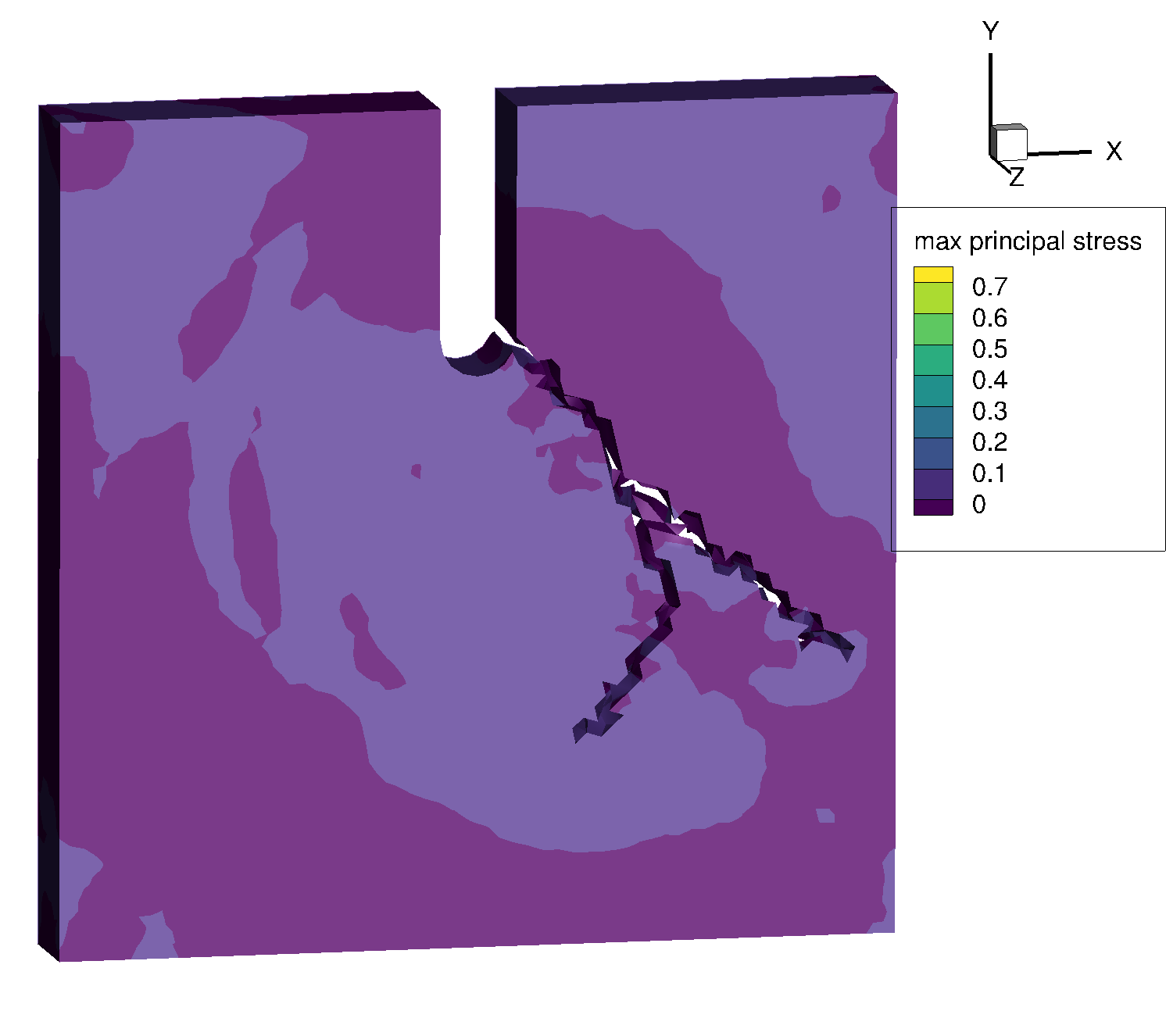}
            (c)
            \end{center}
        \end{minipage}
        \hfill
        \begin{minipage}{0.24\linewidth}
            \begin{center}
            \includegraphics[height=1.5in]{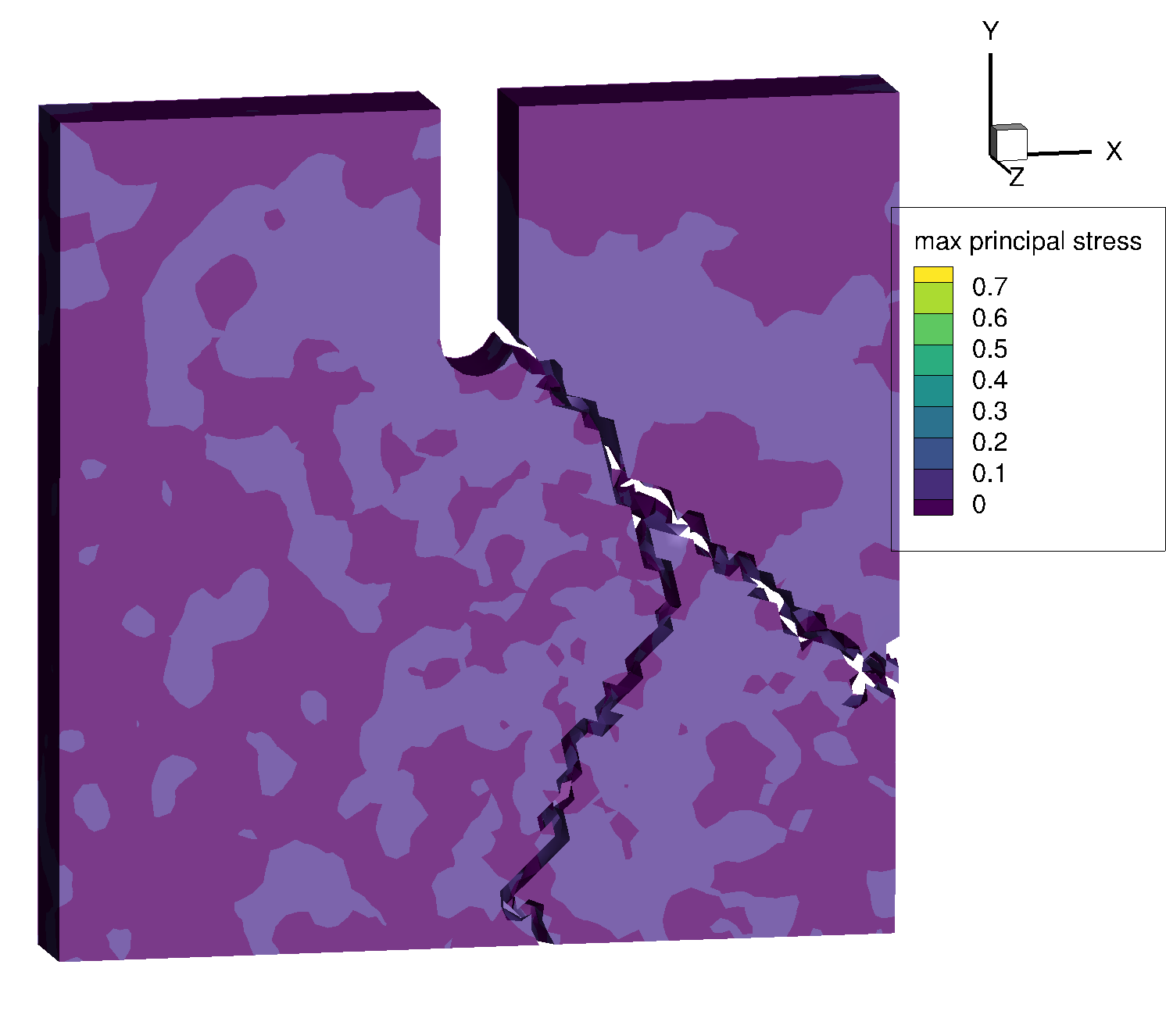}
            (d)
            \end{center}
        \end{minipage}
        \caption{Max. principle stress contour and crack patterns evolution of fine mesh $91939$ elements under the velocity of $v_0=3.318\ m/s$ at times (a). $t=60\ \mu s$, (b). $t=84\ \mu s$, (c). $t=144\ \mu s$, (d). $t=300\ \mu s$.}
        \label{fig29: cbranching-2-crack-evolution-2}
\end{figure}

\begin{figure}[htp]
	\centering
        \begin{minipage}{0.24\linewidth}
            \begin{center}
            \includegraphics[height=1.5in]{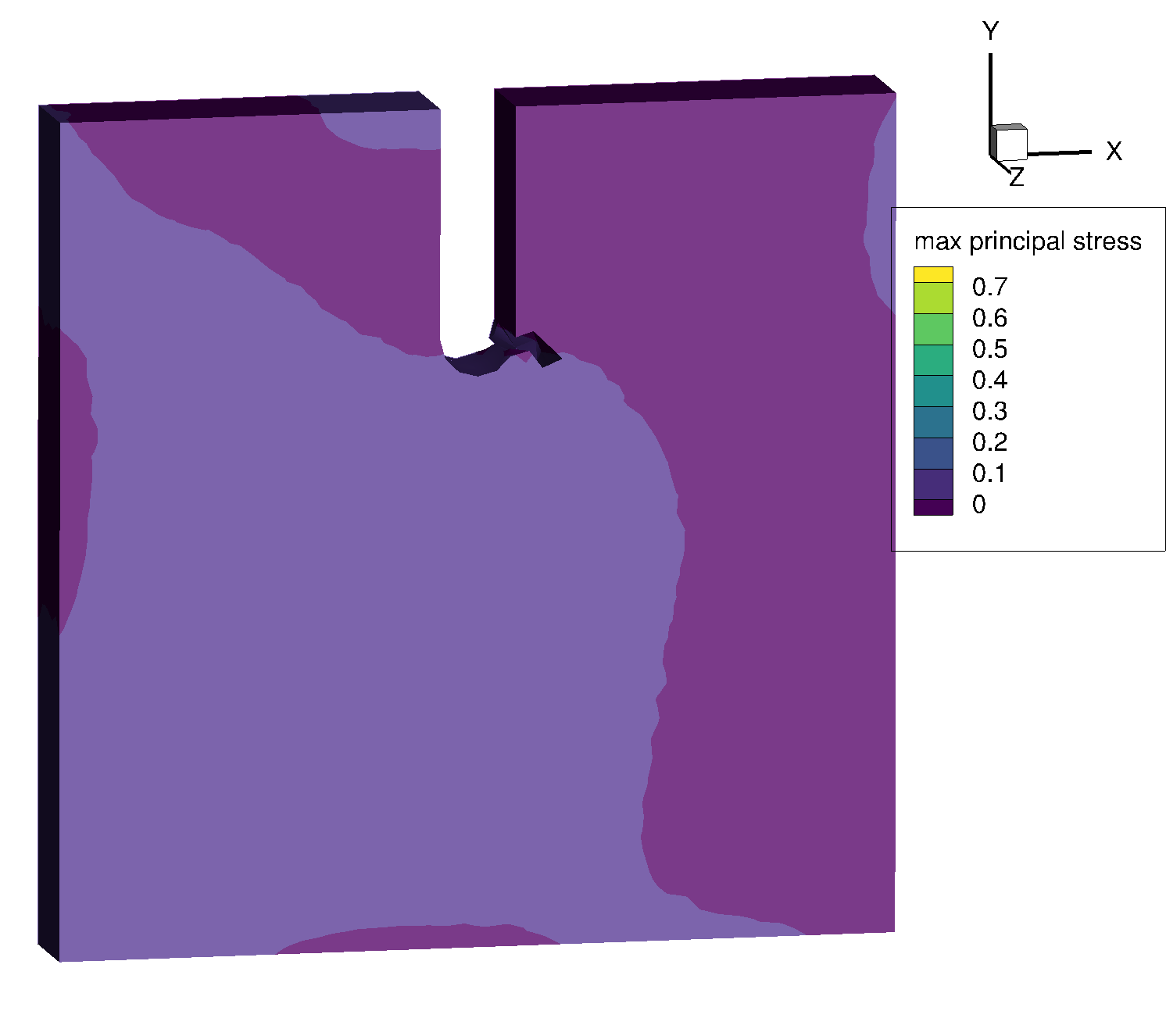}
            (a)
            \end{center}
        \end{minipage}
        \hfill
        \begin{minipage}{0.24\linewidth}
            \begin{center}
            \includegraphics[height=1.5in]{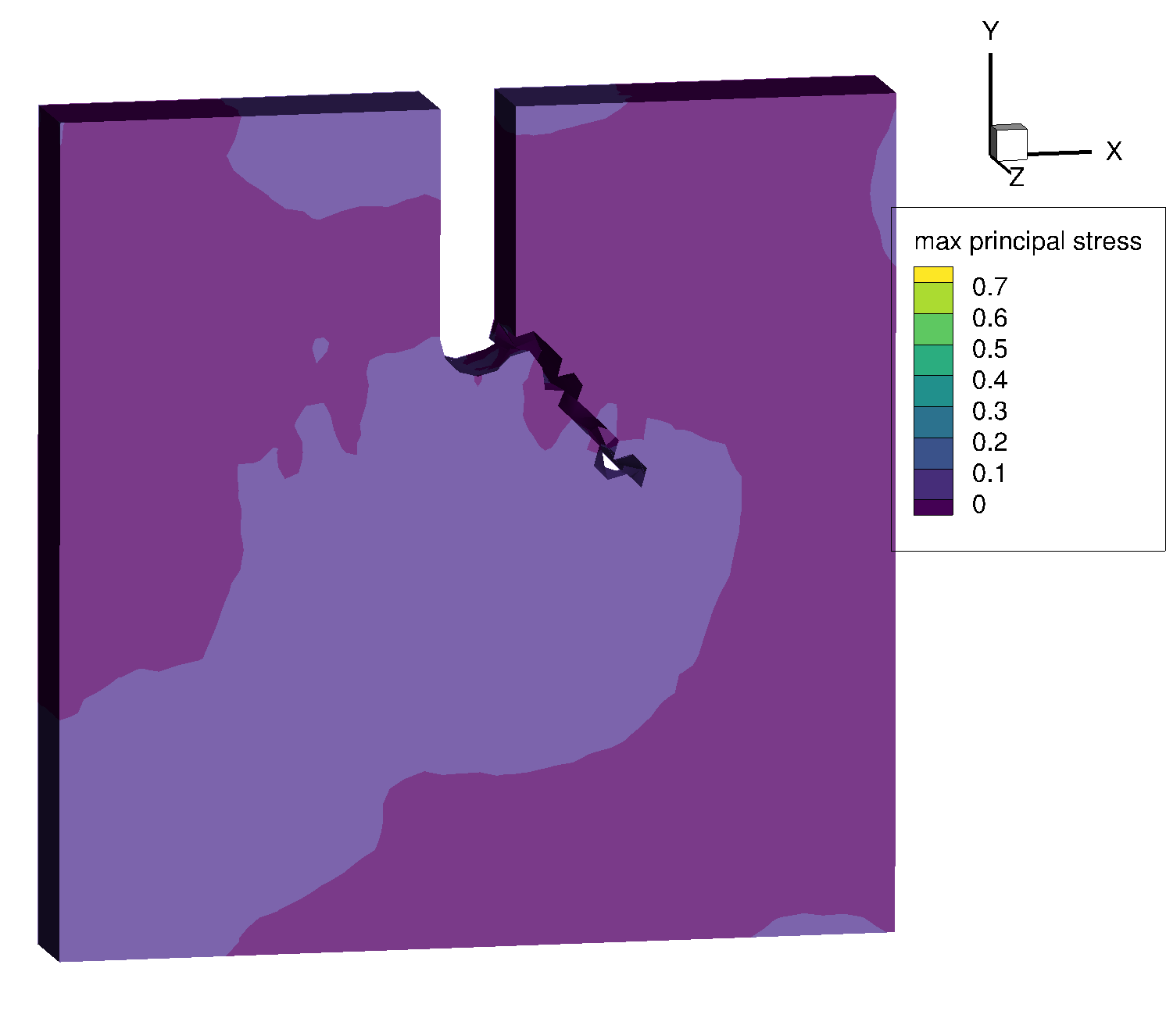}
            (b)
            \end{center}
        \end{minipage}   
        \hfill
        \begin{minipage}{0.24\linewidth}
            \begin{center}
            \includegraphics[height=1.5in]{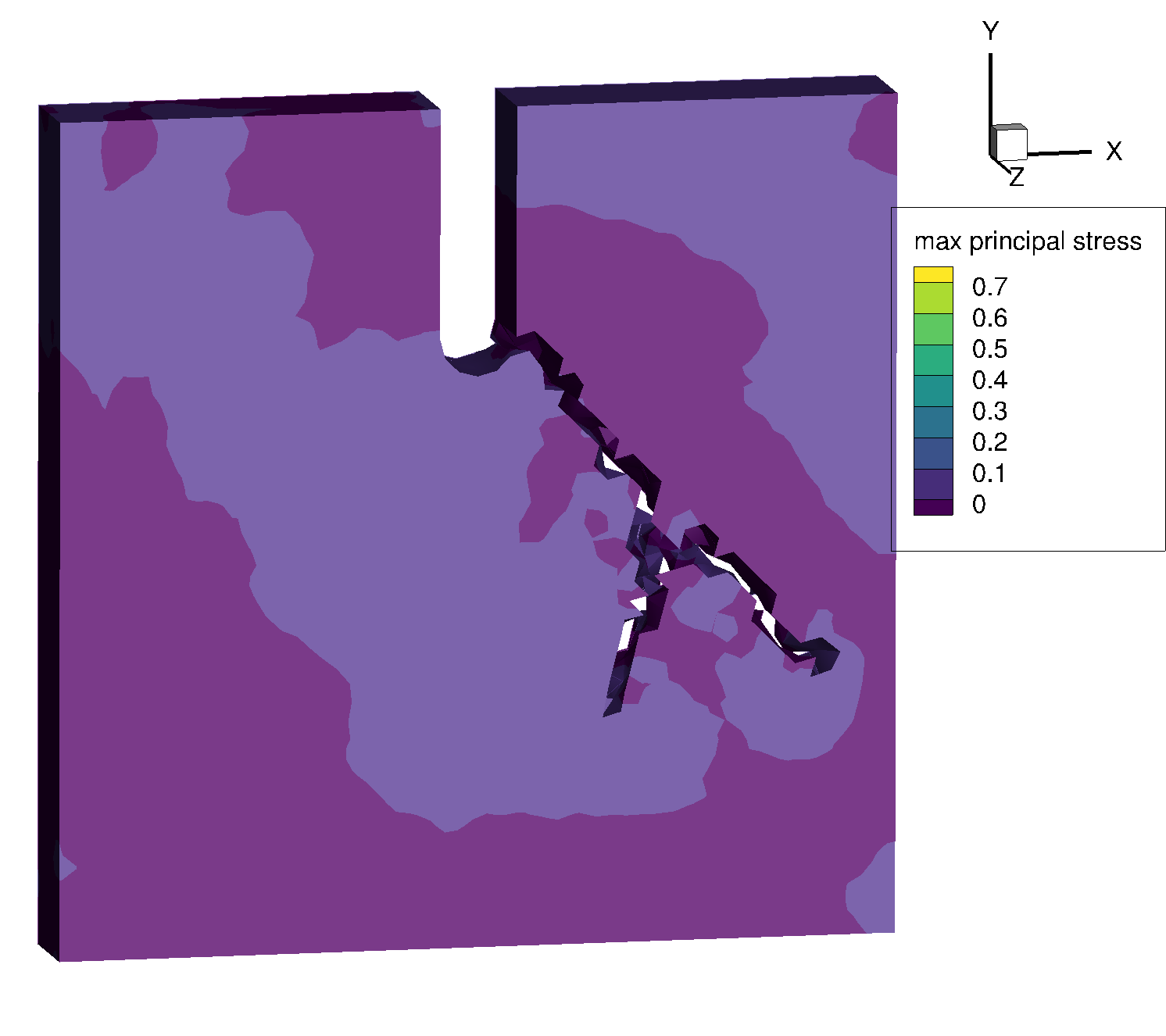}
            (c)
            \end{center}
        \end{minipage}
        \hfill
        \begin{minipage}{0.24\linewidth}
            \begin{center}
            \includegraphics[height=1.5in]{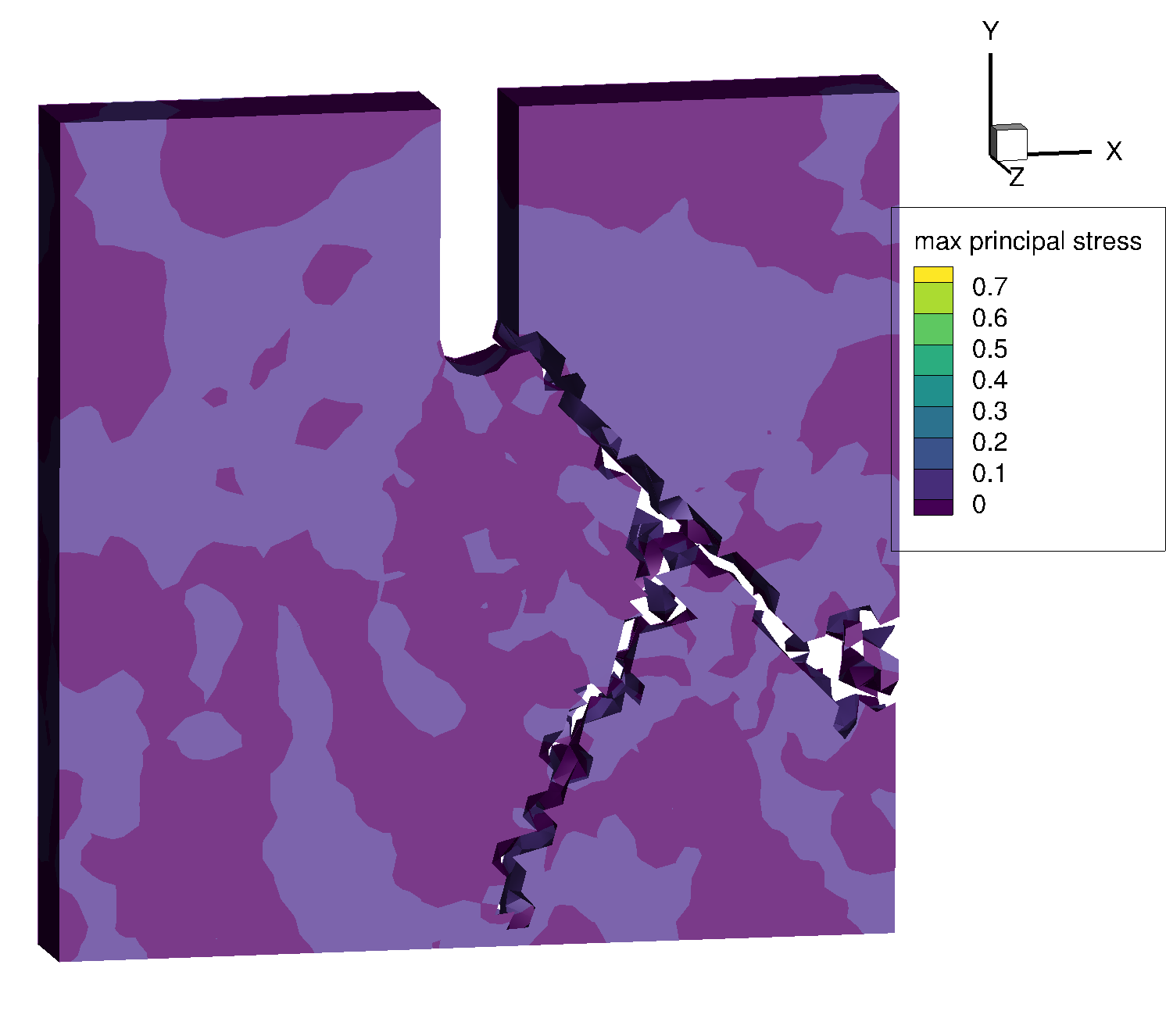}
            (d)
            \end{center}
        \end{minipage}
        \caption{Max. principle stress contour and crack patterns evolution of fine mesh $38960$ elements under the velocity of $v_0=3.318\ m/s$ at times (a). $t=54\ \mu s$, (b). $t=78\ \mu s$, (c). $t=144\ \mu s$, (d). $t=300\ \mu s$.}
        \label{fig30: cbranching-2-crack-evolution-3}
\end{figure}

Figures.\ref{fig28: cbranching-2-crack-evolution-1} - \ref{fig30: cbranching-2-crack-evolution-3} illustrate the evolution of crack patterns at various time steps under an applied velocity of $v_0 = 3.318\ m/s$, using three different meshes: $56048$ nodes, $288693$ elements (Figure.\ref{fig28: cbranching-2-crack-evolution-1}); $19324$ nodes, $91939$ elements (Figure.\ref{fig29: cbranching-2-crack-evolution-2}); and $8819$ nodes, $38960$ elements (Figure.\ref{fig30: cbranching-2-crack-evolution-3}).

\begin{figure}[htp]
	\centering
            \begin{center}
            \includegraphics[height=2.4in]{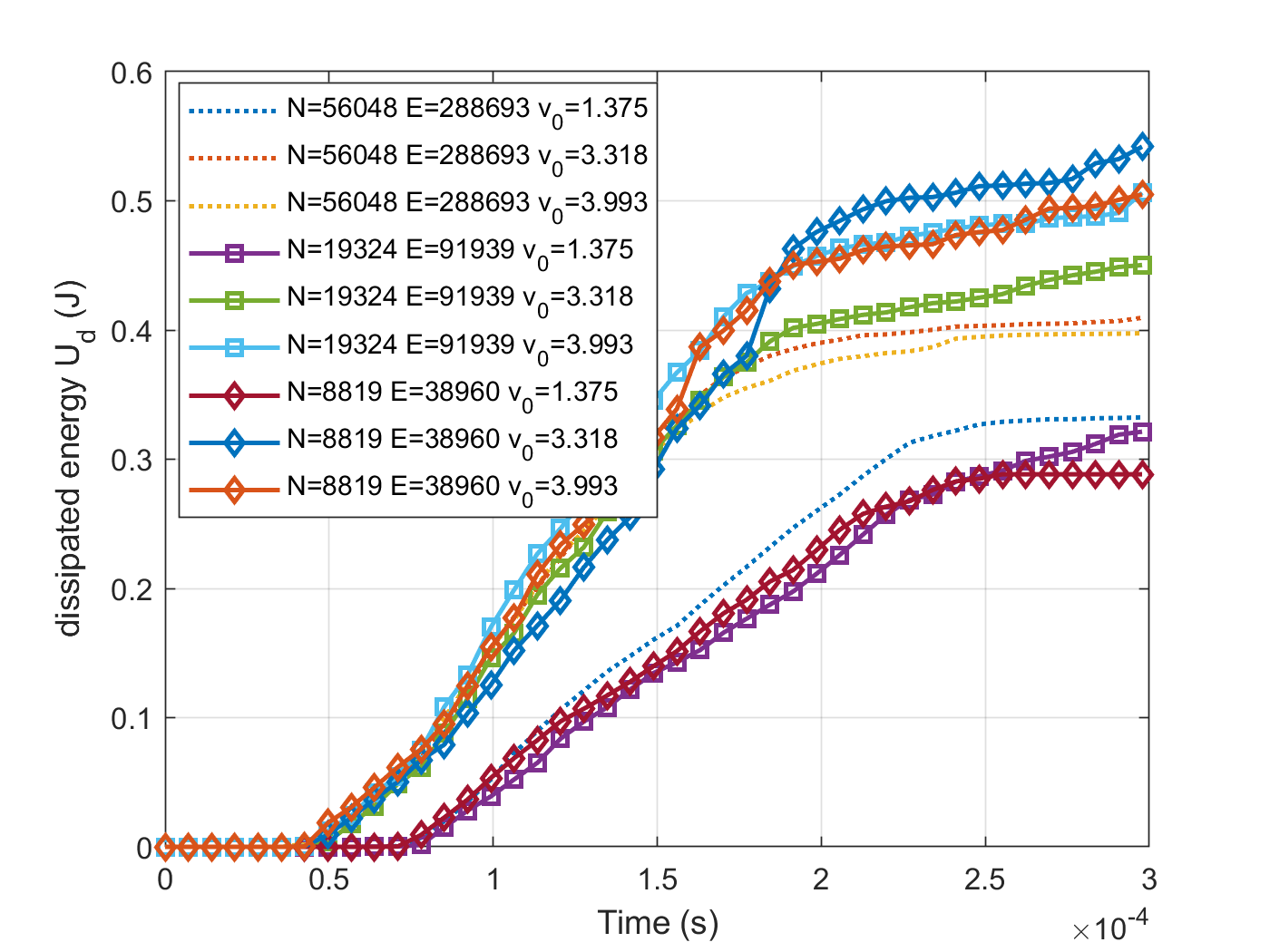}
            \end{center}
        \caption{Comparison of dissipated energy $U_d$ among three representative meshes under three different applied velocities.}
        \label{fig31: cbranching-2-Ud}
\end{figure}

In addition to the qualitative comparison of crack patterns, Figure.\ref{fig31: cbranching-2-Ud} presents a quantitative analysis of dissipated energy for three mesh resolutions, fine ($288693$ elements), medium ($91939$ elements), and coarse ($38960$ elements), under three applied velocities: $1.375$ m/s, $3.318$ m/s, and $3.993$ m/s. It is worth mentioning that no other experimental or numerical results are available for direct comparison. However, for the lower applied velocity $v_0 = 1.375\ m/s$, all three meshes show lower energy dissipation, consistent with the experimental and numerical observation of single crack pattern. The final dissipated energy of all three meshes with lower velocity remain stable around $U_d = 0.3\ J$, indicating consistent single crack behavior across mesh resolutions. Contrary to convergent dissipated energy observed across all three meshes at the velocity $v_0 = 1.375\ m/s$, the higher applied velocities of $v_0 = 3.318\ m/s$ and $v_0 = 3.993\ m/s$ result in more scattered final dissipated energy values. This suggests that crack branching introduces microscale variation in crack paths across the meshes, despite their consistent macroscopic branching behavior. These findings highlights the inherent complexity and challenge of accurately capturing three-dimensional crack branching.

By comparing the performance of the proposed CEM under Neumann and Dirichlet boundary conditions, it can be concluded that more accurate crack patterns are obtained with Dirichlet boundary conditions. This may be attributed to the fact that external forces in Neumann boundary conditions increase stress field variability under transient-dynamic loading, reducing the CEM's predictive accuracy. Additionally, applying surface tractions by converting them into equivalent nodal forces can introduces imbalanced external loads, especially when surface element areas are non-uniform.

\section{Conclusion and Prospects}
Unlike quasi-static crack growth, which assumes steady equilibrium at each step, dynamic crack propagation is characterized by rapidly evolving stress fields, elevated crack tip velocities, and complex stress wave interactions. This time-dependent fracture behavior is particularly critical under extreme loading conditions such as impact, blast, or seismic events. Analytical methods are typically limited to idealized geometries, while experimental techniques often entail significant costs and may offer limited insight into internal fracture processes. In contrast, numerical methods provide a robust framework for in-depth investigation of fracture phenomena, including crack initiation, propagation, branching, and interaction with material heterogeneities, under complex loading and boundary conditions. Consequently, numerical simulation is an essential bridge between theoretical modeling and experimental observation.

Crack branching is one of the most complex and commonly observed phenomena among various fracture behaviors, particularly in brittle materials and alloys subjected to stress corrosion. These branching events can develop symmetrically or asymmetrically and are governed by a range of mechanical and material factors. A comprehensive understanding of crack bifurcation mechanisms, achieved through experimental observation, theoretical modeling, and computational simulation, is essential for accurate failure prediction and the advancement of robust structural design. Despite the development of numerous techniques, a universally accepted method for accurately capturing the full complexity of dynamic crack branching, particularly in three-dimensional settings, remains elusive. This challenge arises primarily from complex crack morphologies, obscure physical mechanisms of branching, and enormous computational demands.

To address the aforementioned limitations, this study presents a novel three-dimensional CEM tailored to efficiently simulate dynamic crack growth and branching. The CEM introduces an advanced element-splitting algorithm that enables element-wise crack growth, including crack branching. Based on the evolving topology of split elements, an original formulation for computing the fracture energy release rate in three dimensions is derived. When the element-wise fracture energy release rate exceeds a predefined critical threshold, the corresponding finite element is deactivated from subsequent computations. This computational procedure of CEM enables a natural and seamless representation of crack evolution in three dimensions. This eliminates the need for complex post-processing procedures or artificial crack-tracking techniques. The method’s accuracy and robustness are validated through a series of benchmark simulations, while its implementation on GPUs demonstrates its capacity to manage the substantial computational demands inherent in high-fidelity 3D fracture analysis. Collectively, the proposed CEM framework offers a scalable, robust, and efficient solution for addressing the complexities of three-dimensional dynamic fracture modeling.

Building on the current advancements in dynamic crack propagation modeling, future research will aim to develop an implicit computational framework specifically tailored for quasi-static fracture problems. In contrast to dynamic scenarios, where inertial effects are predominant, quasi-static crack growth is governed by equilibrium-driven deformation and requires distinct numerical strategies to accurately capture the gradual evolution of cracks under steady or slowly varying loads. Transitioning to an implicit formulation will enable the use of larger time steps and enhance numerical stability, key advantages for simulating long-term degradation phenomena and subcritical crack growth in complex materials. Future developments will focus on extending the current CEM framework to ensure robust convergence characteristics, integrate advanced constitutive models, and accommodate nonlinear material behavior under mixed-mode loading conditions. These enhancements are expected to significantly expand the method's applicability to a wider range of engineering challenges, including structural health monitoring, fracture in composite systems, electronics reliability and fatigue life prediction under realistic service environments.

\newpage

\appendix
\section{Appendix}
\subsection{Appendix I: Co-planarity proof of fracture quadrilateral surface}
\label{Appendix-I: coplanarity}
\begin{figure}[htp]
	\centering
            \begin{center}
            \includegraphics[height=2.5in]{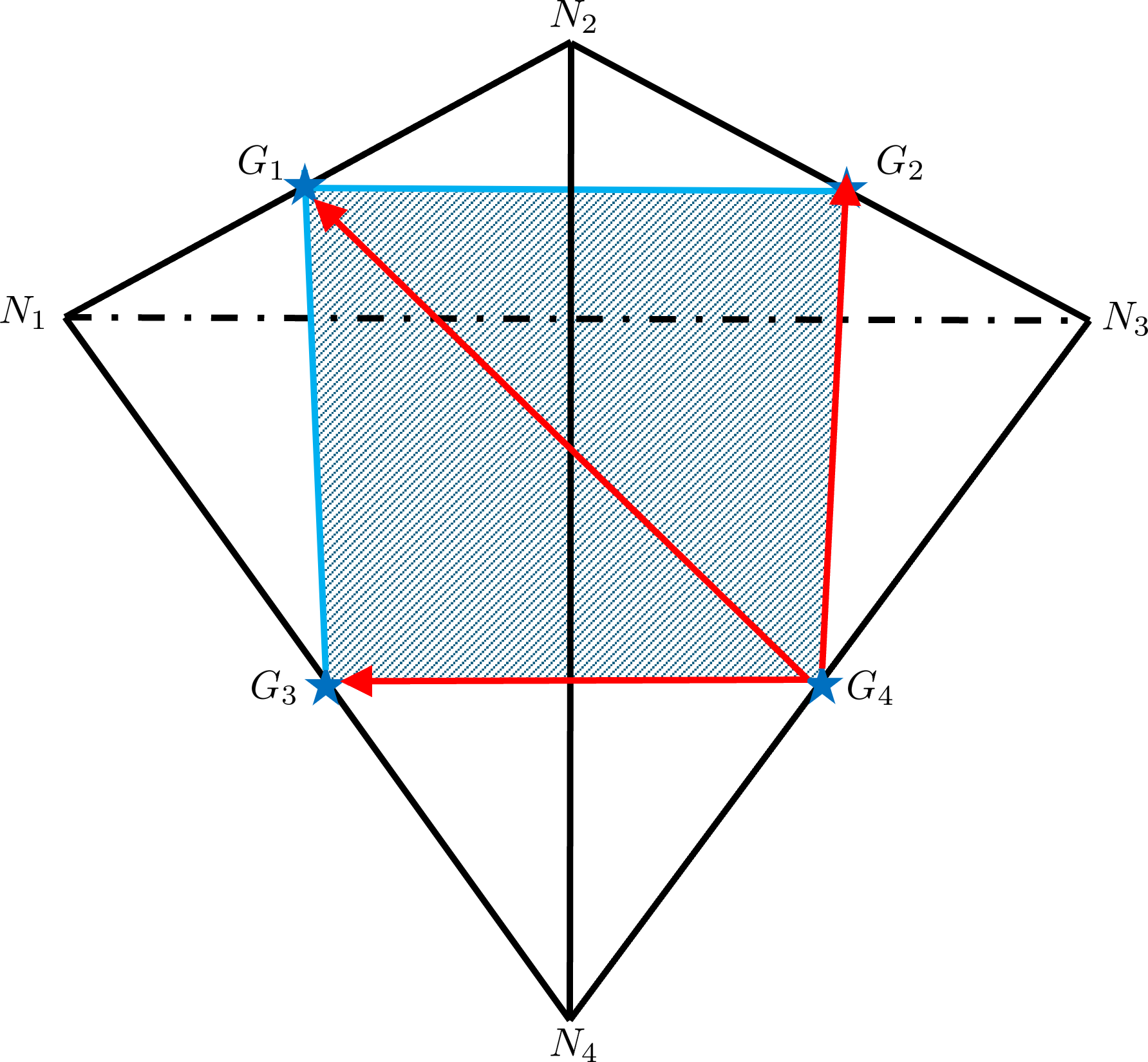}
            \end{center}
        \caption{Illustration of co-planarity proof.}
        \label{fig-app-I: proof-coplanarity}
\end{figure}

\begin{proof}
Given coordinates of four edge quadrature points $G_1$, $G_2$, $G_3$ and $G_4$ as follows,
\begin{eqnarray}
G_1(x_,y,z) &=& \left(x_{G_1}, y_{G_1}, z_{G_1} \right) = \left(\frac{x_{N_1}+x_{N_2}}{2}, \frac{y_{N_1}+y_{N_2}}{2}, \frac{z_{N_1}+z_{N_2}}{2} \right) \nonumber \\
G_2(x_,y,z) &=& \left(x_{G_2}, y_{G_2}, z_{G_2} \right) = \left(\frac{x_{N_2}+x_{N_3}}{2}, \frac{y_{N_2}+y_{N_3}}{2}, \frac{z_{N_2}+z_{N_3}}{2} \right) \nonumber \\
G_3(x_,y,z) &=& \left(x_{G_3}, y_{G_3}, z_{G_3} \right) = \left(\frac{x_{N_1}+x_{N_4}}{2}, \frac{y_{N_1}+y_{N_4}}{2}, \frac{z_{N_1}+z_{N_4}}{2} \right) \nonumber \\
G_4(x_,y,z) &=& \left(x_{G_4}, y_{G_4}, z_{G_4} \right) = \left(\frac{x_{N_3}+x_{N_4}}{2}, \frac{y_{N_3}+y_{N_4}}{2}, \frac{z_{N_3}+z_{N_4}}{2} \right) \nonumber 
\end{eqnarray}

Three vectors formed by $G_4 \rightarrow G_1$, $G_4 \rightarrow G_2$ and $G_4 \rightarrow G_3$ are marked by red in Figure.\ref{fig-app-I: proof-coplanarity}. The mixed product of vectos $\overrightarrow{G_4G_1}$, $\overrightarrow{G_4G_2}$ and $\overrightarrow{G_4G_1}$ are shown as below
\begin{eqnarray}
\left[\overrightarrow{G_4G_1}, \overrightarrow{G_4G_2}, \overrightarrow{G_4G_1} \right] &=& \begin{vmatrix}
x_{G_1}-x_{G_4} & y_{G_1}-x_{G_4} & z_{G_1}-x_{G_4} \\
x_{G_2}-x_{G_4} & y_{G_2}-x_{G_4} & z_{G_2}-x_{G_4} \\
x_{G_3}-x_{G_4} & y_{G_3}-x_{G_4} & z_{G_3}-x_{G_4}
\end{vmatrix} \nonumber \\
&=& \frac{1}{2} \begin{vmatrix}
x_{N_3}-x_{N_4} & y_{N_3}-y_{N_4} & z_{N_3}-z_{N_4} \\
x_{N_1}-x_{N_2}+x_{N_3}-x_{N_4} & y_{N_1}-y_{N_2}+y_{N_3}-y_{N_4} & z_{N_1}-z_{N_2}+z_{N_3}-z_{N_4} \\
x_{N_1}-x_{N_4} & y_{N_1}-y_{N_4} & z_{N_1}-z_{N_4}
\end{vmatrix} \nonumber \\
&=& 0 \nonumber
\end{eqnarray}
Therefore, it is concluded that the rank of matrix is smaller than 3 and the fracture quadrilateral plane is co-plane.
\end{proof}

%



\printcredits

\bibliographystyle{cas-model2-names}

\bibliography{cas-refs}






\end{document}